%% file: main.tex
\documentclass[10pt,journal,compsoc]{IEEEtran}

\ifCLASSOPTIONcompsoc
  \usepackage[nocompress]{cite}
\else
  \usepackage{cite}
\fi

\input{macro}
\begin{document}

\title{Towards Understanding and Mitigating Audio Adversarial Examples for Speaker Recognition}

\author{Guangke Chen, Zhe Zhao, Fu Song, Sen Chen, Lingling Fan, Feng Wang, and Jiashui Wang
}

\input{abstract}

\maketitle

\IEEEdisplaynontitleabstractindextext

\IEEEpeerreviewmaketitle

\input{introduction}

\input{background}

\input{attack-defense}

\input{methodology-of-study}

\input{evaluation}

\input{discussion}

\input{related-work}

\input{conclusion}


\ifCLASSOPTIONcaptionsoff
  \newpage
\fi


\input{biography}

\clearpage
\appendices
\input{appendix}

\end{document}

%% file: macro.tex
\usepackage{tikz}
\usepackage{amsmath}

\usepackage{colortbl}
\usepackage{graphicx,marvosym}
\newcommand{\platformname}{\textsc{SpeakerGuard}\xspace}

\newcommand{\defensename}{\textsc{Feature Compression}\xspace}
\newcommand{\defensenameabbr}{FeCo\xspace}
\usepackage{amsmath}
\usepackage{pifont}
\newcommand{\cmark}{\ding{51}}%
\newcommand{\xmark}{\ding{55}}%
\usepackage{multirow}
\usepackage{url}
\usepackage{hyperref}
\usepackage{xcolor}
\usepackage{bm}
\usepackage{colortbl}
\usepackage{subfigure}
\usepackage{xspace}
\usepackage{enumitem}
\usepackage[many]{tcolorbox}
\usepackage{multirow}
\usepackage{makecell}
\usepackage{threeparttable}

\usepackage{etoolbox}
\makeatletter
\patchcmd{\@makecaption}
  {\scshape}
  {}
  {}
  {}
\makeatletter
\patchcmd{\@makecaption}
  {\\}
  {:\ }
  {}
  {}
\makeatother

\usepackage[switch]{lineno}

\usepackage{algorithm}
\usepackage[noend]{algpseudocode}

\DeclareMathOperator\argmin{argmin}
\usepackage{diagbox}

\usepackage{amsfonts}

%% file: abstract.tex
\IEEEtitleabstractindextext{%
\begin{abstract}
Speaker recognition systems (SRSs) have recently been shown to be vulnerable to adversarial attacks,
raising significant security concerns. In this work, we systematically investigate transformation and 
adversarial training based defenses for securing SRSs. 
According to the characteristic of SRSs, we present 22 diverse transformations 
and thoroughly evaluate them using 7 recent promising adversarial attacks (4 white-box and 3 black-box) on speaker recognition.
With careful regard for best practices in defense evaluations, we analyze the strength of transformations to withstand adaptive attacks.
We also evaluate and understand their effectiveness against adaptive attacks when combined with adversarial training.
Our study provides lots of useful insights and findings, many of them
are new or inconsistent with the conclusions in the image and speech recognition domains,
e.g., variable and constant bit rate speech compressions have different performance,
and some non-differentiable transformations remain effective against current promising evasion techniques which often work well in the image domain.
We demonstrate that the proposed novel feature-level transformation combined with adversarial training is rather effective compared to the sole adversarial training
in a complete white-box setting, e.g., increasing the accuracy  by 13.62\%
and attack cost by two orders of magnitude, while other transformations do not necessarily improve the overall defense capability.
This work sheds further light on the research directions in this field.
We also release our evaluation platform \platformname to foster further research.

\end{abstract}

}

%% file: introduction.tex
\section{{Introduction}}\label{sec:introduction}
Speaker recognition (SR) is the process of
automatically verifying or identifying individual speakers
by extracting and analyzing their unique acoustic characteristics~\cite{Homayoon11}.
State-of-the-art speaker recognition systems (SRSs), based on machine learning (including deep learning), 
 have been adopted by open-source platforms (e.g., Kaldi~\cite{kaldi}) 
and commercial products (e.g., Microsoft Azure~\cite{microsoft-azure-vpr} and Amazon Alexa~\cite{Alexa}), 
and used in safety-critical applications such as remote voice authentication in financial transaction~\cite{TD-Bank}
and device access control in smart home~\cite{RenSYS16}.

The popularity of SRSs has brought new security
concerns. Recent studies have shown that
both open-source and commercial SRSs are vulnerable to adversarial attacks~\cite{abs-1801-03339,li2020adversarial,jati2021adversarial,zhang2021attack,
LiZJXZWM020,xie2020enabling,WangGX20,shamsabadi2021foolhd,chen2019real,du2020sirenattack,DBLP:journals/corr/abs-1711-03280}.
To thwart adversarial attacks,
five input transformations~\cite{chen2019real,du2020sirenattack,voting-defense,HPF-defense}
and two adversarial training~\cite{jati2021adversarial}, derived from other domains, have been studied.
However, these defenses are only evaluated against few non-adaptive attacks.
Thus, it is impossible to fairly compare their performance and also may lead to a false sense of robustness improvement~\cite{tramer2020adaptive}, limiting their usage in practice.
Indeed, these defenses become ineffective against adaptive attacks using evasion techniques from the image domain.

In this work, to secure SRSs against adversarial attacks,
we systematically investigate transformation and adversarial training based defenses
and thoroughly evaluate their effectiveness using both non-adaptive and adaptive attacks under
the same settings.

We study transformations according to the characteristic of audio signals and SRS's architecture.
Different from images and image recognition systems, audio can be transformed at both waveform-level and feature-level,
where at the waveform-level, audio can be transformed in the time- and frequency-domain
while at the feature-level, different types of features in acoustic feature
extraction pipeline can be transformed.
To be diverse and comprehensive, we consider 22 diverse transformations (4 time-domain and 3 frequency-domain transformations,
7 audio compressions that transform audio at both time- and frequency-domains,
and 8 novel feature compressions), covering all the 5 transformations 
studied in~\cite{chen2019real,du2020sirenattack,voting-defense,HPF-defense}.
Furthermore, from the respective of adaptive attacks for evasion,
these transformations cover all the differentiable, non-differentiable, deterministic, and randomized types.

To thoroughly evaluate the defenses,
we extend and implement all the recent promising adversarial attacks~\cite{DBLP:journals/corr/abs-1711-03280,abs-1801-03339,li2020adversarial,zhang2021attack,jati2021adversarial,chen2019real,du2020sirenattack,ASGBWYST19},
including 4 white-box attacks and 3 black-box attacks.
The evaluation on 22 concrete attacks shows that the effectiveness of transformations does not
necessarily decrease with increase of both distortion and attack strength,
and their effectiveness varies with attacks, e.g., two time-domain transformations are more effective than others against $L_\infty$ attacks (i.e., perturbations are limited in $L_\infty$ norm)
and feature-level transformations are often more effective than others against $L_2$ white-box attacks.

However, this evaluation does not provide security guarantees against a future adaptive
adversary who has knowledge of defenses.
To evaluate the robustness of defenses against adaptive attacks,
we design adaptive attacks following the most important lessons of~\cite{tramer2020adaptive},
incorporating evasion techniques (Backward Pass Differentiable
Approximation (BPDA)~\cite{athalye2018obfuscated}, Expectation over Transformation (EOT)~\cite{athalye2018synthesizing},
and Natural Evolution Strategy (NES)~\cite{wierstra2014natural}) that have been shown effective against non-differentiable or randomized transformations in the image domain,
and specific techniques targeting feature-level transformations.
We remark that these evasion techniques have never been considered in the speaker recognition domain except that NES was adopted to estimate gradients by the black-box attack FAKEBOB~\cite{chen2019real}.
The evaluation shows that (1) most transformations including the ones from~\cite{chen2019real,du2020sirenattack,voting-defense,HPF-defense}
  become  \emph{ineffective},
(2) some non-differentiable audio compressions \emph{cannot} be broken by BPDA which is promising in the image domain,
(3) AAC and MP3 with \emph{variable bit rate} are more difficult (resp. easier) to be bypassed than them with \emph{constant bit rate} in the black-box (resp. white-box) setting;
and (4) most of the \emph{randomized} transformations remain resistant to black-box adaptive attacks.

To explore the effectiveness of transformations combined with adversarial training,
we consider the promising adversarial training of \cite{jati2021adversarial} 
and evaluate the combined defenses under adaptive attacks.
The evaluation shows that while the combination of a transformation and
adversarial training
does not necessarily bring the best of both worlds,
the proposed novel feature-level transformation combined with adversarial training
is very effective, improving the accuracy of both benign and adversarial examples in a complete white-box setting.
We further evaluate this combined defense
by varying various attack parameters.
The results show that it is still effective,
improving the accuracy by 13.62\%, the attack cost by two orders of magnitude,
and the distortion of adversarial examples,
compared over vanilla adversarial training.

In summary, we make the following main contributions.
 \begin{itemize}[leftmargin=*]
   \item We perform the most comprehensive investigation of transformation based defenses for securing SRSs
    according to the characteristic of audio signals and SRS's architecture
    and study the impact of their hyper-parameters 
   for mitigating adversarial voices without incurring too much negative impact on the benign voices.

  \item We thoroughly evaluate the proposed transformations
   for mitigating recent promising adversarial attacks
   on SRSs.  With regard for best practices in defense evaluations, we
   carefully analyze their strength, on both models trained naturally and adversarially, to withstand adaptive attacks.
   
\item  Our study provides lots of useful insights and findings, either newly reported or inconsistent with existing findings in other domains, which
        could advance research on adversarial examples in this domain
    and assist the maintainers of SRSs to deploy suitable defense solutions to enhance their systems.
Particularly, we find that our novel feature-level transformations combined
with adversarial training is the most robust one against adaptive attacks. 

    \item We develop the first platform \platformname for systematic and comprehensive
    evaluation of adversarial attacks and defenses on SRSs.
    It features mainstream SRSs, datasets, white- and black-box attacks, widely-used evasion techniques for
    adaptive attacks, evaluation metrics, and diverse defense solutions.
    We release our platform  to foster further research in this direction (\url{https://speakerguard.github.io}).
\end{itemize}

%% file: background.tex
\section{Background}\label{sec:background}


\noindent{\bf Speaker Recognition Systems (SRSs)}.
State-of-the-art SRSs use speaker embedding~\cite{wang2020simulation} to represent
acoustic characteristics of speakers as fixed-dimensional vectors.
The typical
speaker embedding is identity-vector (ivector)~\cite{ivector-2011}
based on the Gaussian Mixture Model (GMM)~\cite{reynolds2000speaker}.
Recently, deep embedding was also proposed to compete with ivector.
It uses deep learning to train a deep neural network
from which speaker characteristics are
extracted and
%
represented
as vectors, e.g. AudioNet~\cite{becker2018interpreting,jati2021adversarial}
and x-vector~\cite{snyder2018x}.

\begin{figure}[t]
    \centering
    \includegraphics[width=0.47\textwidth]{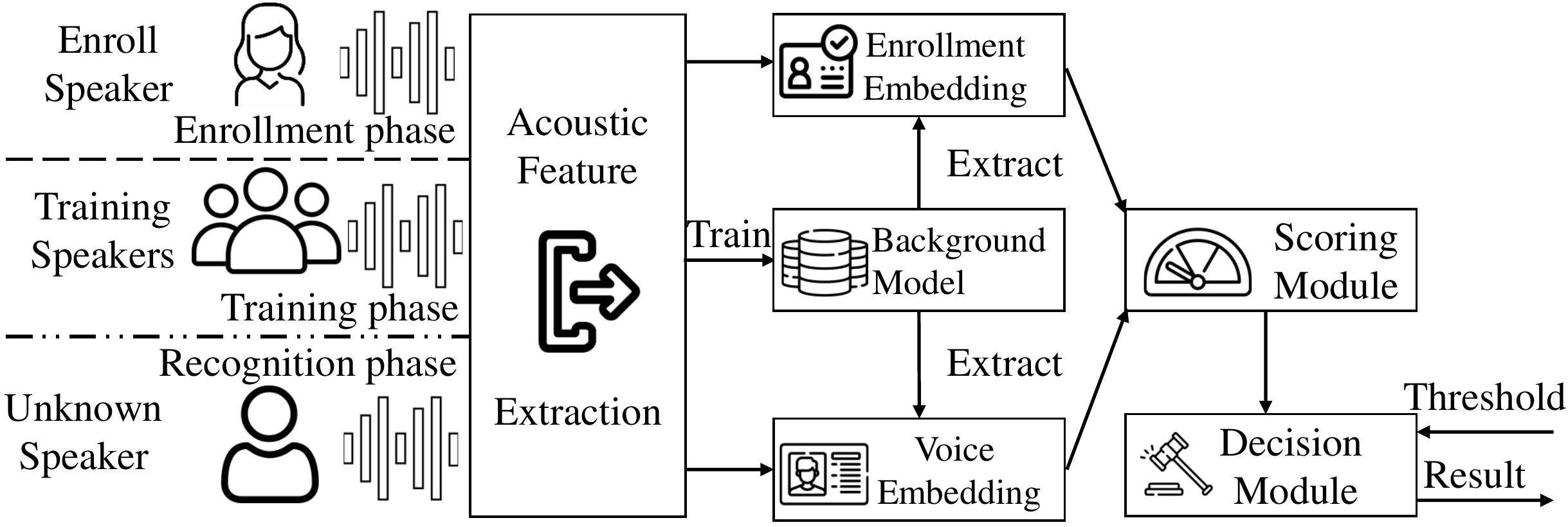}
    \vspace*{-2mm}
    \caption{Architecture of SRSs.}
    \label{fig:typical-SRSs}\vspace*{-4mm}
\end{figure}

A generic architecture of SRSs is shown in \figurename~\ref{fig:typical-SRSs}, consisting of: training, enrollment, and recognition phases.
In the training phase, a background model is trained using tens of thousands of voices from thousands of training speakers,
representing the speaker-independent distribution of acoustic features.
In the enrollment phase, the background model maps the voice uttered by each enrolling speaker to an \emph{enrollment embedding},
regarded as the unique identity.
In the recognition phase, given a voice of an unknown speaker, the \emph{voice embedding} is extracted from the background model. The scoring module measures the similarity between the \emph{enrollment embedding} and \emph{voice embedding} based on which the decision module outputs the result.
There are two typical scoring approaches: Probabilistic Linear Discriminant Analysis (PLDA)~\cite{PLDA}
and cosine similarity~\cite{dehak2010cosine}, where
PLDA works well in most situations but needs to be trained using voices~\cite{wang2020simulation}
while cosine similarity is a reasonable substitution of PLDA
without requiring training.


The acoustic feature extraction module converts the raw audio signals to acoustic features
carrying characteristics of the raw audio signals.
Common feature extraction algorithms include Mel-Frequency Cepstral Coefficients (MFCC)~\cite{muda2010voice} and
Filter-Bank~\cite{FilterBanks}.

\smallskip
\noindent{\bf Recognition task}.
There are three main tasks: close-set identification (CSI), speaker verification (SV), and open-set identification (OSI).
CSI identifies a speaker from
a group of speakers.
SV verifies if an input voice is uttered by the unique enrolled speaker, according to a preset threshold, where
the input voice may be rejected by regarding the speaker as an imposter.
OSI utilizes the scores and a preset threshold to identify which enrolled speaker utters the input voice, where
if the highest score is less than the threshold, the input voice is rejected by regarding the speaker as an imposter.
Moreover, CSI could be classified into two sub-tasks:
CSI with enrollment (CSI-E) and  CSI without enrollment (CSI-NE).
CSI-E exactly follows the above description.
In contrast, CSI-NE does not have the enrollment phase
and the background model is directly utilized to identify speakers.
Thus, ideally,  
   a recognized speaker in CSI-NE task is involved in the training phase,
   while a recognized speaker in the CSI-E task should have enrolled in the enrollment phase but may not be involved in the training phase.

\smallskip\noindent
{\bf Threat model}. 
According to the adversary's knowledge about the SRS and deployed defense, we classify attacks into:
\textit{white-box non-adaptive}, \textit{black-box non-adaptive}, \textit{white-box adaptive}, and \textit{black-box adaptive attacks}.
The adversary for white-box attacks has full access to model architecture, parameters, etc.,
while the adversary for black-box attacks has no knowledge about the model but can access the target model as an oracle, i.e.,
providing a series of carefully crafted inputs to the model and observing its outputs.
Under both white-box and black-box settings, the adversary may be
unaware of the deployed defense, or has complete knowledge of it
(e.g., its implementation detail and concrete values for any tunable parameter)
and intends to bypass it.
We consider non-adaptive attacks for the former adversary
and adaptive attacks for the latter adversary.

%
%



%% file: attack-defense.tex
\section{Defenses}\label{sec:defense} 

\subsection{Motivation}
Recently, adversarial attacks on speaker recognition have been extensively studied~\cite{abs-1801-03339,li2020adversarial,jati2021adversarial,zhang2021attack,
LiZJXZWM020,xie2020enabling,WangGX20,shamsabadi2021foolhd,chen2019real,du2020sirenattack,DBLP:journals/corr/abs-1711-03280}.
Results show that both state-of-the-art open-source and commercial SRSs can be fooled by adding
small perturbations to the original voice, even playing over the air in the physical world.

In the image and speech recognition domains, studies have proposed transformation based
defenses that are able to recover benign counterparts from adversarial examples, e.g.,~\cite{yang2018characterizing,Xu0Q18,MengC17}.
While such defenses are effective for defending against non-adaptive attacks, they may be evaded by adaptive attacks~\cite{tramer2020adaptive}.
Nevertheless, some transformations (but not all) achieve promising results when combined with adversarial training even in a complete white-box setting~\cite{data-aug-advt-effective-image,tramer2020adaptive}.
However, the same conclusion cannot be drawn on speaker recognition without a careful and rigorous evaluation,
because of the difference between speaker recognition and image/speech recognitions.
Compared with image recognition systems, SRSs have complicated architectures and individual components, in particular, the acoustic feature extraction pipeline.
Also, while the well-trained vision model is directly exploited to classify input images into one of the training classes,
the well-trained background model of SRSs is adapted to speaker-specific models during enrollment and used to map input utterances into identity embeddings during recognition,
since the enrolled and inference speakers are not necessarily involved in the training phase.
While speech recognition minimizes speaker-dependent variations
to determine the underlying text or command, speaker recognition treats the phonetic variations as extraneous noise to determine the source of the speech signal.
All these differences may lead to inconsistent conclusions in the speaker recognition domain with other domains.
In fact, we indeed found such inconsistent findings (cf. Section~\ref{sec:adaptive-attack}).

Therefore, in the speaker recognition domain,
five input transformation~\cite{chen2019real,du2020sirenattack,voting-defense,HPF-defense}
and two adversarial training~\cite{jati2021adversarial} based defenses have been studied.
Though promising, these defenses are only evaluated against \emph{few} attacks on different models, recognition tasks, and datasets, let alone adaptive attacks~\cite{tramer2020adaptive}
and combinations of transformation and adversarial training.
Thus, it is impossible to fairly compare their performance and also may lead to a false sense of robustness improvement brought by defenses without considering
adaptive attacks, limiting their usage in practice.
It is also unclear if combining a transformation with adversarial training
results in a more effective defense, as many existing defenses combined with adversarial training result
in lower robustness than adversarial training on its own in the image domain~\cite{tramer2020adaptive}.
Therefore, \emph{there is a lack of comprehensive investigation and rigorous quantitative understanding of defenses on speaker recognition,
in particular, effective defenses}.
This work is aimed at filling this gap.

\subsection{Design Philosophy}
According to the architecture of SRSs (cf. \figurename~\ref{fig:typical-SRSs}),
 we should consider both robust training and input transformation, 
where the former is conducted during the training phase and the latter takes effect in the recognition phase.
When combined, they may lead to a more robust defense.
For input transformation, we design audio transformations based on the following
two key characteristics of speaker recognition, compared over image recognition.

\smallskip
\noindent
{\bf Architecture characteristic}. For state-of-the-art neural network based image recognition,
an image is directly fed to a system without feature engineering.
Due to the time-varying non-stationary property of voices, voices are not resilient enough to noises and other variations,
and audio waveform signals themselves cannot effectively represent speaker characteristics~\cite{voice-feature-review}.
Hence, to achieve better feature representative capacity and system performance~\cite{xiao2016speech},
a modern SRS has an acoustic feature
extraction pipeline for extracting acoustic feature from waveforms (cf. \figurename~\ref{fig:typical-SRSs}).
This gives rise to waveform-level input transformations (W-transformations) and feature-level input transformations (F-transformations).

\smallskip
\noindent
{\bf Audio signal characteristic}.
While images are naturally two-dimensional, raw audio samples form a one-dimensional time series signal~\cite{PurwinsLVSCS19}.
Even though audio signals are often transformed into two-dimensional time-frequency representations,
the two axes, time and frequency, fundamentally differ from the horizontal and vertical axes in an image.
Furthermore, images are commonly analyzed as a whole or in patches
with little order constraints while audio signals have to be
analyzed sequentially in chronological order.  These properties give rise to audio-specific W-transformations
that can be performed either in time-domain or frequency-domain.

Based on the above characteristics, to be diverse and comprehensive, we investigate
both W-transformations and F-transformations, while
for the former, we consider both time-domain and frequency-domain ones.
When necessary and possible, we also evaluate the effectiveness of transformations combined with robust training.
When devising an input transformation based defense,  it is also important to consider
if it is differentiable\footnote{Differentiable here means that a transformation can be implemented in frameworks (e.g., Pytorch)
that supports auto-differentiation~\cite{paszke2017automatic}. 
Auto-differentiation enables the gradients to be back-propagated, 
providing informative gradients for adversarial example generation.
Though it is non-rigorous,
we use it to keep consistent with~\cite{tramer2020adaptive}.} and deterministic, due to the fact that most white-box attacks leverage
gradient to craft adversarial examples. In general, non-differentiable (resp. randomized) input transformations
are more difficult to be evaded than differentiable (resp. deterministic) ones.
Thus, all the types should be addressed to understand their effectiveness.
All the transformations we considered are summarized in Table~\ref{tab:input-transform}, covering  differentiable, non-differentiable,
deterministic, and randomized types.

\subsection{Robust Training}\label{sec:review-defense-adver-train}
Robust training strengthens the resistance of a model to adversarial examples during training. We adopt adversarial training, one of the most effective techniques in the image domain, which augments the training data with adversarial examples.
Formally, adversarial training intends to find the model parameter $\theta$ which minimizes the following loss:
$$\mathbb{E}_{(x,y)\sim \mathcal{D}}[{\tt max}_{\delta\in S}f(\theta,x+\delta,y)]\approx \frac{1}{n}\sum_{i=1}^{n}{\tt max}_{\delta\in S}f(\theta,x_i+\delta,y_i)$$
 where 
$S$ is the set of allowed perturbations,
$\mathcal{D}$ is the underlying data distribution over pairs of samples $x$ and corresponding labels $y$,
$\{(x_i,y_i)\}_{i=1}^{n}$ is the training dataset that mimics the data distribution $\mathcal{D}$,
and $f$ is the training loss function, typically the cross-entropy loss.
Efficient adversarial attacks such as FGSM~\cite{goodfellow2014explaining} and PGD~\cite{madry2017towards} are widely used to solve the above maximization problem.

\begin{table}[t]
  \centering
  \footnotesize\setlength\tabcolsep{2.5pt}
  \caption{Transformations}
  \vspace{-3mm}
  \begin{threeparttable}
  \resizebox{0.48\textwidth}{!}{%
  \begin{tabular}{c|c|c|c|c|c}
  \hline
  \multicolumn{2}{c|}{} & {\bf Name} & {\bf Parameters} & {\bf D} & {\bf R} \\ \hline
 \multirow{17}{*}{\rotatebox{270}{{\bf Waveform Level}}} & \multirow{4}{*}{\rotatebox{270}{\makecell[c]{{\bf Time} \\  {\bf Domain} }  } }
  & {\bf Quantization (QT)} &    $q$: quantized factor   & \xmark & \xmark \\ \cline{3-6}
  & & {\bf Audio Turbulence (AT)} & SNR: signal-to-noise ratio  & \cmark & \cmark \\ \cline{3-6}
 &  & {\bf Average Smoothing (AS)} & $k$: kernel size & \cmark & \xmark \\ \cline{3-6}
 & & {\bf Median Smoothing (MS)} & $k$: kernel size& \cmark & \xmark \\ \cline{2-6}
  & \multirow{4}{*}{\rotatebox{270}{\makecell[c]{{\bf Frequency} \\  {\bf Domain}}}}
  & {\bf Down Sampling (DS)} & $\tau$: downsampling  freq.  & \cmark & \xmark \\ \cline{3-6}
  & &  {{\bf Low Pass Filter (LPF)}} & \makecell[c]{$f_p$: passband edge  freq. \\ $f_s$: stopband edge  freq.}& \cmark & \xmark \\ \cline{3-6}
  & & {{\bf Band Pass Filter (BPF)}} & \makecell[c]{$f_{pl},f_{pu}$: passband edge freq. \\ $f_{sl},f_{su}$: stopband edge freq. } & \cmark & \xmark \\ \cline{2-6}
  &   \multirow{7}{*}{\rotatebox{270}{\makecell[c]{{\bf Speech} \\  {\bf Compression}}}}
  &{{\bf OPUS}} & $b_o$: compression  bitrate  & \xmark & \xmark \\ \cline{3-6}
   & &  {{\bf SPEEX}} & $b_s$: compression  bitrate& \xmark & \xmark \\ \cline{3-6}
   & &  {{\bf AMR}} & $b_r$: compression  bitrate & \xmark & \xmark \\ \cline{3-6}
   & &  {{\bf AAC-V}} & $q_c$: quality & \xmark & \xmark \\ \cline{3-6}
    & &  {{\bf AAC-C}} & $b_c$: compression  bitrate & \xmark & \xmark \\ \cline{3-6}
    & &  {{\bf MP3-V}} & $q_m$: quality  & \xmark & \xmark \\ \cline{3-6}
    & &  {{\bf MP3-C}} & $b_m$: compression  bitrate & \xmark & \xmark \\ \hline
  \multicolumn{2}{c|}{\makecell[c]{{\bf Feature} \\ {\bf Level}}}  & \makecell[c]{{\bf \defensename}  {\bf (\defensenameabbr)}\\ 4 feature types $\times$ 2 compression alg.} &
   \makecell[c]{$cl_m$: cluster method \\ $cl_r$: cluster ratio}& \cmark & \cmark \\ \hline
  \end{tabular}
  }
   \begin{tablenotes}
        \scriptsize
        \item Note: D=Differentiable and R=Randomized.
      \end{tablenotes}
  \end{threeparttable}
  \label{tab:input-transform}
  \vspace*{-2mm}
\end{table}

\subsection{W-Transformations}\label{sec:input-transformation}
For W-transformations, we consider both
time-domain and frequency-domain ones. 
We also consider various speech compression which can be seen as W-transformations
performed both in the time- and  frequency-domains.

\smallskip
\noindent
{\bf Time-domain W-transformations}. 
We study four  time-domain W-transformations,
inspired by image input transformations~\cite{yang2018characterizing}.
(1) Quantization (QT) rounds the amplitude of each sample point of a voice to the nearest integer multiple of a factor $q$,
intended to disrupt the adversarial perturbation since its amplitude is usually small in the input
space. (2) Audio turbulence (AT) adds random noise to an input voice in an element-wise way to disrupt the adversarial perturbation which is assumed to be sensitive to noise.
The magnitude of the noise is adjusted by
signal-to-noise ratio (SNR) $10\log_{10}\frac{P_I}{P_n}$ where $P_I$ (resp. $P_n$) is the power of input voice (resp. random noise).
(3) Average smoothing (AS) and (4) median smoothing (MS) mitigate adversarial examples by smoothing the waveform of the input voice. A mean (resp. median) smooth with kernel size $k$ (must be odd) replaces each element $x_k$ with the \emph{mean} (resp. \emph{median}) value of its $k$ neighbors.
We remark that QT is non-differentiable due to the round operation while the others are differentiable,
and AT is randomized while the others are deterministic.
 
\smallskip
\noindent
{\bf Frequency-domain W-transformations}.
We consider three W-transformations in frequency-domain, all of which are differentiable and
deterministic.
(1) Down sampling (DS) down-samples voices and applies signal
recovery to disrupt adversarial perturbations. The down-sample frequency is determined by the ratio, denoted by $\tau$, between the new and original sampling frequencies.
(2) Low pass filter (LPF) assumes that human voices are within relatively lower frequencies than adversarial perturbation, and applies a low-pass filter to remove the high-frequent perturbations.
A low-pass filter has two  parameters:
the edge frequencies of the passband ($f_p$) and the stopband ($f_s$).
(3) Band pass filter (BPF) combines LPF with a high-pass filter to remove both high-frequent and low-frequent perturbations.
BPF has four parameters: the lower and upper edge frequencies of the passband ($f_{pl}$ and $f_{pu}$),
the lower and upper cutoff frequencies of the stopband ($f_{sl}$ and $f_{su}$).
We remark that these transformations are derived from the speech recognition domain \cite{yang2018characterizing,kwon2019poster,rajaratnam2018speech},
but only DS has been applied in the speaker recognition against two black-box attacks FAKEBOB~\cite{chen2019real} and SirenAttack~\cite{du2020sirenattack}.

\smallskip\noindent
{\bf Speech compression}.
Based on the psychoacoustic principle, 
speech compression aims to suppress redundant information within a speech to improve storage or transmission efficiency.
When an adversarial perturbation is redundant, it can be eliminated by speech compression.
Speech compression achieves the aforementioned purpose by reducing the bit rate, thus can be seen as transformations performed both in the time- and frequency-domains.
We investigate 7 standard lossy speech compression techniques,
grouped into two categories: Constant Bit Rate (CBR) and Variable Bit Rate (VBR).
The former uses a fixed bit rate and the latter exploits a dynamic bit rate schedule controlled by the quality parameter.
We consider OPUS~\cite{vos2013voice}, SPEEX~\cite{valin2016speex},
AMR~\cite{ekudden1999adaptive}, AAC-C~\cite{bosi1997iso}, and MP3-C~\cite{hacker2000mp3} for CBR, and
AAC-V~\cite{bosi1997iso} and MP3-V~\cite{hacker2000mp3} for VBR.
These transformations are non-differentiable and deterministic.

\subsection{F-Transformations}\label{sec:our-approach}
The design of F-transformations is motivated by the following research questions: (Q1) \emph{What kind of acoustic features can be transformed?}
and (Q2) \emph{How to transform them?}.

\begin{figure}[t]
    \centering
    \includegraphics[width=0.4\textwidth]{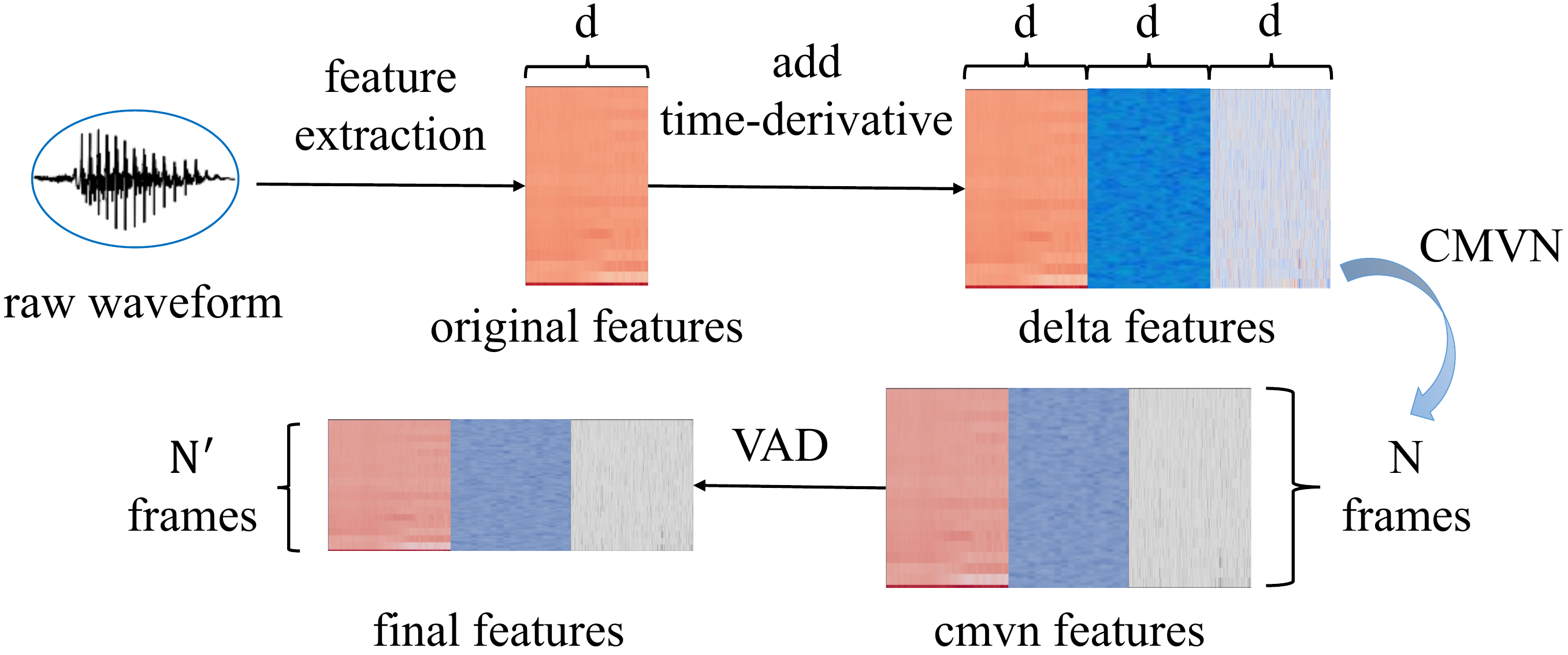}\vspace*{-2mm}
    \caption{A typical flow of feature processing.}
    \label{fig:different-features}\vspace*{-2mm}
\end{figure}

To address Q1, we have to understand what kind of features are used in SRSs.
\figurename~\ref{fig:different-features} shows a typical flow of feature processing.
First, the \emph{original features} (e.g., MFCC or Filter-Bank) are extracted from an input raw waveform. 
Next, to capture temporal information, time-derivative features~\cite{xiao2016speech} 
are 
successively
extracted from and added into the original features, leading to the \emph{delta features}.
After that, cepstral mean and variance normalization
(CMVN)~\cite{speech-processing-book}
is applied to reduce channel and reverberation effects, resulting in \emph{cmvn features}.
Finally, voice activity detection (VAD)~\cite{sohn1999statistical} is utilized to remove the unvoiced frames, resulting in  \emph{final features}.
Therefore,  four types of features could be transformed.

To address Q2, a straightforward idea is to extend W-transformations.
However,  (1) W-transformations work on audio waveforms in two-dimensional time-frequency representations,
while acoustic features
are represented by a matrix,
one row of features per frame.
It prevents frequency-domain W-transformations and speech compression from being extended.
(2) The mapping from waveforms to features is not linear,
and a small perturbation in the input voice may lead to a large
perturbation at the feature level.
This difference refuses time-domain W-transformations
where adversarial perturbations are assumed to be small and/or sensitive to noise.

We propose \defensename (\defensenameabbr)
to disrupt adversarial perturbations at the feature level.
We regard each feature matrix
$\mathcal{M}$ with $N$ frames
and each frame $\mathbf{a}_i$ consisting of $d$ features as $N$ data points in $d$-dimensional space
and compute a compressed feature matrix with $K$ frames for $K< N$.
Our idea is described in Algorithm~\ref{al:FeCo-2}.
The number $K$ of clusters is first computed according to 
the given cluster ratio $cl_r$ (line~\ref{alg:1}).
Then, we partition $N$ frames into $K$ clusters by invoking the  cluster oracle $\mathcal{O}$ (line~\ref{alg:2}),
which returns a list of
indices $b_1,\cdots,b_N$ such that
each frame $\mathbf{a}_i$ is assigned to the $b_i$-th cluster.
Next,  each cluster $C_i$ is represented by a representative vector $\mathbf{m}_i$ (line~\ref{alg:3}).
Finally, $K$ representative vectors are combined to form the new feature matrix $\mathcal{M}'$.

To partition $N$ frames into $K$ clusters, various clustering methods, e.g., kmeans~\cite{hartigan1979algorithm} and soft-kmeans~\cite{mackay2003information},
could be leveraged. In this work, we use kmeans and its variant warped-kmeans~\cite{LeivaV13}
and leave others as future work.
Compared to kmeans, warped-kmeans preserves the temporal dependency of the data by imposing some constraints on the partition operation,
thus is more suitable to cluster sequential data.
Both kmeans and warped-kmeans use the average of all the frames in one cluster as the representative.

Algorithm~\ref{al:FeCo-2} could be applied to any of original, delta, cmvn, and final features.
We use \defensenameabbr-o, \defensenameabbr-d, \defensenameabbr-c, and \defensenameabbr-f to denote
these four concrete F-transformations, all of which are randomized and differentiable.
The randomness of \defensenameabbr lies in the initialization of kmeans and warped-kmeans algorithms.
At the beginning, they \textit{randomly} select $K$ vectors from $N$ vectors as the initial cluster centers,
which will be used in the later clustering operations.
Different initialization may produce different clustering results (Line 2),
thus leading to different feature matrix $\mathcal{M}'$.

\begin{figure}[t]
\vspace*{-3mm}
\begin{algorithm}[H]
\footnotesize
	\caption{\defensenameabbr} 
	\label{al:FeCo-2}
	\begin{algorithmic}[1]
    \Require feature matrix $\mathcal{M}=[\mathbf{a}_1,\cdots,\mathbf{a}_N]$;
    cluster ratio $0<cl_r<1$;  cluster oracle $\mathcal{O}=$ kmeans or warped-kmeans
	\Ensure  compressed feature matrix $\mathcal{M}'$
    \State $K\gets \lceil N\times cl_r\rceil$ \Comment{$K=$ number of clusters} \label{alg:1}
    \State $[b_1,\cdots,b_N]\gets \mathcal{O}(\mathcal{M}, K)$ \Comment{$\mathbf{a}_i$ is assigned to the $b_i$-th cluster}  \label{alg:2}
	\For{$(i=1;i\leq K;i++)$}
        \State $C_i\gets \{\mathbf{a}_k\mid b_k=i\}$   \Comment{compute the $i$-th cluster}
        \State $\mathbf{m}_i\gets \frac{1}{|C_i|}\sum_{\mathbf{a}\in C_i}\mathbf{a}$ \Comment{compute the representative vector}  \label{alg:3}
    \EndFor
    \State $\mathcal{M}'\gets [\mathbf{m}_1,\cdots,\mathbf{m}_K]$ \Comment{concatenate the representative vectors}  \label{alg:4}
    \State \Return{$\mathcal{M}'$}
	\end{algorithmic}
\end{algorithm}
\vspace*{-6mm}
\end{figure}

%% file: methodology-of-study.tex
\begin{table}[t]
    \centering\setlength\tabcolsep{8pt}
     \caption{SR models}
     \vspace{-2.5mm}
  \begin{threeparttable}
 \scalebox{0.85}{
    \begin{tabular}{c|c|c}
    \hline
     & {\bf ivector-PLDA}~\cite{kaldi-ivector-plda} & {\bf AudioNet}~\cite{becker2018interpreting} \\ \hline
      \begin{tabular}[c]{@{}c@{}}{\bf Embedding \& Feature types}\end{tabular} & T \& MFCC & D \& Filter-Bank \\ \hline
        {\bf Add {1{st}} \& 2nd time-derivative} & \cmark & \xmark \\ \hline
      {\bf Apply CMVN \& VAD} & \cmark & \xmark  \\ \hline
      {\bf \#Feature dim} & 72 & 32 \\ \hline
      {\bf Training algorithm} & US & S \\ \hline
    {\bf Scoring method} & PLDA &  - \\ \hline
    \end{tabular}
    }
     \begin{tablenotes}
        \scriptsize
        \item Note: T/D means GMM/deep model and (U)S  means
          (un)supervised learning.
    \end{tablenotes}
   \end{threeparttable}
    \label{tab:MC} \vspace*{-2mm}
\end{table}

\begin{table}[t]
    \centering
\setlength\tabcolsep{6pt} 
    \caption{Voice datasets}  
    \vspace{-3.5mm}
    \begin{threeparttable}
    \scalebox{0.83}{ 
    \begin{tabular}{c|c|c|c|c}
    \hline
         & {\bf Spk$_{10}$-enroll} & {\bf Spk$_{10}$-test} & {\bf Spk$_{251}$-train} & {\bf Spk$_{251}$-test} \\ \hline
         {\bf Task} & \multicolumn{2}{c|}{CSI-E/SV/OSI } & \multicolumn{2}{c}{CSI-NE} \\ \hline
         {\bf \#Speakers} & 10 (5M,5F) & 10 (5M,5F) &
        {\begin{tabular}[c]{@{}c@{}} 251 (126M,125F)\end{tabular}}
         &
        {\begin{tabular}[c]{@{}c@{}} 251 (126M,125F)\end{tabular}}
         \\ \hline
         {\bf \#Voices} & 10$\times$10 & 100$\times$10 & 25652 & 2887 \\ \hline
         {\bf Length} & 3--21s (7.2s) & 1--15s (4.3s) & 1--24s (12.3s) & 1--19s (11.7s) \\ \hline
    \end{tabular}
    }
       \begin{tablenotes}
        \scriptsize
        \item Note: $x$-$y$ ($z$) denotes that the minimal, maximal and average length of voices,
        \item  and $n$M/$m$F denotes that the number of male/female speakers is $n$/$m$.
      \end{tablenotes}
  \end{threeparttable}
    \label{tab:DAC} \vspace*{-3mm}
\end{table}

\section{Evaluation Setup and Metrics}
\subsection{Main Evaluation Setup}

To evaluate defenses against adversarial voices on SRSs, we developed a platform, named \platformname.

\noindent {\bf Models}.
We use two mainstream SRSs:
a pre-trained model ivector-PLDA~\cite{kaldi-ivector-plda} from the popular open-source platform KALDI
having 11.5k stars and 4.9k forks on GitHub~\cite{kaldi},
and a one-dimension convolution neural network based model AudioNet \cite{becker2018interpreting,jati2021adversarial}.
Details of two models are shown in \tablename~\ref{tab:MC}.
Due to the massive experiments, we only target the CSI task (i.e., CSI-E and CSI-NE). 
The results on the SV and OSI tasks could be similar, as demonstrated in \cite{chen2019real}.

\noindent{\bf Datasets}. We use four datasets derived from Librispeech~\cite{panayotov2015librispeech}:
Spk$_{10}$-enroll, Spk$_{10}$-test, Spk$_{251}$-train, and Spk$_{251}$-test.
The datasets are summarized in \tablename~\ref{tab:DAC} (details refer to Appendix~\ref{sec:dataset-detail} in the supplemental material.)

\noindent{\bf Attacks}.
To thoroughly evaluate the defenses, we implement 
4 promising white-box attacks (i.e., FGSM~\cite{goodfellow2014explaining}, PGD~\cite{madry2017towards},
CW$_{\bm{\infty}}$, and CW$_2$~\cite{carlini2017towards}), and
3 state-of-the-art black-box attacks (i.e., FAKEBOB~\cite{chen2019real}, SirenAttack~\cite{du2020sirenattack},
and Kenansville~\cite{ASGBWYST19}). All of them craft adversarial voices via solving optimization problems using
$L_\infty$ norm to limit  perturbations, except that Kenansville is a signal processing-based decision-only attack
and  CW$_2$ minimizes adversarial perturbations in the loss function using $L_2$ norm.
To solve the optimization problems,
FGSM, PGD, CW$_{\bm{\infty}}$, and CW$_2$ use gradients,
FAKEBOB uses gradient-estimation, and SirenAttack uses the gradient-free particle swarm optimization.
Note that CW$_{\bm{\infty}}$ is implemented using the loss function of the CW attack but optimized by PGD,
the same as \cite{madry2017towards}, to improve the attack efficiency.
Details refer to Appendix~\ref{sec:review-attacks}.

To avoid fake adversarial voices due to the discretization problem~\cite{bu2019taking}, i.e.,
adversarial voices become benign after {being} transformed into concrete voices,
they are evaluated after storing back into the 16-bit PCM form.
We only consider untargeted attacks which are more challenging to be defeated than targeted attacks~\cite{athalye2018obfuscated}.

We use a machine with Ubuntu 18.04, an Intel Xeon E5-2697 v2 2.70GHz CPU, 376GiB memory, and a GeForce RTX 2080Ti GPU.
 
\subsection{Evaluation Metrics}  
\noindent
{\bf Attack effectiveness}.
To evaluate the effectiveness of an attack,
we use model accuracy on adversarial examples ($A_a$), i.e.,
the proportion of adversarial examples that are correctly classified by the model.
Thus,  smaller $A_a$ indicates better attack.
Note that $100\%-A_a$ is the attack success rate.

\noindent
{\bf Defense effectiveness}.
A usable defense should not only improves resistance to adversarial examples,
but also sacrifices accuracy on benign examples as little as possible.
Thus, we measure the effectiveness of a defense using
model accuracy on adversarial examples ($A_a$) and model accuracy on benign examples ($A_b$), respectively,
where the larger  $A_a$ (resp. $A_b$) is, the better the defense is.
We also use the R1 score,  $R1=\frac{2\times A_b\times A_a}{A_b+A_a}$~\cite{rajaratnam2018speech}, which assigns equal importance to $A_b$ and $A_a$, to quantify the usability of a defense.

\noindent
{\bf Imperceptibility}.
To measure the imperceptibility, we use Signal-to-Noise Ratio (SNR)~\cite{yuan2018commandersong} and
Perceptual Evaluation of Speech Quality (PESQ)~\cite{rix2001perceptual}. SNR is defined as $10\log_{10}\frac{P_x}{P_\delta}$, where $P_x$ (resp. $P_\delta$) is the power of the original voice (resp. perturbation).
PESQ is one of the objective perceptual measures, simulating human auditory system~\cite{xiang2017digital}.
The calculation of PESQ is more involved.
It first applies an auditory transform to obtain the loudness spectra of the original and adversarial voices, and then compares two loudness spectra to obtain a metric score whose value is in the range of -0.5 to 4.5.
We refer readers to \cite{rix2001perceptual} for more details.
Larger SNR and higher PESQ indicate better imperceptibility.

%% file: evaluation.tex
\input{non-adaptive-transformation}

\input{adaptive-attack-algo}

\input{adaptive-transformation}

\input{adaptive-transformation-AdvT}

%% file: non-adaptive-transformation.tex
\section{Evaluation of Transformations}\label{sec:non-adaptive-attack}
\subsection{Evaluation Setup}

We limit the
perturbation budget
$\epsilon$ to 0.002 for $L_\infty$ attacks, the same as \cite{chen2019real,jati2021adversarial}, unless explicitly stated.
The number of steps for PGD and CW$_{\bm{\infty}}$ range from 10 to 50 with step\_size $\alpha=\frac{\epsilon}{5}=0.0004$ for each step.
For CW$_2$, we set the initial trade-off constant $c$ to 0.001, use 9 binary search steps to minimize perturbations, run 900-9000 iterations to converge, and vary the confidence parameter $\kappa$ from 0, 2, 5, 10, 20 to 50.
For FAKEBOB, we limit the number of iterations to 200 with the parameter samples\_per\_draw of NES $m=50$ and $\kappa=0.5$.
For SirenAttack, we use the optimal parameters reported in \cite{du2020sirenattack},
i.e., the maximum number of epochs $epoch_{max}=300$,
the iteration limit of the PSO subroutine $iter_{max}=30$,
and the number of particles $n\_particles=25$.
For Kenansville, we use the SSA method to perturb a voice and set the maximal attack factor to 100
and maximal number of iterations to 30, which is sufficient for the attack to converge according to our experiments.
FFT method is not considered since it is much less effective than the SSA method~\cite{AWBPT20}.

We consider the ivector-PLDA model for the CSI-E task which is enrolled with 10 speakers using the Spk$_{10}$-enroll dataset. 
We use the Spk$_{10}$-test dataset  to test the model, resulting in 99.8\% accuracy on benign examples.
We also use the Spk$_{10}$-test dataset to craft adversarial examples.
Though the ivector-PLDA model is pre-trained without any transformations,
it still produces sufficient accuracy on benign examples, as shown in column ($A_b$) of \tablename~\ref{tab:evaluate-defense-non-adaptive}.
Thus, we do not re-train it when transformations are deployed.
As each transformation contains at least one tunable parameter
which may affect the effectiveness, we tune parameters and choose the best ones according to their R1 scores for the remaining experiments.
Details are given in Appendix~\ref{sec:parameter}.

\begin{table*}
  \caption{Results of transformations against non-adaptive attacks}\vspace*{-3.5mm}
  \label{tab:evaluate-defense-non-adaptive}\centering\setlength\tabcolsep{3pt}
 \begin{threeparttable}
 \scalebox{0.63}{
  \begin{tabular}{c|c|c||c|c|c|c|c|c|c|c|c|c|c|c|c|c|c|c|c|c|c||c|c|c}
  \hline
  \multirow{4}{*}{\bf Defense} &  \multirow{4}{*}{\begin{tabular}[c]{@{}c@{}}{\bf {R1}} \\ {\bf {Score}} \end{tabular}}  & \multirow{4}{*}{$\mathbf{A_b}$} &  \multicolumn{22}{c}{$\mathbf{A_a}$} \\ \cline{4-25}
    &  &  &  \multicolumn{13}{c|}{\bf L$_\infty$ white-box attacks} & \multicolumn{6}{c||}{\bf L$_2$ white-box attacks} & \multicolumn{3}{c}{\bf black-box attacks} \\ \cline{4-25}
    & & & \multirow{2}{*}{\bf FGSM} & \multicolumn{6}{c|}{\bf PGD} & \multicolumn{6}{c|}{\bf CW$_\infty$} & \multicolumn{6}{c||}{\bf CW$_2$} & \multicolumn{2}{c|}{\bf Score-based (L$_\infty$)} & {\bf Decision-only}  \\ \cline{5-25}
    & & & & {\bf 10}& {\bf 20} & {\bf 30} & {\bf 40} & {\bf 50} & {\bf 100} & {\bf 10}& {\bf 20} & {\bf 30} & {\bf 40} & {\bf 50} & {\bf 100} & {\bf 0} & {\bf 2} & {\bf 5} &{\bf 10} & {\bf 20} & {\bf 50} & {\bf FAKEBOB} & {\bf SirenAttack} & {\bf Kenansville} \\ \hline \hline
  {\bf Baseline} & 8.3\% & 99.8\% & 42.3\% & 0\% & 0\% & 0\% & 0\% & 0\% & 0\% & 0\% & 0\% & 0\% & 0\% & 0\% & 0\% & 6.5\% & 0\% & 0\% & 0\% & 0\% & 0\% & 19.8\% & 18.7\% & 8.0\% \\ \hline\hline
  \textbf{QT} & \textcolor{blue}{\bf 76.0\%} & \textcolor{red}{\bf 86.8\%} & \textcolor{blue}{\bf 76.8\%} & \textcolor{blue}{\bf 61.2\%} & \textcolor{blue}{\bf 55.4\%} & \textcolor{blue}{\bf 56.6\%} & \textcolor{blue}{\bf 62.5\%} & \textcolor{blue}{\bf 59.8\%} & \textcolor{blue}{\bf 67.2\%} & \textcolor{blue}{\bf 60.7\%} & \textcolor{blue}{\bf 55.0\%} & \textcolor{blue}{\bf 55.8\%} & \textcolor{blue}{\bf 62.2\%} & \textcolor{blue}{\bf 57.4\%} & \textcolor{blue}{\bf 65.0\%} & 86.8\% & 86.1\% & 86.4\% & 86.2\% & \textcolor{blue}{\bf 84.9\%} & \textcolor{blue}{\bf 49.9\%} & \textcolor{blue}{\bf 91.3\%} & 88.2\% & 31.7\% \\ \hline
  \textbf{AT} & \textcolor{blue}{\bf 84.5\%} & {89.2\%} & \textcolor{blue}{\bf 82.9\%} & \textcolor{blue}{\bf 77.8\%} & \textcolor{blue}{\bf 75.9\%} & \textcolor{blue}{\bf 75.6\%} & \textcolor{blue}{\bf 78.5\%} & \textcolor{blue}{\bf 76.6\%} & \textcolor{blue}{\bf 81.2\%} & \textcolor{blue}{\bf 77.9\%} & \textcolor{blue}{\bf 76.7\%} & \textcolor{blue}{\bf 74.2\%} & \textcolor{blue}{\bf 77.8\%} & \textcolor{blue}{\bf 75.5\%} & \textcolor{blue}{\bf 81.2\%} & 89.1\% & 89.0\% & 89.1\% & \textcolor{blue}{\bf 89.2\%} & \textcolor{blue}{\bf 88.9\%} & \textcolor{blue}{\bf 78.5\%} & \textcolor{blue}{\bf 95.4\%} & 94.0\% & \textcolor{blue}{\bf 40.6\%} \\ \hline
  \textbf{AS} & 39.8\% & 98.1\% & \textcolor{red}{\bf 46.0\%} &  \textcolor{red}{\bf 0.0\%} & \textcolor{red}{\bf 0.0\%} & \textcolor{red}{\bf 0.0\%} & \textcolor{red}{\bf 0.0\%} & \textcolor{red}{\bf 0.0\%} & \textcolor{red}{\bf 0.0\%} & \textcolor{red}{\bf 0.0\%} & \textcolor{red}{\bf 0.0\%} &\textcolor{red}{\bf 0.0\%} & \textcolor{red}{\bf 0.0\%} & \textcolor{red}{\bf 0.0\%} & \textcolor{red}{\bf 0.0\%} & \textcolor{blue}{\bf 96.8\%} & \textcolor{blue}{\bf 95.0\%} & 87.4\% & 65.5\% & \textcolor{red}{\bf 20.1\%} & \textcolor{red}{\bf 0.0\%} & \textcolor{red}{\bf 47.5\%} & \textcolor{red}{\bf 69.3\%} & 21.9\% \\ \hline
  \textbf{MS} & 53.9\% & \textcolor{red}{\bf 83.9\%} & 65.6\% & \textcolor{blue}{\bf 21.3\%} & \textcolor{blue}{\bf 17.1\%} & \textcolor{blue}{\bf 17.3\%} & \textcolor{blue}{\bf 22.1\%} & \textcolor{blue}{\bf 18.3\%} & \textcolor{blue}{\bf 24.5\%} & \textcolor{blue}{\bf 21.2\%} & \textcolor{blue}{\bf 17.9\%} & \textcolor{blue}{\bf 17.1\%} & \textcolor{blue}{\bf 23.6\%} & \textcolor{blue}{\bf 18.9\%} & \textcolor{blue}{\bf 24.4\%} & \textcolor{red}{\bf 77.1\%} & 76.4\% & 73.2\% & 68.8\% & 57.9\% & 26.9\% & 71.5\% & \textcolor{red}{\bf 70.6\%} & \textcolor{blue}{\bf 41.8\%} \\ \hline \hline
  \textbf{DS} & 38.3\% & 91.8\% & 57.2\% & 0.3\% & 0.2\% & 0.2\% & 0.2\% & 0.1\% & 0.2\% & 0.2\% & 0.3\% & 0.3\% & 0.1\% & 0.2\% & 0.3\% & \textcolor{red}{\bf 77.2\%} & \textcolor{red}{\bf 73.4\%} & \textcolor{red}{\bf 68.1\%} & 59.9\% & 39.3\% & 0.7\% & 67.3\% & \textcolor{red}{\bf 66.8\%} & 20.2\% \\ \hline
  \textbf{LPF} & 38.2\% & 96.9\% & 59.8\% & \textcolor{red}{\bf 0.0\%} & \textcolor{red}{\bf 0.0\%} & \textcolor{red}{\bf 0.0\%} & \textcolor{red}{\bf 0.0\%} & \textcolor{red}{\bf 0.0\%} & \textcolor{red}{\bf 0.0\%} & \textcolor{red}{\bf 0.0\%} & \textcolor{red}{\bf 0.0\%} & \textcolor{red}{\bf 0.0\%} &\textcolor{red}{\bf 0.0\%} & \textcolor{red}{\bf 0.0\%} & \textcolor{red}{\bf 0.0\%} & 84.6\% & 78.2\% & 71.7\% & 59.7\% & 22.2\% & \textcolor{red}{\bf 0.0\%} & 54.3\% & 81.5\%  & \textcolor{red}{\bf 10.6\%} \\ \hline
  \textbf{BPF} & \textcolor{red}{\bf 36.1\%} & 91.0\% & 51.4\% & \textcolor{red}{\bf 0.0\%} & \textcolor{red}{\bf 0.0\%} & \textcolor{red}{\bf 0.0\%} & \textcolor{red}{\bf 0.0\%} & \textcolor{red}{\bf 0.0\%} & \textcolor{red}{\bf 0.0\%} & \textcolor{red}{\bf 0.0\%} & 0.1\% & \textcolor{red}{\bf 0.0\%}& \textcolor{red}{\bf 0.0\%} & \textcolor{red}{\bf 0.0\%} & \textcolor{red}{\bf 0.0\%} & \textcolor{red}{\bf 79.0\%} & 75.9\% & 68.5\% & \textcolor{red}{\bf 52.9\%} & 21.2\% & 0.2\% & 58.5\% & 76.8\%  & 11.6\% \\ \hline \hline
  \textbf{OPUS} & 56.8\% & \textcolor{red}{\bf 88.6\%} & 67.9\% & 17.4\% & 14.1\% & 15.0\% & 17.9\% & 17.1\% & 23.3\% & 17.0\% & 14.2\% & 15.3\% & 18.5\% & 17.9\% & 22.4\% & 84.0\% & 82.9\% & 81.0\% & 78.8\% & 71.8\% & 31.5\% & 87.5\% & 86.0\% & \textcolor{blue}{\bf 37.2\%} \\ \hline
  \textbf{SPEEX} & 53.5\% &  93.8\% & \textcolor{blue}{\bf 71.8\%} & 7.2\% & 6.6\% & 7.9\% & 11.9\% & 10.6\% & 21.8\% & 6.7\% & 6.8\% & 7.8\% & 11.3\% & 9.6\% & 22.5\% & 88.1\% & 87.5\% & 84.0\% & 77.4\% & 59.6\% & 18.3\% & 87.9\% & 89.0\% & 30.0\% \\ \hline
  \textbf{AMR} & 55.4\% & 96.8\% & 67.4\% & 6.4\% & 7.0\% & 7.7\% & 11.0\% & 8.1\% & 15.9\% & 5.8\% & 8.0\% & 7.8\% & 11.4\% & 8.7\% & 17.3\% & 94.8\% & 93.7\% & \textcolor{blue}{\bf 92.3\%} & \textcolor{blue}{\bf 88.6\%} & 67.2\% & 24.6\% & \textcolor{blue}{\bf 94.2\%} & 93.6\% & 22.9\% \\ \hline
  \textbf{AAC-V} & \textcolor{red}{\bf 29.8\%} & \textcolor{blue}{\bf 99.8\%} & \textcolor{red}{\bf 47.1\%} & \textcolor{red}{\bf 0.0\%} & \textcolor{red}{\bf 0.0\%} & \textcolor{red}{\bf 0.0\%} & \textcolor{red}{\bf 0.0\%} & \textcolor{red}{\bf 0.0\%}  & \textcolor{red}{\bf 0.0\%} & \textcolor{red}{\bf 0.0\%}  &\textcolor{red}{\bf 0.0\%}  &\textcolor{red}{\bf 0.0\%}  & \textcolor{red}{\bf 0.0\%}  & \textcolor{red}{\bf 0.0\%} &  \textcolor{red}{\bf 0.0\%} & 89.7\% & \textcolor{red}{\bf 72.4\%} & \textcolor{red}{\bf 37.3\%} & \textcolor{red}{\bf 5.9\%} & \textcolor{red}{\bf 0.0\%}  & \textcolor{red}{\bf 0.0\%} & \textcolor{red}{\bf 34.6\%} & 87.5\% & \textcolor{red}{\bf 10.4\%} \\ \hline
  \textbf{AAC-C} & 44.7\% & 92.7\% & 64.2\% & 2.8\% & 2.3\% & 1.8\% & 2.5\% & 2.4\% & 2.7\% & 3.2\% & 2.3\% & 1.6\% & 2.6\% & 2.2\% & 2.6\% & 83.6\% & 82.3\% & 78.5\% & 71.8\% & 51.1\% & 8.1\% & 76.4\% & 89.8\% & 12.6\% \\ \hline
  \textbf{MP3-V} & \textcolor{red}{\bf 27.1\%} & \textcolor{blue}{\bf 99.6\%} & \textcolor{red}{\bf 48.0\%} & \textcolor{red}{\bf 0.0\%} & \textcolor{red}{\bf 0.0\%} & \textcolor{red}{\bf 0.0\%} & \textcolor{red}{\bf 0.0\%} & \textcolor{red}{\bf 0.0\%} & \textcolor{red}{\bf 0.0\%} & \textcolor{red}{\bf 0.0\%} & \textcolor{red}{\bf 0.0\%} & \textcolor{red}{\bf 0.0\%} & \textcolor{red}{\bf 0.0\%} & \textcolor{red}{\bf 0.0\%} & \textcolor{red}{\bf 0.0\%} & 87.4\% & \textcolor{red}{\bf 62.2\%} & \textcolor{red}{\bf 15.9\%} & \textcolor{red}{\bf 0.3\%} & \textcolor{red}{\bf 0.0\%} & \textcolor{red}{\bf 0.0\%} & \textcolor{red}{\bf 32.0\%} & 90.5\% & \textcolor{red}{\bf 8.9\%} \\ \hline
   \textbf{MP3-C} & 40.9\% & 96.4\% & 53.1\% &\textcolor{red}{\bf 0.0\%} & \textcolor{red}{\bf 0.0\%}  & 0.1\% & 0.1\% & \textcolor{red}{\bf 0.0\%} & \textcolor{red}{\bf 0.0\%} & \textcolor{red}{\bf 0.0\%} & \textcolor{red}{\bf 0.0\%} & 0.1\% & 0.1\% & \textcolor{red}{\bf 0.0\%} & \textcolor{red}{\bf 0.0\%} & 87.6\% & 84.3\% & 79.0\% & 63.9\% & 29.3\% & 0.4\% & 71.1\% & 91.1\% & 11.4\% \\ \hline \hline
  \textbf{\defensenameabbr-o(k)} & \textcolor{blue}{\bf 58.0\%} & 94.0\% & 70.4\% & 16.3\% & 13.8\% & 13.0\% & 17.0\% & 12.7\% & 20.8\% & 14.1\% & 14.2\% & 15.2\% & 15.1\% & 10.7\% & 22.5\% & 91.4\% & 86.1\% & 86.5\% & 83.4\% & \textcolor{blue}{\bf 74.0\%} & \textcolor{blue}{\bf 42.0\%} & 85.5\% & 92.1\% & 26.7\% \\ \hline
  \textbf{\defensenameabbr-d(k)} &  49.6\% & \textcolor{blue}{\bf 99.4\%} & 70.5\% & 0.2\% &\textcolor{red}{\bf 0.0\%}  & 0.2\% & 0.9\% & 0.3\% & 1.1\% & 0.1\% & 0.1\% & 0.5\% & 0.6\% & 0.8\% & 1.0\% & \textcolor{blue}{\bf 97.1\%} & \textcolor{blue}{\bf 94.4\%} & \textcolor{blue}{\bf 94.1\%} & \textcolor{blue}{\bf 87.3\%} & 62.8\% & 14.7\% & 85.6\% & \textcolor{blue}{\bf 94.4\%} & 20.1\% \\ \hline
  \textbf{\defensenameabbr-c(k)} & 48.2\% & 98.8\% & 68.8\% & \textcolor{red}{\bf 0.0\%}  & 0.2\% & 0.1\% & 0.1\% & 0.1\% & 0.5\% & 0.1\% & 0.3\% & 0.1\% & 0.2\% & 0.3\% & 1.0\% &  \textcolor{blue}{\bf 96.3\%} & \textcolor{blue}{\bf 93.8\%} & \textcolor{blue}{\bf 91.0\%} & 82.1\% & 55.0\% & 11.2\% & 84.0\% & \textcolor{blue}{\bf 95.1\%} & 20.8\% \\ \hline
  \textbf{\defensenameabbr-f(k)} & 47.2\% & 98.2\% & 67.1\% & 0.3\% & 0.3\% & 0.5\% & 0.3\% & 0.4\% & 0.9\% & 0.5\% & 0.5\% & 0.6\% & 0.6\% & 0.7\% & 1.0\% & 93.4\% & 90.3\% & 86.6\% & 78.7\% & 51.2\% & 10.8\% & 83.6\% & \textcolor{blue}{\bf 94.8\%} & 21.0\% \\ \hline
  \textbf{\defensenameabbr-o(wk)} & 50.5\% & 96.7\% & 66.6\% & 3.9\% & 3.5\% & 3.7\% & 4.3\% & 4.0\% & 6.5\%  & 3.6\% & 3.2\% & 4.4\% & 4.8\% & 3.9\% & 7.0\% & 91.3\% & 88.5\% & 84.4\% & 77.5\% & 58.5\% & 26.8\% & 89.6\% & 91.7\% & 24.0\% \\ \hline
  \textbf{\defensenameabbr-d(wk)} & 49.8\% & 98.2\% & 70.2\% & 1.7\% & 1.1\% & 1.1\% & 3.0\% & 1.8\% & 3.5\% & 1.4\% & 0.6\% & 1.1\% & 2.7\% & 2.0\% & 2.7\% & 93.9\% & 90.9\% & 88.3\% & 82.9\% & 64.0\% & 23.4\% & 88.1\% & 88.7\% & 19.9\% \\ \hline
  \textbf{\defensenameabbr-c(wk)} & 48.5\% & 98.0\% & 68.3\% & 1.4\% & 0.7\% & 0.7\% & 2.4\% & 1.3\% & 2.5\% & 1.1\% & 0.6\% & 1.0\% & 2.3\% & 1.8\% & 1.8\% & 93.0\% & 89.0\% & 87.1\% & 79.4\% & 58.6\% & 20.1\% & 87.6\% & 87.9\% & 21.2\% \\ \hline
  \textbf{\defensenameabbr-f(wk)} & 49.0\% & 97.6\% & 68.5\% & 2.0\% & 1.2\% & 0.8\% & 3.0\% & 1.5\% & 3.0\% & 2.2\% & 1.5\% & 1.2\% & 2.0\% & 2.2\% & 3.2\% & 91.6\% & 88.7\% & 85.7\% & 79.5\% & 60.4\% & 22.1\% & 88.7\% & 88.8\% & 22.1\% \\ \hline
  \end{tabular}%
  }
  \begin{tablenotes}
        \scriptsize
        \item Note: k (resp. wk) denotes kmeans (resp. warped-kmeans).   The top-3 highest/lowest results are highlighted in {blue}/{red} color except for Baseline where no
  \item   defense is deployed.  The accuracy $A_a$ used for computing  R1 Score is the average of all the attacks.   
     \end{tablenotes}
  \end{threeparttable}
  \vspace*{-5mm}
  \end{table*}

\subsection{Results}
The results are reported in \tablename~\ref{tab:evaluate-defense-non-adaptive}, where row (Baseline)
shows the accuracy 
without any defense, indicating
the effectiveness of attacks. In general, the effectiveness of transformations
significantly
varies with attacks.
The results provide many interesting
and useful findings, including but not limited to the following ones.

\noindent {\bf Effectiveness versus level/domain.}
Time-domain W-transformations (e.g., QT, AT and MS) are often more effective than others on L$_\infty$ attacks,
while F-transformations are often more effective than others on L$_2$ attacks.
Among W-transformations, 
\defensenameabbr-o and \defensenameabbr-d often perform slightly better than others,
as transformation on preceding features also affects succeeding features.
Between kmeans and warped-kmeans,
the effectiveness varies with attacks
and in general they are almost comparable.
In terms of R1 score, \defensenameabbr-o with kmeans, i.e., \defensenameabbr-o(k), ranks the first place.

\begin{tcolorbox}[size=title,breakable,arc=1mm, boxsep=0.4mm, left = 1pt, right = 1pt, top = 1pt, bottom = 1pt] 
  \textbf{Findings 1.}
  Time-domain (resp. feature-level) transformations
  are often more effective than others on L$_\infty$ (resp. L$_2$) attacks.
\end{tcolorbox}

\begin{table}[t]
  \centering\setlength\tabcolsep{12pt}
  \caption{Imperceptibility and strength of non-adaptive attacks}\vspace*{-3mm}
   \begin{threeparttable}
 \scalebox{0.8}{
  \begin{tabular}{c|c|c|c|c|c}
  \hline
    \multicolumn{2}{c|}{\multirow{2}{*}{\bf Attack}} & \multicolumn{2}{c|}{\bf Imperceptibility} & \multicolumn{2}{c}{\bf Loss} \\ \cline{3-6}
     \multicolumn{2}{c|}{} & {\bf SNR} & {\bf PESQ} & $\mathbf{\mathcal{L}_{CE}}$ & $\mathbf{\mathcal{L}_{M}}$ \\ \hline
     \multicolumn{2}{c|}{\bf FGSM} & 28.53 & 2.23 & 3.91 & -1.66 \\ \hline
     \multirow{6}{*}{\bf {PGD-$\mathbf{x}$}} & $\mathbf{x}${\bf =10} & 32.77 & 2.85 & 45.88 & -45.87 \\ \cline{2-6}
     & $\mathbf{x}${\bf =20}  & 31.57 & 2.72 & 54.50 & -54.50 \\ \cline{2-6}
     & $\mathbf{x}${\bf =30}  & 31.42 & 2.70 & 58.38 & -58.38 \\ \cline{2-6}
     & $\mathbf{x}${\bf =40}  & 31.45 & 2.71 & 60.52 & -60.52 \\ \cline{2-6}
     & $\mathbf{x}${\bf =50}  & 31.31 & 2.69 & 62.23 & -62.23  \\ \cline{2-6}
     & $\mathbf{x}${\bf =100} & 31.29 & 2.70 & 67.10 & -67.10 \\ \hline
      \multirow{6}{*}{\bf CW\bm{$\infty$}-$\mathbf{x}$} & $\mathbf{x}${\bf =10} & 32.74 & 2.85 & 44.59 & -44.56 \\ \cline{2-6}
          & $\mathbf{x}${\bf =20} & 31.88 & 2.76 & 53.21 & -53.19  \\ \cline{2-6}
          & $\mathbf{x}${\bf =30} & 31.62 & 2.73 & 57.36 & -57.35 \\ \cline{2-6}
          & $\mathbf{x}${\bf =40} & 31.51 & 2.72 & 59.94 & -59.93 \\ \cline{2-6}
          & $\mathbf{x}${\bf =50} & 31.45 & 2.71 & 61.04 & -61.03 \\ \cline{2-6}
          & $\mathbf{x}${\bf =100} & 31.38 & 2.71 & 66.36 & -66.36 \\ \hline
     \multirow{5}{*}{\bf CW$_2$-$\kappa$} & $\mathbf{\kappa}${\bf =0} & 52.99 & 4.24 & 1.54 & -1.12 \\\cline{2-6}
     & $\mathbf{\kappa}${\bf =2} & 51.42 & 4.19 & 2.94 & -2.87 \\ \cline{2-6}
     & $\mathbf{\kappa}${\bf =5}  & 49.73 & 4.10 & 6.42 & -6.35 \\ \cline{2-6}
     & $\mathbf{\kappa}${\bf =10}  & 47.09 & 3.95 & 11.28 & -11.31 \\ \cline{2-6}
  & $\mathbf{\kappa}${\bf =20}  & 42.14 & 3.60 & 21.70 & -21.25 \\ \cline{2-6}
    & $\mathbf{\kappa}${\bf =50}  & 30.44 & 2.46 & 51.88 & -51.43 \\ \hline
  \multicolumn{2}{c|}{\bf FAKEBOB}  & 31.40 & 2.71 & 0.91 & -0.10 \\ \hline
  \multicolumn{2}{c|}{\bf SirenAttack} & 31.03 & 2.66 & 0.91 & -0.10 \\ \hline
  \multicolumn{2}{c|}{\bf Kenansville} & 8.73 & 1.87 & 3.32 & -2.82 \\ \hline
  \end{tabular}
  }  
  \begin{tablenotes}
        \scriptsize
        \item Note:  $\mathbf{\mathcal{L}_{CE}}$ and $\mathbf{\mathcal{L}_{M}}$ respectively denote cross entropy loss and margin 
        \item loss. The larger $\mathbf{\mathcal{L}_{CE}}$ (resp. the smaller $\mathbf{\mathcal{L}_{M}}$), the stronger the attack.
     \end{tablenotes}
 \end{threeparttable}
  \label{tab:imper-non-adaptive-atatck} \vspace*{-5mm}
\end{table}

\noindent {\bf Effectiveness versus distortion.}
Almost all the transformations perform better
against FGSM, FAKEBOB, Kenansville and SirenAttack attacks
than PGD, CW$_\infty$, and CW$_2$-50 attacks.
To find out the reason for this difference, we report the imperceptibility and strength of non-adaptive attacks in \tablename~\ref{tab:imper-non-adaptive-atatck}.
According to the imperceptibility metrics SNR and PESQ,
we observe that FGSM, SirenAttack, and Kenansville (resp. FAKEBOB) attacks introduce larger (resp. comparable)
levels of distortion than PGD, CW$_\infty$, and CW$_2$ attacks.
This indicates that there is no direct correlation between the distortion of adversarial voices and the effectiveness of input transformations.
In contrast, according to the loss values of $\mathbf{\mathcal{L}_{CE}}$ and $\mathbf{\mathcal{L}_{M}}$,
we observe that the single-step attack FGSM and the black-box attacks (i.e., FAKEBOB, SirenAttack, and Kenansville)
are much weaker than PGD, CW$_\infty$, and CW$_2$ attacks.
In fact, FGSM is a single-step attack, FAKEBOB and SirenAttack adopt an early-stop strategy,
and Kenansville is a decision-based attack,
so adversarial examples crafted by them are weak (i.e., close to the decision boundary),
while PGD, CW$_{\bm{\infty}}$, and CW$_2$-50 continue searching for
strong adversarial examples (i.e., far from the decision boundary)
even if an adversarial example has been found.

\begin{tcolorbox}[size=title,breakable,arc=1mm, boxsep=0.4mm, left = 1pt, right = 1pt, top = 1pt, bottom = 1pt] 
  \textbf{Findings 2.}
  The effectiveness of input transformations does not necessarily
  decrease with increase of  distortion, 
  since large distortion does not imply stronger adversarial voices.
\end{tcolorbox}

\begin{figure}[t]
  \centering
  \includegraphics[width=0.3\textwidth]{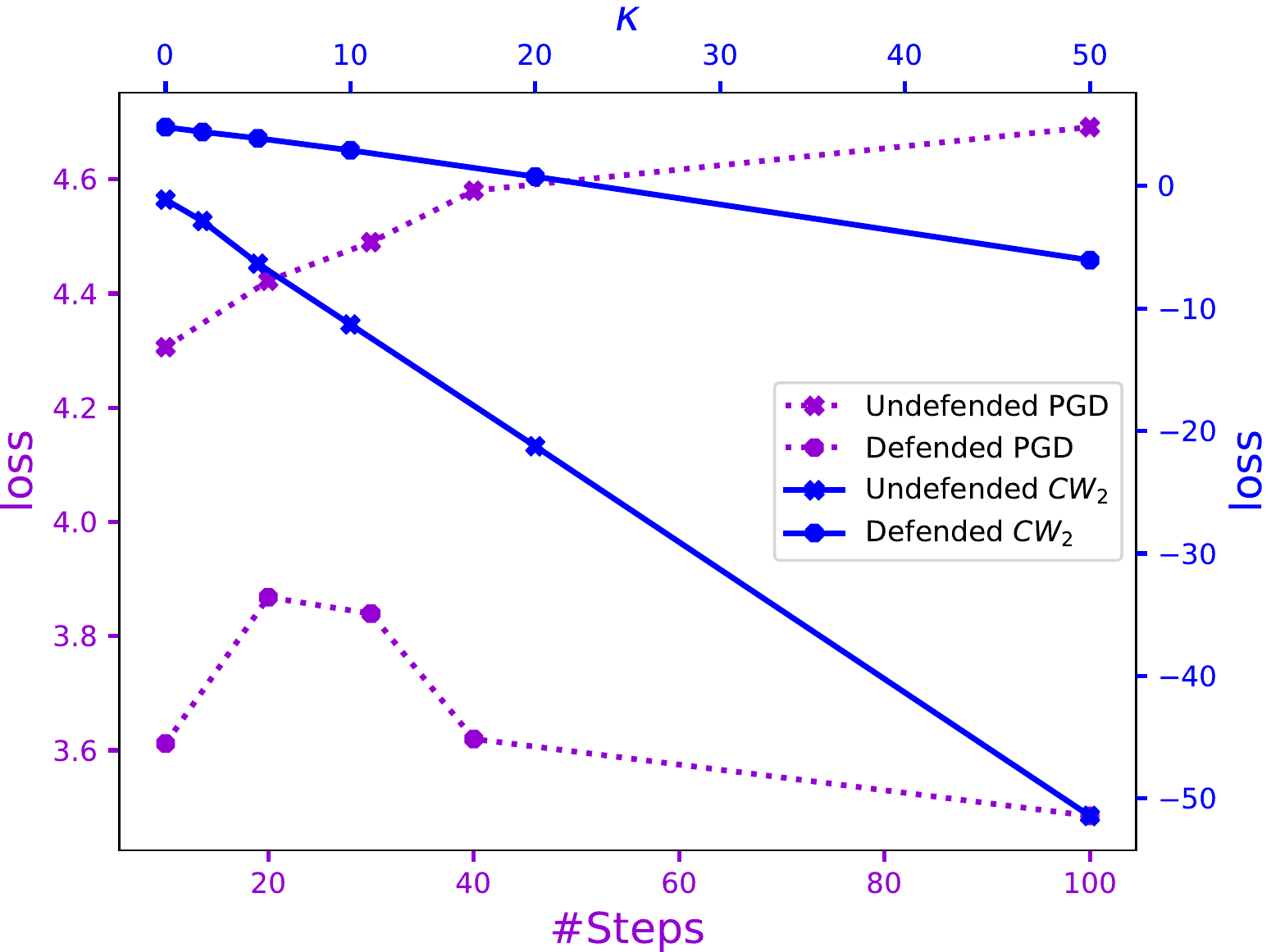}\vspace*{-3mm}
  \caption{The loss values (i.e., strength) of the adversarial voices
  on the model without/with the MS input transformation versus \#Steps of PGD and $\kappa$ of CW$_2$.
  The larger the loss of PGD (resp. the smaller the loss of CW$_2$),
the stronger the adversarial examples.
  The loss of PGD is scaled for better visualization.}\vspace*{-5mm}
  \label{fig:loss-step-kappa}
\end{figure}

\noindent {\bf Effectiveness versus attack strength.}
With increase of $\kappa$ in CW$_2$ (i.e., attack strength), unsurprisingly, the effectiveness of all the transformations decreases.
However, though the attack strength of PGD and CW$_\infty$ attacks increase with \#Steps (cf. \tablename~\ref{tab:imper-non-adaptive-atatck}), 
the effectiveness of the input transformations (e.g., QT, AT, MS, OPUS, SPEEX and \defensenameabbr-o) does \emph{not} decrease monotonically.
To understand this, we analyze the strength of adversarial voices before/after applying MS in \figurename~\ref{fig:loss-step-kappa}
and find that the strength of the adversarial examples crafted by CW$_2$ remains monotonic after applying MS with increase of $\kappa$,
while the strength of the adversarial examples crafted by PGD
becomes non-monotonic after applying
MS with increase of \#Steps.
This may be because CW$_2$ introduces larger distortion with increase of $\kappa$,
but PGD does not introduce obviously larger distortion with increase of \#Steps, as shown in \tablename~\ref{tab:imper-non-adaptive-atatck}.

Since the step size $\alpha$ may impact the capacity of the PGD attack,
we also adopt another three dynamic strategies $\alpha=\frac{\epsilon}{5\times\text{\#Steps}}$,
$\alpha=\frac{\epsilon}{\text{\#Steps}}$, and $\alpha=\frac{10\times\epsilon}{\text{\#Steps}}$
which reduces the step size $\alpha$ with increase of \#Steps (Recall that previously we set $\alpha=\frac{\epsilon}{5}$).
The same phenomenon also occurs (cf. \tablename~\ref{tab:PGD-step-size-fractional} in Appendix~\ref{sec:pgd-cwinf-step-size}), indicating this phenomenon is not due to unsuitable step size.

\begin{tcolorbox}[size=title,breakable,arc=1mm, boxsep=0.4mm, left = 1pt, right = 1pt, top = 1pt, bottom = 1pt]
\textbf{Findings 3.}
The effectiveness of input transformations does not necessarily
decrease with  increase of  attack strength.
\end{tcolorbox}

\noindent {\bf Overall effectiveness.}
Transformations are often more effective against L$_2$ white-box, L$_\infty$ black-box, and signal processing attacks
than L$_\infty$ white-box attacks.
For instance, AS, LPF, AAC-V, and MP3-V
cannot improve any robustness against the PGD and CW$_\infty$ attacks regardless of \#Steps, and the CW$_2$-50 attack.
By analyzing the strength
of adversarial voices in \tablename~\ref{tab:imper-non-adaptive-atatck},
we found that:

\begin{tcolorbox}[size=title,breakable,arc=1mm, boxsep=0.4mm, left = 1pt, right = 1pt, top = 1pt, bottom = 1pt]
  \textbf{Findings 4.}
AS, LPF, AAC-V, and MP3-V are \emph{completely} ineffective against attacks that craft high-confidence adversarial voices (i.e., PGD, CW$_\infty$ and CW$_2$ with $\kappa=50$), in non-adaptive setting.
\end{tcolorbox}

\noindent {\bf  VBR and CBR in speech compression.}
We noticed significant difference of effectiveness between VBR speech compression (e.g., AAC-V and MP3-V)
and CBR speech compression (OPUS, SPEEX, AMR, AAC-C, and MP3-C).
For instance, the accuracy of MP3-C (resp. AAC-C) against CW$_2$-10
is 212 (resp. 11) times larger than that of MP3-V (resp. AAC-V).
Compared to CBR speech compression, VBR speech compression dynamically adjusts the bit rate of the voices
to better fit to the psychoacoustic perception of the human ear and thus achieves better quality.
As a result, although they incur less side effect on the benign voices
($A_b$ of AAC-V and MP3-V only drops by 0\% and 0.2\% compared to the Baseline),
they are limited in disrupting the adversarial perturbation.

\begin{tcolorbox}[size=title,breakable,arc=1mm, boxsep=0.4mm, left = 1pt, right = 1pt, top = 1pt, bottom = 1pt]
\textbf{Findings 5.}
  VBR speech compression has less side-effect,
  but are less effective in mitigating adversarial voices.
\end{tcolorbox}

More findings in the non-adaptive setting refer to Appendix~\ref{sec:non-adaptive-more-findings}.

%% file: adaptive-attack-algo.tex
\section{Adaptive Attacks}\label{sec:adaptiveattack-algo}
To evaluate the robustness of
transformations in the adaptive setting where the adversary has complete knowledge of defense
and attempts 
to bypass the defense,
we design adaptive attacks tailored to input transformations, following the
suggestions of~\cite{tramer2020adaptive}, i.e.,
being as simple
as possible while resolving any potential optimization difficulties.

To bypass a certain input transformation $g(\cdot)$,
the adversary attempts to find an adversarial voice $x^{adv}$ from a benign voice $x$
such that $x^{adv}$ remains adversarial after being transformed by $g(\cdot)$, namely,
 solving the following optimization problem:
\begin{center}
    $\argmin_{x^{adv}}  \mathcal{L}(g(x^{adv}),y) \quad
    \text{such that }  \quad \|x^{adv}-x\|_p\leq \epsilon$
\end{center}
where $\mathcal{L}$ is the loss function used in non-adaptive attack
(cross-entropy loss for FGSM, PGD, and margin loss for CW$_\infty$, CW$_2$, FAKEBOB, and SirenAttack),
$p=2,\infty$ is the L$_p$ norm-based distance, and $y$ is the ground-truth label of $x$ for untargeted attack.

FAKEBOB, SirenAttack, and Kenansville  solve the optimization problem without gradient back-propagation, thus can be directly mounted,
except that the adaptive version goes through the deployed transformation when querying the model, while the non-adaptive one does not.
For differentiable and deterministic transformations (i.e., AS, MS, DS, LPF, and BPF) on which reliable and informative gradients can be computed via back-propagation,
the optimization problem can be easily solved by white-box attacks
using gradient descents.
However, the gradient of the loss function $\mathcal{L}$ w.r.t. $x^{adv}$ cannot be back-propagated for non-differentiable transformations (e.g., QT and speech compressions)
while the gradient is less reliable and informative for randomized transformations  (e.g., AT and FeCo).
To address this issue, we adopt evasion techniques for white-box attacks (i.e., FGSM, PGD, CW$_\infty$, and CW$_2$ attacks).

\subsection{Bypassing W-Transformations}\label{sec:bypass-non-diff}
To enable backpropagation of the gradient from a non-differentiable but deterministic W-transformation $g$,
the adversary may utilize Backward Pass Differentiable Approximation (BPDA)~\cite{athalye2018obfuscated}.
Specifically, during the forward pass, the adversary directly uses $g$ to compute the loss,
while uses a differentiable function $\hat{g}$ in the backward pass,
i.e., approximating $\nabla_x g(x)$ with $\nabla_x \hat{g}(x)$.
We set $\hat{g}(x)=x$, i.e., the identity function,
which has been shown effective for breaking non-differentiable input transformations
in the image domain~\cite{tramer2020adaptive}.

To tackle randomized transformations,
the adversary may exploit Expectation over Transformation (EOT)~\cite{athalye2018synthesizing}, i.e.,
the loss function is reformulated as $\mathbb{E}_r[\mathcal{L}(g_r(x),y)]\approx \frac{1}{R}\sum_{i=1}^{R}\mathcal{L}(g_{r_i}(x),y)$
where $r$ denotes the randomness of $g$, $r_i$ is an independent draw of the randomness, and
$R$ is the number of independent draws.
Intuitively, a randomized transformation is independently sampled multiple times
and the average of the loss function is used during gradient descent.
It reduces the variance of the gradient and enables a more stable search direction.
We remark that four differentiable and randomized transformation based defenses have been
broken using EOT in the image domain ~\cite{tramer2020adaptive,athalye2018obfuscated}.

\subsection{Bypassing F-Transformations}
Since \defensenameabbr is differentiable and randomized,
one could use EOT to bypass \defensenameabbr (cf. Section~\ref{sec:bypass-non-diff}).
 Below, we design more specific evasion techniques for white-box attacks, tailored to \defensenameabbr,
 called Replicate attack, including Replicate-F(feature) and Replicate-W(ave).

\noindent {\bf Replicate-F.}
To bypass \defensenameabbr, 
the adversary first crafts an adversarial voice $x'$ on the model {\it without} FeCo,
and then builds a new
feature matrix $\mathcal{M}'$ from the feature matrix $\mathcal{M}$ of $x'$ with the goal $\text{\defensenameabbr}(\mathcal{M}')=\mathcal{M}$, i.e., when $\mathcal{M}'$
is fed to the model defended by \defensenameabbr,
$\mathcal{M}'$ is likely compressed to $\mathcal{M}$, leading to a successful attack.

\begin{figure}[t]
\vspace*{-3mm}
\begin{algorithm}[H]
\footnotesize
	\caption{Replicating features}
	\label{alg:replicate-F}
	\begin{algorithmic}[1]
    \Require feature matrix $\mathcal{M}=[\mathbf{a}_1,\cdots,\mathbf{a}_N]$;
    cluster ratio $0<cl_r<1$;  cluster oracle $\mathcal{O}=$ kmeans or warped-kmeans
	\Ensure  replicated feature matrix $\mathcal{M}'$
    \State $k\gets \lfloor\frac{1}{cl_r}\rfloor$ \label{alg2:1}
    \For{$(i=1;i\leq N;i++)$}
        $\mathcal{A}_i\gets$ matrix that replicats the vector $\mathbf{a}_i$ $k$ times
    \EndFor
    \For{$(i=1;\lceil (N\times k+i-1)\times cl_r\rceil\neq N;i++)$}
       append the vector $\mathbf{a}_i$ to $\mathcal{A}_i$
    \EndFor
    \State $\mathcal{M}_1\gets [\mathcal{A}_1,\cdots,\mathcal{A}_N]$ \Comment{concatenate the replicated vectors} \label{alg2:2}
    \State $[b_1,\cdots,b_{|\mathcal{M}_1|}]\gets \mathcal{O}(\mathcal{M}_1, N)$  \label{alg2:3}
    \State Let $i_1,\cdots,i_N$ be a permutation of $1,\cdots,N$ s.t. for each $1\leq j\leq N$, most of vectors of
    $\mathcal{A}_{i_j}$ are divided into the $b_{i_j}$-cluster
    \State $\mathcal{M}'\gets [\mathcal{A}_{i_1},\cdots,\mathcal{A}_{i_N}]$
    \State \Return{$\mathcal{M}'$}
	\end{algorithmic}
\end{algorithm}
\vspace*{-7mm}
\end{figure}

The desired feature matrix $\mathcal{M}'$
is built by applying Algorithm~\ref{alg:replicate-F}.
Suppose $\mathcal{M}=[\mathbf{a}_1,\cdots,\mathbf{a}_N]$
where $\mathbf{a}_i$ is the feature vector of the $i$-th frame. It first replicates each feature vector $\mathbf{a}_i$
of $\mathcal{M}$ by $k=\lfloor\frac{1}{cl_r}\rfloor$ times
and then appends vectors to the replicated vectors $\mathcal{A}_i$'s
until the concatenated matrix $\mathcal{M}_1$
of $[\mathcal{A}_1,\cdots,\mathcal{A}_N]$
will lead to a feature matrix with $N$ frames after applying
\defensenameabbr.
It is expected that $\text{\defensenameabbr}(\mathcal{M}_1)$
has the same frames as $\mathcal{M}$.
However, the order of frames of $\text{\defensenameabbr}(\mathcal{M}_1)$
may differ from that of $\mathcal{M}$.
To overcome this problem, we run the clustering algorithm
with the parameter $cl_r$ on the matrix $\mathcal{M}_1$
to get the order of the frames of  $\text{\defensenameabbr}(\mathcal{M}_1)$.
This order is used to permute the replicated vectors
$\mathcal{A}_i$'s intended to make  $\text{\defensenameabbr}(\mathcal{M}')=
[\frac{\sum\mathcal{A}_{i_1}}{|\mathcal{A}_{i_1}|},\cdots,\frac{\sum\mathcal{A}_{i_N}}{|\mathcal{A}_{i_N}|}]$ being $\mathcal{M}$.

\smallskip\noindent {\bf Replicate-W.}
Replicate-F is infeasible in practice, as the API exposed by a SRS  {\it only} accepts waveforms.
Therefore, we introduce
Replicate-W, which is similar to Replicate-F except that the adversarial voice $x^{adv}$ is reconstructed from $\mathcal{M}'$ using Griffin-Lim algorithm~\cite{GL-algo}
and fed to SRS defended with FeCo.

%% file: adaptive-transformation.tex
\begin{table*}
  \centering\setlength\tabcolsep{4pt}
  \caption{Results ($A_a$, SNR, PESQ) of transformations against adaptive attacks
  }\vspace*{-3.5mm}
\begin{threeparttable}
\scalebox{0.72}{ 
  \begin{tabular}{c||c|c|c|c|c|c||c|c|c|c|c|c|c|c|c||c|c|c}
  \hline
  \multirow{3}{*}{\bf Defense} & \multirow{3}{*}{\begin{tabular}[c]{@{}c@{}} {\bf Adaptive} \\ {\bf Techniques} \end{tabular}} & \multicolumn{5}{c||}{\bf L$_\infty$ white-box attacks} & \multicolumn{9}{c||}{\bf L$_2$ white-box attacks} &  \multicolumn{3}{c}{\bf black-box attacks} \\ \cline{3-19}
   & & {\bf FGSM} & {\bf PGD-10} &
   {\bf PGD-100} & {\bf CW$_{\bm{\infty}}$-10} & {\bf CW$_{\bm{\infty}}$-100} & \multicolumn{3}{c|}{\bf CW$_2$-0} & \multicolumn{3}{c|}{\bf CW$_2$-2} & \multicolumn{3}{c||}{\bf CW$_2$-50} & {\bf FAKEBOB} & {\bf SirenAttack} & {\bf Kenansville} \\  \cline{3-19}
   & & $\mathbf{A_a}$ & $\mathbf{A_a}$ & $\mathbf{A_a}$ & $\mathbf{A_a}$ & $\mathbf{A_a}$ & $\mathbf{A_a}$ & {\bf SNR} & {\bf PESQ} & $\mathbf{A_a}$ & {\bf SNR} & {\bf PESQ} & $\mathbf{A_a}$ & {\bf SNR} & {\bf PESQ} & $\mathbf{A_a}$ & $\mathbf{A_a}$ & $\mathbf{A_a}$ \\ \hline \hline
    \cellcolor{gray!20} {\bf QT} & BPDA & 18.6\% & 0\% & 0\% & 0\% & 0\% & 14.6\% & 46.81 & 3.86 & 0\% & 44.04 & 3.71 & \multicolumn{3}{c||}{-} & 40.1\% & 75.0\% & 9.9\% \\ \hline
   \cellcolor{green!20} {\bf AT} & EOT & 18.7\%  & 4.3\%  & 1.8\%  & 4.5\% & 1.9\% & 64.4\%  & 37.47 & 3.03 & 26.2\% & 35.45 & 2.88 & 0\% & 20.71 & 1.70 & {\textcolor{red}{96.67\%}} & \textcolor{red}{95.0\%} & 18.5\% \\ \hline
    {\bf AS} & \xmark & 31.5\%  & - & - & - & - & 19.0\% & 49.70 & 4.16 & 0\% & 48.49 & 4.11 & \multicolumn{3}{c||}{-} & 14.5\% & \textcolor{red}{93.0\%} & 9.8\% \\ \hline
  {\bf MS} & \xmark & 1.6\% & 0\% & 0\% & 0\% & 0\% & 4.7\% & 61.76 & 4.45 & \multicolumn{3}{c|}{-} & \multicolumn{3}{c||}{-} & 0.3\% & 23.0\% & 6.5\% \\ \hline \hline
{\bf DS} & \xmark &  24.2\%  & - & - & - & - & 18.2\% & 57.28 & 4.35 & 0\% & 55.02 & 4.29  & \multicolumn{3}{c||}{-} & 15.0\% & \textcolor{red}{93.0\%} & 8.5\% \\ \hline
    {\bf LPF} & \xmark & 32.6\%  & - & - & - & - & 20.2\% & 55.34 & 4.35 & 0\% & 53.46 & 4.29 & \multicolumn{3}{c||}{-} & 18.8\%  &  \textcolor{red}{95.9\%}  & 7.1\% \\ \hline
   {\bf BPF} & \xmark & 26.4\%  & - & - & - & - & 17.3\% & 57.98 & 4.37 & 0\% & 55.99 & 4.31 & \multicolumn{3}{c||}{-} & 12.3\% & \textcolor{red}{82.7\%} & 6.8\% \\ \hline \hline
    \cellcolor{gray!20} {\bf OPUS} & BPDA & \textcolor{red}{89.1\%} & \textcolor{red}{86.8\%} & \textcolor{red}{84.4\%} & 86.5\% & \textcolor{red}{84.0\%}  & 25.1\% & 20.97 & 1.89 & 0\% & 15.94 & 1.71 & \multicolumn{3}{c||}{-} & 82.3\% & 73.2\% & 8.7\% \\ \hline
    \cellcolor{gray!20} {\bf SPEEX} & BPDA & \textcolor{red}{89.7\%} & \textcolor{red}{80.6\%} &  \textcolor{red}{75.4\%} & 80.0\% & \textcolor{red}{75.2\%} & 1.9\% & 24.33 & 1.92 & \multicolumn{3}{c|}{-} & \multicolumn{3}{c||}{-} & 87.7\% & 72.0\% & 7.2\% \\ \hline
 \cellcolor{gray!20} {\bf AMR} & BPDA & \textcolor{red}{90.4\%} & \textcolor{red}{73.2\%} & \textcolor{red}{63.4\%} & 73.5\% & \textcolor{red}{63.5\%} & 2.1\% & 24.30 & 1.96 & \multicolumn{3}{c|}{-} & \multicolumn{3}{c||}{-} & 92.0\% & 80.1\% & 6.3\% \\ \hline
    \cellcolor{gray!20} {\bf AAC-V} & BPDA &  \textcolor{red}{51.9\%} &  0\%  & 0\%  & 0\% & 0\% & 2.3\% & 48.96 & 4.06 & \multicolumn{3}{c|}{-} & \multicolumn{3}{c||}{-} & \textcolor{red}{44.9\%} & \textcolor{red}{97.0\%} & 9.1\% \\ \hline
    \cellcolor{gray!20} {\bf AAC-C} & BPDA & \textcolor{red}{88.8\%} & \textcolor{red}{43.2\%}  & \textcolor{red}{6.2\%}  & 44.5\% & \textcolor{red}{6.7\%} & 19.9\% & 32.67 & 2.59 & 0\% & 29.23 & 2.36 & \multicolumn{3}{c||}{-} & 23.1\% & 65.0\% & 8.3\% \\ \hline
    \cellcolor{gray!20} {\bf MP3-V} & BPDA & \textcolor{red}{52.2\%} & 0\% & 0\% & 0\%  & 0\% & 2.4\% & 49.95 & 4.12 & \multicolumn{3}{c|}{-} & \multicolumn{3}{c||}{-} & \textcolor{red}{46.4\%} & \textcolor{red}{96.1\%} & 6.9\% \\ \hline
 \cellcolor{gray!20} {\bf MP3-C} & BPDA &  \textcolor{red}{89.4\%}  & \textcolor{red}{10.2\%}  & \textcolor{red}{0.9\%} & 10.5\% & \textcolor{red}{1.2\%} & 15.5\% & 34.70 & 2.88 & 0\% & 31.11 & 2.64 & \multicolumn{3}{c||}{-} & 54.2\% & 64.2\% & 7.3\% \\ \hline \hline
 \cellcolor{green!20} & EOT & 54.1\%  &  0\%   & 0\% & 0\% & 0\% & 90.4\% & 56.20 & 4.14  & 88.0\% & 53.54 & 4.05 & 1.2\% & 18.38 & 1.57 & \textcolor{red}{92.17\%} & \textcolor{red}{96.4\%} & \textcolor{red}{31.0\%} \\ \cline{2-19}
 \cellcolor{green!20} & Replicate-W & 68.0\%  &  39.4\%   & 49.0\%  &  39.3\% & 49.9\% & 82.7\% & - & -  & 78.7\% & - & - & 58.6\% & - & - & \textcolor{red}{87.8\%} & 83.9\% & 20.0\% \\ \cline{2-19}
  \multirow{-3}{*}{\cellcolor{green!20} \bf \defensenameabbr-o(k)} & Replicate-F & 72.4\%  &  7.9\%   & 15.6\%  &  7.3\% & 14.5\% & 92.8\% & - & -  & 88.6\% & - & - & 36.7\% & - & - & \textcolor{red}{98.1\%} & \textcolor{red}{93.2\%} & 22.6\% \\ \hline
 \end{tabular}
  }
   \begin{tablenotes}
        \scriptsize
        \item Note: The accuracy in {red} indicates that an adaptive attack is not stronger than its non-adaptive version.  The cells with {gray} (resp. {green})  color indicate that the 
  \item transformations are non-differentiable (resp. randomized).  Distortion levels of $L_\infty$ attacks are not reported since they are similar. The distortion levels of Replicate
  \item  attacks are not reported  since the benign and adversarial voices do not align with each other  due to the replication operation.
      \end{tablenotes}
  \end{threeparttable}
  \vspace*{-3mm}
  \label{tab:evaluate-defense-adaptive-iv}
\end{table*}
 
\section{Evaluation of Adaptive Attacks}\label{sec:adaptive-attack}

\subsection{Evaluation Setup}
We evaluate 
transformations
in the same setup
as in Section~\ref{sec:non-adaptive-attack}
against adaptive attacks
 derived from a subset of representative attacks according to Section~\ref{sec:adaptiveattack-algo}.
For adaptive attacks derived from FGSM, CW$_2$-0, FAKEBOB, SirenAttack and Kenansville, we consider all the transformations,
as they are effective in the non-adaptive setting,
but the effectiveness varies.
For adaptive attacks derived from PGD-10, PGD-100, CW$_{\bm{\infty}}$-10, and CW$_{\bm{\infty}}$-100,
we do not consider AS, DS, LPF, and BPF, as they are differentiable, deterministic,
and almost completely ineffective in the non-adaptive setting.
The CW$_2$-2 (resp. CW$_2$-50) attack is considered only when a transformation is effective (i.e., at least 5\% accuracy)
against CW$_2$-0 (resp. CW$_2$-2).
We do not consider all the
combinations of attacks and transformations,
as the current experiments
already require substantial effort.
 
\subsection{Results}
The results are shown in \tablename~\ref{tab:evaluate-defense-adaptive-iv}.
Overall, the effectiveness varies with transformations and attacks.
Below, we compare the results with those obtained in the non-adaptive setting (i.e., \tablename~\ref{tab:evaluate-defense-non-adaptive}), by distinguishing if
the transformations are differentiable or not.

\noindent
{\bf Results of non-differentiable transformations} (\textcolor{gray!100}{gray} color in \tablename~\ref{tab:evaluate-defense-adaptive-iv}).
First, QT becomes less effective against both white-box and black-box attacks,
indicating both BPDA and adaptive black-box attacks are able to circumvent QT.
 
Second, against adaptive white-box attacks, the effectiveness of CBR speech compressions (i.e., OPUS, SPEEX, AMR, AAC/MP3-C)
does not decrease, 
indicating that BPDA is not able to circumvent them.
Indeed,
(1) BPDA cannot reduce the accuracy of speech CBR compressions on the adversarial examples crafted by FGSM, PGD, and CW$_{\bm{\infty}}$-0
when compared with the results in \tablename~\ref{tab:evaluate-defense-non-adaptive}.
(2) Though BPDA can reduce the accuracy on the adversarial examples crafted by CW$_2$-0 and CW$_2$-2,
much more distortions are introduced than the non-adaptive CW$_2$ attack,
e.g., the SNR of the adaptive CW$_2$-0 (with BPDA) on AAC-C (resp. MP3-C) is 32.67 dB (resp. 34.70 dB),
20 dB (resp. 18 dB) smaller than that of the non-adaptive CW$_2$-0 (52.99 dB, cf. \tablename~\ref{tab:imper-non-adaptive-atatck}).
Recall that CW$_2$ does not have any perturbation threshold,
while other attacks have. Thus, adaptive CW$_2$ attacks still achieve high attack success rate at the cost of distortion.

In contrast, we found that BPDA with the identity function is effective in breaking VBR speech compression (i.e., AAC/MP3-V).
Compared with the result of non-adaptive CW$_2$-0 attack in \tablename~\ref{tab:evaluate-defense-non-adaptive},
the adaptive CW$_2$-0 attack equipped with BPDA reduces the accuracy of AAC-V (resp. MP3-V) by 70.1\% (resp. 59.8\%)
with no more than 0.2 and 4.1 dB decrease in PESQ and SNR, respectively.

To understand why BPDA has different effectiveness between QT, CBR and VBR speech compressions,
we checked the appropriateness of approximating non-differentiable transformations by the identity function and found that QT and VBR speech compressions are much closer to the identity function
than CBR speech compressions (cf. Appendix~\ref{sec:evaapptrans}),
indicating that BPDA with the identity function is not strong enough to bypass CBR speech compressions,
and better approximation functions are required to circumvent them.
We leave this as future work (cf. Section~\ref{sec:discuss-findings} for discussion).

\begin{tcolorbox}[size=title,breakable,arc=1mm, boxsep=0.4mm, left = 1pt, right = 1pt, top = 1pt, bottom = 1pt]
\textbf{Findings 6.}
  BPDA with identity function can evade non-differentiable QT and VBR speech compressions,
  but fail to evade CBR speech compressions.
\end{tcolorbox}
We highlight that in the image domain, \cite{athalye2018obfuscated} and \cite{tramer2020adaptive} successfully evade all the seven input transformation-based adversarial defenses using BPDA with the identity function,
which is inconsistent with our Findings 6.
Also, while \cite{Carlini018} showed MP3 robust audio adversarial examples against speech recognition models can be crafted with BPDA
at the cost of approximately 15dB larger distortion (close to our result of MP3-C), Findings 6 shows that
MP3-V can be easily evaded with BPDA without obvious distortion increase.

Third, CBR speech compressions become less effective against adaptive FAKEBOB and SirenAttack,
especially, AAC-C and MP3-C
reduce 53.3\% and 16.90\% accuracy against adaptive FAKEBOB, respectively.
However, AAC/MP3-V achieve higher accuracy, indicating that adaptive FAKEBOB and SirenAttack
are limited in circumventing VBR speech compressions.
It is because the gradients estimated by NES of FAKEBOB for AAC/MP3-V are not informative enough,
and the particles moving direction of PSO in SirenAttack is not stable,
due to the variable bit rate of AAC/MP3-V.

\begin{tcolorbox}[size=title,breakable,arc=1mm, boxsep=0.4mm, left = 1pt, right = 1pt, top = 1pt, bottom = 1pt]
\textbf{Findings 7.}
  Variable bit rate (VBR)  makes speech compressions
  more resistant against adaptive black-box attacks.
\end{tcolorbox}

\noindent
{\bf Results of differentiable transformations} (\textcolor{gray!100}{non-gray} color in \tablename~\ref{tab:evaluate-defense-adaptive-iv}).
All the deterministic transformations become less effective against white-box and black-box adaptive attacks,
except for AS, DS, LPF, and BPF against SirenAttack because the perturbation budget $\epsilon=0.002$ is not sufficient enough
for SirenAttack to evade these transformations.
When $\epsilon=0.02$, the adaptive SirenAttack becomes stronger than the non-adaptive one,
reducing at least 16\% accuracy, on these transformations (cf. Appendix~\ref{sec:siren-0.02}).

Randomized transformations (i.e., AT and \defensenameabbr-o(k)) can also be evaded
by the white-box adaptive attacks with EOT or larger parameter $\kappa$.
However, AT and \defensenameabbr-o(k) remain effective on the adversarial examples
crafted by the black-box adaptive attacks FAKEBOB, SirenAttack, and Kenansville
(except for AT due to the larger distortion
introduced by Kenansville which suffices to overcome the randomness of AT).
This is because: their randomness makes the estimated gradients of NES uninformative for
FAKEBOB, the moving direction of PSO unreliable
for SirenAttack,
and randomized decision
for Kenansville.

\begin{tcolorbox}[size=title,breakable,arc=1mm, boxsep=0.4mm, left = 1pt, right = 1pt, top = 1pt, bottom = 1pt]
\textbf{Findings 8.}
Differentiable transformations become less effective against the white-box adaptive attacks,
but randomized transformations remain resistant to the black-box adaptive attacks.
\end{tcolorbox}

\noindent
{\bf Replicate attack versus EOT}.
We observe that EOT is more effective than
the Replicate attack
to bypass \defensenameabbr-o(k).
To understand the reason, we analyze if the expectation (i.e., $\text{FeCo}(\mathcal{M}')=\mathcal{M}$) of the Replicate attack is satisfied.
We found that $\text{FeCo}(\mathcal{M}')$ has almost the same frames (i.e., feature vectors) as $\mathcal{M}$, but their orders are not the same,
due to the randomness of \defensenameabbr.
Indeed, it is impossible to
ensure the same orders, even if a brute-force adversary can enumerate the randomness, where
the adversary has to craft and submit an adversarial voice for each randomness, would result in a low success rate (cf. Appendix~\ref{sec:more-detail-replicate-attack}).
In contrast, EOT allows to craft an adversarial voice that remains adversarial against the randomness of \defensenameabbr
by taking average of the loss functions conditioned at multiple randomness during the gradient descent.

Besides,  Relicate attack
replicates the speech content of each frame,
and the lossy reconstruction of voices from features introduce additional noise,
making the adversarial voices more perceptible (visit our website for listening audios) and less robust (i.e., Replicate-W is worse than Replicate-F for strong attacks).

\begin{tcolorbox}[size=title,breakable,arc=1mm, boxsep=0.4mm, left = 1pt, right = 1pt, top = 1pt, bottom = 1pt]
\textbf{Findings 9.}
Against \defensenameabbr, EOT is more effective than Replicate attack  in terms of both attack success rate and imperceptibility.
\end{tcolorbox}
  
\begin{table*}[t]
  \centering\setlength\tabcolsep{8pt}
  \caption{Results ($A_a$, SNR, PESQ) on {Standard}, {Vanilla-AdvT}, and {AdvT+Transformation}}\vspace*{-2mm}
 \begin{threeparttable} \scalebox{0.68}{ 
  \begin{tabular}{c|c||c||c|c|c|c|c||c|c|c||c|c|c}
  \hline
  \multirow{3}{*}{} & \multirow{3}{*}{\begin{tabular}[c]{@{}c@{}}{\bf R1} \\ {\bf Score} \end{tabular}} &  \multirow{3}{*}{$\mathbf{A_b}$} & \multicolumn{5}{c||}{{\bf L$_\infty$ white-box attacks}} & \multicolumn{3}{c||}{\bf L$_2$ white-box attacks} & \multicolumn{3}{c}{\bf black-box attacks} \\ \cline{4-14}
   & & & {\bf FGSM} & {\bf PGD-10} & {\bf PGD-100} & {\bf CW$_{\bm{\infty}}$-10} & {\bf CW$_{\bm{\infty}}$-100} & \multicolumn{3}{c||}{\bf CW$_2$-1} & {\bf FAKEBOB} & {\bf SirenAttack} & {\bf Kenansville} \\ \cline{4-14}
   & & & $\mathbf{A_a}$ & $\mathbf{A_a}$ & $\mathbf{A_a}$ & $\mathbf{A_a}$ & $\mathbf{A_a}$ & $\mathbf{A_a}$ & {\bf SNR} & {\bf PESQ} & $\mathbf{A_a}$ & $\mathbf{A_a}$ & $\mathbf{A_a}$ \\ \hline \hline
   {\bf Standard} & 6.54 & {99.06\%} & 19.61\% & 0\%  & 0\%  & 0\%  & 0\% & 0\% & 55.87 & 4.47 & 0.35\% & 0.38\% & 0.03\% \\ \hline \hline
   {\bf Vanilla-AdvT} & 61.48 & 95.67\% & 75.20\% & 58.19\%  & 53.83\%  & 58.95\% & 55.56\% & 0\% & 36.96 & 3.91 & 85.63\% & 86.73\% & 0.03\% \\ \hline \hline
   {\bf AdvT+QT} & 67.68 & 95.74\% & 88.19\%  & 72.12\%  & 64.08\%  & 73.20\% & 65.43\% & 0.7\% & \cellcolor{green!20}46.59 & 3.86 & \cellcolor{green!20}79.84\% & 88.81\% & 0.31\% \\ \hline
   {\bf AdvT+AT} & 71.11 & \cellcolor{green!20}95.57\% & \cellcolor{green!20}71.10\%  & 61.10\%  & 59.22\% & 61.47\% & 59.89\% & 9.3\% & 36.21 & 3.90 & 94.69\% & 95.39\% & 39.80\% \\ \hline
   {\bf AdvT+AS} & \cellcolor{green!20}58.35 & \cellcolor{green!20}93.59\% & 82.72\% & \cellcolor{green!20}53.83\% & \cellcolor{green!20}43.12\% & \cellcolor{green!20}54.10\% & \cellcolor{green!20}45.24\% & 0\% & 35.46 & 3.45 & \cellcolor{green!20}83.55\% & 87.08\% & 0.03\% \\ \hline
   {\bf AdvT+MS} & \cellcolor{green!20}54.66 & \cellcolor{green!20}92.76\% & \cellcolor{green!20}65.85\% & \cellcolor{green!20}49.77\%  & \cellcolor{green!20}44.13\%  & \cellcolor{green!20}50.33\% & \cellcolor{green!20}46.66\% & 0\% & \cellcolor{green!20}37.85 & 3.66 & \cellcolor{green!20}76.38\% &  \cellcolor{green!20}77.24\% & 0.17\% \\ \hline \hline
   {\bf AdvT+DS} & \cellcolor{green!20}56.41 & \cellcolor{green!20}95.32\% & \cellcolor{green!20}70.14\% & \cellcolor{green!20}51.44\%  & \cellcolor{green!20}44.06\%  & \cellcolor{green!20}52.13\% & \cellcolor{green!20}45.41\% & 0\% & 36.23 & 3.91 & \cellcolor{green!20}79.91\% & \cellcolor{green!20}85.04\% & 0.69\% \\ \hline \hline
   {\bf AdvT+\defensenameabbr-o(k)} & \textcolor{blue}{\bf 88.03} & \textcolor{blue}{\bf 97.81\%} & \textcolor{blue}{\bf 95.06\%}  & \textcolor{blue}{\bf 93.65\%}  & \textcolor{blue}{\bf 85.50\%} & \textcolor{blue}{\bf 94.14\%} & \textcolor{blue}{\bf 86.11\%} &
   \textcolor{blue}{\bf 96.0\%} & \textcolor{blue}{\bf 29.89} & \textcolor{blue}{\bf 2.53} &
   \textcolor{blue}{\bf 98.08\%} & \textcolor{blue}{\bf 97.42\%} & \textcolor{blue}{\bf 39.94\%} \\ \hline
  \end{tabular}
  }
    \begin{tablenotes}
        \scriptsize
        \item Note: The top-1 is highlighted in {blue} excluding Standard. The results in {green} background indicate that the transformation worsens adversarial training.
      \end{tablenotes}
  \end{threeparttable}
   \label{tab:evaluate-defense-adaptive-cnn}
  \vspace*{-4mm}
  \end{table*}

%% file: adaptive-transformation-AdvT.tex
\section{Evaluation of Transformations on Adversarially Trained Model}\label{sec:advTadaptive-attack}
\subsection{Evaluation Setup}
As ivector-PLDA cannot be adversarially trained due to unsupervised learning,
we adversarially train AudioNet for the CSI-NE task using the datasets Spk$_{251}$-train and Spk$_{251}$-test for training and testing, respectively.
The training uses a minibatch of size 128 for 300 epoches,
cross-entropy loss as the objective function,
and Adam~\cite{kingma2014adam} to optimize trainable parameters.
The naturally trained model is denoted by {Standard}.
For adversarial training, we use PGD with 10 steps (i.e., PGD-10) to generate adversarial examples.
The model is denoted by {Vanilla-AdvT}.

For each chosen transformation {X},
we implement it as a proper layer
in AudioNet. Note that this layer does not involve any trainable parameter, similar to the ReLU activation layer~\cite{eckle2019comparison}.
The resulting network is adversarially trained the same as above,
except that BPDA is leveraged for training the network with non-differentiable transformations
and EOT with $R$=10 is leveraged for training the network with randomized transformations.
The resulting model is denoted by {AdvT+X}.
We do not consider speech compressions, LPF and BPF,
as BPDA is not effective for estimating the gradients of speech compressions,
and the accuracy of the resulting model with LPF/BPF
is extreme low on both training dataset (i.e., 24.10\%/23.65\%) and testing dataset (i.e., 2.04\%/2.25\%).

The adaptive attacks are derived from 
FGSM, PGD-10, PGD-100, CW$\mathbf{_\infty}$-10, CW$\mathbf{_\infty}$-100, CW$_2$-1, FAKEBOB, SirenAttack, and Kenansville,
armed with EOT ($R$=50) and BPDA to evade
randomized and non-differentiable transformations. 
To improve the attack capability of FAKEBOB, 
we increase the parameter samples\_per\_draw $m$ to 300, 
allowing more precise gradient estimation at the cost of increased attack overhead.
Since adversarially trained models tend to yield smaller loss than naturally trained one,
we increase the initial trade-off constant $c$ of CW$_2$ attack from $0.001$ to $0.1$ when attacking Vanilla-AdvT and AdvT+X.
This helps finding adversarial examples with better imperceptibility according to our experiments.

\subsection{Results}
The results are reported in \tablename~\ref{tab:evaluate-defense-adaptive-cnn}.
We observe that the sole adversarial training
(i.e., {Vanilla-AdvT})
is effective for defeating adversarial examples compared over
{Standard} except for Kenansville,
at the cost of slightly sacrificing accuracy on benign examples (i.e., $A_b$ reduces from 99.06\% to 95.67\%). 
Adversarial training either significantly improves the accuracy by more than 53\% on the adversarial examples crafted by $L_\infty$ attacks,
or amplifies the distortions of the adversarial examples crafted by CW$_2$-1 (the SNR of Vanilla-AdvT is 18 dB smaller than that of Standard).
However, adversarial training does not improve the model accuracy on the adversarial examples crafted by Kenansville.
This is not surprising since Kenansville is a signal processing-based attack
while the adversarial examples used for adversarial training is generated by the optimization-based attack PGD-10.
We also tried to improve the model robustness against Kenansville by incorporating  Kenansville in adversarial training,
but the result is not promising (cf. Section~\ref{sec:discuss-findings} for discussion).

While sole adversarial training is often effective compared over 
{Standard},
the combination of adversarial training with a transformation, highlighted in {green} color in \tablename~\ref{tab:evaluate-defense-adaptive-cnn}, does not necessarily bring the best of both worlds,
which also exists in image domain~\cite{tramer2020adaptive}.

Interestingly, we found that adversarial training combined with \defensenameabbr-o(k), i.e., AdvT+\defensenameabbr-o(k), is very effective,
achieving higher accuracy on both the adversarial and benign examples compared with {Vanilla-AdvT}.
This improvement is brought by the randomness of \defensenameabbr.
In fact, during the training of AdvT+\defensenameabbr-o(k), the training data are randomly transformed by \defensenameabbr,
which enhances the quantity and diversity of the training data, similar to data augmentation in the image domain~\cite{data-aug-advt-effective-image}.
Consequently, the distribution mimicked by the training dataset $\{(x_i,y_i)\}_{i=1}^{B}$
becomes closer to the underlying data distribution $\mathcal{D}$ (cf. Section~\ref{sec:review-defense-adver-train}),
on which AdvT+\defensenameabbr-o(k) encounters more diverse adversarial examples during training.
Thus, it becomes more robust than Vanilla-AdvT.

Compared to the other transformations, 
\defensenameabbr enjoys larger randomness space than AT (cf. Section~\ref{sec:ablation-study-eot-size})
and other deterministic transformations (without randomness), 
hence AdvT+\defensenameabbr-o(k) outperforms other AdvT+X.
 
\begin{figure}[t]
    \centering

    \subfigure[\#Steps=1]{
        \includegraphics[width=0.14\textwidth]{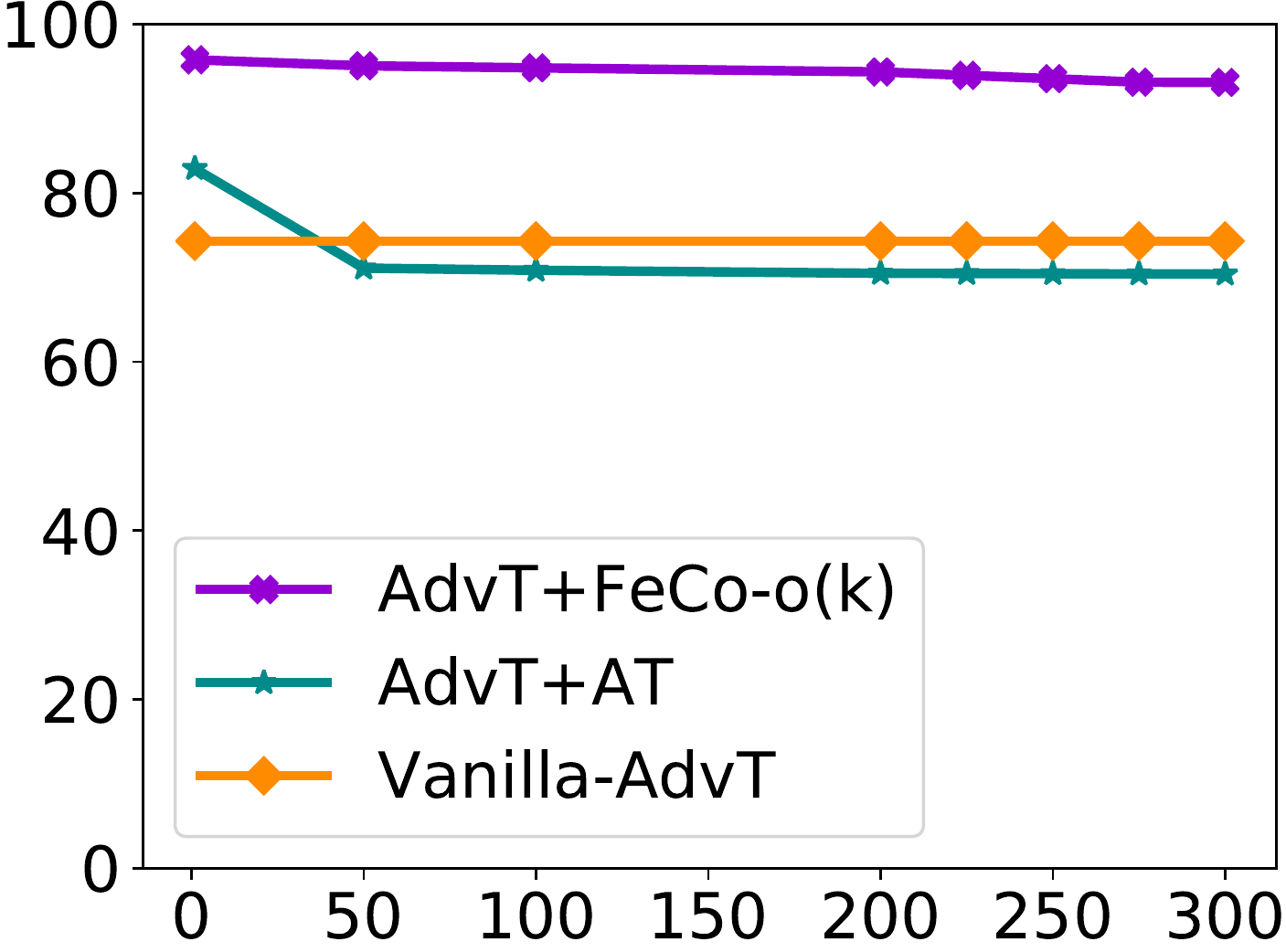}
    }
    \subfigure[\#Steps=100]{
        \includegraphics[width=0.14\textwidth]{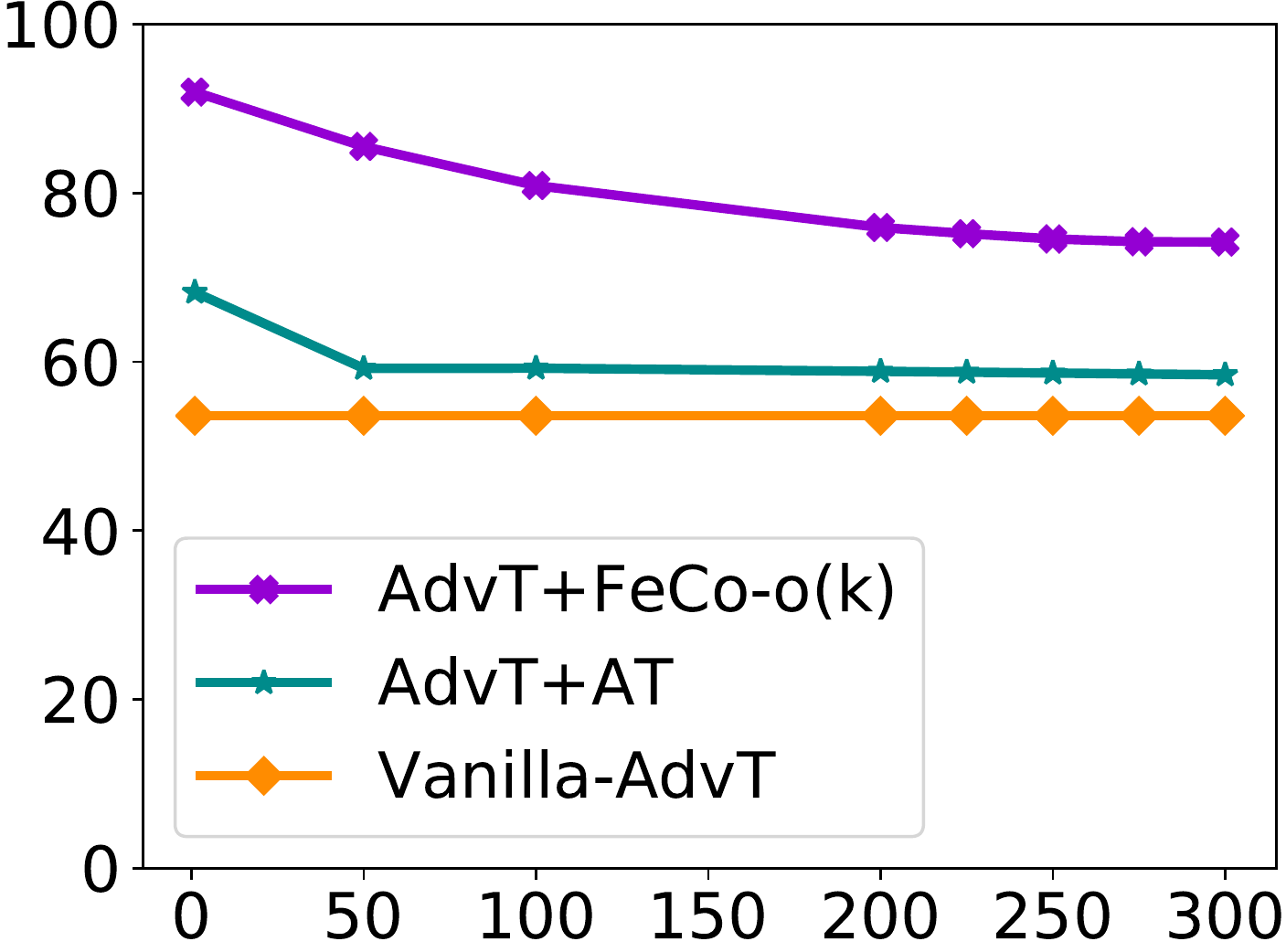}
    }
    \subfigure[\#Steps=200]{
        \includegraphics[width=0.14\textwidth]{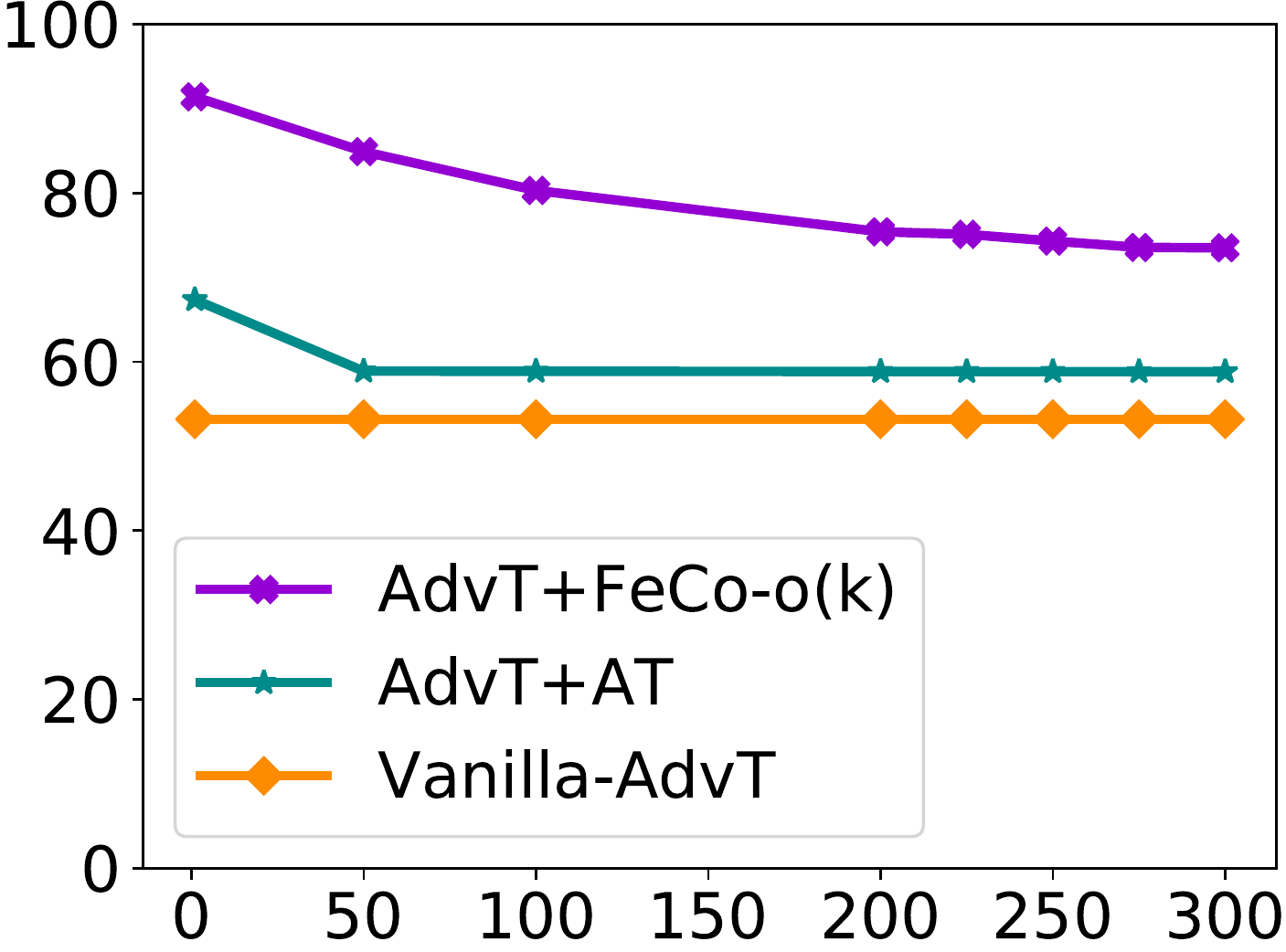}
    }
    \vspace*{-3mm} 
    \caption{x-axis is EOT\_size ($R$) and y-axis is $A_a$.}
    \label{fig:acc-vs-r}
    \vspace*{-3mm}
\end{figure}
 
\subsection{Attack Parameters Tuning}\label{sec:ablation-study-eot-size}
To thoroughly evaluate the robustness of AdvT+\defensenameabbr-o(k) against adaptive versions of the PGD and CW$_2$  attacks,
we further conduct a series of experiments by tuning the attack parameters,
including EOT\_size ($R$), number of steps (\#Steps), step\_size ($\alpha$), and confidence ($\kappa$).
Since these experiments on the entire Spk$_{251}$-test dataset require huge effort,
we randomly select 1,000 voices out of 2,887 voices in Spk$_{251}$-test from which  adversarial examples are crafted.

\noindent
{\bf EOT\_size ($R$)}.
We study the impact of EOT\_size ($R$) on the effectiveness of AdvT+\defensenameabbr-o(k).
We set PGD's step\_size $\alpha=\epsilon/5=0.0004$ (the same as previous experiments) and \#Steps$=$1, 100, 200.
For each number of steps (\#Steps), EOT\_size ($R$) ranges from 1 to 300.
The results are shown in \figurename~\ref{fig:acc-vs-r}.
We observe that with the increase of EOT\_size ($R$),  the accuracy of both AdvT+\defensenameabbr-o(k) and AdvT+AT decreases.
This is because larger EOT\_size ($R$) allows EOT to more accurately approximate the distributions of randomized transformations,
enabling the PGD attack to obtain more reliable gradient
and thus more stable search direction for adversarial examples.
However, when $R\geq 275$
(resp. $R\geq 50$), further increasing $R$ has negligible effect on AdvT+\defensenameabbr-o(k) (resp. AdvT+AT), i.e., the accuracy becomes stable.
Note that AdvT+\defensenameabbr-o(k) converges at a larger EOT\_size ($R$) than AdvT+AT,
i.e., 275 vs. 50.
Recall that EOT is exploited to overcome the randomness of a transformation.
Thus, EOT\_size ($R$) is a reasonable metric for quantifying the degree of randomness that a transformation introduces.
Accordingly,
we can conclude that \defensenameabbr introduces larger randomness than AT.

\begin{figure}[t]
    \centering

    \subfigure[r=1]{
        \includegraphics[width=0.14\textwidth]{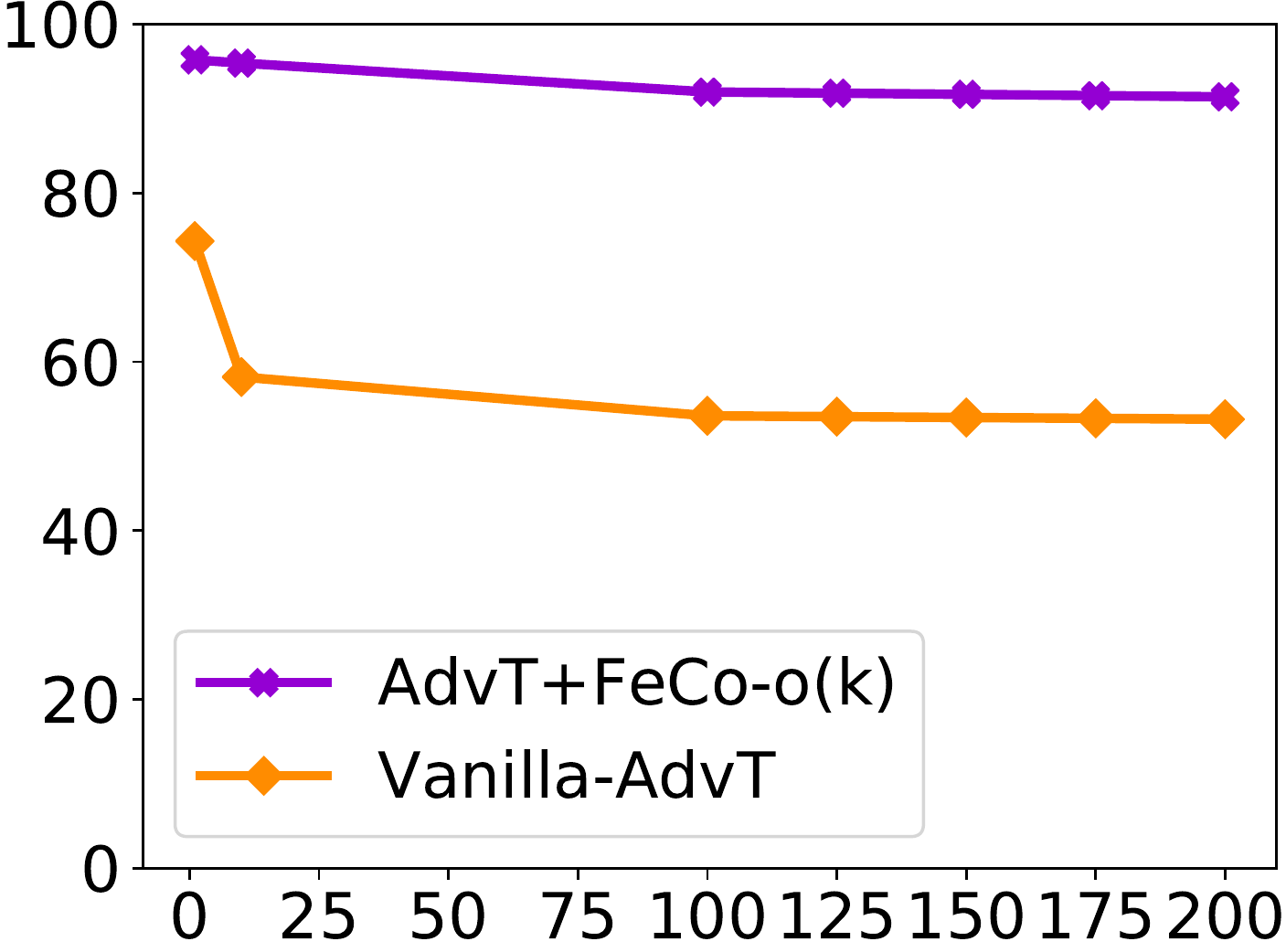}
    }
    \subfigure[r=100]{
        \includegraphics[width=0.14\textwidth]{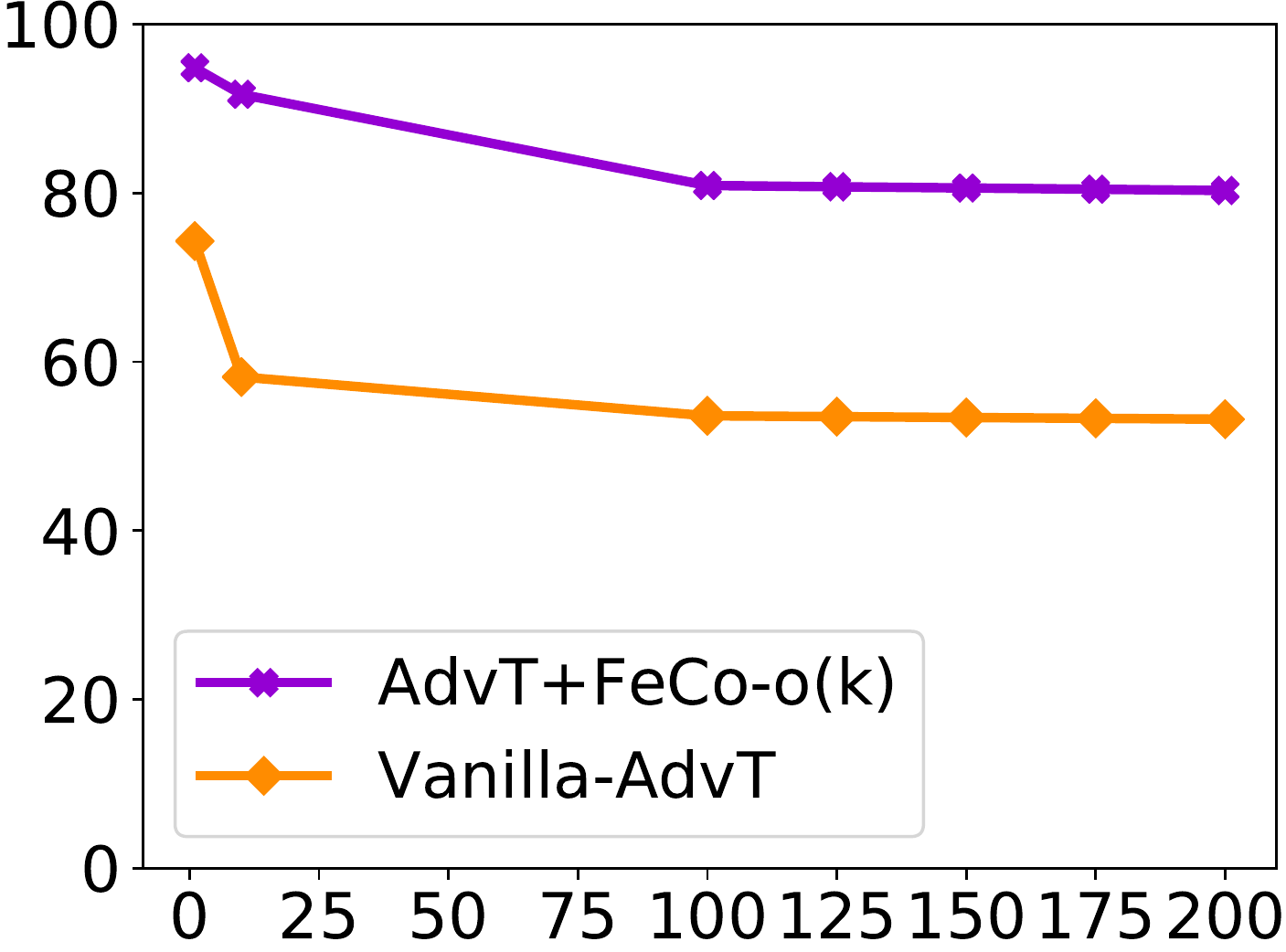}
    }
    \subfigure[r=300]{
        \includegraphics[width=0.14\textwidth]{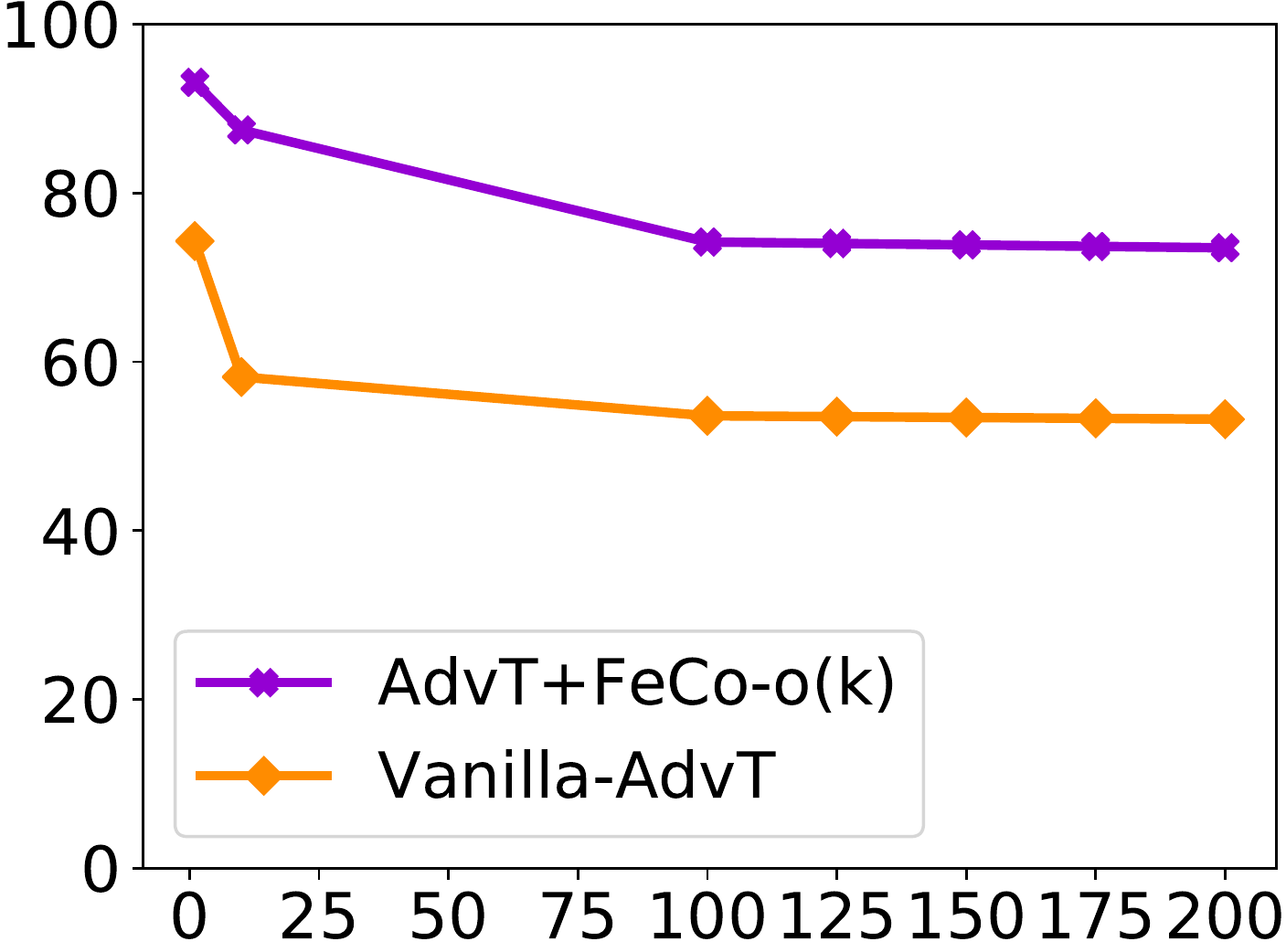}
    }
    \vspace*{-3mm}
    \caption{x-axis is the number of steps (\#Steps), and y-axis is $A_a$, where \#Steps=1 is indeed the FGSM attack.}
    \label{fig:acc-vs-steps}
    \vspace*{-4mm}
\end{figure}

\noindent
{\bf Number of steps (\#Steps)}.
We study the impact of the number of steps (\#Steps) in the PGD attack on the effectiveness of AdvT+\defensenameabbr-o(k).
We set PGD's step\_size $\alpha=\epsilon/5=0.0004$ and EOT\_size $R=1,100,300$.
The number of steps (\#Steps) ranges from 1 to 200 for every EOT\_size ($R$).
The results are shown in \figurename~\ref{fig:acc-vs-steps}.
We observe that the accuracy of AdvT+\defensenameabbr-o(k) decreases gradually when \#Steps increase from 1 to 100.
This is not surprising as increasing \#Steps improves the strength of adversarial examples (cf. \figurename~\ref{fig:loss-step-kappa} in Appendix).
However, when \#Steps$>$100, the accuracy of AdvT+\defensenameabbr-o(k)
remains almost unchanged
with the increase of the number of steps (\#Steps). 

\begin{figure}[t]
    \centering
    \subfigure[x-axis is step\_size ($\alpha$) and y-axis is $A_a$.]{ 
        \includegraphics[width=0.19\textwidth]{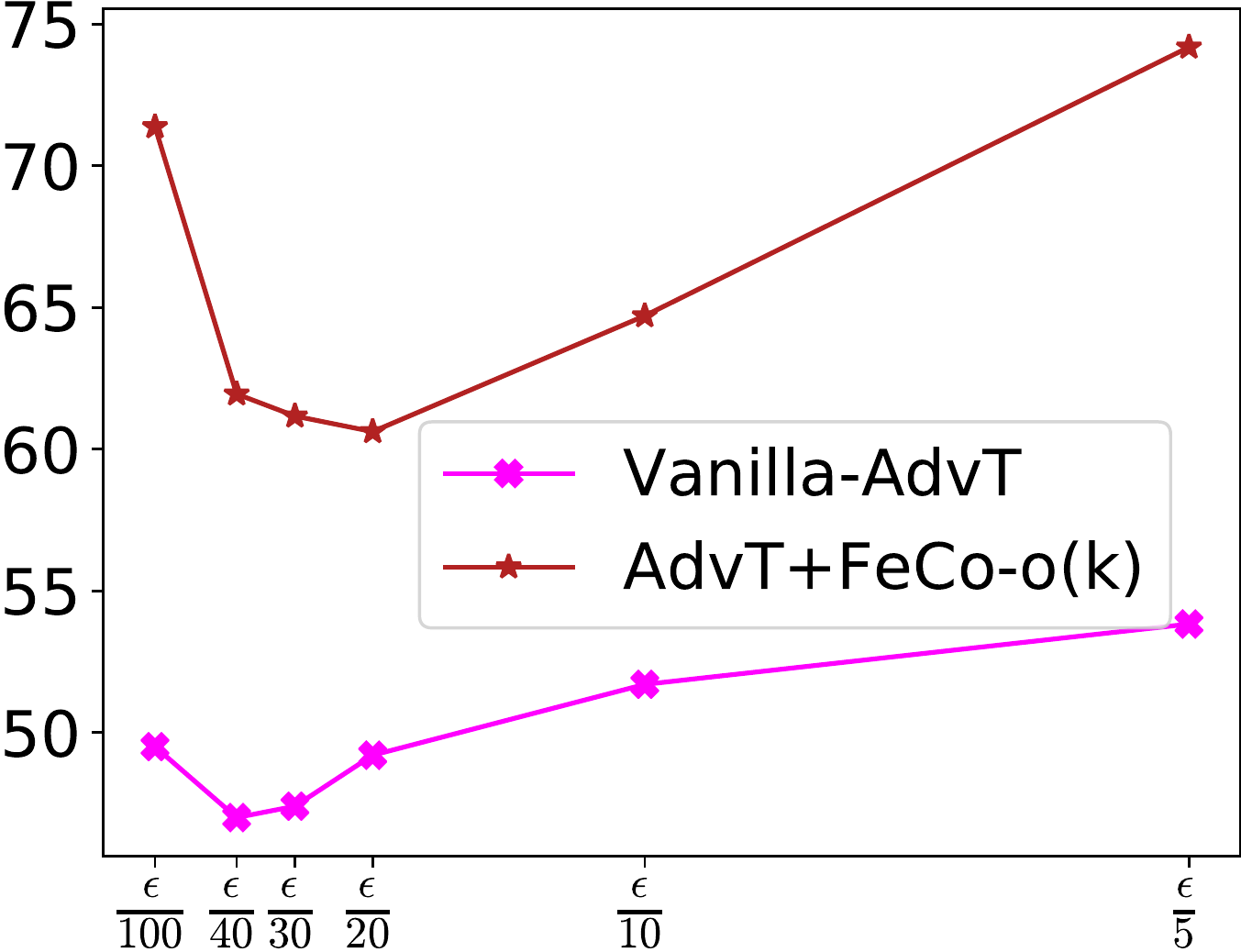}
        \label{fig:acc-vs-step-size}
    }
    \subfigure[Accuracy and distortion of adversarial voices with the increase of $\kappa$ for CW$_2$.]{ 
        \includegraphics[width=0.23\textwidth]{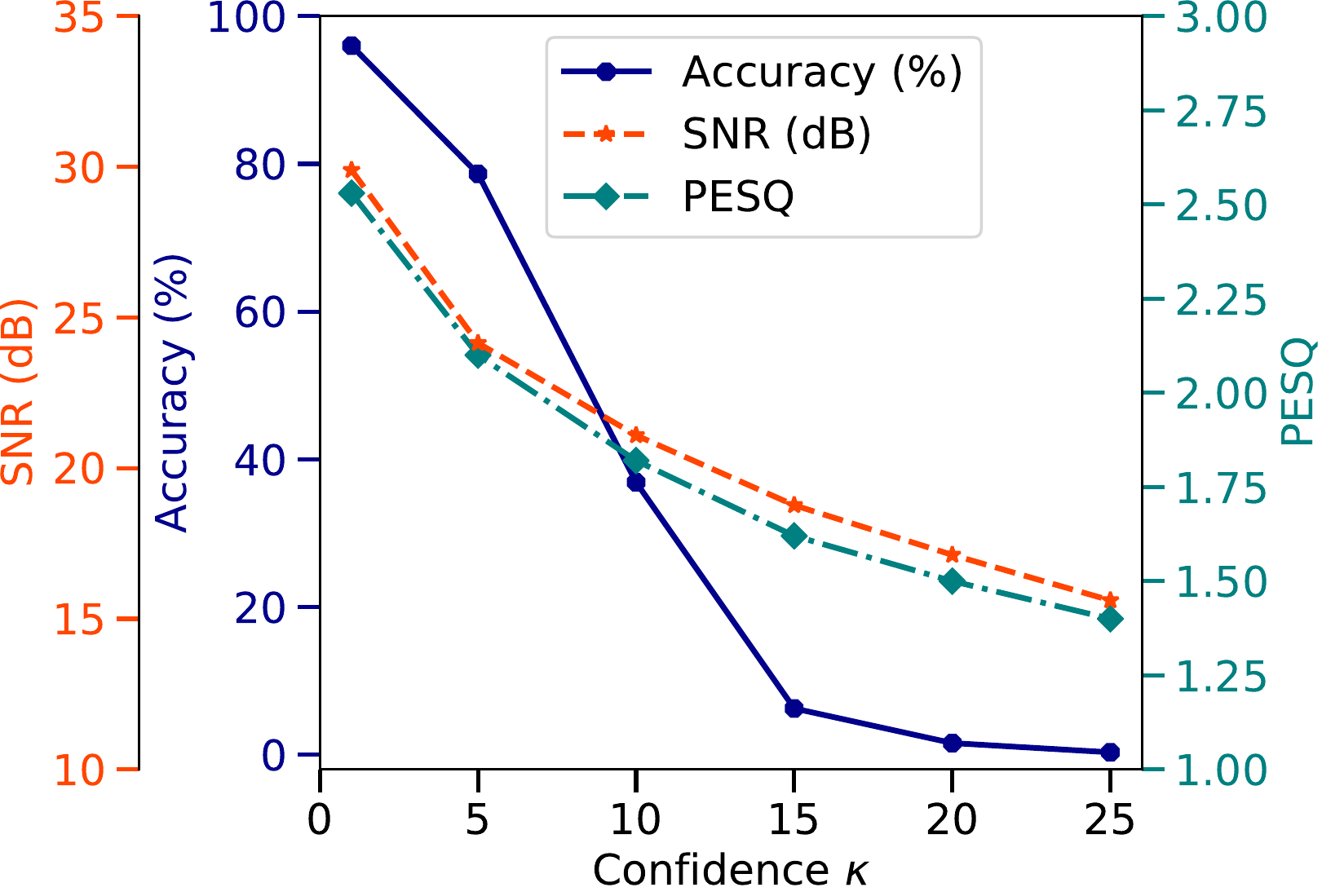}
        \label{fig:AdvT-FF-CW2-kappa}
    }
    \vspace*{-2mm} 
    \caption{Tuning the step\_size ($\alpha$) and confidence $(\kappa)$.}
    \vspace*{-4mm}
\end{figure}

\noindent
{\bf Step\_size ($\alpha$)}.  
Based on the above results, we fix \#Steps=100 and 
EOT\_size $R=275$
when studying the impact of step\_size ($\alpha$) on the effectiveness of AdvT+\defensenameabbr-o(k) 
by setting $\alpha=\epsilon/100$, $\epsilon/40$, $\epsilon/30$, $\epsilon/20$, $\epsilon/10$, $\epsilon/5$.
The results are shown in \figurename~\ref{fig:acc-vs-step-size}.
We found that decreasing step\_size
reduces the accuracy of both Vanilla-AdvT and AdvT+\defensenameabbr-o(k).
We conjecture that 
the PGD attack with small step\_size is less likely to oscillate across different directions,
thus can search for adversarial examples in a more stable way.
However, when $\alpha\leq \epsilon/20$ (resp. $\alpha\leq \epsilon/40$),  decreasing step\_size ($\alpha$)
reduces the attack success rate on AdvT+\defensenameabbr-o(k) (resp. Vanilla-AdvT).
 
From the above three stuides, we can observe that
the accuracy of AdvT+\defensenameabbr-o(k) plateaus at 60.62\% 
with $R=275$,
\#Steps=100, and $\alpha=\epsilon/20$,
while the accuracy of Vanilla-AdvT plateaus at 47.0\% with $R=1$, \#Steps=100, and $\alpha=\epsilon/40$.
Thus, AdvT+\defensenameabbr-o(k) achieves 13.62\% higher accuracy than Vanilla-AdvT.
Furthermore, the attack has to query the AdvT+\defensenameabbr-o(k) model 
$275\times 100=27,500$ times,
while it only has to query the Vanilla-AdvT model $1\times 100=100$ times.
This indicates that \defensenameabbr-o(k) significantly improves the attack cost by two orders of magnitude.

\noindent
{\bf Confidence ($\kappa$)}.  
We launch the CW$_2$ attack by setting the parameter $\kappa=1,5, 10, 15, 20, 25$, 
where the larger $\kappa$, the stronger the attack. 
As shown in \figurename~\ref{fig:AdvT-FF-CW2-kappa}, though the accuracy on the adversarial examples decreases with the increase of
$\kappa$, 
the distortion also increases. 
For instance, when $\kappa=25$, the attack success rate is nearly 100\%, but the SNR (resp. PESQ) is
15.61 dB (resp. 1.40), 3 times (resp. 2 times) smaller
than that of 
{Standard}, 
indicating that the adversarial examples become much less imperceptible. This demonstrates the
effectiveness of AdvT+\defensenameabbr-o(k) against powerful attacks. 
\begin{tcolorbox}[size=title,breakable,arc=1mm, boxsep=0.4mm, left = 1pt, right = 1pt, top = 1pt, bottom = 1pt]
\textbf{Findings 10.} 
    Among the adversarially trained models combined with transformations,
    AdvT+\defensenameabbr-o(k) is the unique one that is effective against all the adaptive attacks.
    Compared with Vanilla-AdvT, it improves the accuracy on both benign examples and adversarial examples
    against $L_\infty$, $L_2$ and signal processing-based adaptive attacks,
    largely increases the attack cost of the PGD based adaptive attack,
    and significantly worsens the imperceptibility of adversarial examples crafted by the CW$_2$ based adaptive attack. 
    \end{tcolorbox}

%% file: discussion.tex
\section{Discussion}
We discuss some key findings and the limitations of our study,
interspersed with possible future works motivated by them. 
\subsection{Discussion of Findings}\label{sec:discuss-findings}
\noindent
{\bf Combination of different transformations}.
According to Findings 1, \tablename~\ref{tab:evaluate-defense-non-adaptive}, and \tablename~\ref{tab:evaluate-defense-adaptive-iv},
the effectiveness of transformations varies with attacks.
Moreover,
different types of transformations operate on different domains (time vs. frequency), different levels (waveform vs. acoustic feature) and
own different properties (differentiable vs. non-differentiable, deterministic vs. randomized).
Therefore, it is interesting to study if the combinations of transformations (e.g., AT and \defensenameabbr)
could improve adversarial robustness.

\noindent
{\bf Attacks against speech compression defenses}. 
Findings 6 and Findings 7 reveal that BPDA, FAKEBOB and SirenAttack
are hard to circumvent non-differentiable CBR and VBR speech compression, respectively.
BPDA cannot succeed since replacing speech compression with the identity function in the backward pass is not precise enough
(cf. \figurename~\ref{fig:BPDA-differ} in Appendix).
Diving deeper into speech compression, we found that its bit allocation would assign unequal number of bits to voice sample points,
according to their contribution to human perception of the voices.
Consequently, the transformed voice by speech compression does not align with the original one in time axis,
making speech compression far from the identity function.
To improve BPDA, we may utilize time sequence alignment techniques, e.g., dynamic time warping~\cite{DTW},
to align the original and transformed voices to make speech compression close to the identity function as much as possible,
or adopt more accurate approximation functions than the identity function.
The failure of FAKEBOB and SireAttack may be attributed to the large non-smoothness introduced
by the variable bit rate of speech compression.
The smoothness assumption of NES and PSO does not hold anymore~\cite{NATTACK},
making the estimated gradient of NES and the search direction of PSO not reliable and informative enough for gradient descent.
$\mathcal{N}$ATTACK~\cite{NATTACK}, which will not be impeded by the non-smoothness of models,
and gradient-free decision-only attacks from
the image domain, e.g., boundary attack~\cite{boundary-attack} and evolutionary attack~\cite{CMA-ES},
may be good alternatives to evade speech compression.

\noindent
{\bf Black-box attacks against randomized defenses}.
According to Findings 8,
all the black-box attacks (FAKEBOB, SirenAttack, and Kenansville) have limited attack success rate on the models with randomized transformations (e.g., AT and \defensenameabbr).
This is probably because NES of FAKEBOB becomes ineffective for estimating gradients,
PSO of SirenAttack becomes unstable for searching better particle locations,
and Kenansville gets misled in updating the attack factor, in presence of randomness.
To bypass such randomized transformations, one may use $\mathcal{N}$ATTACK 
which is effective in breaking the randomized defenses in the image domain.
Adapting $\mathcal{N}$ATTACK to speaker recognition is an interesting future work.

\noindent
{\bf Robust training against Kenansville}. 
The results in \tablename~\ref{tab:evaluate-defense-adaptive-cnn} show that adversarial training
fails to improve robustness against Kenansville. 
The reason is that the adversarial training uses the optimization-based attack PGD,
while Kenansville is a signal processing-based attack.
We also tried to incorporate Kenansville into adversarial training but found that it
not only fails to increase adversarial robustness against Kenansville,
but also significantly degrades accuracy on benign voices.
The former may be due to low-confidence of adversarial examples crafted
by Kenansville that are not suitable for solving the inner maximization problem in adversarial training (cf. in Section~\ref{sec:review-defense-adver-train})
while the latter may be due to large distortion
introduced by Kenansville. Details refer to \tablename~\ref{tab:imper-non-adaptive-atatck} in Appendix.
Since adversarial training does not work well for Kenansville, we may turn to other robustness training techniques,
e.g., Lipschitz regularization~\cite{jati2021adversarial}. We leave this as future work.

\subsection{Discussion of Limitations}

\noindent
{\bf Suitability of audio imperceptibility metrics}.
We use $L_\infty$ and $L_2$ norms to quantify the perturbation magnitude in adversarial example generation,
and adopt SNR and PESQ to measure the imperceptibility of crafted adversarial voices.
These metrics have been widely adopted in the literature~\cite{chen2019real,du2020sirenattack,jati2021adversarial,LiZJXZWM020,shamsabadi2021foolhd,WangGX20,zhang2021attack}
and in general, can consistently reflect the degree of distortions according to our experimental results.
Moreover, PESQ is an objective perceptual measure simulating the human auditory system~\cite{xiang2017digital}.
However, it remains unknown to what extent do these metrics correlate with human hearing perception.
In the image domain, the proximity of two images measured by $L_p$ norm is neither necessary nor sufficient for them to be visually indistinguishable by humans~\cite{suitibility-lp}.
Therefore,  it is worthy to explore in future the sufficiency and necessity of these metrics in quantifying
the audio perceptual similarity.
 
\noindent
{\bf Securing commercial SRSs}.
We did not directly target commercial SRSs, although they are also vulnerable to black-box attacks \cite{chen2019real,Occam}.
The reason is that it is more important to consider the most powerful adversaries when evaluating defenses, 
while the adversaries are not able to mount white-box attacks without having access to the internal structures of commercial SRSs.
Instead, we directly evaluate defenses against the black-box attacks FAKEBOB~\cite{chen2019real}, SirenAttack~\cite{du2020sirenattack} and Kenansville~\cite{ASGBWYST19}
which could be used to attack commercial SRSs and FAKEBOB is able to fool commercial SRSs.
Investigating and evaluating if our findings are applicable to commercial SRSs is left for future work.

\noindent
{\bf Detection of adversarial voices}.
While we focus on adversarial training and transformation based defenses against adversarial attacks, 
effective transformations could be leveraged to detect adversarial voices
by comparing the degree-of-change of benign and adversarial voices before and after transformations~\cite{Xu0Q18,MengC17}.
This is reasonable as benign voices are generally more robust~\cite{zhao2021attack},
their results are less likely to change after transformations, 
which is validated by our Findings 11 in Appendix~\ref{sec:non-adaptive-more-findings}.

\noindent
{\bf Defending against over-the-air attacks}.
Our evaluation focuses on digital attacks
where adversarial voices are directly fed to the SRS via exposed API, as
it is more important to evaluate defenses against powerful adversaries while over-the-air attack will be compromised by various sources of distortions~\cite{AS2T}.
We emphasize that input transformations are also applicable to over-the-air attacks
where the adversarial voices are played and recorded by hardware and transmitted in the air.
Transformations can back-up liveness detection~\cite{liveness-detection-wisec,liveness-detection-usenix} when liveness detection has false negatives,
where liveness detection detects over-the-air attacks by
exploiting the different characteristics of the voices generated by human vocal tract and electronic loudspeaker.
Evaluating the effectiveness of these transformations in defending against over-the-air attacks is left for future work.
 
\noindent
{\bf Input transformations against other attacks}.
This work focuses on defending against adversarial attacks.
There are other attacks against SRSs which have different attack goals and scenarios from adversarial attack.
Thus, it is interesting to investigate whether input transformations can defend against those attacks.
As a first attempt, we carry out a preliminary evaluation against hidden voice attack~\cite{AbdullahGPTBW19} and speech synthesis attack~\cite{wenger2021hello} (cf. Appendix~\ref{sec:hidden-spoofing}).
We found that input transformations are also effective in mitigating these two attacks
and speech synthesis attack is more difficult to defeat than the other two attacks.
More thorough evaluations against more other attacks are needed in the future.

%% file: related-work.tex
\section{Related Work}\label{sec:relate}
Adversarial attacks and defenses in the speech and speaker recognition domains recently have attracted intensive attention.
Though both of them share a similar feature extraction pipeline, they perform different tasks
and speaker recognition owns unique enrollment phase and decision making mechanism~\cite{chen2021sok,chen2019real}.
Thus, in this section, we do not discuss adversarial attacks and defenses that focus on speech recognition~\cite{Carlini018,
qin2019imperceptible,yuan2018commandersong,psychoacoustic-hiding-attack,247642,LiW00020,TaoriKCV19,dolphin-attack,yang2018characterizing,Dompteur} (cf.~\cite{AWBPT20,chen2021sok} for survey).
There are other voice attacks in the speaker recognition domain,
such as hidden voice attacks~\cite{AbdullahGPTBW19} and spoofing attacks~\cite{hautamaki2013vectors,shirvanian2019quantifying,mukhopadhyay2015all,shirvanian2014wiretapping,shirvanian2018short,wenger2021hello}.
Though these attacks have different attack goals and scenarios from adversarial attacks~\cite{chen2019real},
our preliminary evaluation shows that it is possible to mitigate hidden voice attack~\cite{AbdullahGPTBW19} and speech synthesis attack~\cite{wenger2021hello}
via input transformations.
Below, we discuss adversarial attacks and defenses in the speaker recognition domain.

\noindent
{\bf Adversarial attacks}.
Existing white-box attacks in the speaker recognition domain are derived from the attacks in the image recognition domain.
The FGSM method was adopted to attack the CSI-NE task~\cite{DBLP:journals/corr/abs-1711-03280}
and the SV task~\cite{abs-1801-03339,li2020adversarial}.
Zhang at al. used PGD to attack the CSI-NE task~\cite{zhang2021attack}.
Jati at al. attacked the CSI-NE task by leveraging FGSM, PGD, CW$\mathbf{_\infty}$ and CW$_2$~\cite{jati2021adversarial} methods.
However, these attacks have not been thoroughly evaluated on the systems with various defenses
and it is difficult to conclude which one is better due to inconsistent benchmarks (e.g., models and datasets).
We consider all these white-box attacks and adaptive variants thereof in this work.
Though our main goal is to investigate and evaluate transformation
and adversarial training based defenses, our results also provide
a fair comparison of these attacks under the same settings when various defenses are deployed.

There are also some specific white-box attacks, aiming at
crafting universal perturbations~\cite{LiZJXZWM020,xie2020enabling}
or improving the imperceptibility of adversarial voices~\cite{WangGX20, shamsabadi2021foolhd},
yet these works did not consider any defense.
We do not incorporate these attacks into our study,
as all of them are not publicly available and non-trivial to reproduce.

FAKEBOB~\cite{chen2019real}, SirenAttack~\cite{du2020sirenattack}, Kenansville~\cite{ASGBWYST19}, and Occam~\cite{Occam} 
are four black-box adversarial attacks targeting SRSs, 
where FAKEBOB, SirenAttack, and Occam are optimization-based attacks,
and Kenansville is a signal processing-based attack. 
All of them, except for Occam which is not publicly available and non-trivial to reproduce,
have been used to evaluate defenses in this work.

\noindent
{\bf Adversarial defenses: mitigation and detection}.
Robust training is one way to mitigate adversarial examples.
\cite{du2020sirenattack,jati2021adversarial} showed that adversarial training can enhance the robustness of models.
\cite{jati2021adversarial} also proposed another  technique
which adds a regularization term using Lipschitz smoothness to the loss function for model training.
This technique performs better than FGSM based adversarial training, but worse
than PGD based adversarial training. This motivated us to evaluate PGD based adversarial training in this work.

The transformations (QT, MS and DS) and (DS and AS)
have been evaluated against FAKEBOB and SirenAttack respectively.
But, they were neither combined with adversarial training nor thoroughly evaluated under various attacks.
Our evaluation shows that these transformations are \emph{not} effective against adaptive attacks
and \emph{cannot} improve the adversarial robustness of adversarially trained models.
Furthermore, we investigate and evaluate significantly more defenses against both non-adaptive and adaptive attacks.
We note that AT, Auto-Encoder~\cite{MengC17,zhao2021attack}, and GAN~\cite{Defend-GAN} have been evaluated against four white-black attacks in \cite{TIFS-defense}.
Compared to the transformations considered in this work, Auto-Encoder and GAN are data-dependent methods which require additional overhead for training from benign examples
to model the distribution of unperturbed voices, thus may exhibit different performance on difference datasets.
Although BPDA was used to solve the non-differentiability of GAN in \cite{TIFS-defense}, the randomness of AT was not properly addressed, leading to false sense of adversarial robustness.
Our findings show that AT becomes ineffective against adaptive attack armed with EOT to address the randomness.
Moreover, \cite{TIFS-defense} did not consider black-box attacks, while we did and found some useful related findings (Findings 6-8).

Detection is another way to defend against adversarial voices. 
\cite{XuLidetection} proposed to detect adversarial examples by training a CNN-based binary classifier, while \cite{pairing-weak-strong} checks the consistence of results of twin models. 
However, these approaches have not been evaluated against adaptive attacks and may be evaded by incorporating the detector into loss functions~\cite{CW17a}.
Another direction is liveness detection~\cite{liveness-detection-wisec,liveness-detection-usenix}
which detects malicious audios by exploiting the different characteristics of the voices
generated by human vocal tract and electronic loudspeaker.
Liveness detection is a promising approach for defeating physical adversarial attacks.
However, it is not suitable for API attacks
where adversarial voices are directly fed to the SRSs in the form of audio file via exposed API.

%% file: conclusion.tex
\section{Conclusion}
We have systematically investigated diverse transformations for mitigating adversarial voices in the speaker recognition domain,
including waveform-level transformations in both time-domain and frequency-domain,  speech compression, and feature-level transformations, 
and covering all the differentiable, non-differentiable,
deterministic, and randomized types.
We have thoroughly evaluated those transformations on both naturally trained and adversarially trained models against promising
white-box and black-box attacks, as well as carefully designed adaptive variants for circumventing
different types of transformations.
Our study revealed lots of interesting and useful findings for both researchers and practitioners.

Among all the transformations, we showed that our novel feature-level transformation  \defensenameabbr is rather effective against black-box attacks and
improves the robustness of adversarially trained models against both white-box and black-box adaptive attacks in terms of accuracy, attack cost, and distortion level.
This opens up a new research direction on transformations for mitigating adversarial examples.
We pointed out many possible future works in both adversarial attacks and defenses  in the speaker recognition domain, and
released our evaluation platform \platformname to foster  further research.

%% file: biography.tex
\begin{IEEEbiography}[{\includegraphics[width=1in,clip,keepaspectratio]{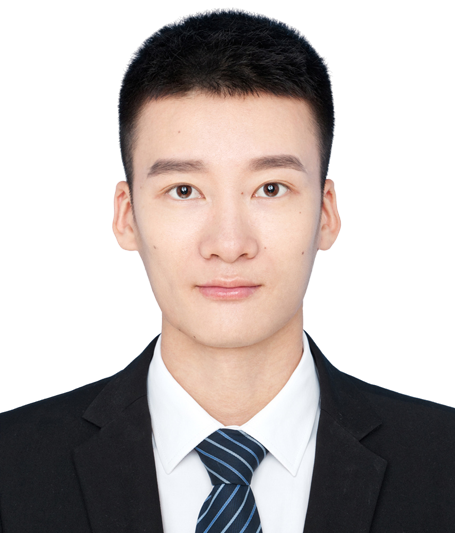}}]{Guangke Chen}
received his BEng degree from South China University of Technology, Guangzhou, China, in 2019.
He is currently pursuing the Ph.D. degree with ShanghaiTech University, advised by Dr. Song. 
His research interest lies in the area of multimedia and machine learning security and privacy.
He is currently doing research on the security issues of speaker and speech recognition systems.
More information is available at \url{http://guangkechen.site/}.
\end{IEEEbiography}

\begin{IEEEbiography}[{\includegraphics[width=1in,clip,keepaspectratio]{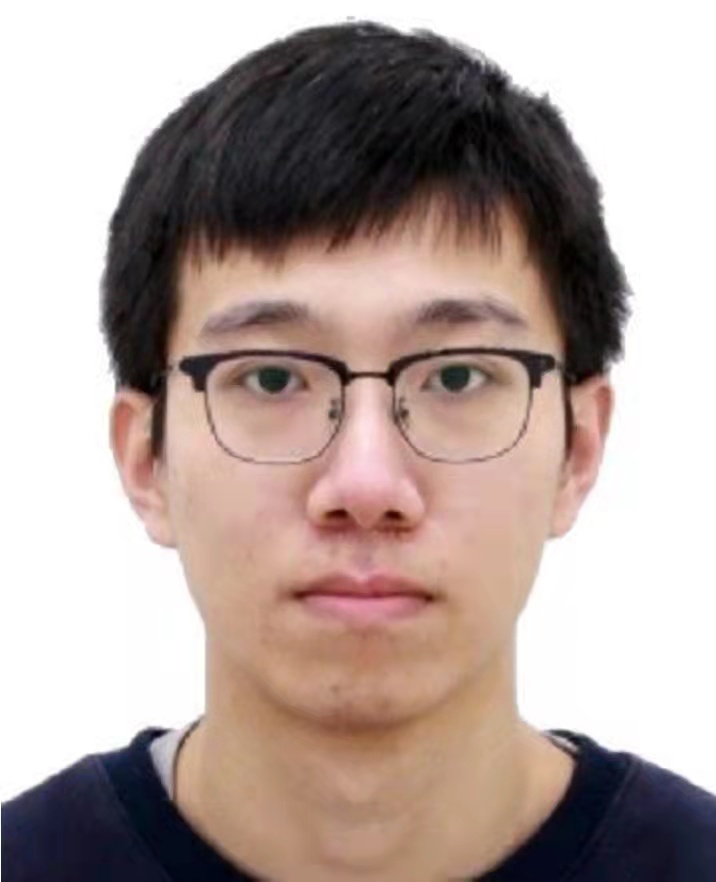}}]{Zhe Zhao}
received his B.S. degree from Ocean University of China, Tsingtao, China, in 2016.
From 2016 to 2018, he was a software engineer at Hewlett-Packard Company.
Now he is a Ph.D. student at School of Information Science and Technology, ShanghaiTech University.
His research interest lies in the area of software engineering and testing.
He is currently doing research in trusted artificial intelligence.
His supervisor is Dr. Song.
\end{IEEEbiography}

\begin{IEEEbiography}[{\includegraphics[width=1in,height=1.25in,clip,keepaspectratio]{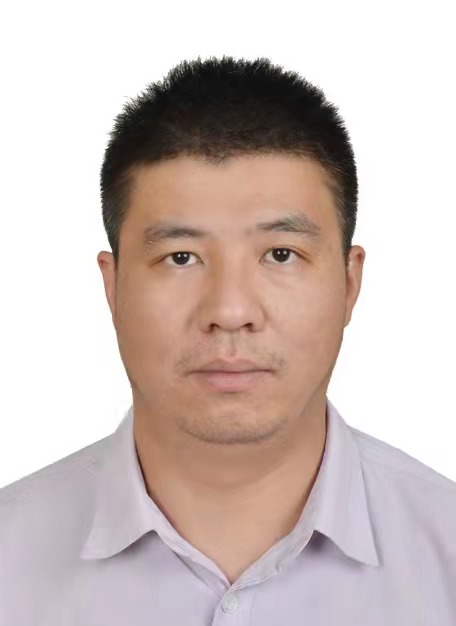}}]{Fu Song}
    received the B.S. degree from Ningbo
    University, Ningbo, China, in 2006, the M.S. degree from East China Normal University, Shanghai,
    China, in 2009, and the Ph.D. degree in computer science from University Paris-Diderot,
    Paris, France, in 2013.
    From 2013 to 2016, he was a Lecturer and Associate Research Professor at East China Normal University.
    From August 2016 to July 2021, he is an Assistant Professor with ShanghaiTech University, Shanghai, China.
    Since July 2021, he is an Associate Professor with ShanghaiTech University.
    His research interests include formal methods and computer/AI security. 
    \end{IEEEbiography}

\begin{IEEEbiography}[{\includegraphics[width=1.12in,clip,keepaspectratio]{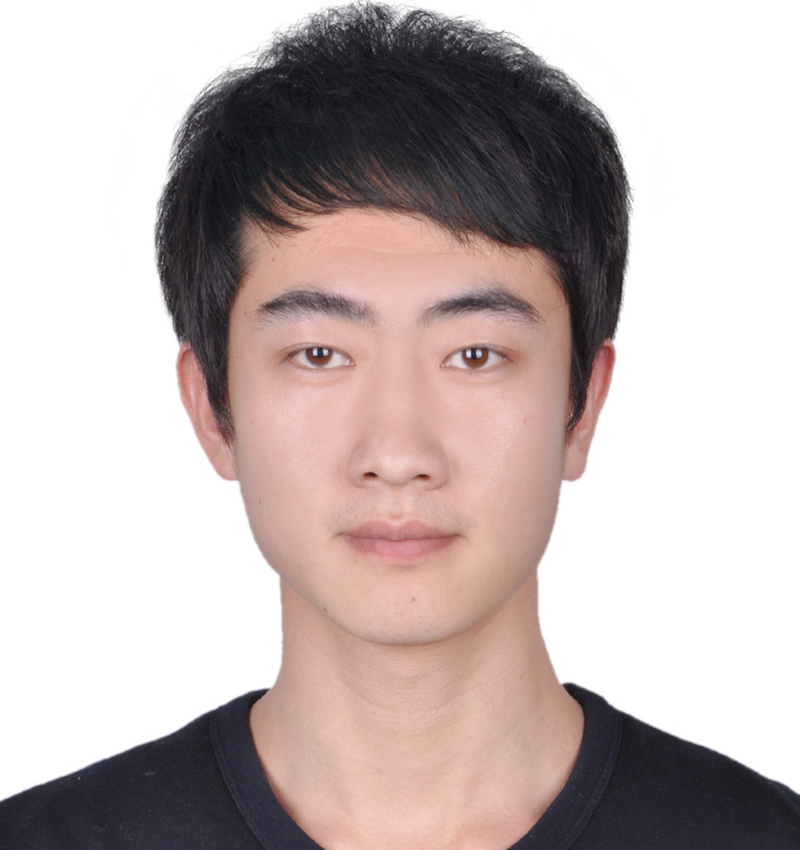}}]{Sen Chen} (Member, IEEE) is an Associate Professor at Tianjin University, China. Before that, he was a Research Assistant Professor at Nanyang Technological University (NTU), Singapore, and a Research Assistant of NTU from 2016 to 2019 and a Research Fellow from 2019-2020. He received his Ph.D. degree in Computer Science  East China Normal University, China, in 2019. His research focuses on Security and Software Engineering. 
More information is available on {\url{https://sen-chen.github.io/}.}
\end{IEEEbiography}

\begin{IEEEbiography}[{\includegraphics[width=1.12in,clip,keepaspectratio]{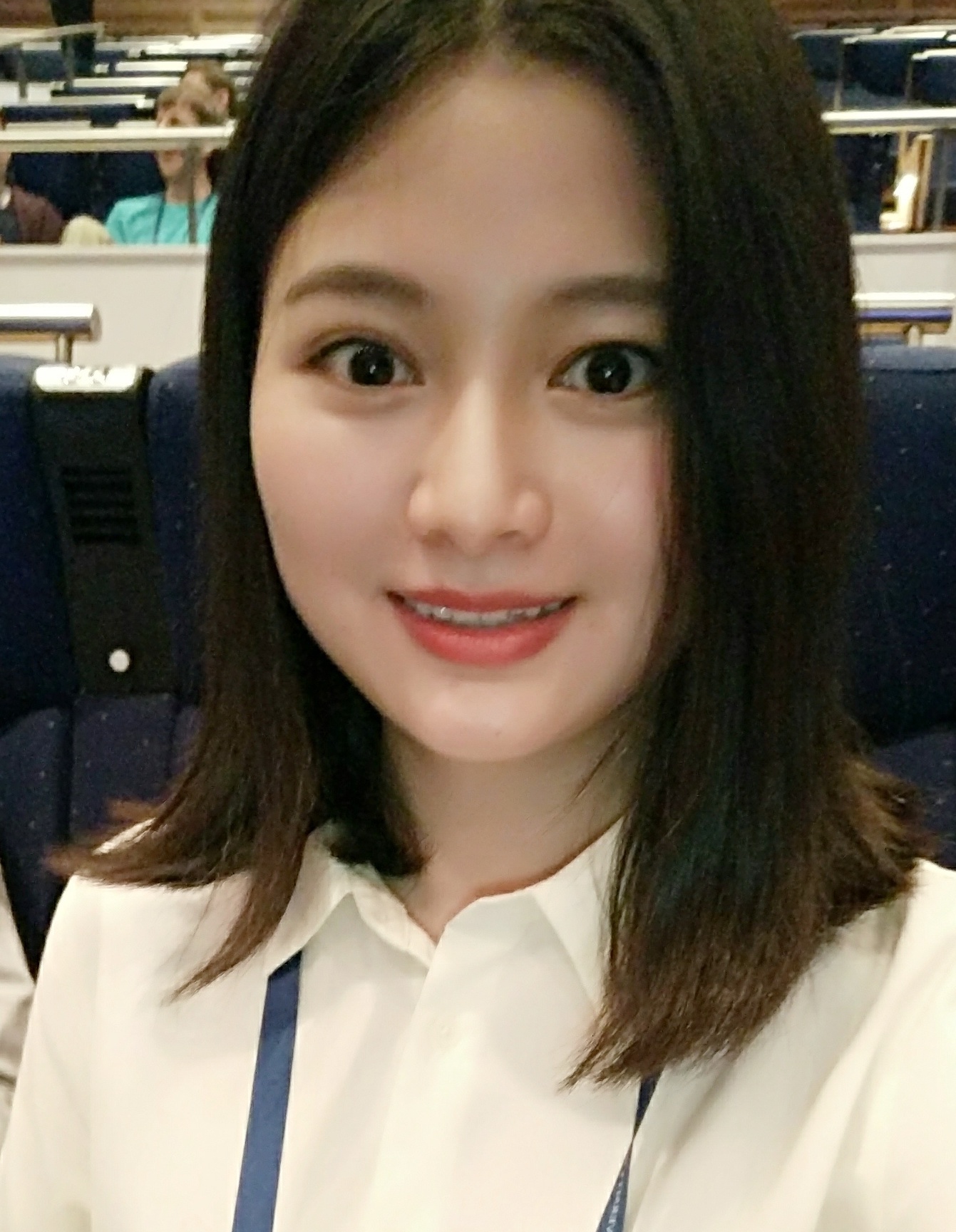}}]{Lingling Fan} is an Associate Professor at Nankai University, China. She received her Ph.D and BEng degrees in computer science from East China Normal University, Shanghai, China in June 2019 and June 2014, respectively. In 2017, she joined Nanyang Technological University (NTU), Singapore as a Research Assistant and then had been as a Research Fellow of NTU since 2019. Her research focuses on program analysis and testing, software security. She got an ACM SIGSOFT Distinguished Paper Award at ICSE 2018.
\end{IEEEbiography}


\begin{IEEEbiography}[{\includegraphics[width=1.12in,clip,keepaspectratio]{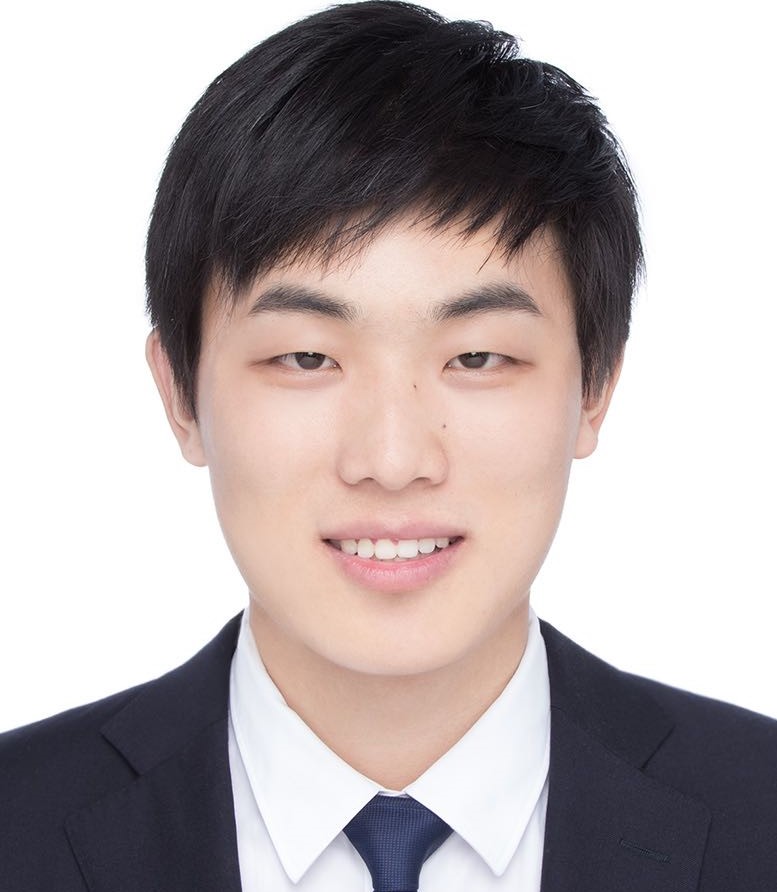}}]{Feng Wang} 
    received the bachelor's degree in network engineering from the Nanjing University of Post and Telecommunication, Nanjing, China, in 2016.  
    He obtained master's degree from University of Chinese Academy of Sciences, under the supervision of Prof. Fu Song from ShanghaiTech University in 2019. 
    He is now working at Security Countermeasure Technology Department of Ant Group.
\end{IEEEbiography}

\begin{IEEEbiography}[{\includegraphics[width=1.12in,clip,keepaspectratio]{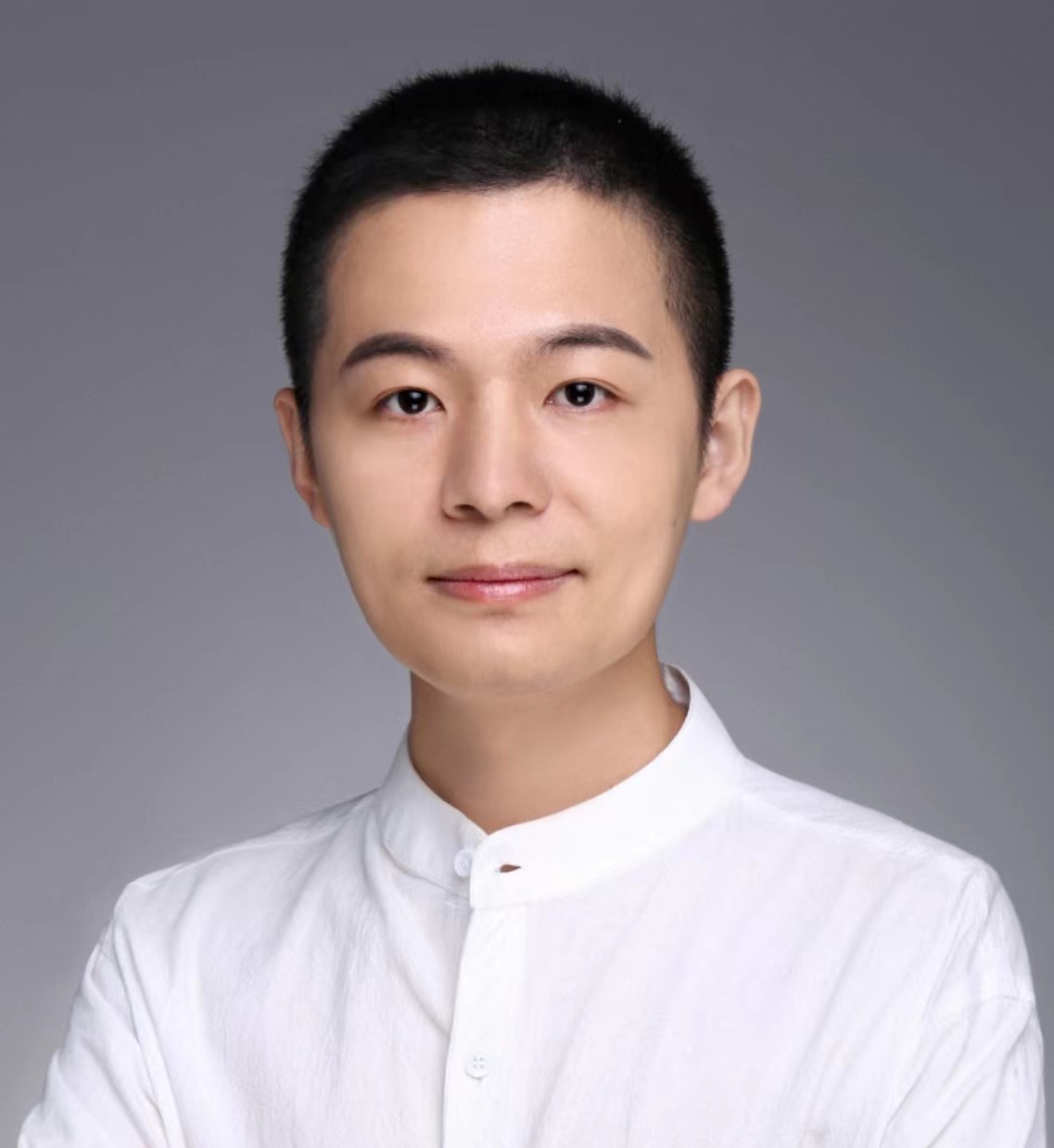}}]{Jiashui Wang} 
    is the head of Security Countermeasure Technology Department of Ant Group and the main founder of Ant Security Light-Year Lab.
\end{IEEEbiography}

%% file: appendix.tex
\section{Supplemental Material} 

\subsection{Details of the Datasets}\label{sec:dataset-detail}\vspace*{-1mm}  
Spk$_{10}$-enroll consists of 10 speakers (5 males and 5 females), 10 voices per speaker.
The speakers are randomly selected from the ``test-other'' and ``dev-other'' subsets of the popular dataset Librispeech~\cite{panayotov2015librispeech,LiZJXZWM020,zhang2021attack,jati2021adversarial,chen2019real}.
For each speaker, we select the top-10 longest voices in order to have better enrollment embedding~\cite{ParkYKKA17, app9183697}. The voices in Spk$_{10}$-enroll are used for speaker enrollment
of the CSI-E, SV, and OSI tasks.
Spk$_{10}$-test consists 10 speakers (5 males and 5 females), 100 randomly selected voices per speaker.
Spk$_{10}$-test has the same speakers as Spk$_{10}$-enroll, but distinct voices.

Both Spk$_{251}$-train and Spk$_{251}$-test
are taken from the ``train-clean-100'' subset of Librispeech,
each of which has the same 251 speakers (126 males and 125 females).
Following \cite{jati2021adversarial}, for each speaker, 90\% of his/her voices are added into Spk$_{251}$-train,
and the remaining 10\% are added into Spk$_{251}$-test.
Spk$_{251}$-train is used to train background
models while Spk$_{251}$-test is used for adversarial attacks
on the CSI-NE task.
Note that there are no overlapping speakers among Spk$_{251}$-train, Spk$_{10}$-enroll and Spk$_{10}$-test.

\subsection{Attacks}\label{sec:review-attacks}

A plethora of adversarial attacks have been proposed, most of which are primarily studied in computer vision~\cite{akhtar2018threat}.
It is largely unknown if they can successfully be ported to the speaker recognition domain.
Thus, we only consider the attacks that have been demonstrated to be effective on at least one speaker recognition task,
including four white-box attacks: Fast Gradient Sign Method (FGSM)~\cite{goodfellow2014explaining},
Projected Gradient Descent (PGD)~\cite{madry2017towards},
Carlini and Wagner's attack (CW)~\cite{carlini2017towards},
an integration of the CW and PGD attacks,
and three black-box attacks:
FAKEBOB~\cite{chen2019real}, SirenAttack~\cite{du2020sirenattack}, and Kenansville~\cite{ASGBWYST19}.

{\bf FGSM} perturbs an input $x$ by performing one-step gradient ascent to maximize a loss
function. Formally, a potential adversarial example is:
$$\hat{x} = x + \epsilon \times \emph{sign}(\nabla_x \mathcal{L}(x,y)),$$
where $\epsilon$ is the step\_size of gradient ascent,
\emph{sign} is the sign function, and
$\mathcal{L}(x,y)$ is the loss function describing the cost of classifying $x$ as label $y$.

{\bf PGD} is an iterative version of FGSM. In each iteration, PGD applies FGSM with a small step\_size $\alpha$
and clips the result to ensure that it stays within an $\epsilon$-neighborhood of the original input $x$.
The intermediate example after the $i$-th iteration is:
$$x^i = \emph{clip}_{x,\epsilon}(x^{i-1} + \alpha \times \emph{sign}(\nabla_x \mathcal{L}(x^{i-1},y))).$$
Note that the PGD attack starts from a randomly perturbed example, which helps the attack find a better local optimum.
We denote by PGD-$n$ the PGD attack with $n$ iteration steps, where
the larger $n$ is, the stronger the attack is.

{\bf CW} is introduced to search for adversarial examples with the small magnitude of perturbations.
It formulates finding adversarial examples as an optimization problem whose objective function is the trade-off (controlled by a factor $c$)
between the effectiveness and imperceptibility of adversarial examples.
The effectiveness is measured by a loss function $\mathcal{L}(x,y)$ such that $\mathcal{L}(x,y)\leq 0$ if and only if the attack succeeds.
The imperceptibility can be instantiated by $L_0$, $L_2$, and $L_\infty$ distance
between adversarial and original examples, leading to three versions of CW attack,
denoted by CW$_0$, CW$_2$, and CW$_{\bm{\infty}}$, respectively.
CW attack is equipped with a parameter $\kappa$, where
the larger $\kappa$ is, the stronger the adversarial examples are.
We consider CW$_2$ and CW$_\infty$ in this works and denote by CW$_2$-$x$ (resp. CW$_\infty$-$n$)
the CW$_2$ (resp. CW$_\infty$) attack with $\kappa=x$ (resp. $n$ iteration steps).

{\bf FAKEBOB} is similar to PGD except that it estimates gradients via Natural Evolution Strategy (NES)~\cite{wierstra2014natural}
that only relies on the output of the model.
NES first creates $m$ noisy examples by adding Gaussian noises onto an example.
Then, the values of the loss function of $m$ examples are obtained by querying the model,
which are finally exploited to approximate the gradient.
FAKEBOB adopts an early-stop strategy to reduce the number of queries, i.e., stop searching once an adversarial example is found.
Similar to the CW attack, FAKEBOB also provides an option to control the confidence of adversarial examples via
a parameter $\kappa$. FAKEBOB also proposed the first algorithm to estimate the threshold for SV and OSI tasks.
One of the crucial parameter of FAKEBOB is the samples\_per\_draw $m$ of NES.

{\bf Integration of the CW and PGD attacks} (i.e., CW$_{\bm{\infty}}$ in our evaluation) uses the loss function of the CW attack but optimized by PGD,
the same as \cite{madry2017towards}, to improve the attack efficiency.

{\bf SirenAttack} is a gradient-free black-box attack based on Particle Swarm Optimization (PSO)~\cite{PSO} that only relies on the output of the model.
PSO maintains a swarm of particles, each of which is a candidate solution to the optimization problem.
They are iteratively updated via the weighted linear combination of three parts,
i.e., inertia, local best solution, and global best solution.
When the algorithm terminates, the global best solution returns an optima.
SirenAttack runs the PSO subroutine multiple times (each run called an epoch) and globally keeps track of the best solution.
One of the crucial parameter of SirenAttack is the number of particles $n\_particles$ used in PSO.

{\bf Kenansville} is a signal-processing based attack which crafts adversarial voices by decomposing benign voices
and then reconstructing voices using part of the decomposing information (other decomposing information is discarded).
The amount of information used in reconstruction is controlled by the attack factor
which will be iteratively updated via a binary search to improve the imperceptibility of the attack.
Kenansville features two signal processing methods, i.e., Fast Fourier Transform (FFT) and Singular Spectrum Analysis (SSA),
where FFT method is not considered in our evaluation since it is much less effective than the SSA method~\cite{AWBPT20}.
 
\begin{table}[t]
    \centering
     \caption{The ranges and optimal values for parameters of transformations.}
    \resizebox{0.49\textwidth}{!}{
    \begin{tabular}{c|c|c}
    \hline
        \makecell[c]{{\bf Transformation}\\ {\bf (Parameter)}} &  {\bf Range} & {\bf Optimal} \\ \hline
        {\bf QT ($q$)} & 128, 256, 512, 1024 & 512 \\ \hline
        {\bf AT ($snr$)} & 2 to 20 dB, step 2 dB & 16 dB \\ \hline
        {\bf AS ($k$)} & 3 to 21, step 2 & 17 \\ \hline
        {\bf MS ($k$)} & 3 to 21, step 2 & 7 \\ \hline \hline
        {\bf DS ($\tau$)} & 0.05 to 0.95, step 0.05 & 0.45 \\ \hline
        {\bf LPF ($f_p$, $f_s$)} & \begin{tabular}[c]{@{}c@{}}$f_p$: 4000 Hz \\ $f_s$: 4500 to 8000 Hz, step 500 Hz\end{tabular} & $f_s$=4500 Hz \\ \hline
        \makecell[c]{{\bf BPF} \\ {($f_{pl}$, $f_{pu}$, $f_{sl}$, $f_{su}$)}} & \begin{tabular}[c]{@{}c@{}}$f_{pl}$: 300 Hz \\ $f_{pu}$: 4000 Hz \\ $f_{sl}$: 50 to 200 Hz, step 50 Hz \\ $f_{su}$: 5000 Hz to 8000 Hz, step 500 Hz \end{tabular} & \begin{tabular}[c]{@{}c@{}}$f_{sl}$=150 Hz \\ $f_{su}$=6000 Hz \end{tabular} \\ \hline \hline
        {\bf OPUS ($b_o$)} & 6-20 kbps, step 1 kbps & 8 kbps \\ \hline
        {\bf SPEEX ($b_s$)} & 4-44 kbps, step 2 kbps & 11 kbps \\ \hline
        {\bf AMR ($b_r$)} & \begin{tabular}[c]{@{}c@{}}6.6, 8.85, 12.65, 14.25, 15.85 \\ 18.25, 19.85, 23.05, 23.85 kbps \end{tabular} & 6.6 kbps \\ \hline
        {\bf AAC-V ($q_c$)} & 1-5, step 1 & 1 \\ \hline
        {\bf AAC-C ($b_c$)} & 15-85 kbps, step 5 kbps & 15 kbps \\ \hline
        {\bf MP3-V ($q_m$)} & 0-9, step 1 & 4 \\ \hline
        {\bf MP3-C ($b_m$)} & \begin{tabular}[c]{@{}c@{}}8, 16, 24, 32, 40, 48, \\ 64, 80, 96, 112, 128, 160 kbps \end{tabular} & 24 kbps \\ \hline \hline
        {\bf \defensenameabbr ($cl_m$, $cl_r$)} & \begin{tabular}[c]{@{}c@{}} $cl_{m}$: kmeans/warped-kmeans \\ $cl_{r}$: 0.05 to 0.95, step 0.05 \end{tabular} & \begin{tabular}[c]{@{}c@{}} \defensenameabbr-o(k): $cl_{r}$=0.2 \\ \defensenameabbr-o(wk): $cl_{r}$=0.35 \\ \defensenameabbr-d: $cl_{r}$=0.1 \\ \defensenameabbr-c: $cl_{r}$=0.1 \\ \defensenameabbr-f: $cl_{r}$=0.1 \end{tabular} \\ \hline
    \end{tabular}
    }
    \label{tab:range-optimal}
    \vspace*{-3mm}
\end{table}

\subsection{Tuning the Parameters of Transformations}\label{sec:parameter}
To tune the parameters of the transformations,
we vary the parameters as shown in Table~\ref{tab:range-optimal}
and conduct all the attacks mentioned in
Section~\ref{sec:non-adaptive-attack}.

The results are depicted as curves  in \figurename~\ref{fig:parameter-1}, \figurename~\ref{fig:parameter-2}, and \figurename~\ref{fig:parameter-3}.
We choose the optimal parameters according to the R1 score on FGSM, as R1 score assigns equal importance to the accuracy on benign examples
and the accuracy on adversarial examples.
We consider FGSM as it is the weakest one among all the attacks, as shown in
the (Baseline) row of \tablename~\ref{tab:evaluate-defense-non-adaptive},
and a good parameter should provide strong resilience to the weakest attack.
Although these optimal parameters may not be the optimal ones against the other attacks,
they are still very promising.

\input{tuningparameters}

\subsection{More Details of Section 7}  
\label{sec:moredetails}

In this section, we report more detailed results of the evaluation of transformations against non-adaptive attacks.

\begin{table*}[htbp]
    \centering
    \caption{The effectiveness of input transformations in terms of accuracy (\%) against non-adaptive PGD attack
        when the step\_size is fractional to the number of steps (\#Steps).}\vspace*{-3mm}
    \scalebox{.8}{ 
      \begin{tabular}{c||c|c|c|c|c|c||c|c|c|c|c|c||c|c|c|c|c|c}
      \hline
      \multirow{2}[4]{*}{} & \multicolumn{6}{c||}{\boldmath{}\textbf{$\alpha=\frac{\epsilon}{5\text{\#Steps}}$}\unboldmath{}} & \multicolumn{6}{c||}{\boldmath{}\textbf{$\alpha=\frac{\epsilon}{\text{\#Steps}}$}\unboldmath{}} & \multicolumn{6}{c}{\boldmath{}\textbf{$\alpha=\frac{10\epsilon}{\text{\#Steps}}$}\unboldmath{}} \\
  \cline{2-19}          & \textbf{10} & \textbf{20} & \textbf{30} & \textbf{40} & \textbf{50} & \textbf{100} & \textbf{10} & \textbf{20} & \textbf{30} & \textbf{40} & \textbf{50} & \textbf{100} & \textbf{10} & \textbf{20} & \textbf{30} & \textbf{40} & \textbf{50} & \textbf{100} \\
      \hline
      \textbf{QT} & 52.4  & 60.3  & 64.1  & 66.3  & 68.4  & 71.7  & 72.1  & 76.0  & 77.2  & 78.4  & 79.1  & 81.2  & 48.7  & 55.3  & 58.3  & 61.7  & 63.7  & 70.2  \\
      \hline
      \textbf{AT} & 76.3  & 81.9  & 85.1  & 86.8  & 88.2  & 90.3  & 86.4  & 90.5  & 91.0  & 91.8  & 92.2  & 92.6  & 68.6  & 75.0  & 78.3  & 81.0  & 81.9  & 87.6  \\
      \hline
      \textbf{AS} & 0.0   & 0.0   & 0.0   & 0.0   & 0.0   & 0.0   & 0.0   & 0.0   & 1.3   & 5.4   & 9.5   & 24.8  & 0.0   & 0.0   & 0.0   & 0.0   & 0.0   & 0.0  \\
      \hline
      \textbf{MS} & 13.6  & 22.3  & 29.1  & 35.8  & 38.6  & 50.4  & 31.2  & 39.4  & 46.6  & 50.0  & 52.1  & 58.0  & 22.5  & 17.9  & 15.1  & 15.1  & 14.8  & 28.8  \\
      \hline \hline
      \textbf{DS} & 0.0   & 0.0   & 0.1   & 0.6   & 1.2   & 4.1   & 0.8   & 0.6   & 1.5   & 2.2   & 3.9   & 9.5   & 6.5   & 0.7   & 0.0   & 0.0   & 0.0   & 0.1  \\
      \hline
      \textbf{LPF} & 0.0   & 0.0   & 0.0   & 0.0   & 0.0   & 0.1   & 0.0   & 0.1   & 0.3   & 0.6   & 1.2   & 3.8   & 0.8   & 0.0   & 0.0   & 0.0   & 0.0   & 0.1  \\
      \hline
      \textbf{BPF} & 0.0   & 0.0   & 0.0   & 0.0   & 0.0   & 0.0   & 0.1   & 0.5   & 1.2   & 2.1   & 3.2   & 9.1   & 1.0   & 0.0   & 0.0   & 0.0   & 0.0   & 0.0  \\
      \hline \hline
      \textbf{OPUS} & 13.7  & 17.0  & 21.6  & 26.2  & 31.5  & 44.6  & 29.9  & 38.2  & 44.6  & 48.2  & 51.9  & 59.3  & 31.8  & 19.1  & 15.2  & 14.3  & 15.4  & 26.9  \\
      \hline
      \textbf{SPEEX} & 1.7   & 2.9   & 6.3   & 9.9   & 14.2  & 28.7  & 6.4   & 14.1  & 21.9  & 27.0  & 31.7  & 44.0  & 16.8  & 7.5   & 4.8   & 4.5   & 5.5   & 14.6  \\
      \hline
      \textbf{AMR} & 5.5   & 8.5   & 14.9  & 19.5  & 24.4  & 38.4  & 11.1  & 21.1  & 29.3  & 34.6  & 39.9  & 48.7  & 18.5  & 10.6  & 5.8   & 6.1   & 7.4   & 17.8  \\
      \hline
      \textbf{AAC-V} & 0.0   & 0.0   & 0.0   & 0.0   & 0.0   & 0.0   & 0.0   & 0.0   & 0.0   & 0.0   & 0.1   & 0.7   & 0.0   & 0.0   & 0.0   & 0.0   & 0.0   & 0.0  \\
      \hline
      \textbf{AAC-C} & 1.5   & 1.7   & 3.0   & 4.5   & 6.0   & 13.1  & 2.3   & 4.0   & 6.3   & 7.3   & 10.1  & 19.2  & 14.3  & 1.5   & 0.3   & 0.4   & 0.4   & 1.3  \\
      \hline
      \textbf{MP3-V} & 0.0   & 0.0   & 0.0   & 0.0   & 0.0   & 0.0   & 0.0   & 0.0   & 0.0   & 0.0   & 0.0   & 0.0   & 0.0   & 0.0   & 0.0   & 0.0   & 0.0   & 0.0  \\
      \hline
      \textbf{MP3-C} & 0.0   & 0.1   & 0.1   & 0.3   & 0.8   & 3.1   & 0.2   & 0.3   & 1.7   & 2.5   & 3.0   & 7.8   & 1.9   & 0.1   & 0.0   & 0.0   & 0.0   & 0.0  \\
      \hline \hline
      \textbf{FeCo-o(k)} & 10.4  & 12.9  & 14.9  & 19.3  & 20.5  & 29.4  & 22.4  & 26.5  & 27.7  & 33.1  & 36.3  & 38.2  & 40.3  & 22.1  & 19.0  & 16.7  & 17.3  & 22.1  \\
      \hline
      \textbf{FeCo-d(k)} & 0.6   & 0.6   & 1.2   & 0.9   & 2.3   & 5.6   & 1.2   & 3.7   & 5.3   & 6.2   & 8.0   & 13.5  & 9.2   & 2.7   & 1.0   & 1.0   & 1.6   & 4.8  \\
      \hline
      \textbf{FeCo-c(k)} & 0.5   & 0.6   & 1.2   & 1.2   & 1.4   & 6.6   & 1.8   & 2.1   & 3.9   & 4.5   & 6.3   & 12.2  & 8.8   & 1.8   & 0.8   & 0.5   & 1.1   & 2.9  \\
      \hline
      \textbf{FeCo-f(k)} & 0.6   & 0.7   & 1.0   & 1.2   & 1.7   & 6.9   & 1.8   & 2.2   & 3.8   & 4.8   & 7.0   & 12.5  & 8.0   & 2.0   & 0.6   & 0.5   & 1.5   & 2.6  \\
      \hline
      \textbf{FeCo-o(wk)} & 2.4   & 2.5   & 2.8   & 3.8   & 3.9   & 8.2   & 3.1   & 3.5   & 5.7   & 6.2   & 8.0   & 12.6  & 18.8  & 9.4   & 5.7   & 4.6   & 4.4   & 5.0  \\
      \hline
      \textbf{FeCo-d(wk)} & 1.9   & 2.5   & 2.4   & 4.1   & 5.4   & 12.7  & 4.3   & 7.8   & 11.2  & 12.5  & 13.6  & 21.4  & 19.2  & 6.3   & 3.5   & 3.0   & 2.9   & 7.7  \\
      \hline
      \textbf{FeCo-c(wk)} & 1.5   & 1.6   & 2.1   & 2.8   & 3.7   & 9.9   & 2.7   & 4.8   & 7.0   & 8.9   & 10.5  & 17.6  & 16.9  & 4.8   & 2.8   & 2.0   & 1.8   & 6.7  \\
      \hline
      \textbf{FeCo-f(wk)} & 1.3   & 1.8   & 1.9   & 2.6   & 3.8   & 11.2  & 2.2   & 5.0   & 6.8   & 9.0   & 10.6  & 18.1  & 16.0  & 5.0   & 3.2   & 2.3   & 2.0   & 7.5  \\
      \hline
      \end{tabular}%
        }
    \label{tab:PGD-step-size-fractional}%
  \end{table*}%

\subsubsection{Impact of Step\_size in PGD Attack}\label{sec:pgd-cwinf-step-size}
\tablename~\ref{tab:PGD-step-size-fractional} reports the effectiveness of input transformations in terms
of accuracy against the PGD attack when step\_size $\alpha$ is fractional to the number \#Steps of steps, namely,
$\alpha=\frac{\epsilon}{5\text{\#Steps}}$, $\alpha=\frac{\epsilon}{\text{\#Steps}}$, and $\alpha=\frac{10\times\epsilon}{\text{\#Steps}}$
(Recall that previously we set $\alpha=\frac{\epsilon}{5}$).
We can observe that increasing the number \#Steps of steps and simultaneously decreasing
step\_size $\alpha$ does not necessarily reduce the effectiveness of input transformations (e.g., QT, AT,
MS, OPUS, SPEEX and FeCo-o).

\subsubsection{More Findings}\label{sec:non-adaptive-more-findings}
In this subsection, we discuss in more detail the side effect of transformations on benign examples
and usability of transformations.

\smallskip \noindent{\bf Side effect on benign examples.}
Most transformations slightly degrade accuracy on benign examples, but the degradation varies.
The accuracy degradation reflects the degree of distortions induced by each transformation, i.e., how well the transformation preserves the speech quality.
Among all the transformations,
QT, AT, MS, and OPUS cause the greatest accuracy degradation ($>10\%$),
indicating that they add more distortions.
AAC-V, MP3-V and
\defensenameabbr-d(k)
almost have no side effects, reducing only
$0\%$, $0.2\%$ and $0.4\%$ accuracy on benign examples, respectively.
We observe that dynamic bit rate based speech compressions
have fewer side effects (e.g., MP3-V vs. MP3-C, and AAC-V vs. AAC-C), as they preserve the better quality of voices.
Among the feature-level transformations, we observe that
\defensenameabbr-d outperforms the others, indicating
that \defensenameabbr
has fewer effects on the delta
features than the others.

\begin{tcolorbox}[size=title,breakable,arc=1mm, boxsep=0.4mm, left = 1pt, right = 1pt, top = 1pt, bottom = 1pt]
\textbf{Findings 11.}
Most of transformations can be freely composed with pre-trained models to defeat adversarial examples with slight accuracy degradation on the benign examples.
Input transformations with dynamic bit rate and delta feature transformation have the least side effects,
while the input transformations QT, AT, MS, and OPUS
have the greatest side effects.
\end{tcolorbox}

\smallskip\noindent{\bf Usability of transformations.}
Though effective transformations against adversarial examples degrade accuracy  on benign examples,
compared to Baseline, all transformations
show good usability
in terms of the R1 score.
The best one (i.e., AT) improves the R1 score by 77.3\% and the worst one (i.e., MP3-V) improves it by 19.4\%.
This is because in general, the accuracy improvements on adversarial examples
are often larger than the accuracy degradation on benign examples.

Among the feature-level transformations, we can observe that
\defensenameabbr-o and \defensenameabbr-d often significantly outperform.
This is because transformation on preceding features also affects succeeding features, which amplifies the effect of the transformation.
Between two clustering algorithms kmeans and warped-kmeans,
the effectiveness varies with attacks
and in general they are almost comparable.
In terms of the R1 score, \defensenameabbr-o with kmeans, i.e., \defensenameabbr-o(k), ranks the first place.

\begin{tcolorbox}[size=title,breakable,arc=1mm, boxsep=0.4mm, left = 1pt, right = 1pt, top = 1pt, bottom = 1pt] 
\textbf{Findings 12.}
All transformations exhibit good usability since they lead to significantly better R1 scores.
 While the transformations QT, AT, and \defensenameabbr-o(k) degrade the accuracy on benign examples, they
 are the three most effective transformations against non-adaptive attacks.
 \end{tcolorbox}

\begin{figure*}[t]
    \centering
    \subfigure[QT, 0.92]{
    \includegraphics[width=0.22\textwidth]{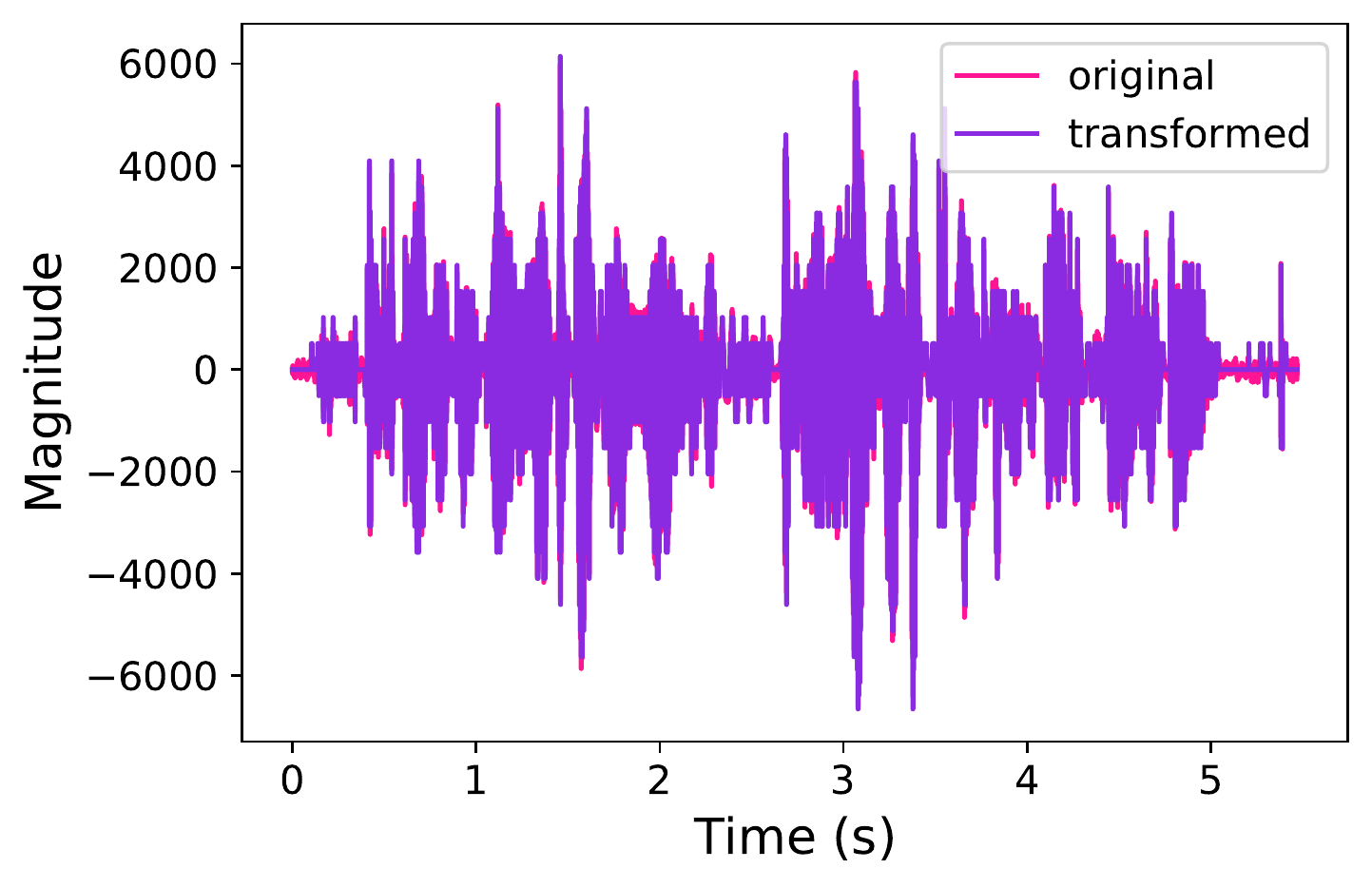}
    \label{fig:BPDA-QT}
    }
    \subfigure[OPUS, 15.24]{
    \includegraphics[width=0.22\textwidth]{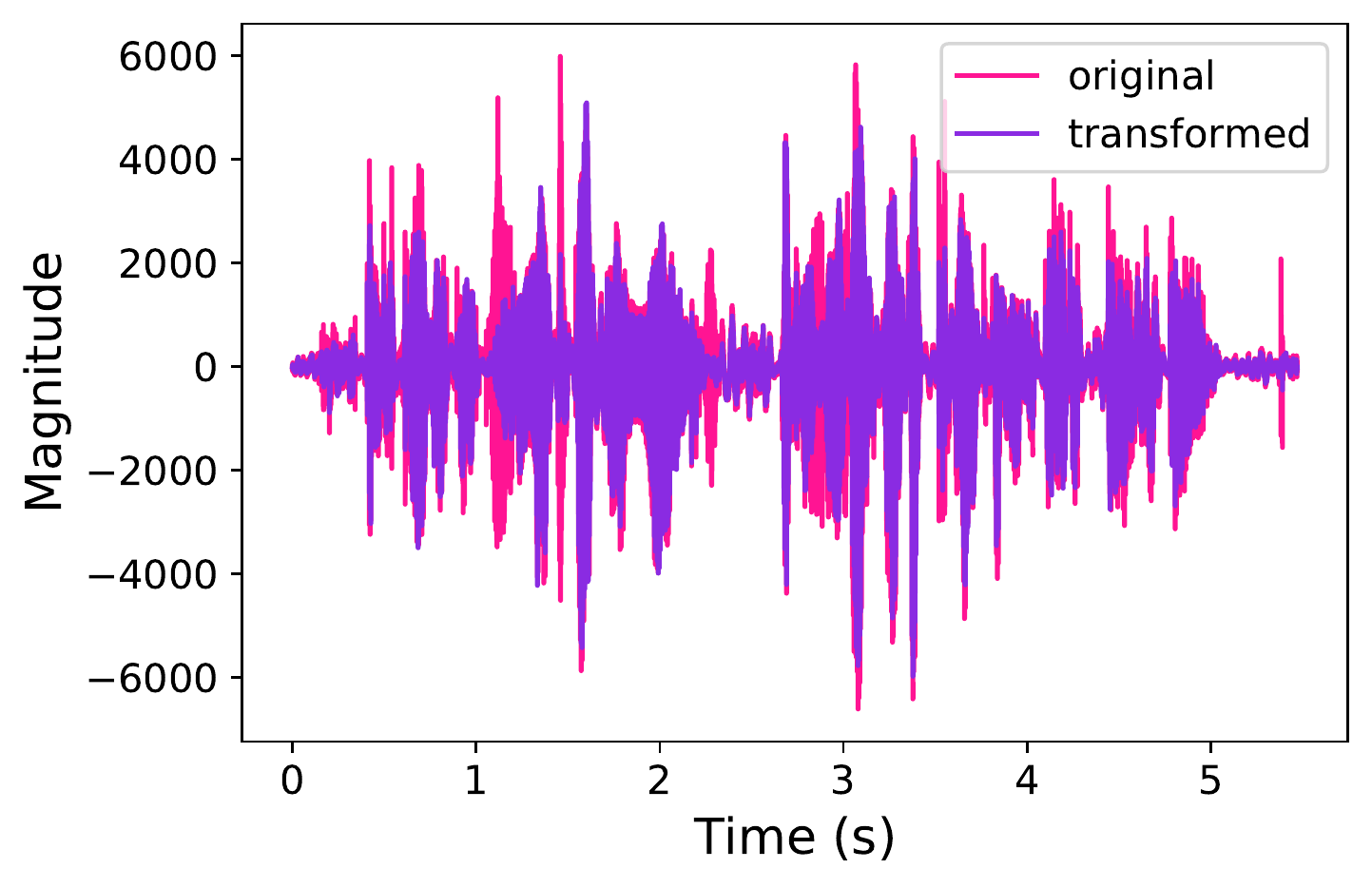}
    }
    \subfigure[SPEEX, 29.57]{
    \includegraphics[width=0.22\textwidth]{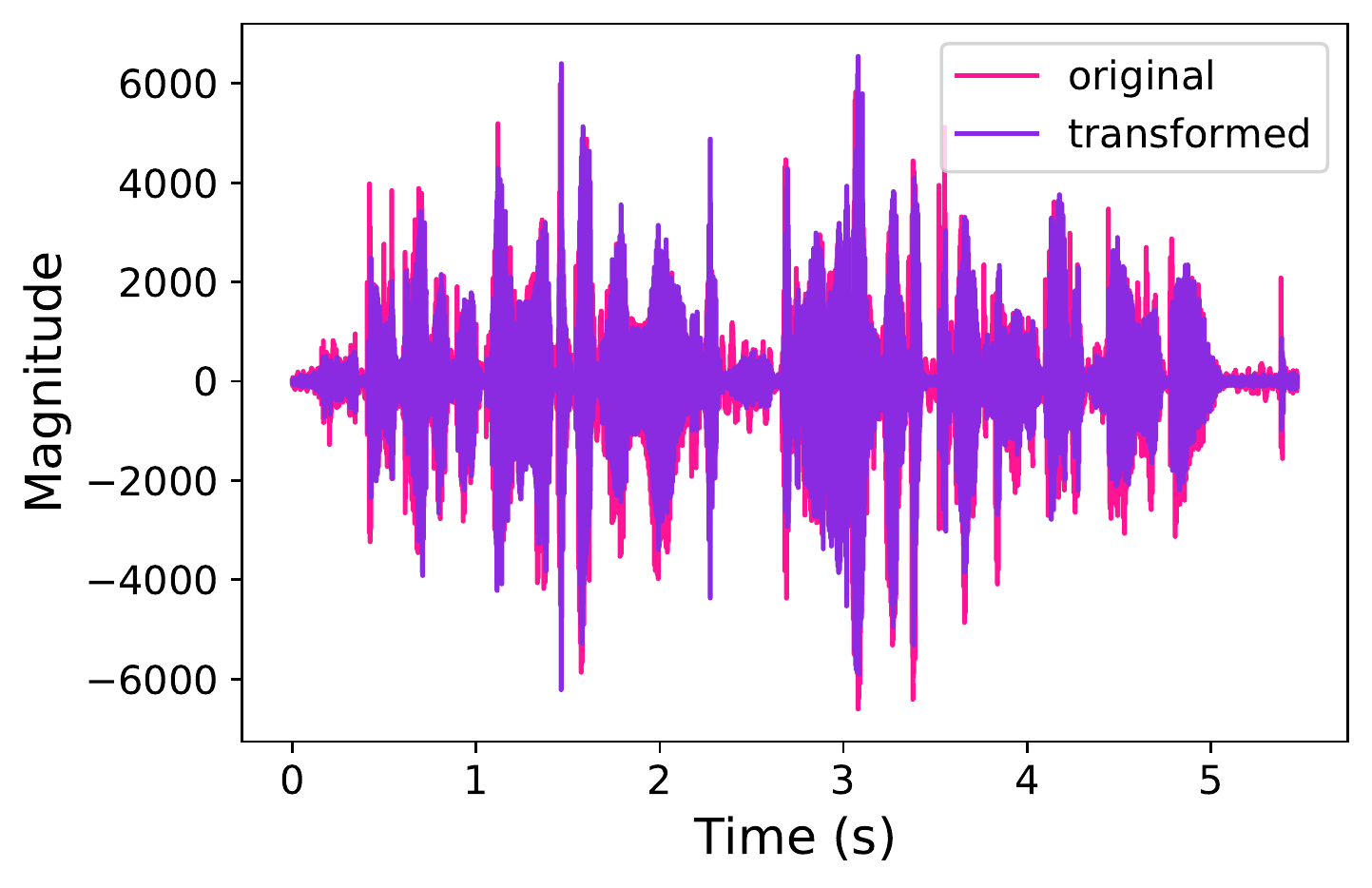}
    }
    \subfigure[AMR, 23.68]{
    \includegraphics[width=0.22\textwidth]{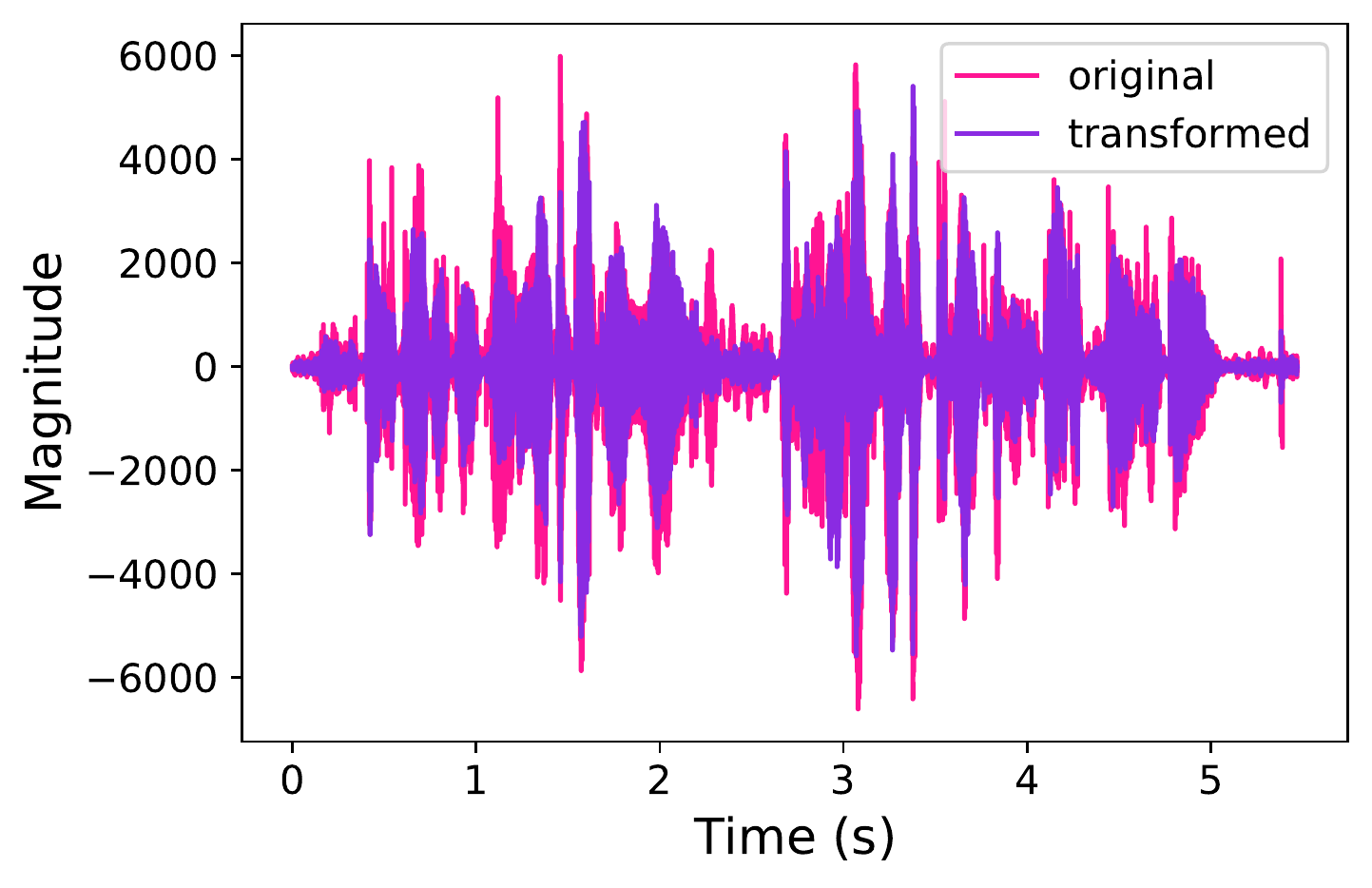}
    }
    \quad
  \subfigure[AAC-V, 3.87]{
    \includegraphics[width=0.22\textwidth]{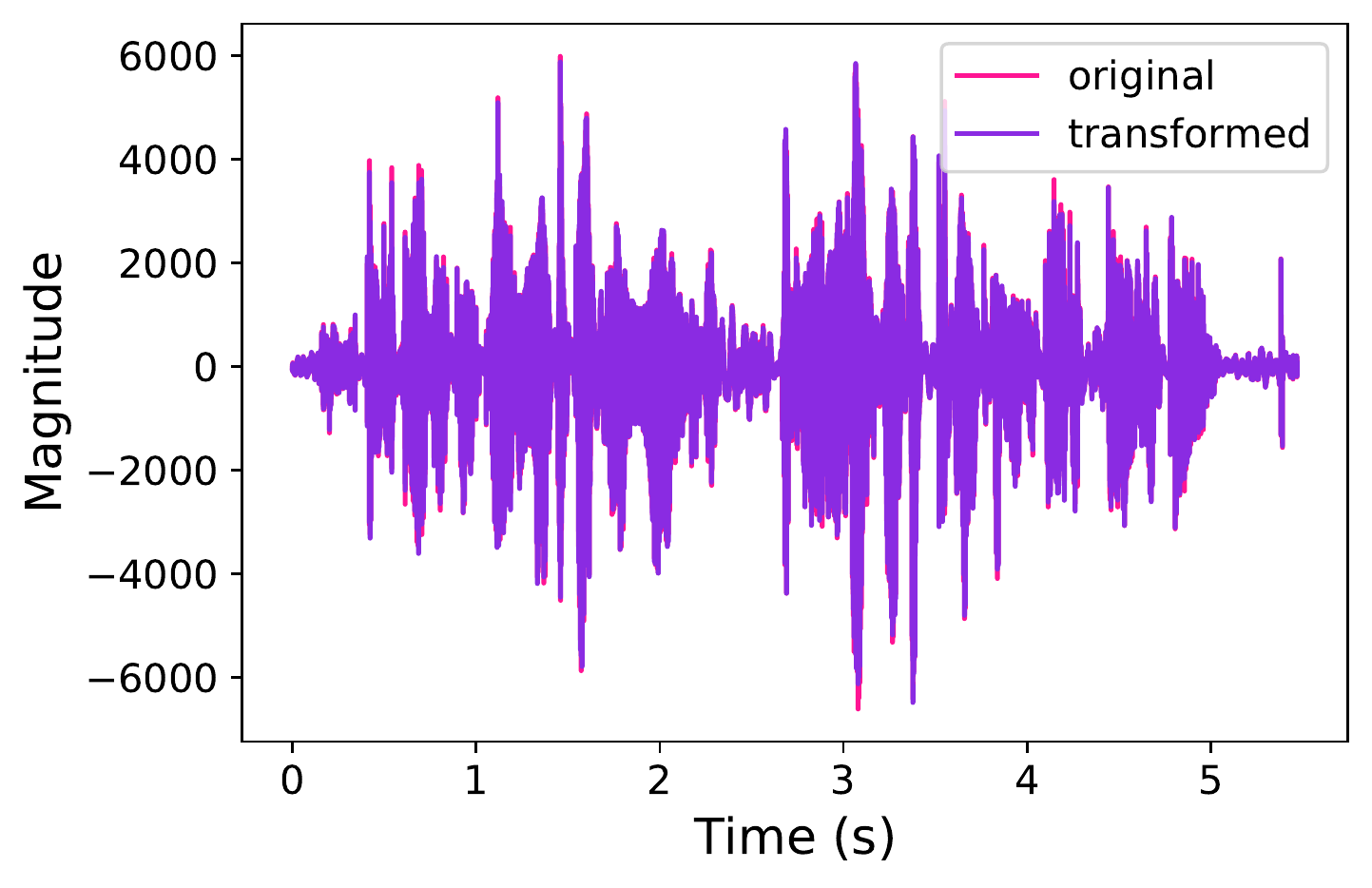}
    }
    \subfigure[AAC-C, 6.92]{
    \includegraphics[width=0.22\textwidth]{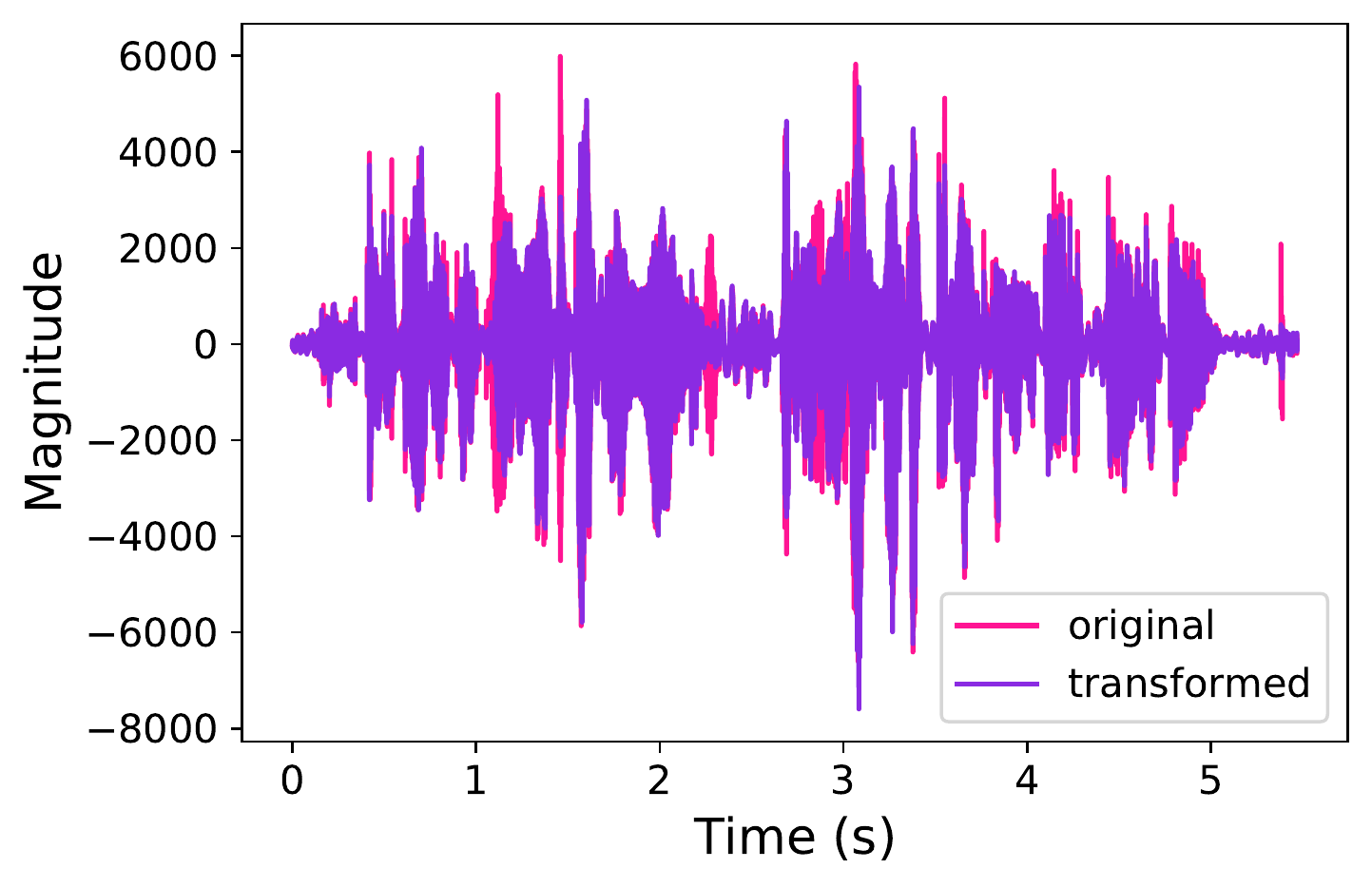}
    }
    \subfigure[MP3-V, 4.19]{
    \includegraphics[width=0.22\textwidth]{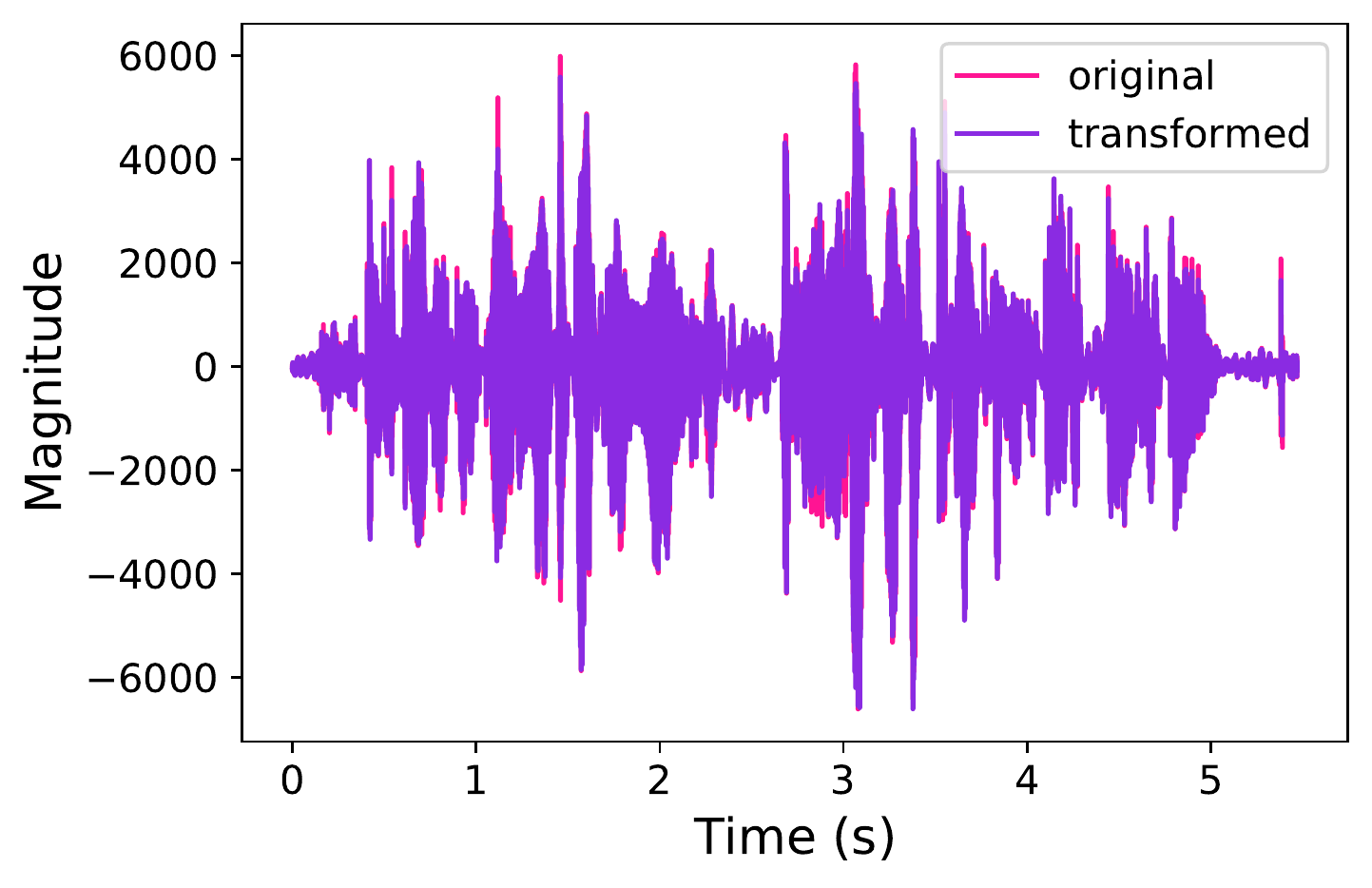}
    }
    \subfigure[MP3-C, 5.92]{
    \includegraphics[width=0.22\textwidth]{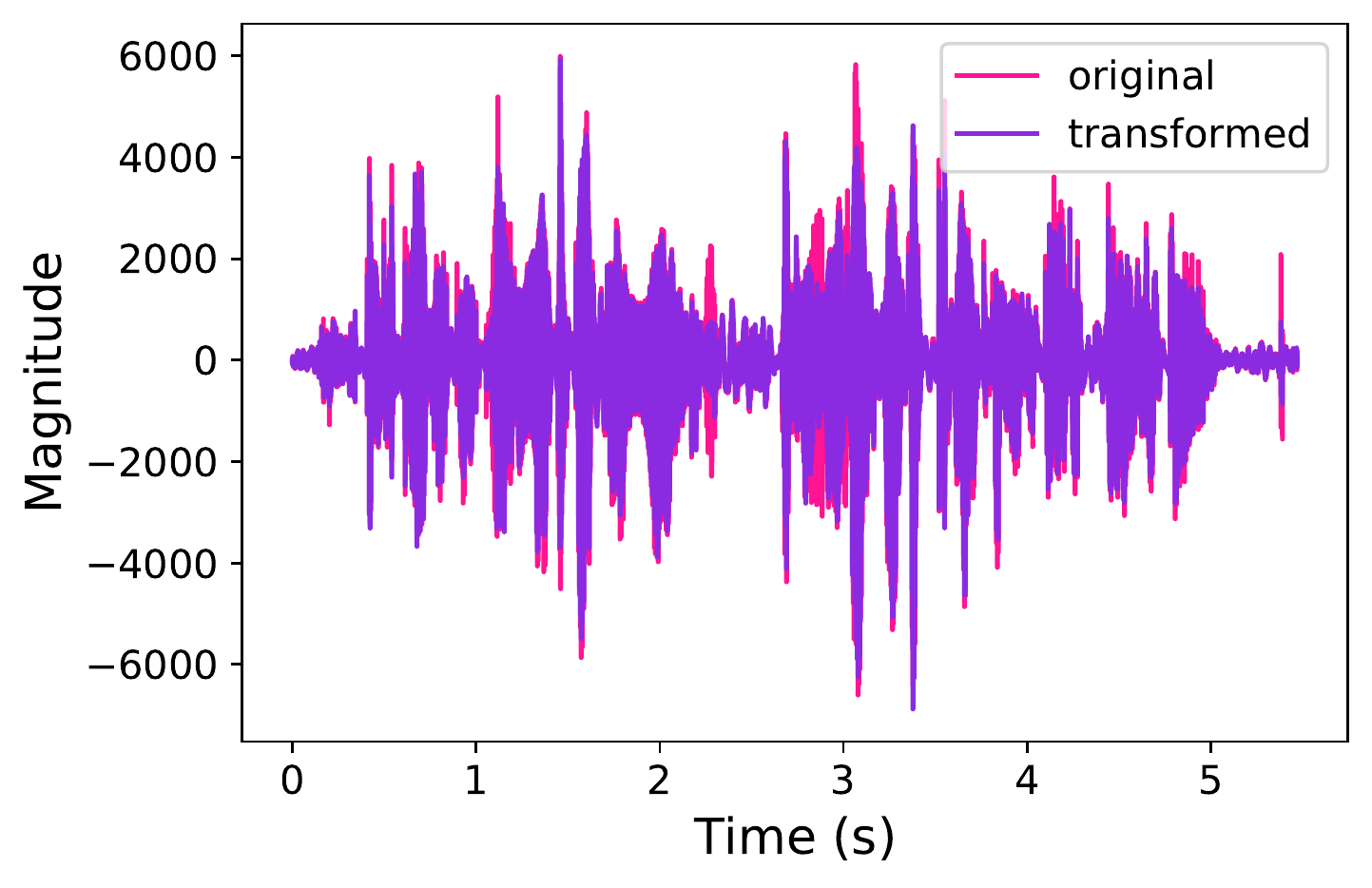}
    }
   \vspace*{-2mm}
    \caption{The visualization of an original voice and transformed voice by different input transformations.
    The average $L_2$ distance between original and transformed voices is listed right of the transformation name.}
    \label{fig:BPDA-differ}\vspace*{-3mm}
\end{figure*}

\subsection{Approximation of Non-differentiable Transformations by the Identity Function}
\label{sec:evaapptrans}
To measure how accurate it is to substitute a non-differentiable transformation with
the identity function, we compute the average $L_2$ distance between the original voices and the voices after the transformation.
The results are shown in \figurename~\ref{fig:BPDA-differ}, where the $L_2$ distance is given in the caption of each sub-figure,
and the curves in each sub-figure are the waveform of a random chosen voice and the voice after transformation.

From \figurename~\ref{fig:BPDA-differ}, we can observe that the $L_2$ distance of QT, AAC-V and MP3-V is much smaller than that of OPUS, SPEEX, AMR, AAC-C and MP3-C,
indicating that QT, AAC-V and MP3-V are much closer to the identity function.
We can also observe that the difference between the original voice and the voice after transformation of CBR speech compressions is more significant than that of QT and VBR speech compressions (i.e., AAC-V and MP3-V).
In conclusion, it seems that it suffices to replace QT and VBR speech compressions with the identity function in the backward pass,
but more accurate approximation functions or more advanced adaptive attacks than BPDA are required to circumvent other speech compressions.

\subsection{SirenAttack to AS, DS, LPF and BPF}\label{sec:siren-0.02}
The results are shown in \tablename~\ref{tab:siren-0.02} when $\epsilon=0.02$ for the adaptive SirenAttack.
We can observe that the adaptive SirenAttack reduces the accuracy of these input transformations by at least 16\% compared to the non-adaptive one.

\begin{table}[t]
    \centering
    \caption{Non-adaptive and adaptive SirenAttack against AS, DS, LPF, and BPF when $\epsilon=0.02$ in terms of model accuracy.}
    \vspace*{-3mm}
    \resizebox{0.35\textwidth}{!}{
    \begin{tabular}{c|c|c|c|c}
        \hline
        & {\bf AS} & {\bf DS} & {\bf LPF} & {\bf BPF} \\ \hline
    {\bf Non-adaptive} & 42.9\% & 53.3\% & 58.0\% & 48.0\% \\ \hline
    {\bf Adaptive} & 0\% & 0\% & 42.0\% & 16.3\% \\ \hline
    \end{tabular}
    }\vspace*{-3mm}
    \label{tab:siren-0.02}
\end{table}
 
\subsection{Brute-force Replicate Attack}\label{sec:more-detail-replicate-attack} 
Replicate attack is not strong 
due to the randomness of FeCo. The adversary may attempt to improve the attack by enumerating the randomness in a brute-force way.
Below we analyze the success probability such a brute-force adversary can achieve.

Suppose the randomness space is $\mathcal{B}=\{B_1,\cdots,B_Q\}$
where $B_i$ denotes a possible clustering result.
In each trail of the brute-force, suppose the adversary samples $B_a$ and the victim model samples $B_r$,
we have:
$Pr[\text{the attack succeeds}]\geq Pr[B_r=B_a]=\frac{1}{Q}.$

The size of the randomness space $Q$ depends on the duration of the voice and the initial method of the clustering algorithm.
For a voice with duration of one second (the minimal duration of the voices in Spk10\_test and Spk251\_test), the number $N$ of frames is nearly $100$.
If the initial method is kmeans++~\cite{kmeans++}, $Q=kN=500$ ($k=\frac{1}{cl_r}=\frac{1}{0.2}=5$ for \defensenameabbr-o(k))
and $Pr[\text{the attack succeeds}]\geq 0.2\%$.
If the initial method is random, $Q=C_{kN}^{N}>2.04\times 10^{107}$ and
$Pr[\text{the attack succeeds}]$ is close to $0\%$ in the worst case.
The success probability of the brute-force attack is very low.

\subsection{Defending against Hidden Voice and Speech Synthesis Attacks}\label{sec:hidden-spoofing}
To be comprehensive, apart from adversarial attacks,
we also evaluate the input transformation-based defenses
against hidden voice and speech synthesis attacks under both the non-adaptive and adaptive settings.

Hidden voice and speech synthesis attacks have different attack purposes from adversarial attacks.
Given a voice uttered by a source speaker, an adversarial attack intends to perturb the voice
such that the perturbed voice is recognized as another speaker by the target SRS,
but still recognized as the source speaker by human.
In contrast, hidden voice attack aims to craft a perturbed voice which is treated as mere noise by human,
but still correctly recognized as the source speaker by the target SRS,
and a speech synthesis attack attempts to produce a voice that contains the desired speech content
and sound as spoken by the source speaker from the perspective of both human and the target SRS.

For hidden voice attack, we consider the signal processing-based attack in \cite{AbdullahGPTBW19}.
It generates incomprehensible voices for human by inverting the speech in the time domain (Time Domain Inversion, TDI),
accelerating the speed of the speech (Time Scaling, TS), adding high-frequency signal, or generating random phases.
We exploit TDI and TS to perturb each speech since they are the two most effective methods~\cite{AbdullahGPTBW19}.
TDI and TS feature the parameters window size $w$ and scaling factor $\beta$, respectively, where
the smaller $w$ (resp. larger $\beta$), the less comprehensible the voices for human and the harder the voices to be correctly recognized by the target SRS.
As suggested in \cite{AbdullahGPTBW19}, we use a linear search to find the optimal parameters
where the attack can produce the least understandable voices for human when ensuring the correct recognition of the target SRS.
Specifically, to find the optimal $w$, the TDI attack starts from $w=1$ milliseconds (ms), gradually increases to $10$ ms with step of $0.5$ ms,
and terminates once the target SRS correctly recognizes the perturbed voice.
To find the optimal $\beta$, the TS attack starts from $\beta=20$, gradually reduces to $1.5$ with step of $0.5$,
and terminates once the target SRS correctly recognizes the perturbed voice.
Since both TDI and TS are black-box attacks, their adaptive versions are similar to the non-adaptive versions
except that the attack terminates once the {\it defended} SRS correctly recognizes the perturbed voice.

For speech synthesis attack, we exploit the deep learning-based speech synthesis tool used in \cite{wenger2021hello}.
The tool takes as input a set of voice samples of the source speaker and the desired speech content.
We use the voices in Spk10\_enroll as the set of voice samples (10 speakers and 10 voices per speaker)
and the ten sentences used in \cite{wenger2021hello} as the desired speech content.
We consider the following adaptive adversary: (1) running the non-adaptive speech synthesis attack with input voice samples $x_1,\cdots,x_N$.
Suppose the output voice is $\hat{x}$;
(2) generating adversarial perturbation $\delta$ for $\hat{x}$ with the objective of minimizing $\frac{1}{N}\sum_{i=1}^{N}d(Enc(g(\hat{x}+\delta))-Enc(x_i))$
where $g$ is the input transformation, $Enc(x)$ extracts the embedding (i.e., the vector representing the speaker characteristic) of the voice $x$,
and $d$ measures the distance between two embeddings;
(3) using $\hat{x}+\delta$ to attack the {\it defended} SRS.
Specifically, in step (2), we use cosine distance to measure distance between two embeddings
and exploit PGD with $\epsilon=0.002$ and \#Steps=50 to craft $\delta$.

\begin{table}[t]
    \centering
    \caption{Results of input transformations against hidden voice and speech synthesis attacks
    under both non-adaptive and adaptive settings
    in terms of attack success rate (\%).} \vspace*{-3mm}
    \resizebox{0.4\textwidth}{!}{
      \begin{tabular}{c|c|c|c|c|c|c}
      \hline
      \multirow{3}[4]{*}{} & \multicolumn{3}{c|}{\bf Non-adaptive} & \multicolumn{3}{c}{\bf Adaptive} \\ \cline{2-7}
      & \multicolumn{2}{c|}{\textbf{Hidden}} & \multirow{2}[1]{*}{\begin{tabular}[c]{@{}c@{}}\textbf{Speech} \\ {\bf Synthesis}\end{tabular}} &  \multicolumn{2}{c|}{\textbf{Hidden}} & \multirow{2}[1]{*}{\begin{tabular}[c]{@{}c@{}}\textbf{Speech} \\ {\bf Synthesis}\end{tabular}} \\ \cline{2-3}\cline{5-6}
      & \textbf{TDI} & \textbf{TS} &  & \textbf{TDI} & \textbf{TS} & \\
      \hline
      \textbf{Baseline} & 96.1 & 96.0 & 96.0 & 96.1 & 96.0 & 96.0 \\
      \hline
      \hline
      \textbf{QT} & 49.2  & \textcolor{blue}{\bf 43.3}  & 76.2 & 77.7 & 73.9 & 85.7 \\
      \hline
      \textbf{AT} & 55.6  & 52.7  & \textcolor{blue}{\bf 72.1} & 79.7 & \textcolor{blue}{\bf 71.8} & \textcolor{blue}{\bf 79.1} \\
      \hline
      \textbf{AS} & \textcolor{red}{\bf 70.0}  & \textcolor{red}{\bf 73.7}  & \textcolor{red}{\bf 91.0} & \textcolor{red}{\bf 93.8} & 89.5 & 92.5 \\
      \hline
      \textbf{MS} & 52.8  & 49.3  & \textcolor{blue}{\bf 66.3} & 81.2 & 84.0 & 84.5 \\
      \hline
      \hline
      \textbf{DS} & 51.2  & 54.0    & \textcolor{blue}{\bf 72.4} & 79.7 & 85.3 & 80.7 \\
      \hline
      \textbf{LPF} & 42.4  & 59.6  & 88.6 & 78.5 & 86.9 & \textcolor{red}{\bf 92.8} \\
      \hline
      \textbf{BPF} & \textcolor{blue}{\bf 31.6}  & 49.0  & 80.0 & 70.8 & 85.1 & 92.6 \\
      \hline
      \hline
      \textbf{OPUS} & 38.0  & 49.8  & 78.3 & 69.4 & 79.2 & \textcolor{blue}{\bf 72.2} \\
      \hline
      \textbf{SPEEX} & 64.0  & 59.0    & 80.2 & 93.2 & 89.6 & \textcolor{blue}{\bf 70.6} \\
      \hline
      \textbf{AMR} & 49.5  & 62.2  & 88.6 & 92.7 & \textcolor{red}{\bf 93.0} & 88.2 \\
      \hline
      \textbf{AAC-V} & \textcolor{red}{\bf 89.9}  & \textcolor{red}{\bf 90.3}  & \textcolor{red}{\bf 94.9} & \textcolor{red}{\bf 96.5} & \textcolor{red}{\bf 95.2} & \textcolor{red}{\bf 97.9} \\
      \hline
      \textbf{AAC-C} & \textcolor{blue}{\bf 31.6}  & 53.8  & 86.9 & 69.1 & 88.8 & 90.2 \\
      \hline
      \textbf{MP3-V} & \textcolor{red}{\bf 88.9}  & \textcolor{red}{\bf 90.7}  & \textcolor{red}{\bf 95.2} & \textcolor{red}{\bf 95.5} & \textcolor{red}{\bf 96.4} & \textcolor{red}{\bf 98.2} \\
      \hline
      \textbf{MP3-C} & \textcolor{blue}{\bf 23.3}  & 56.3  & 88.5 & \textcolor{blue}{\bf 68.0} & 88.4 & 90.9 \\
      \hline
      \hline
      \textbf{FeCo-o(k)} & 43.3  & 57.1  & 83.8 & \textcolor{blue}{\bf 53.8} & \textcolor{blue}{\bf 67.4} & \textcolor{blue}{\bf 79.1} \\
      \hline
      \textbf{FeCo-d(k)} & 52.1  & 49.2  & 88.2 & 72.4 & 74.8 & 83.2 \\
      \hline
      \textbf{FeCo-c(k)} & 52.3  & 47.8  & 86.7 & \textcolor{blue}{\bf 68.8} & 75.3 & 81.0 \\
      \hline
      \textbf{FeCo-f(k)} & 52.2  & 47.6  & 86.5 & 68.2 & 75.0 & 80.8 \\
      \hline
      \textbf{FeCo-o(wk)} & 59.6  & 60.0    & 84.5 & 74.1 & 75.5 & 89.5 \\
      \hline
      \textbf{FeCo-d(wk)} & 49.8  & 46.5  & 87.3 & 71.8 & 74.9 & 87.9 \\
      \hline
      \textbf{FeCo-c(wk)} & 52.8  & \textcolor{blue}{\bf 44.3}  & 86.3 & 70.0 & \textcolor{blue}{\bf 72.5} & 86.6 \\
      \hline
      \textbf{FeCo-f(wk)} & 52.4  & \textcolor{blue}{\bf 44.2}  & 86.8 & 70.0 & 72.3 & 86.6 \\
      \hline
      \end{tabular}%
    }\vspace*{-3mm}
    \label{tab:hidden-spoofing}%
  \end{table}%

The results are shown in \tablename~\ref{tab:hidden-spoofing}.
We can observe that all the input transformations are able to reduce the attack success rate under the non-adaptive setting
compared to Baseline without any defense, indicating they can also be exploited to mitigate hidden voice and speech synthesis attacks.
We also notice that regardless of the attack types,
AS, MP3-V and AAC-V are the least effective ones
while CBR speech compressions are more effective than VBR ones.
However, we should point out some transformations performing differently between adversarial, hidden voice and speech synthesis attacks.
While the time-domain W-transformations, especially QT and AT, are quite effective against adversarial attacks,
they are not as much effective against hidden voice attack.
Input transformations are in general less effective against speech synthesis attack than the other two attacks.
The reason is that speech synthesis attack attempts to synthesize high-quality and natural speeches to deceive both human and the target SRS,
unlike adversarial and hidden command attacks which perturb the original voice to cause inconsistent recognition between human and
the target SRS.

Under the adaptive setting, the adaptive hidden voice attack achieves much higher attack success rate
against all the input transformations than the non-adaptive one,
e.g., the success rate improves by over 43\% on AMR.
However, the adaptive speech synthesis attack does not perform better than the non-adaptive one on some input transformations,
e.g., OPUS, SPEEX, AMR, and some \defensenameabbr.
The reason is that the adaptive speech synthesis attack involves solving an optimization problem,
and these transformations introduce optimization obstacles,
i.e., non-differenetiability of OPUS, SPEEX, and AMR,
and the randomness of \defensenameabbr.

\begin{tcolorbox}[size=title,breakable,arc=1mm, boxsep=0.4mm, left = 1pt, right = 1pt, top = 1pt, bottom = 1pt] 
  \textbf{Findings 13.}
  Input transformations exhibits general defense capability against
  adversarial, hidden voice, and speech synthesis attacks, although with some differences.
  Speech synthesis attack is more difficult to defeat by input transformations than the other two attacks.
\end{tcolorbox}

%% file: tuningparameters.tex
\begin{figure*}
    \centering

    \subfigure[QT]{
    \begin{minipage}[t]{0.22\textwidth}
    \includegraphics[width=1.0\textwidth]{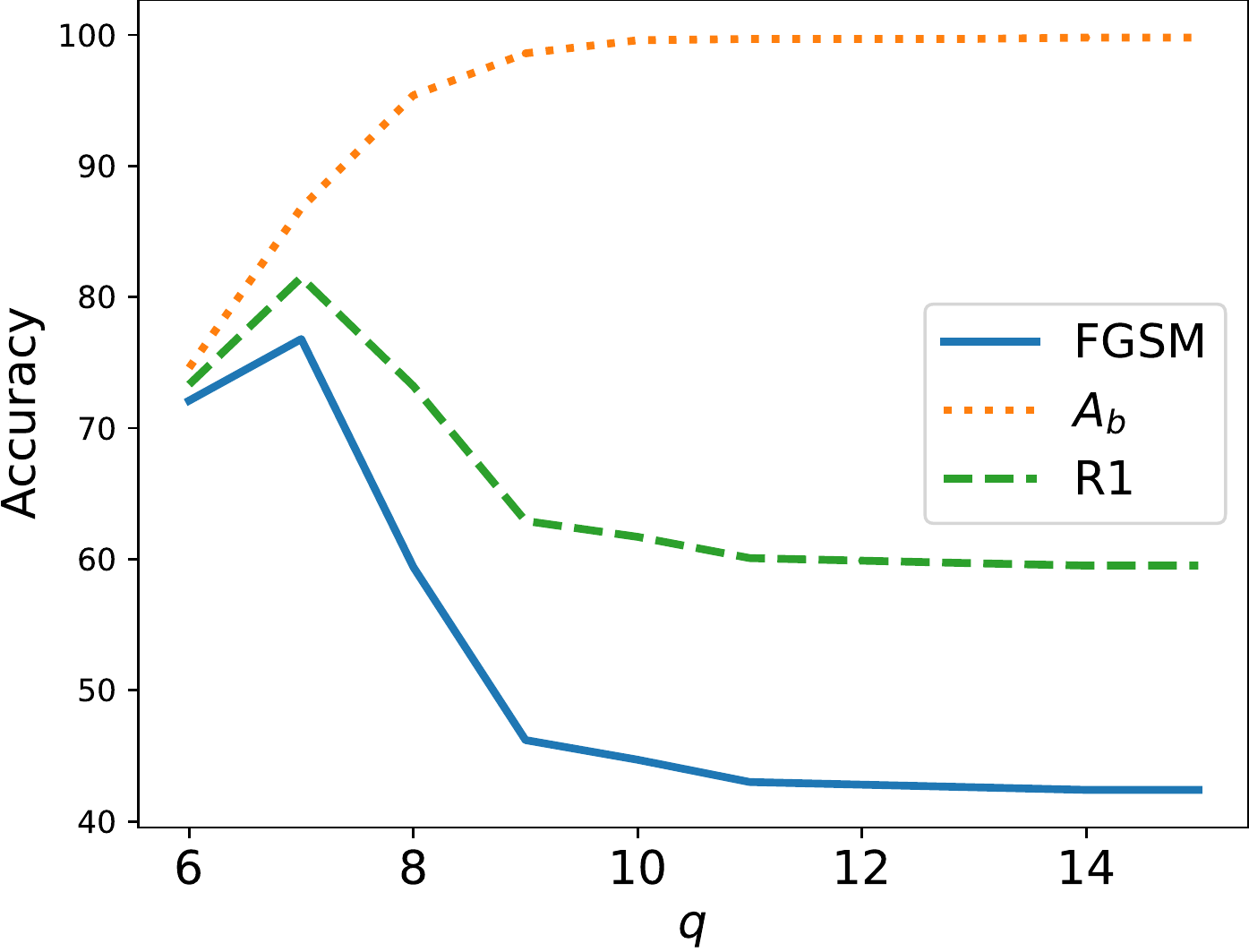}
    \end{minipage}
    \begin{minipage}[t]{0.22\textwidth}
    \includegraphics[width=1.0\textwidth]{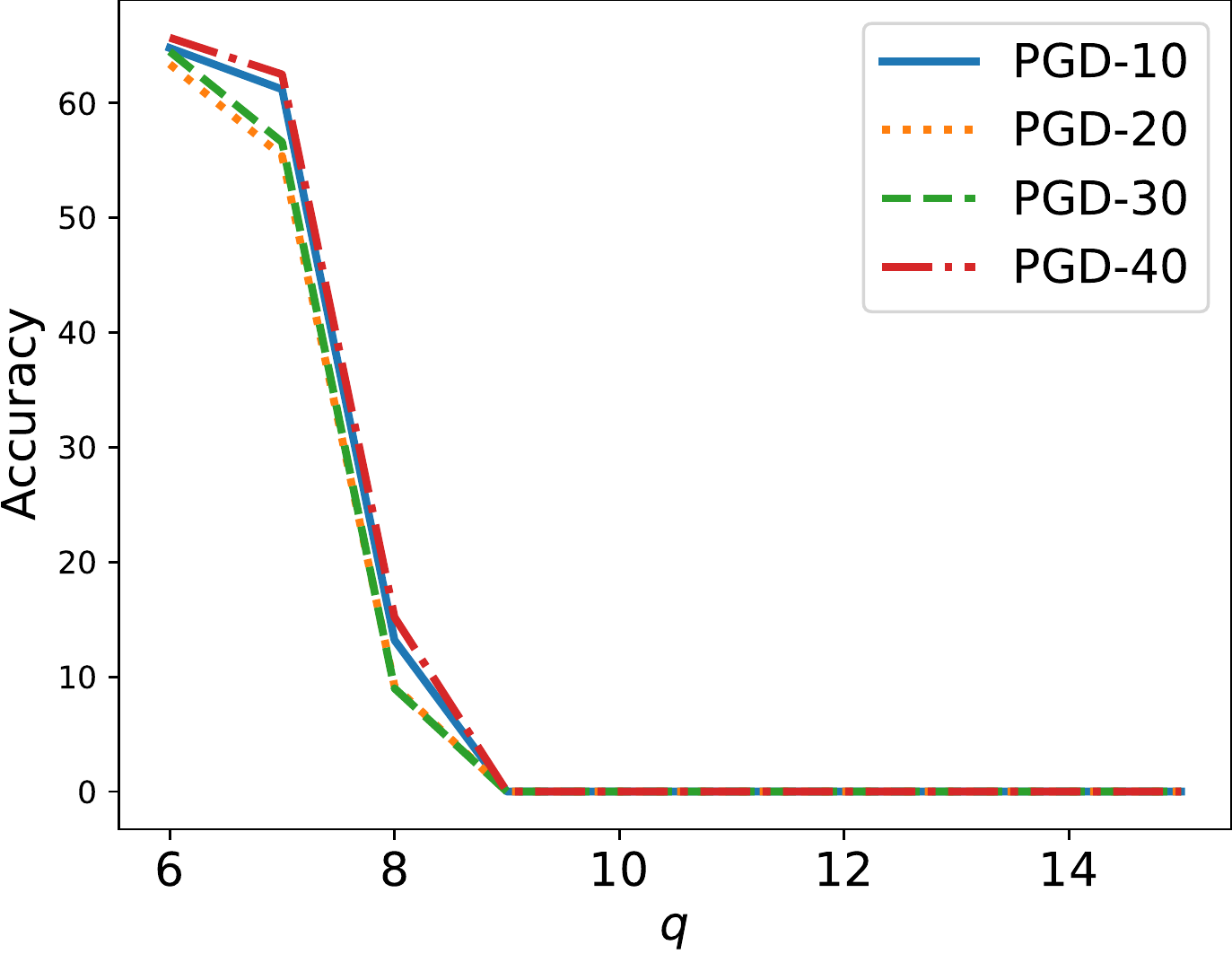}
    \end{minipage}
    \begin{minipage}[t]{0.22\textwidth}
    \includegraphics[width=1.0\textwidth]{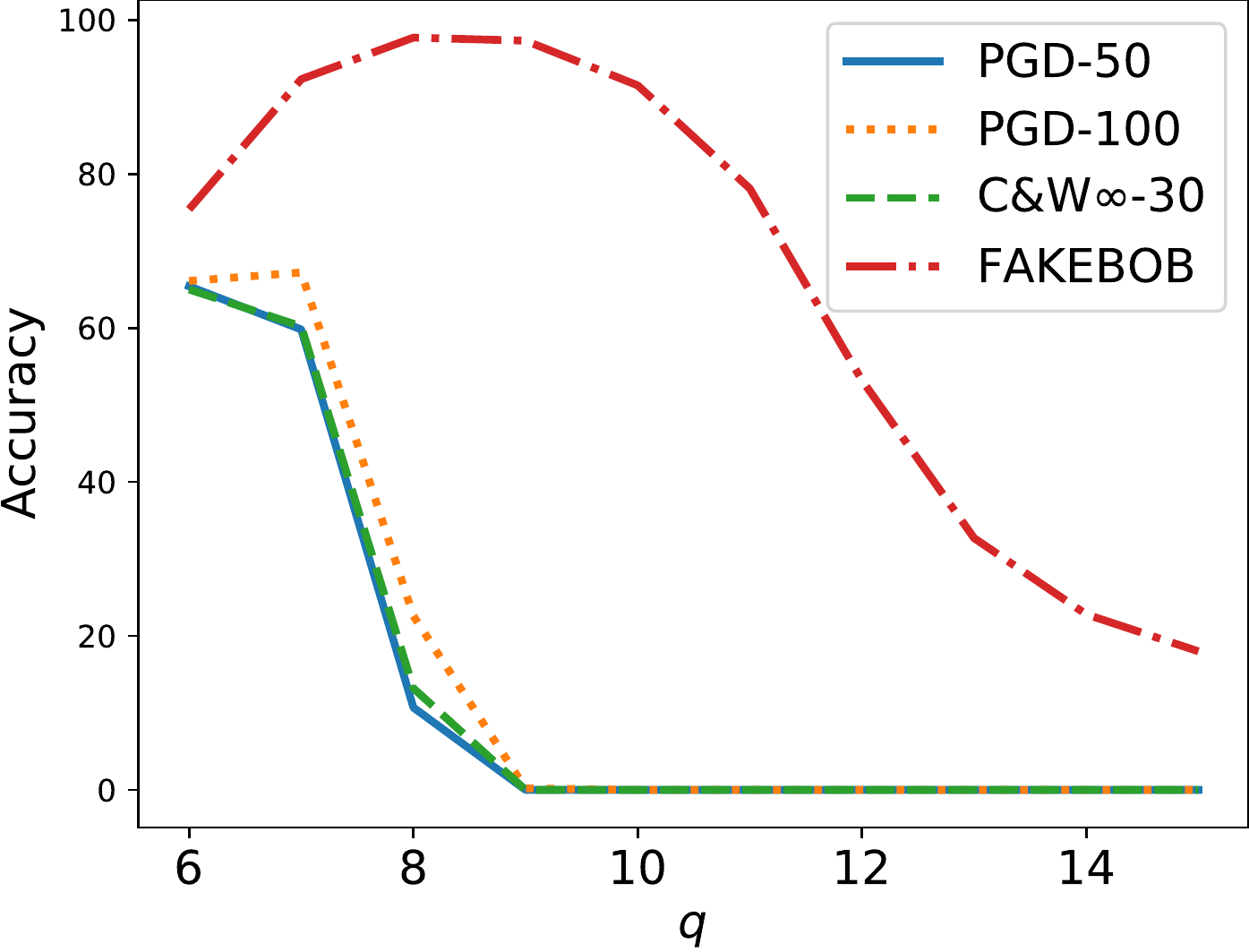}
    \end{minipage}
    \begin{minipage}[t]{0.22\textwidth}
    \includegraphics[width=1.0\textwidth]{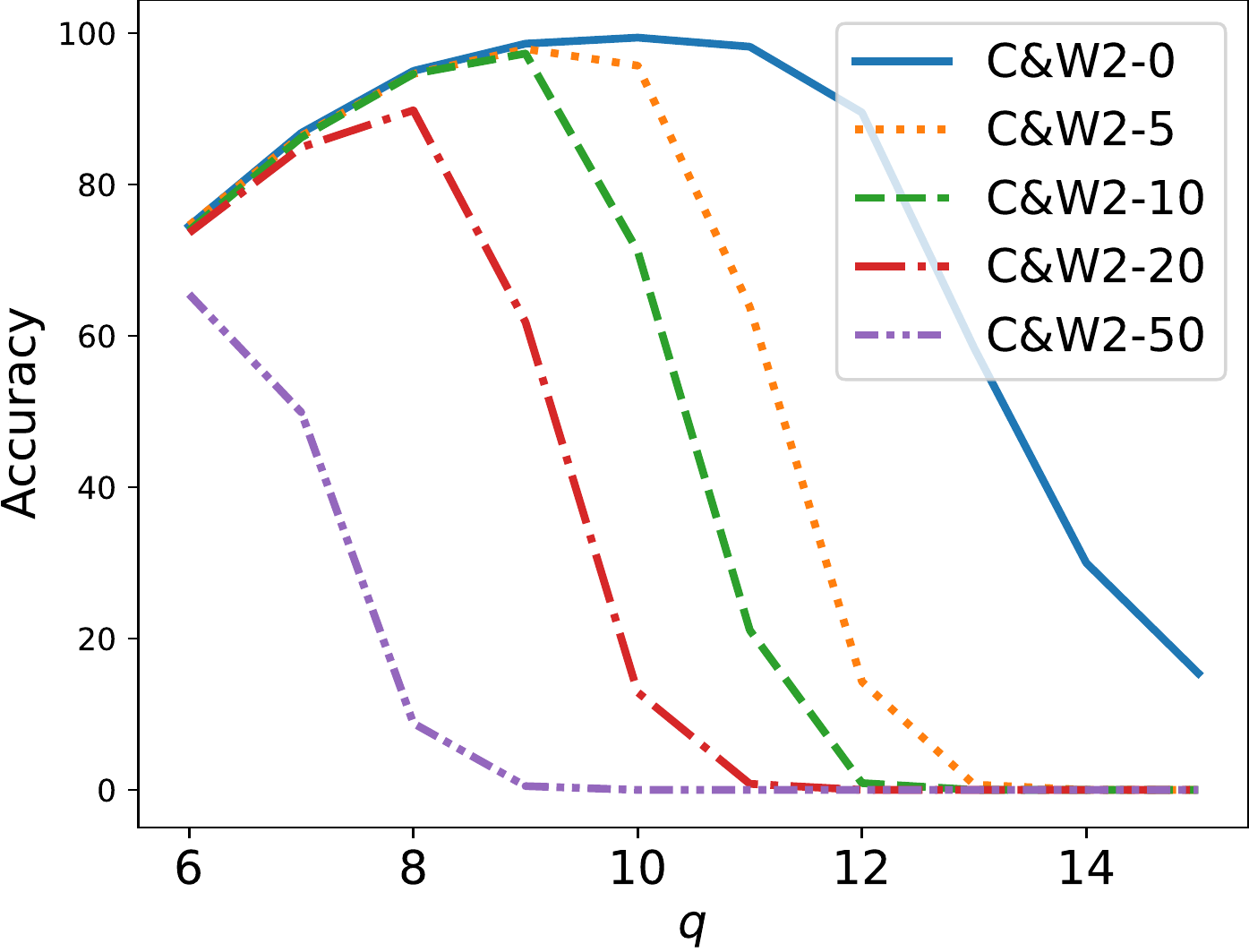}
    \end{minipage}
    }\vspace{-3mm}

    \subfigure[DS]{
    \begin{minipage}[t]{0.22\textwidth}
    \includegraphics[width=1.0\textwidth]{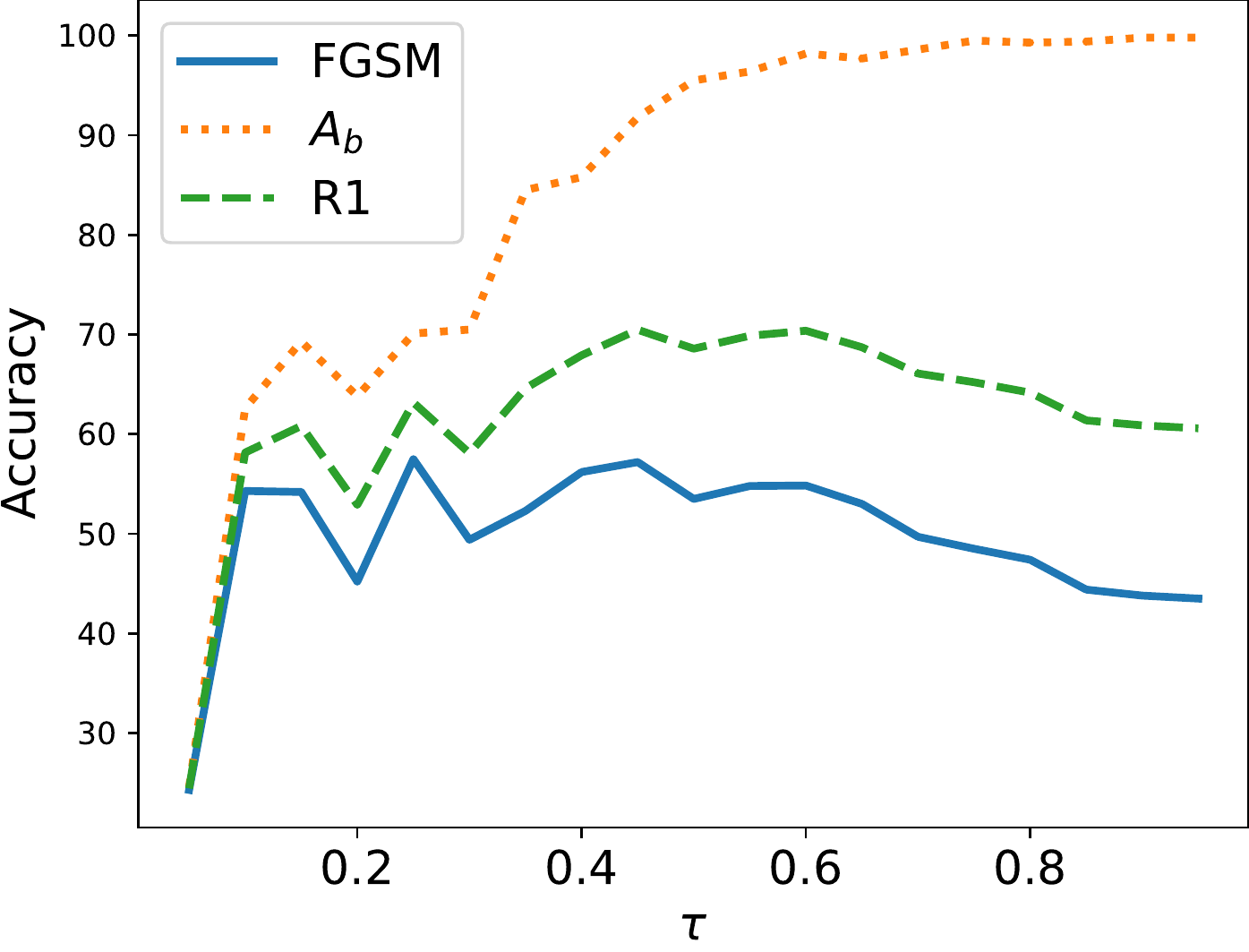}
    \end{minipage}
    \begin{minipage}[t]{0.22\textwidth}
    \includegraphics[width=1.0\textwidth]{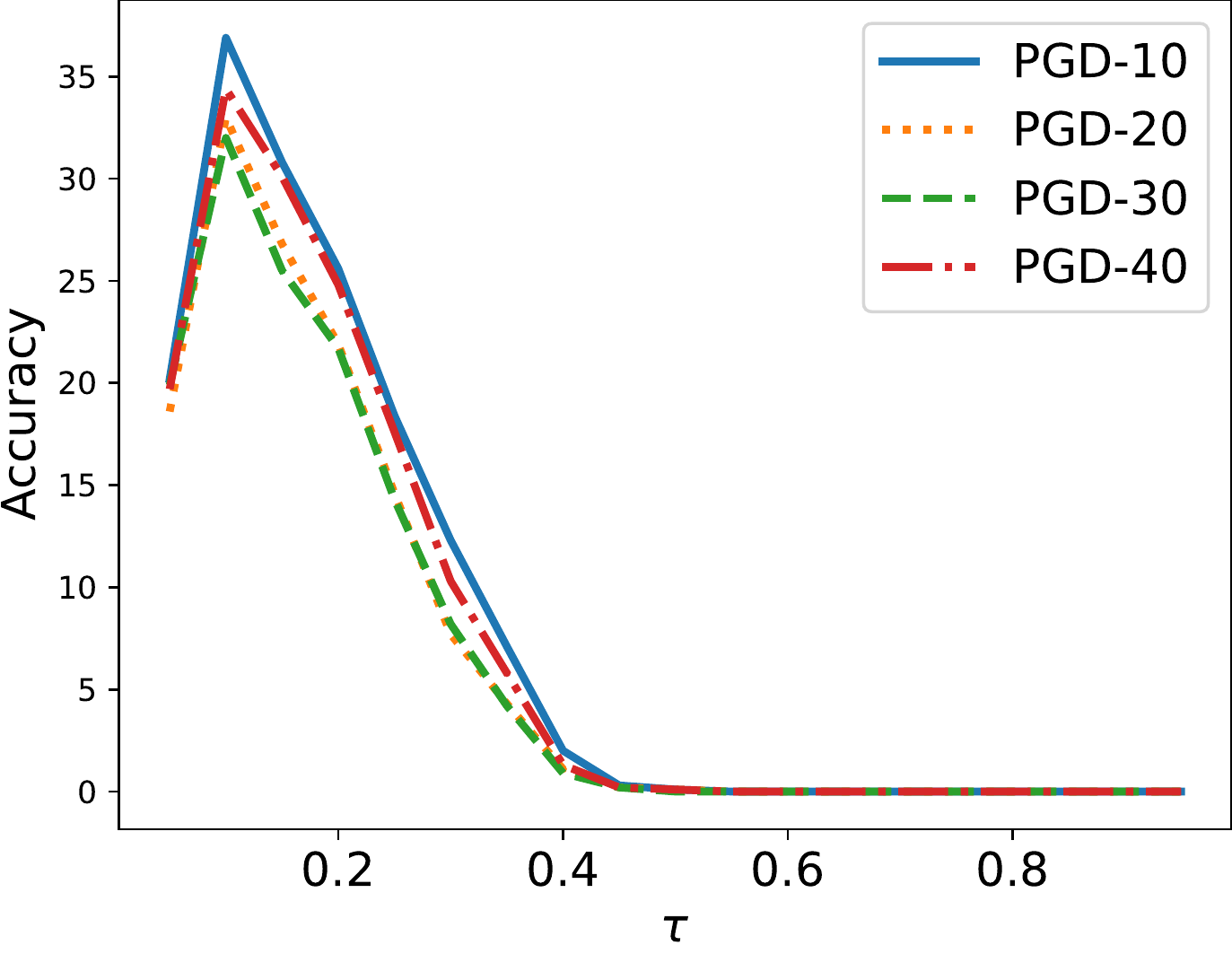}
    \end{minipage}
    \begin{minipage}[t]{0.22\textwidth}
    \includegraphics[width=1.0\textwidth]{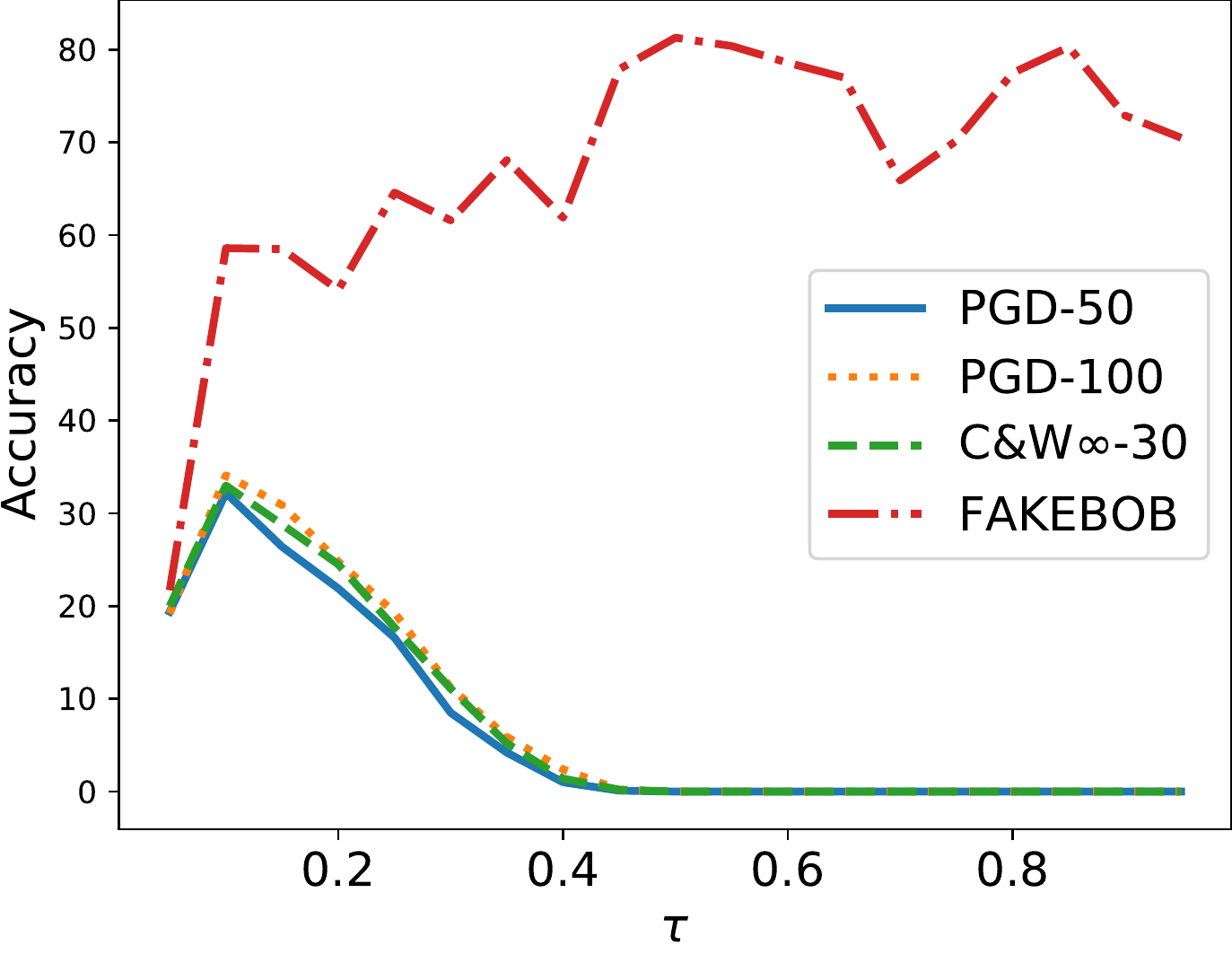}
    \end{minipage}
    \begin{minipage}[t]{0.22\textwidth}
    \includegraphics[width=1.0\textwidth]{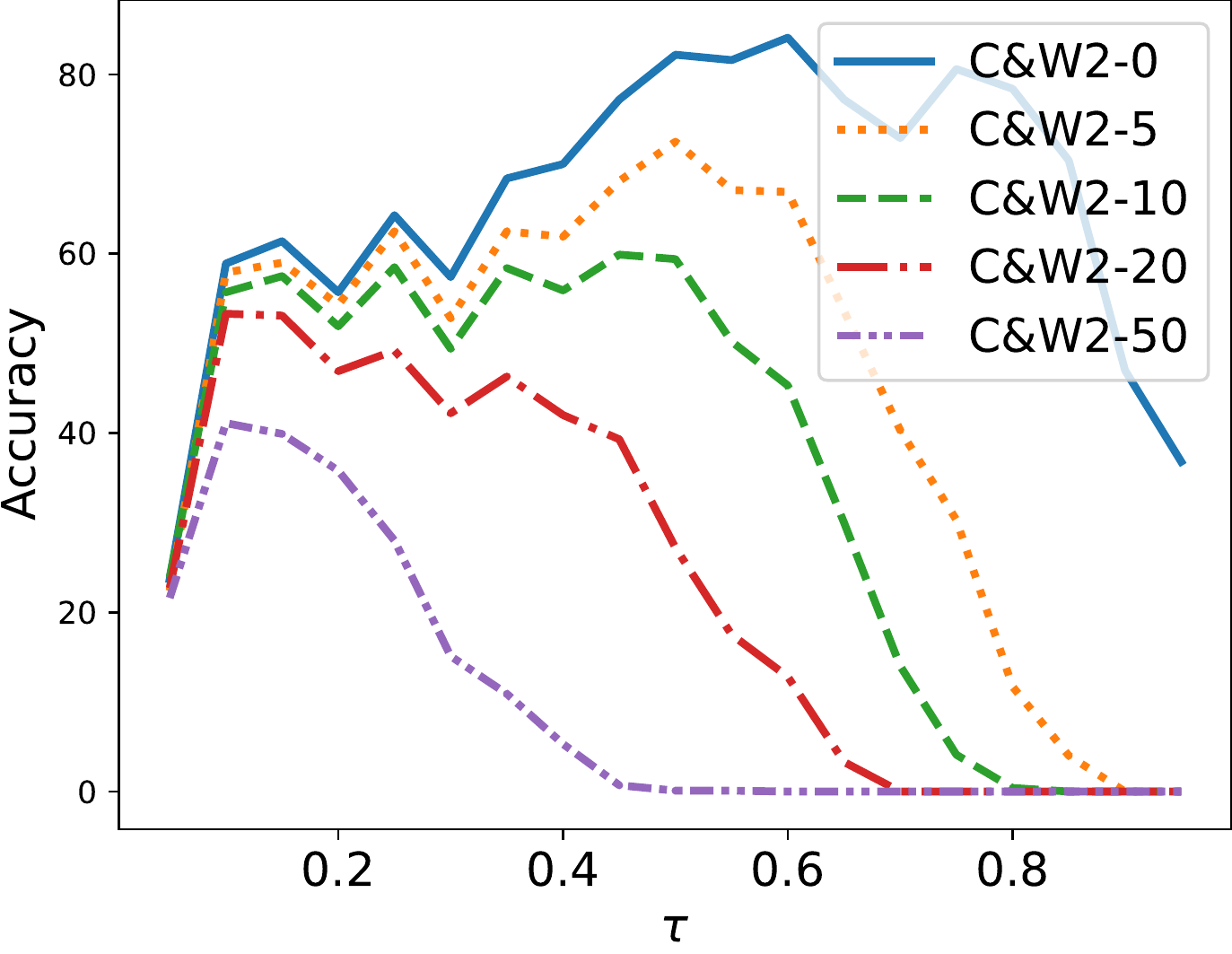}
    \end{minipage}
    }\vspace{-3mm}

    \subfigure[AT]{
    \begin{minipage}[t]{0.22\textwidth}
    \includegraphics[width=1.0\textwidth]{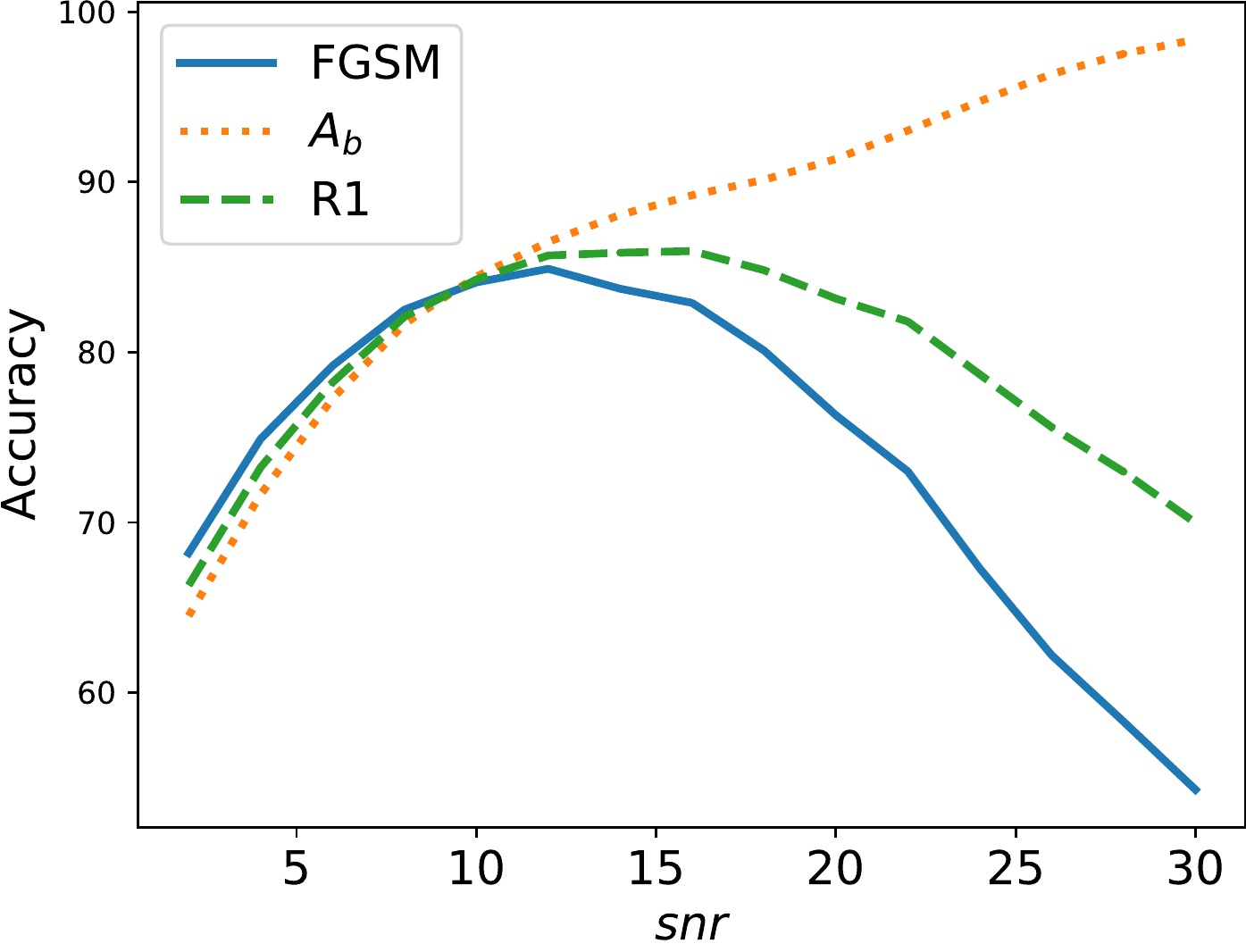}
    \end{minipage}
    \begin{minipage}[t]{0.22\textwidth}
    \includegraphics[width=1.0\textwidth]{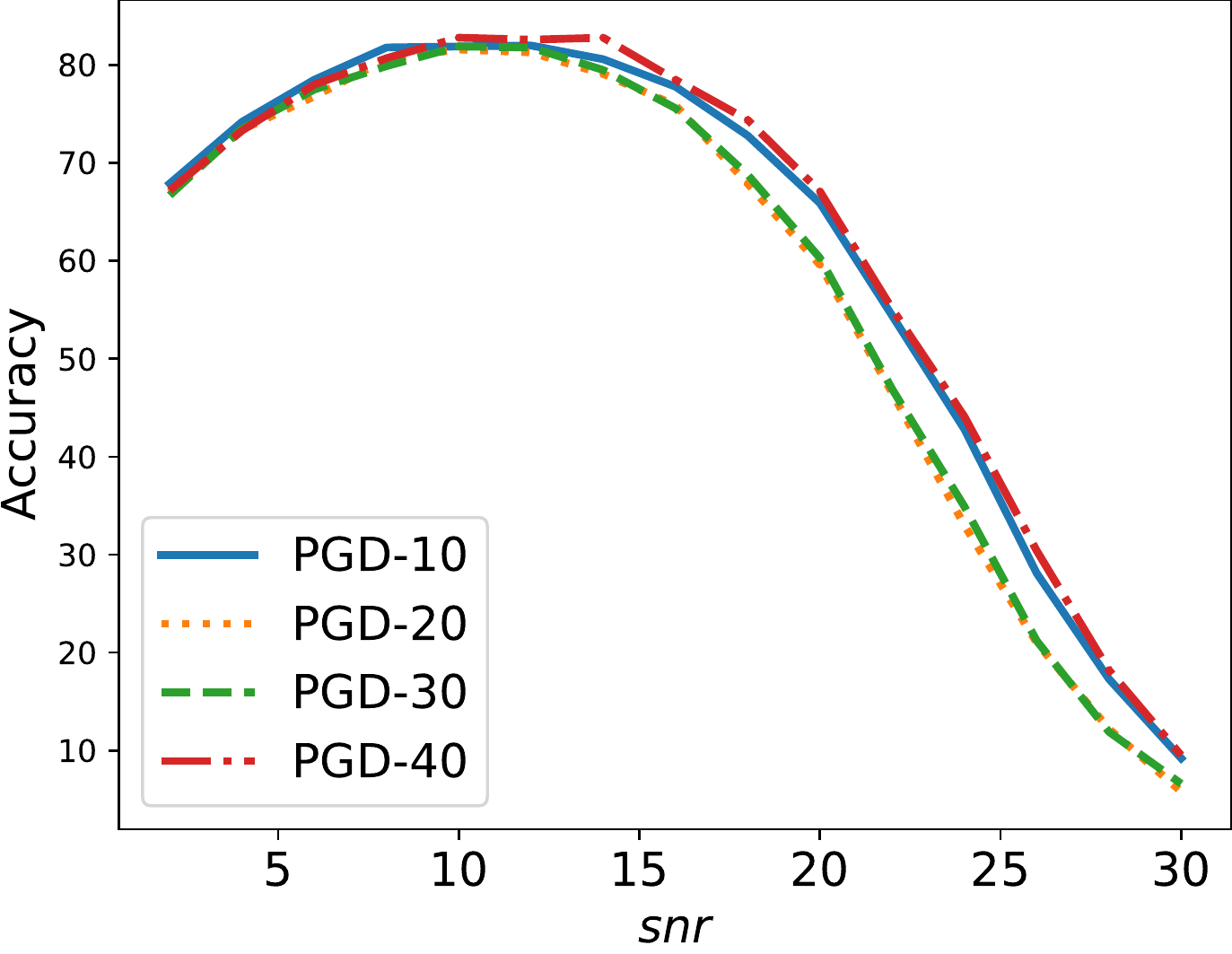}
    \end{minipage}
    \begin{minipage}[t]{0.22\textwidth}
    \includegraphics[width=1.0\textwidth]{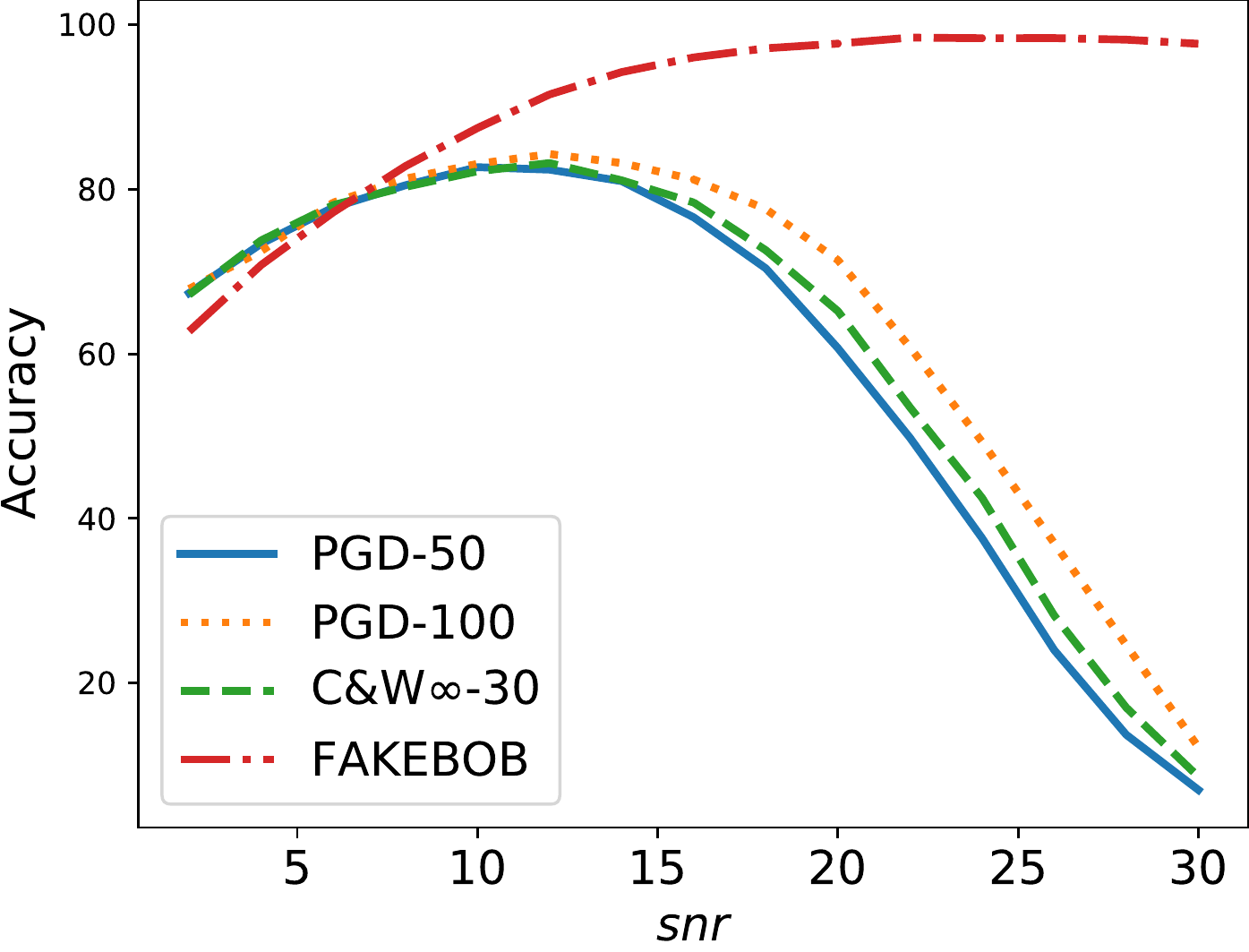}
    \end{minipage}
    \begin{minipage}[t]{0.22\textwidth}
    \includegraphics[width=1.0\textwidth]{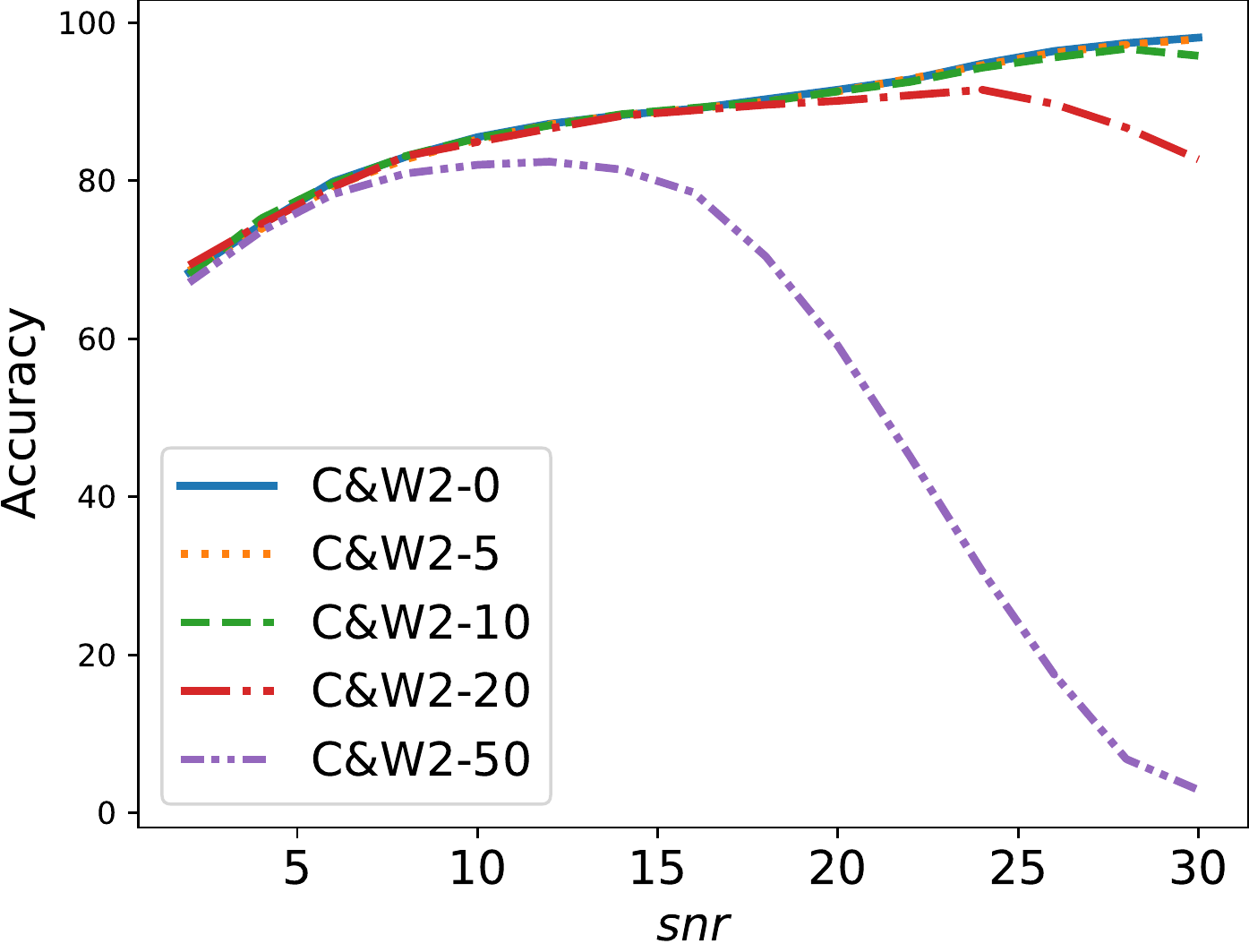}
    \end{minipage}
    }\vspace{-3mm}

    \subfigure[AS]{
    \begin{minipage}[t]{0.22\textwidth}
    \includegraphics[width=1.0\textwidth]{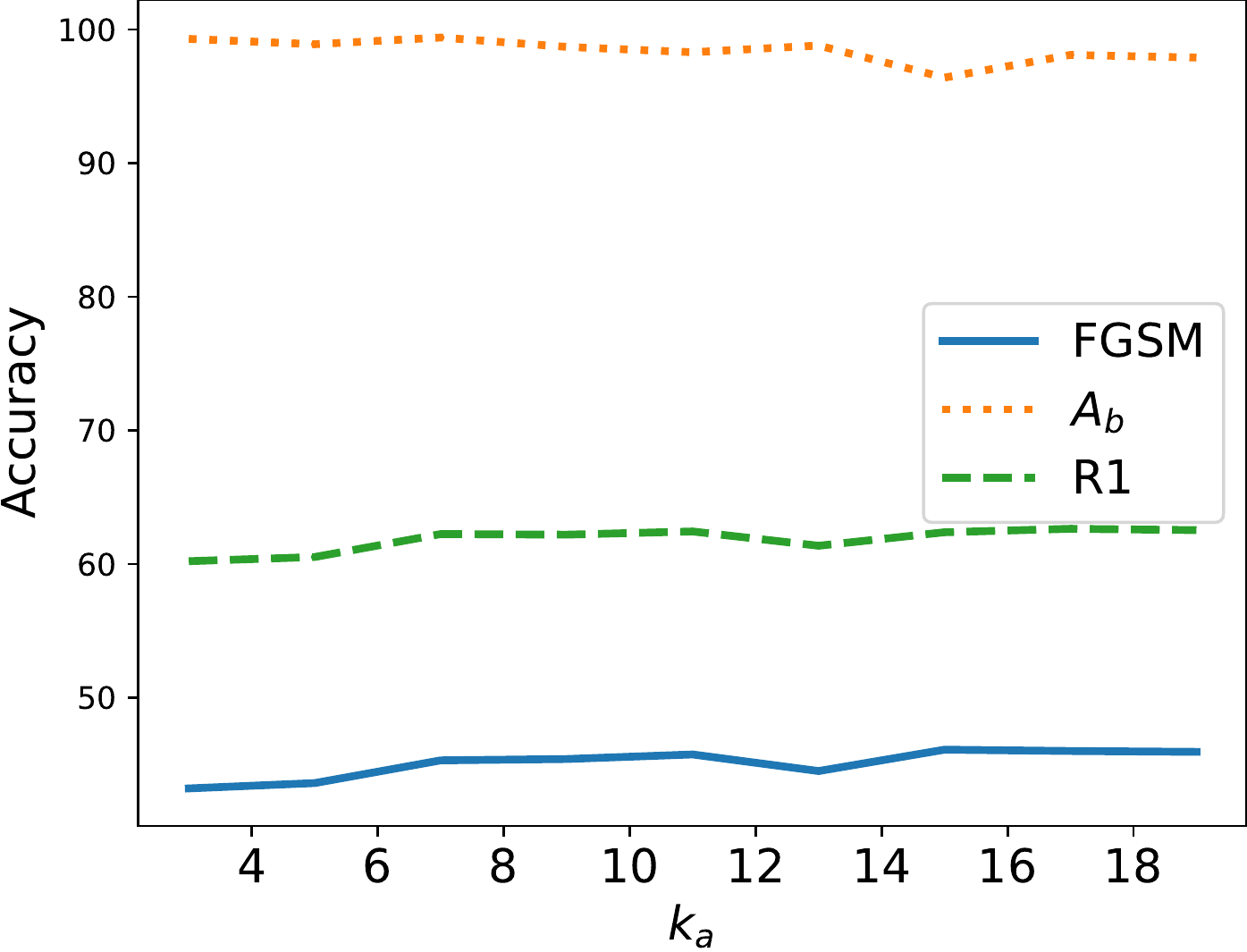}
    \end{minipage}
    \begin{minipage}[t]{0.22\textwidth}
    \includegraphics[width=1.0\textwidth]{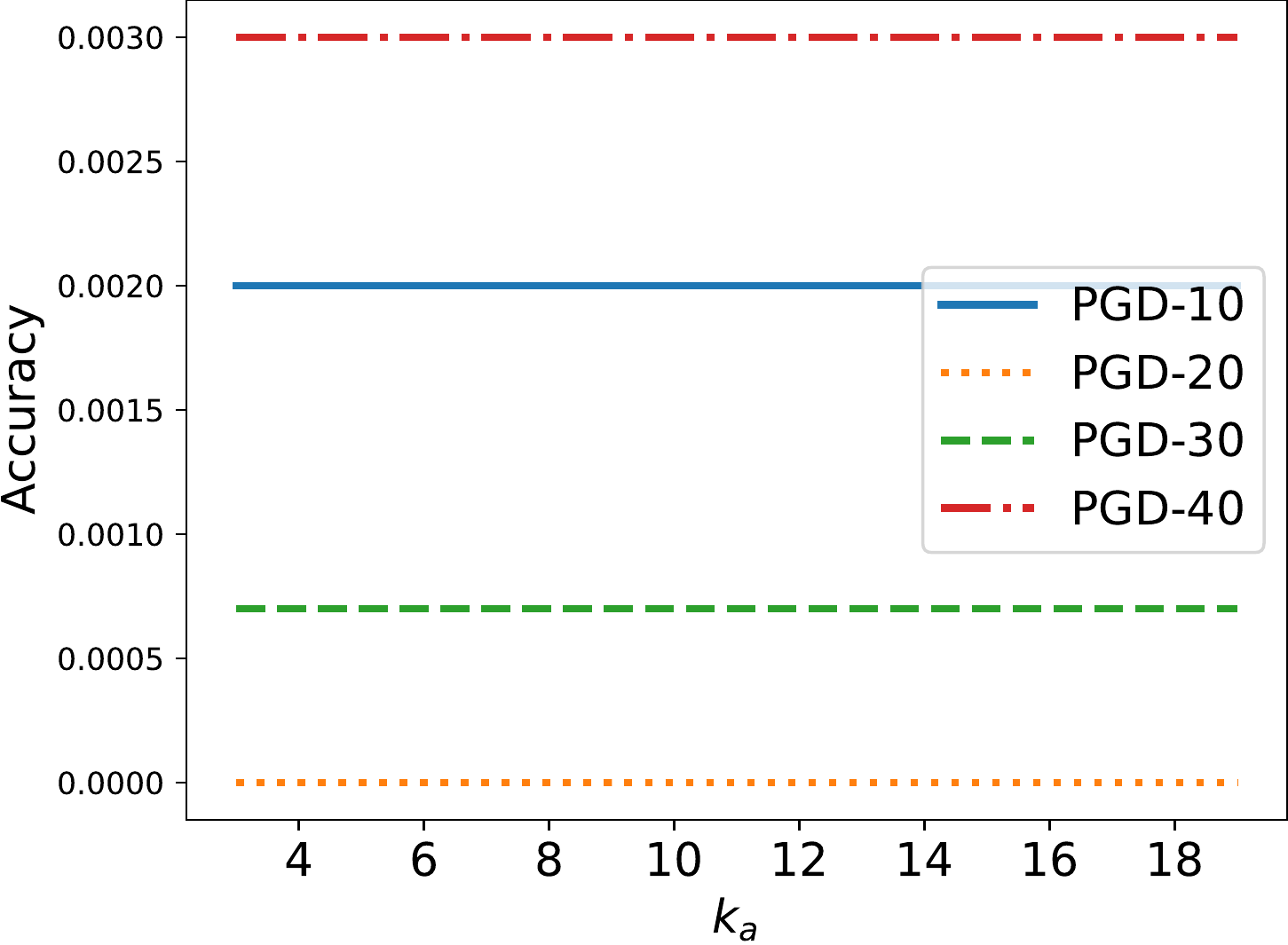}
    \end{minipage}
    \begin{minipage}[t]{0.22\textwidth}
    \includegraphics[width=1.0\textwidth]{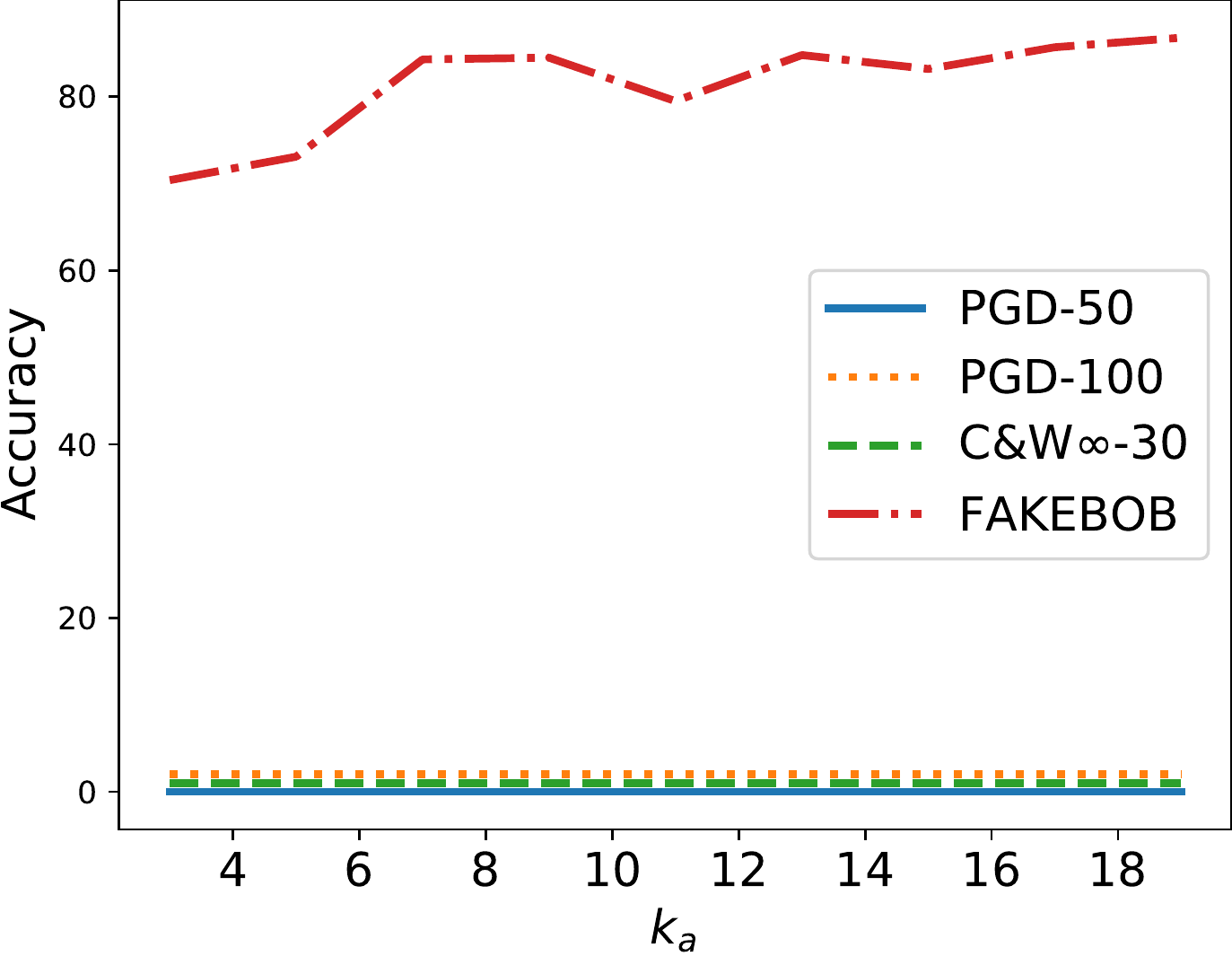}
    \end{minipage}
    \begin{minipage}[t]{0.22\textwidth}
    \includegraphics[width=1.0\textwidth]{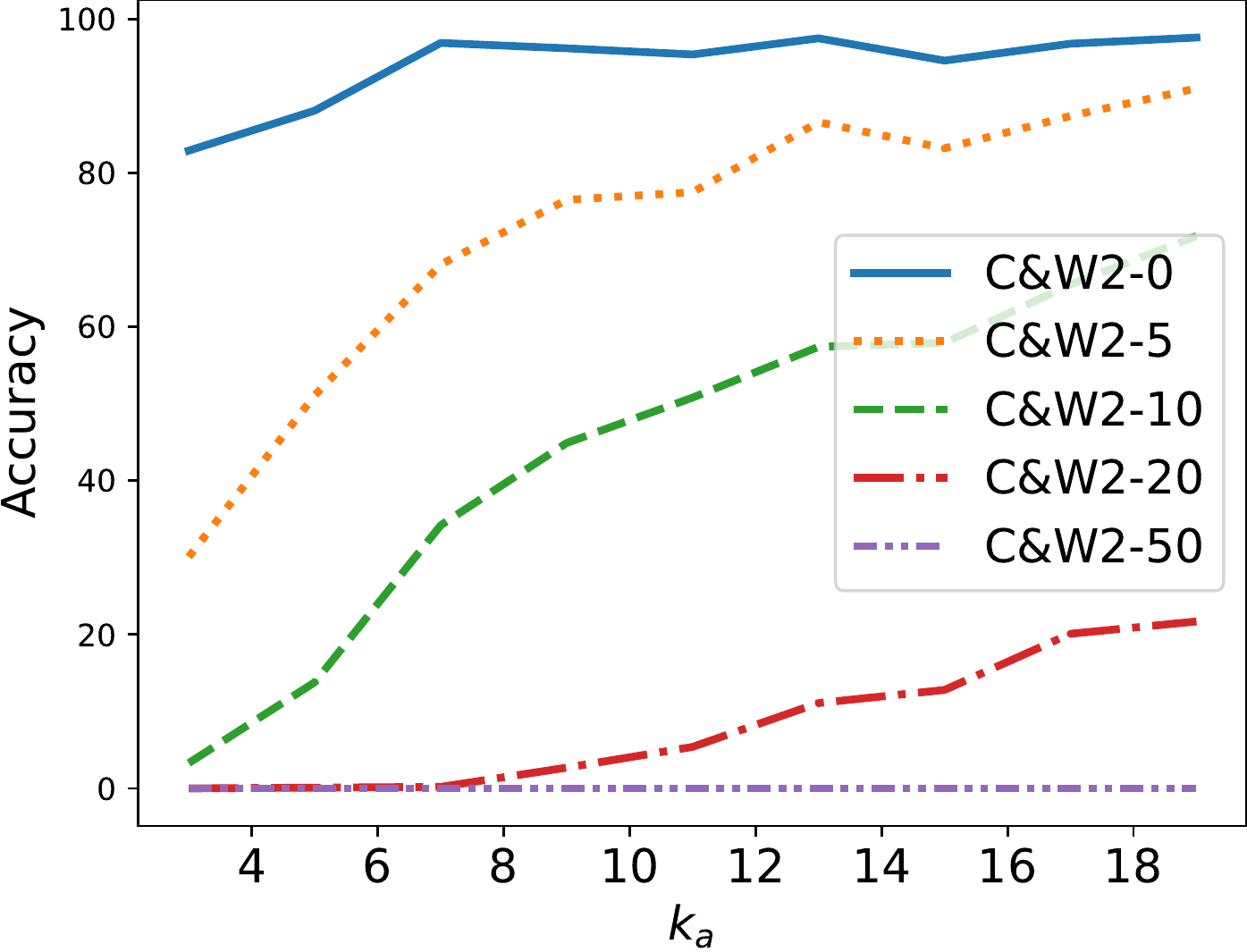}
    \end{minipage}
    }\vspace{-3mm}

    \subfigure[MS]{
    \begin{minipage}[t]{0.22\textwidth}
    \includegraphics[width=1.0\textwidth]{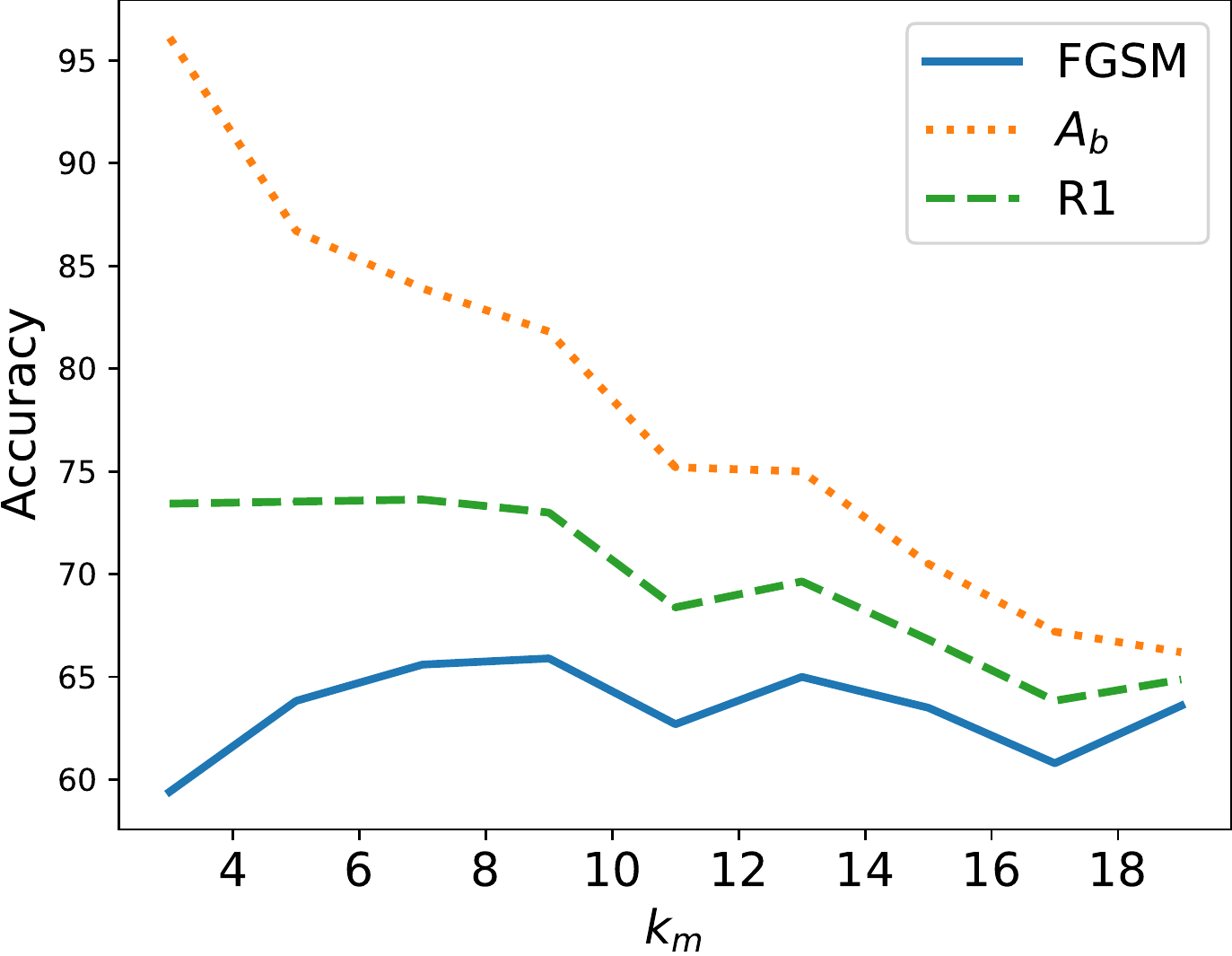}
    \end{minipage}
    \begin{minipage}[t]{0.22\textwidth}
    \includegraphics[width=1.0\textwidth]{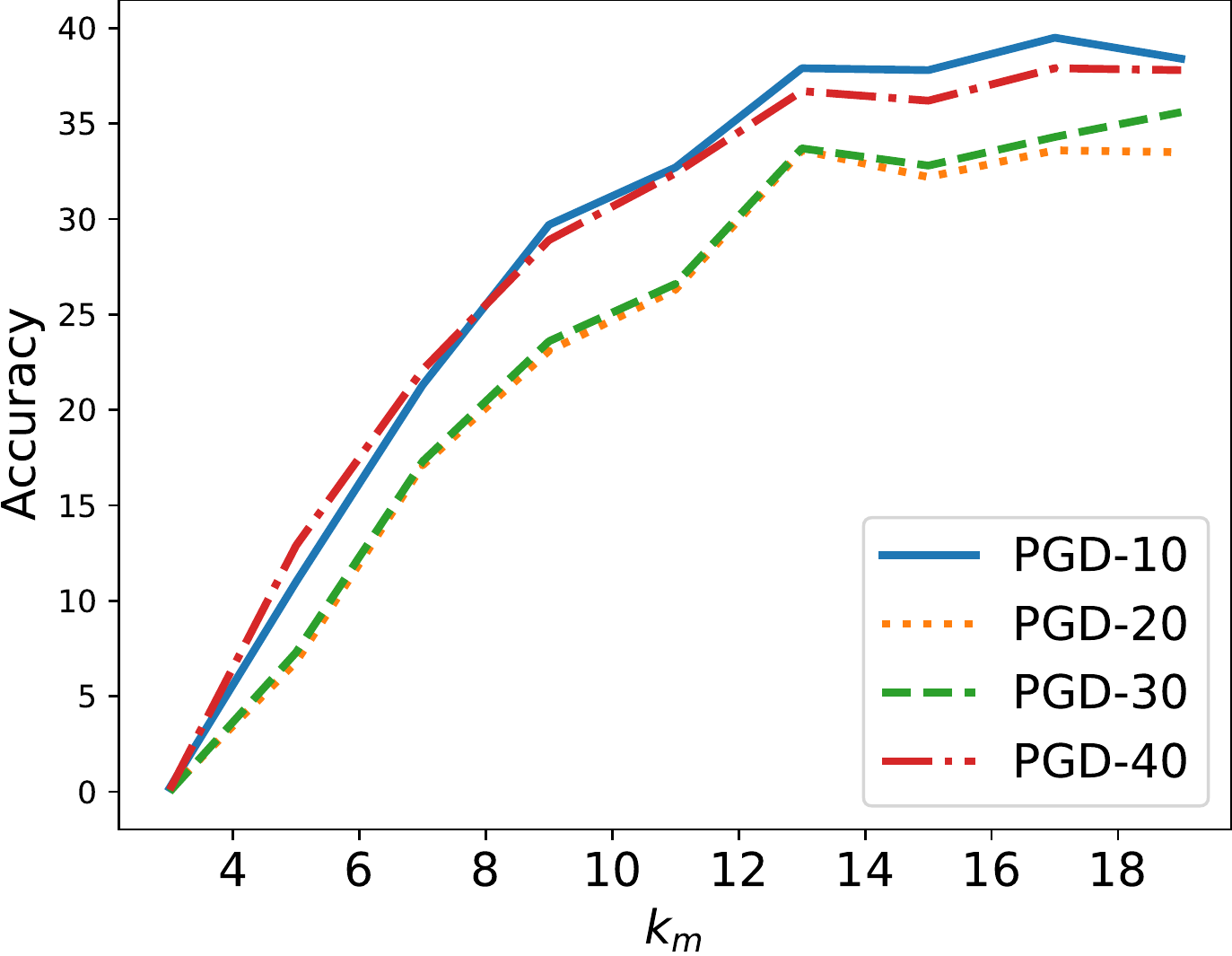}
    \end{minipage}
    \begin{minipage}[t]{0.22\textwidth}
    \includegraphics[width=1.0\textwidth]{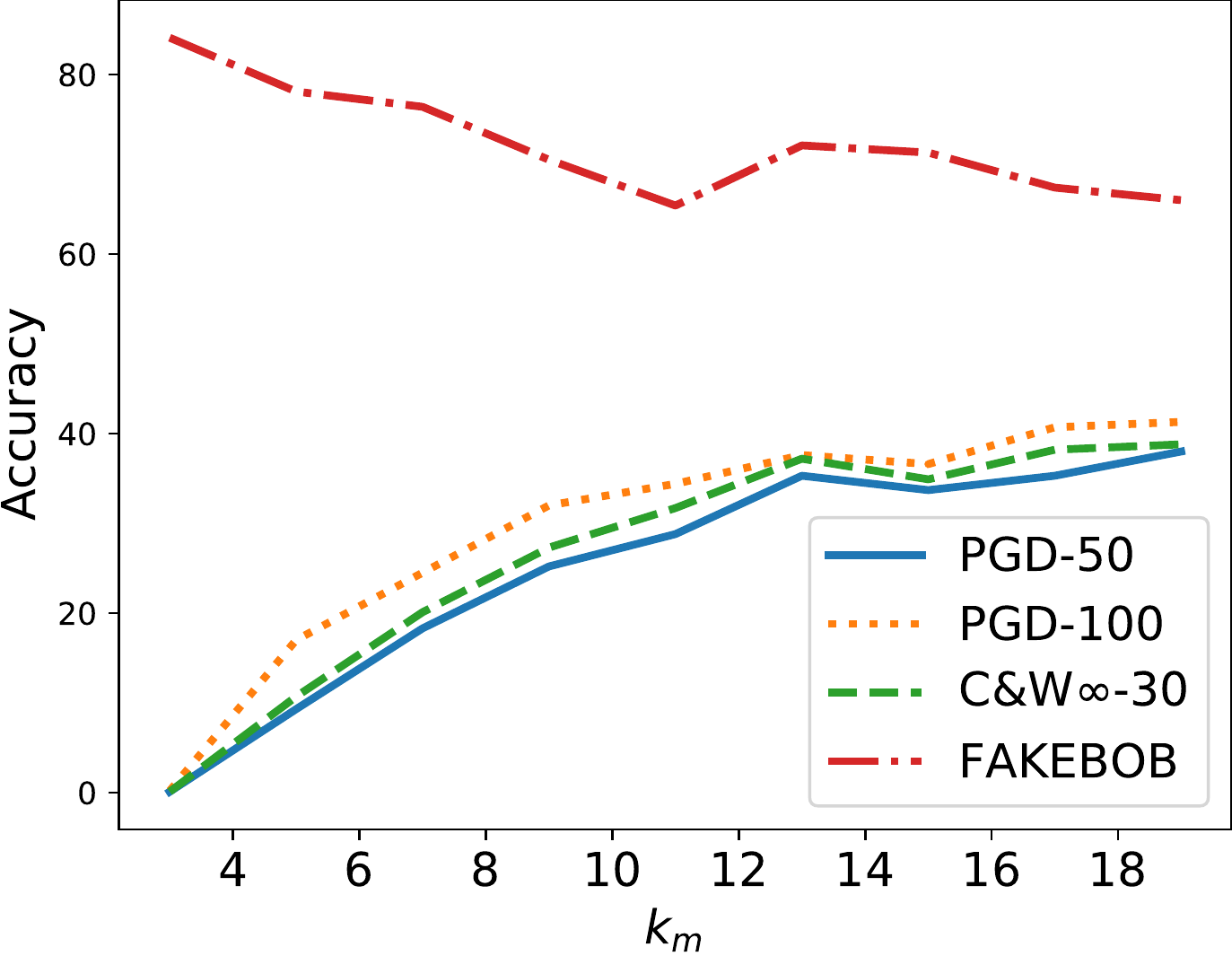}
    \end{minipage}
    \begin{minipage}[t]{0.22\textwidth}
    \includegraphics[width=1.0\textwidth]{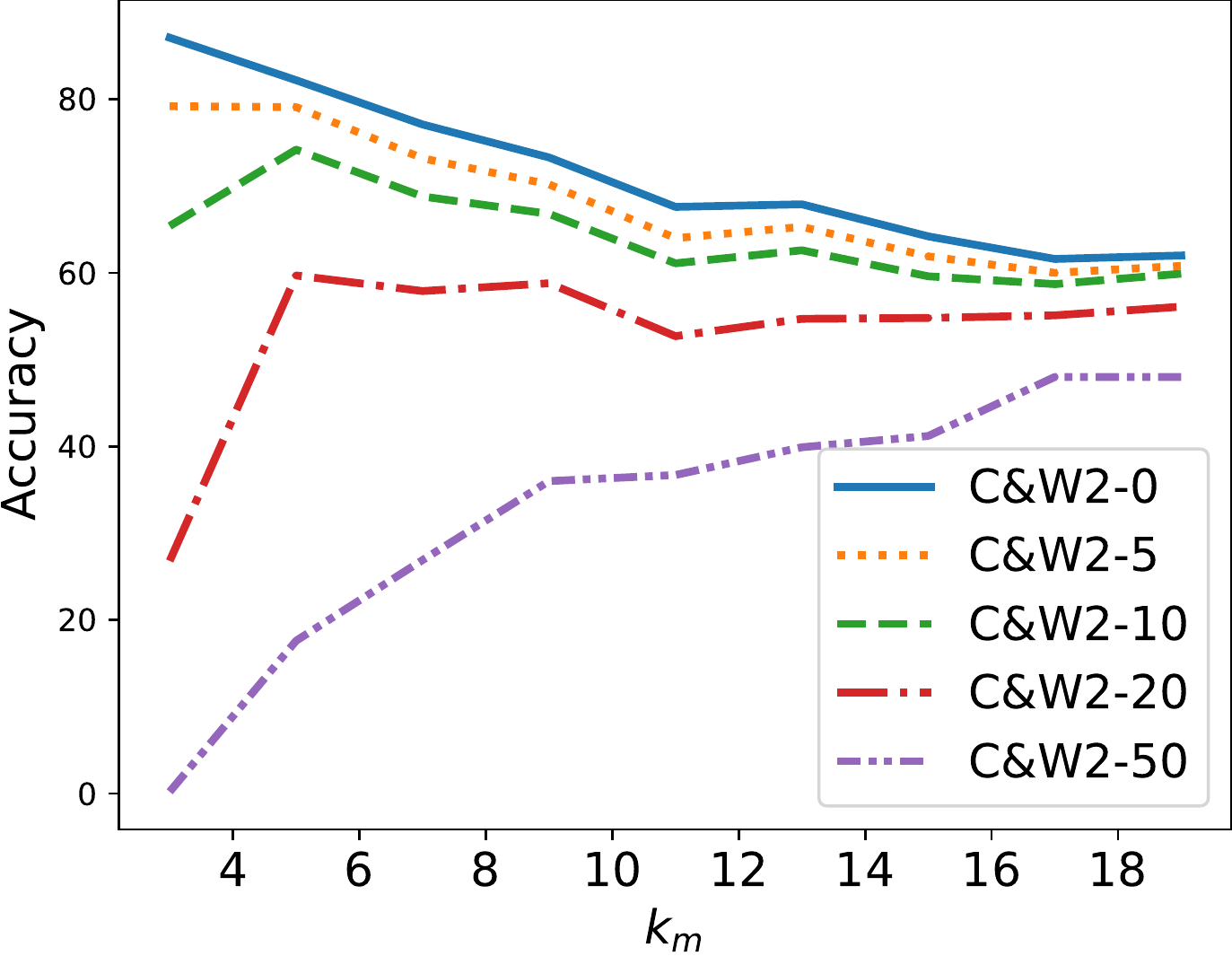}
    \end{minipage}
    }\vspace{-3mm}

    \subfigure[LPF]{
    \begin{minipage}[t]{0.22\textwidth}
    \includegraphics[width=1.0\textwidth]{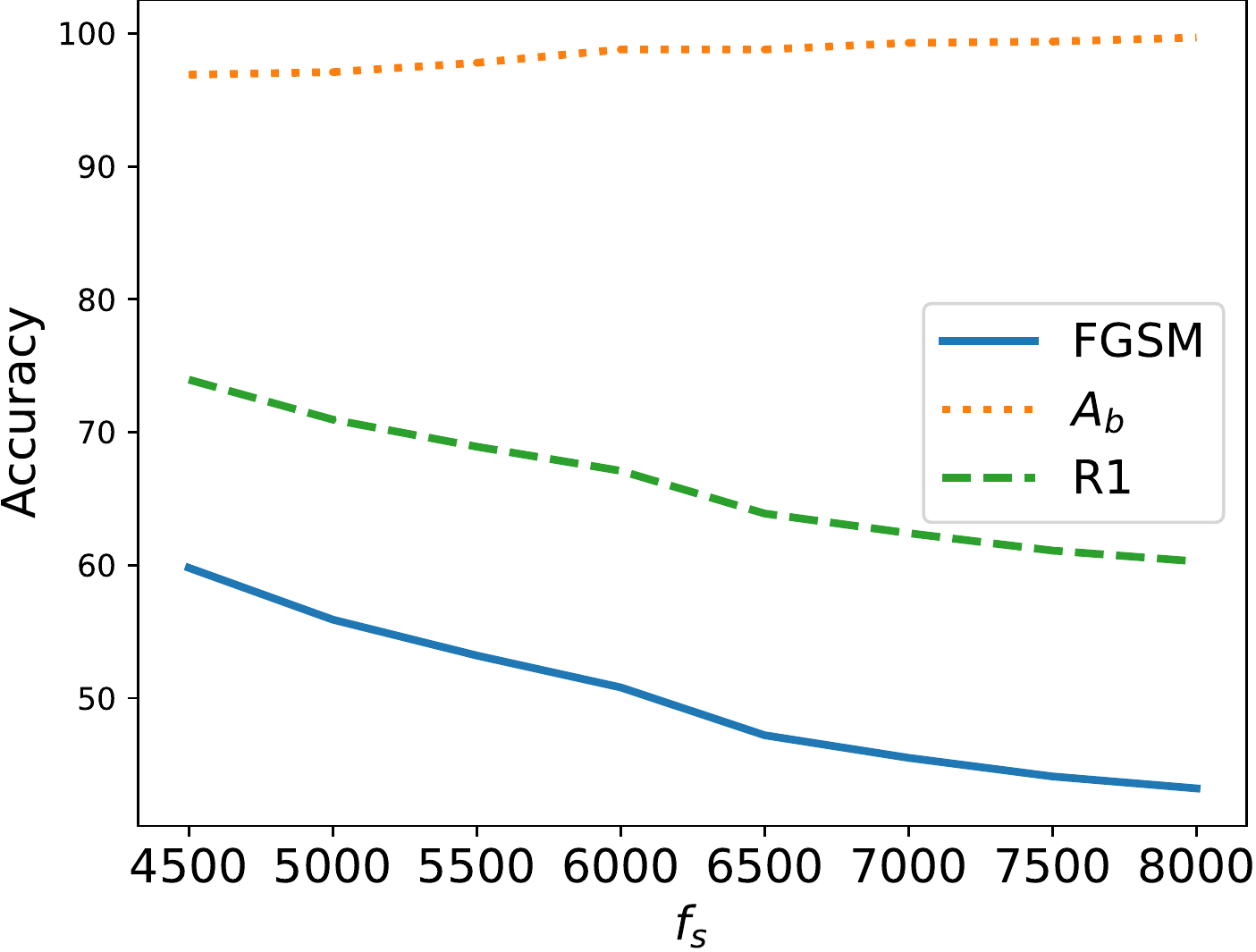}
    \end{minipage}
    \begin{minipage}[t]{0.22\textwidth}
    \includegraphics[width=1.0\textwidth]{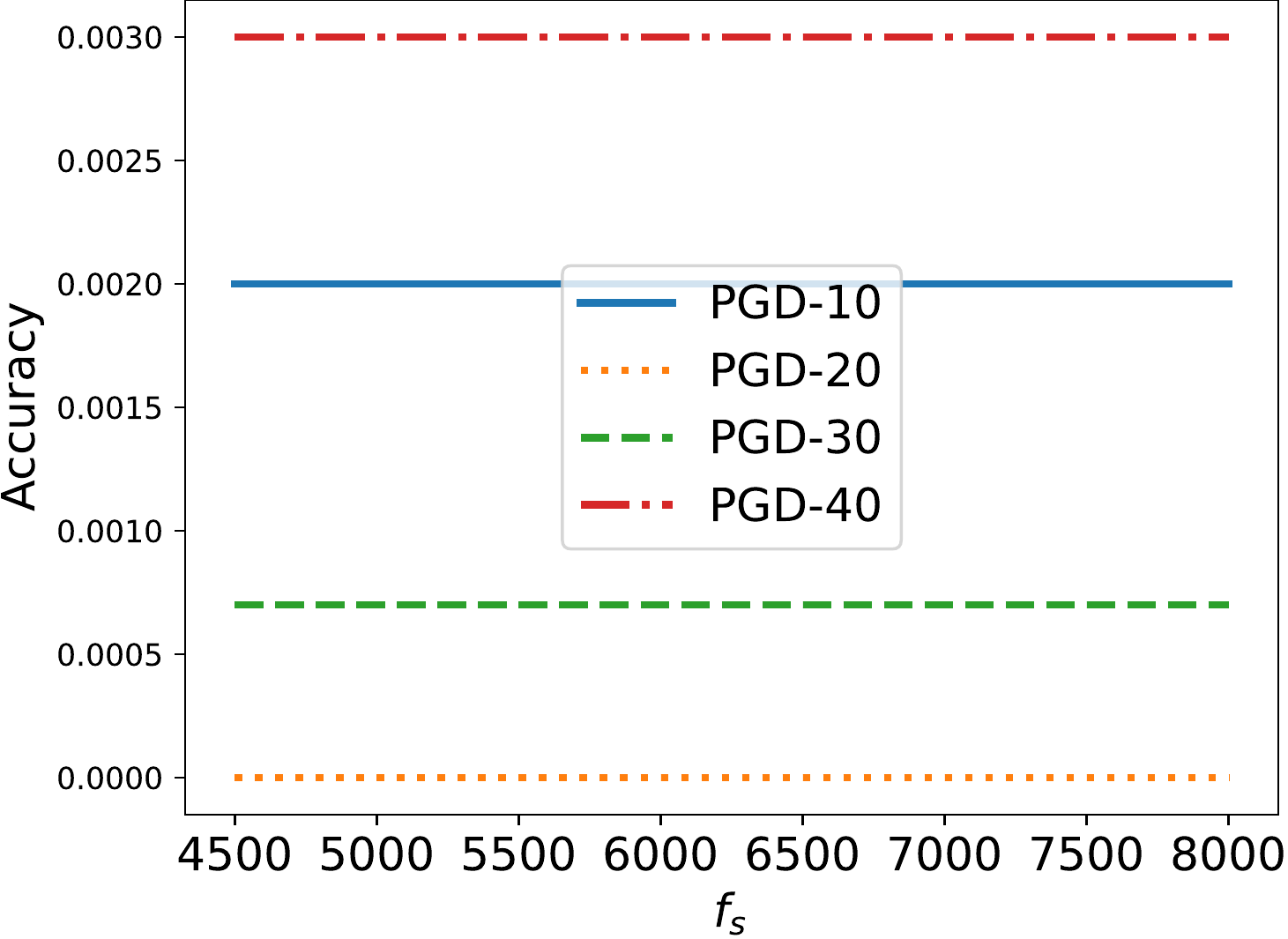}
    \end{minipage}
    \begin{minipage}[t]{0.22\textwidth}
    \includegraphics[width=1.0\textwidth]{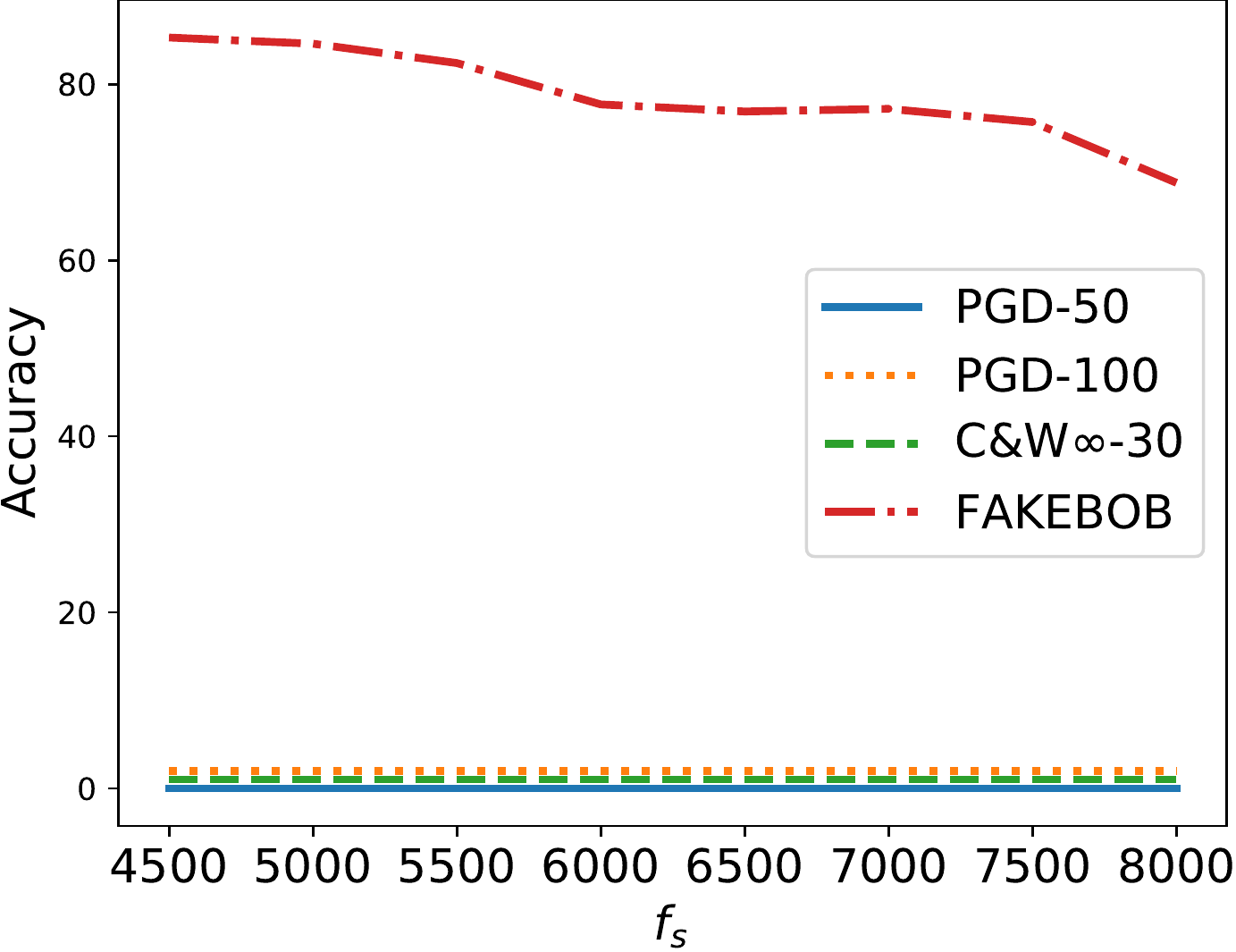}
    \end{minipage}
    \begin{minipage}[t]{0.22\textwidth}
    \includegraphics[width=1.0\textwidth]{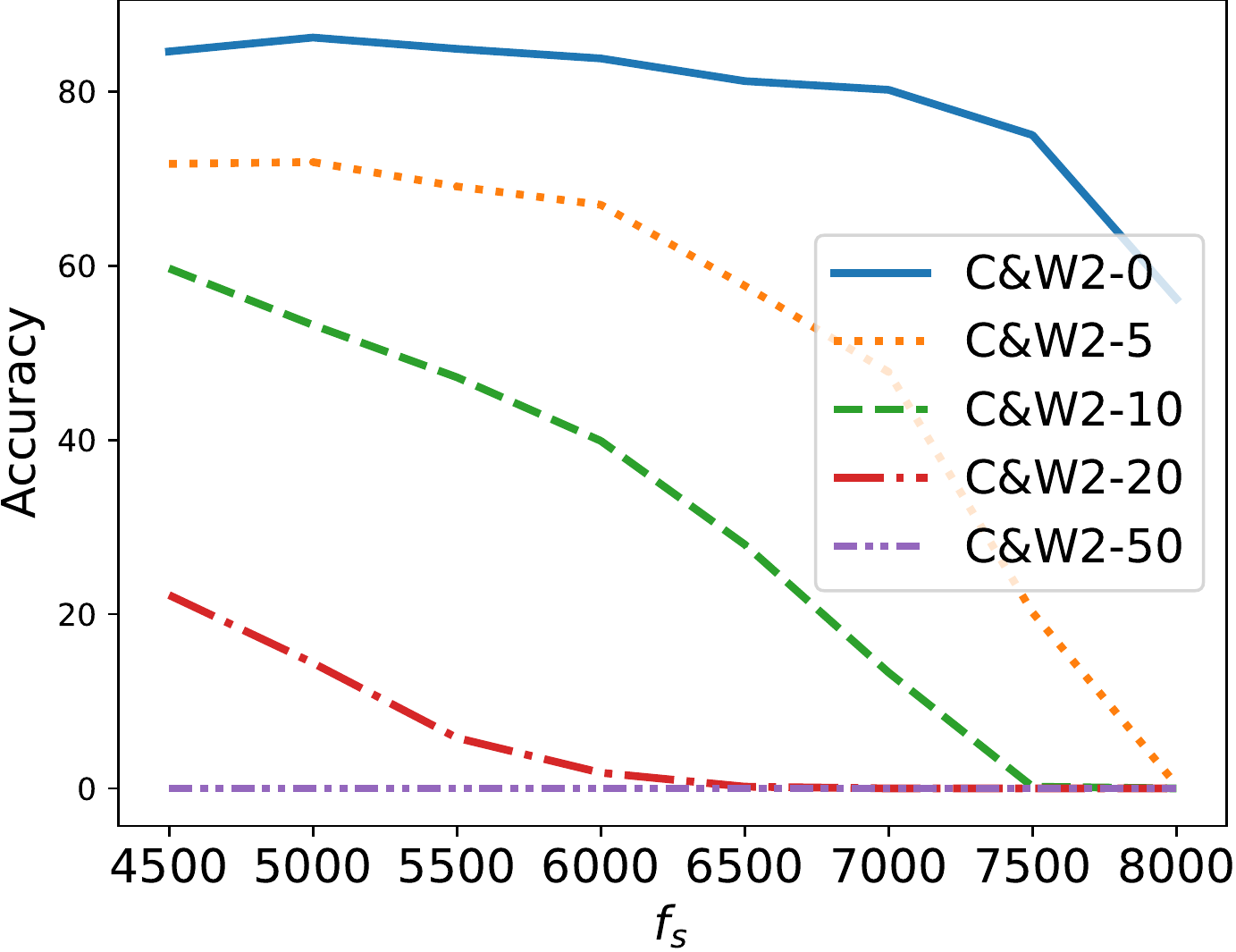}
    \end{minipage}
    }\vspace{-3mm}
    \caption{The performance of input transformations vs. parameter values.}
    \label{fig:parameter-1}
\end{figure*}

\begin{figure*}
  \centering

  \subfigure[BPF]{
  \begin{minipage}[t]{0.22\textwidth}
  \includegraphics[width=1.0\textwidth]{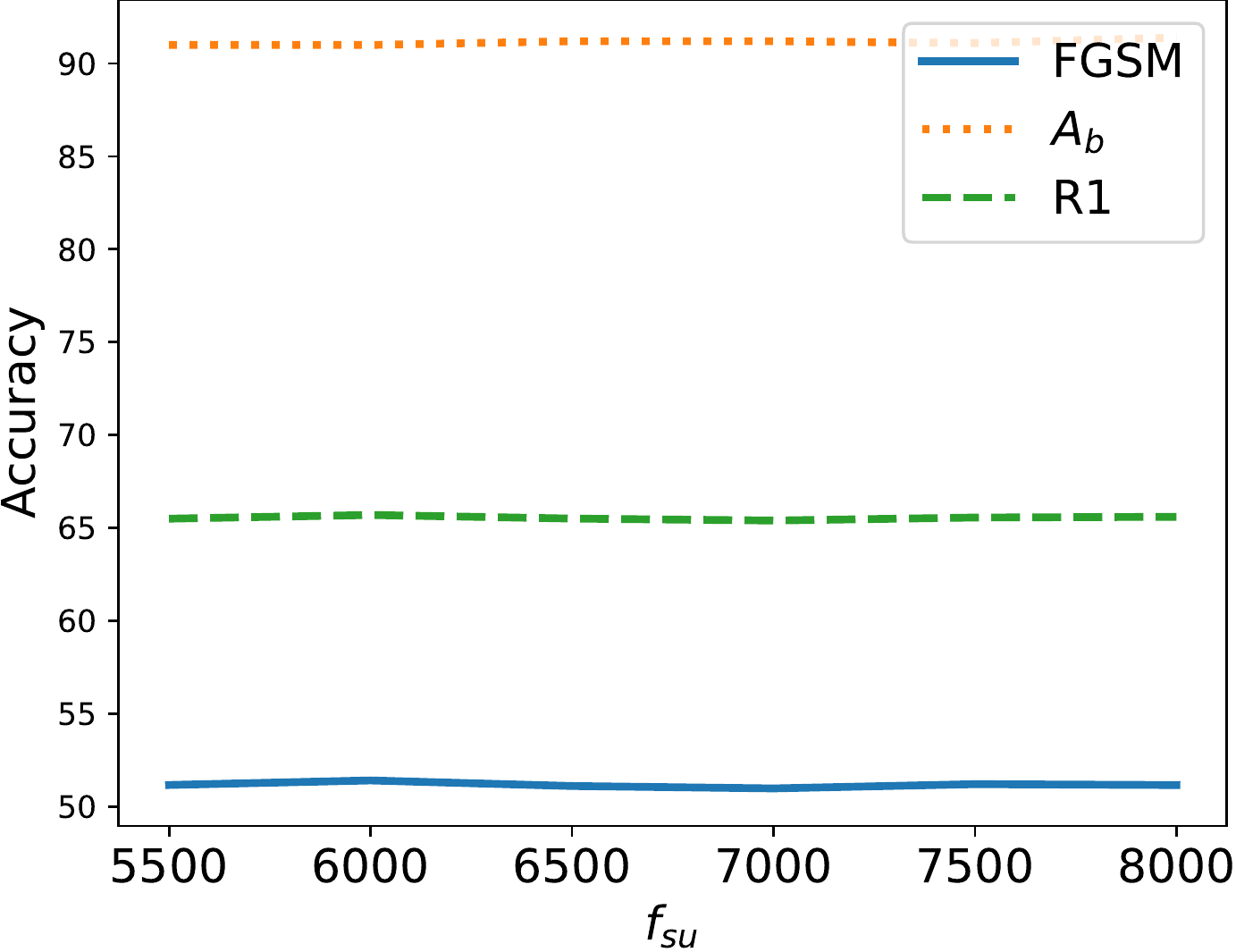}
  \end{minipage}
  \begin{minipage}[t]{0.22\textwidth}
  \includegraphics[width=1.0\textwidth]{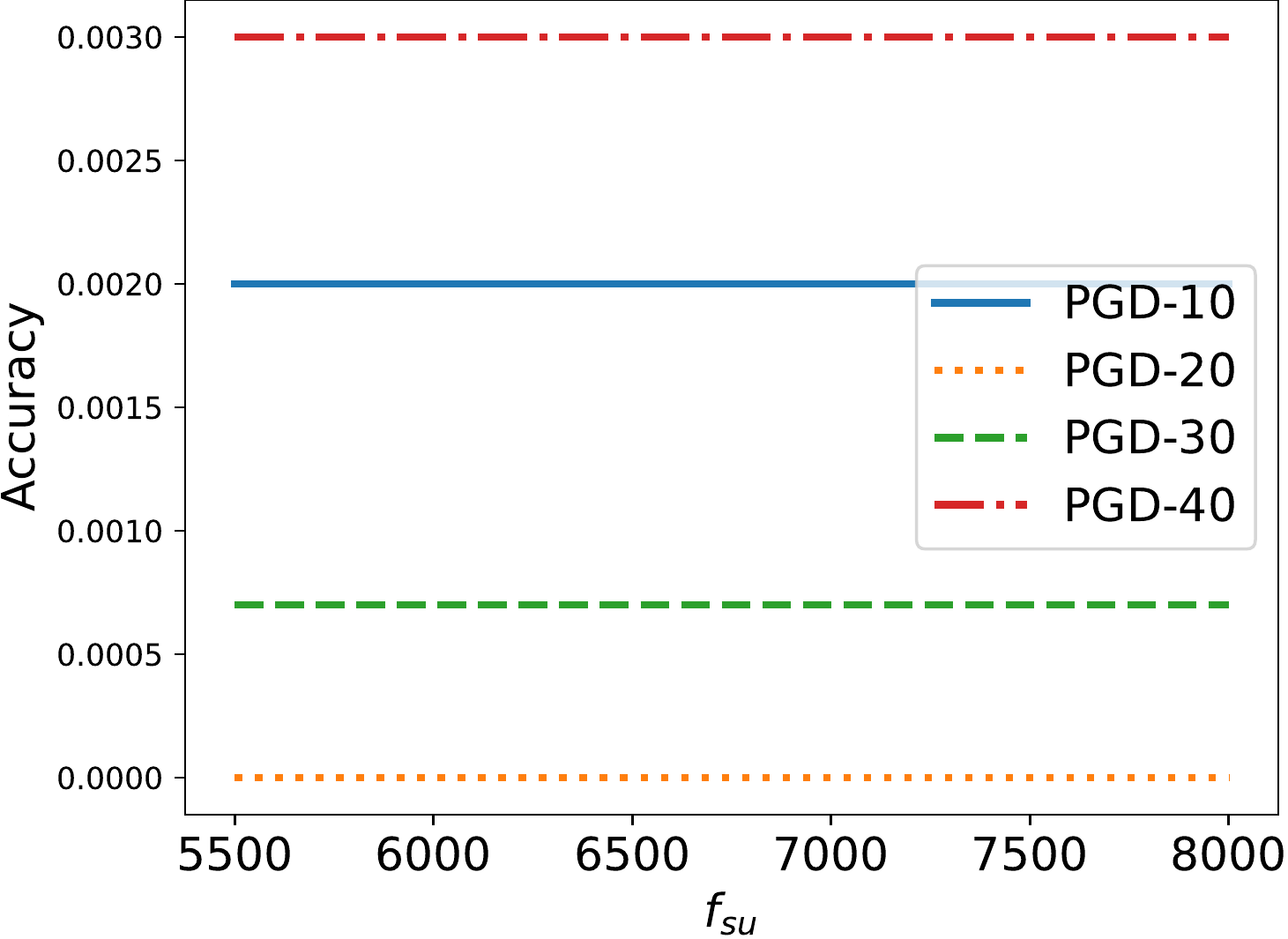}
  \end{minipage}
  \begin{minipage}[t]{0.22\textwidth}
  \includegraphics[width=1.0\textwidth]{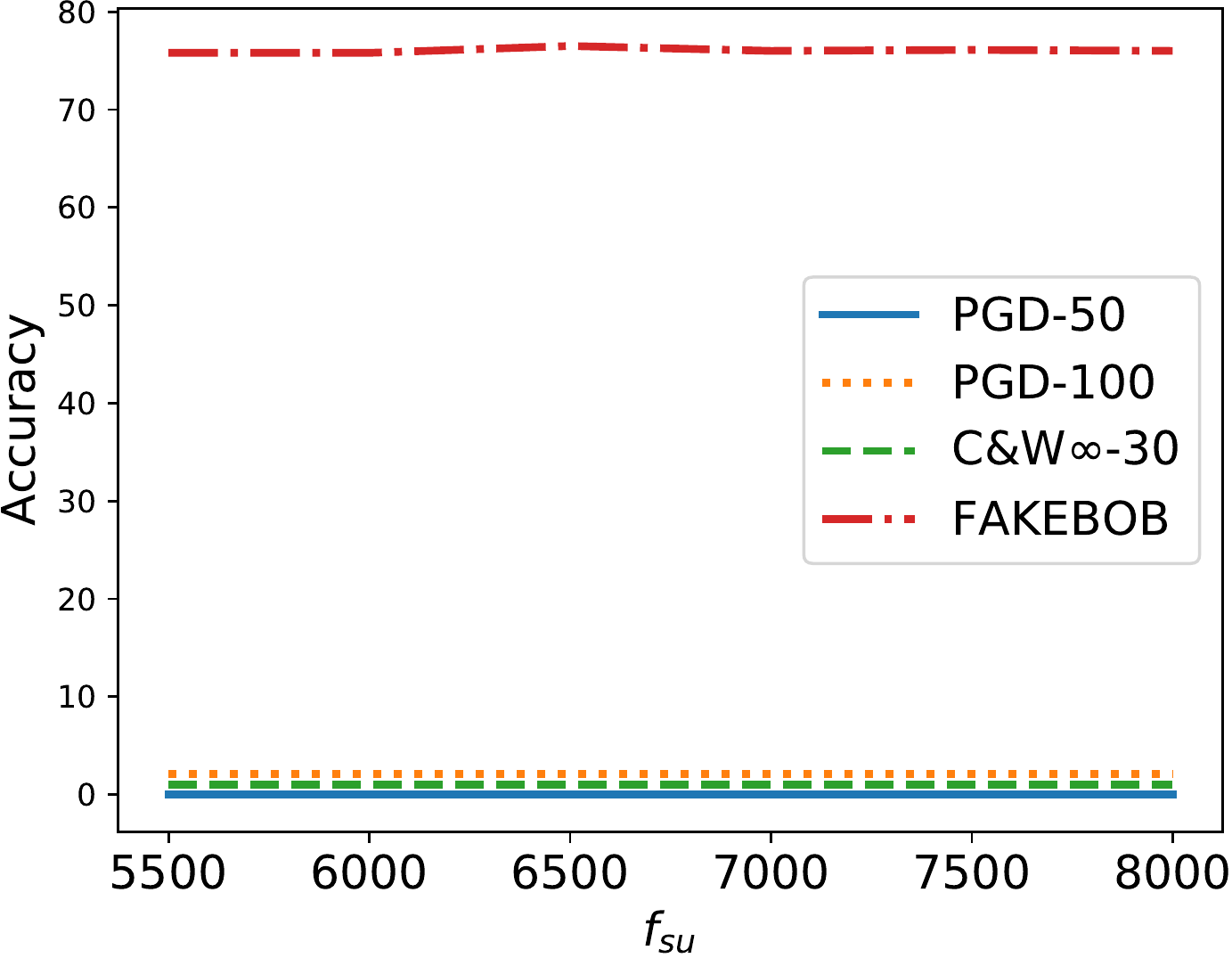}
  \end{minipage}
  \begin{minipage}[t]{0.22\textwidth}
  \includegraphics[width=1.0\textwidth]{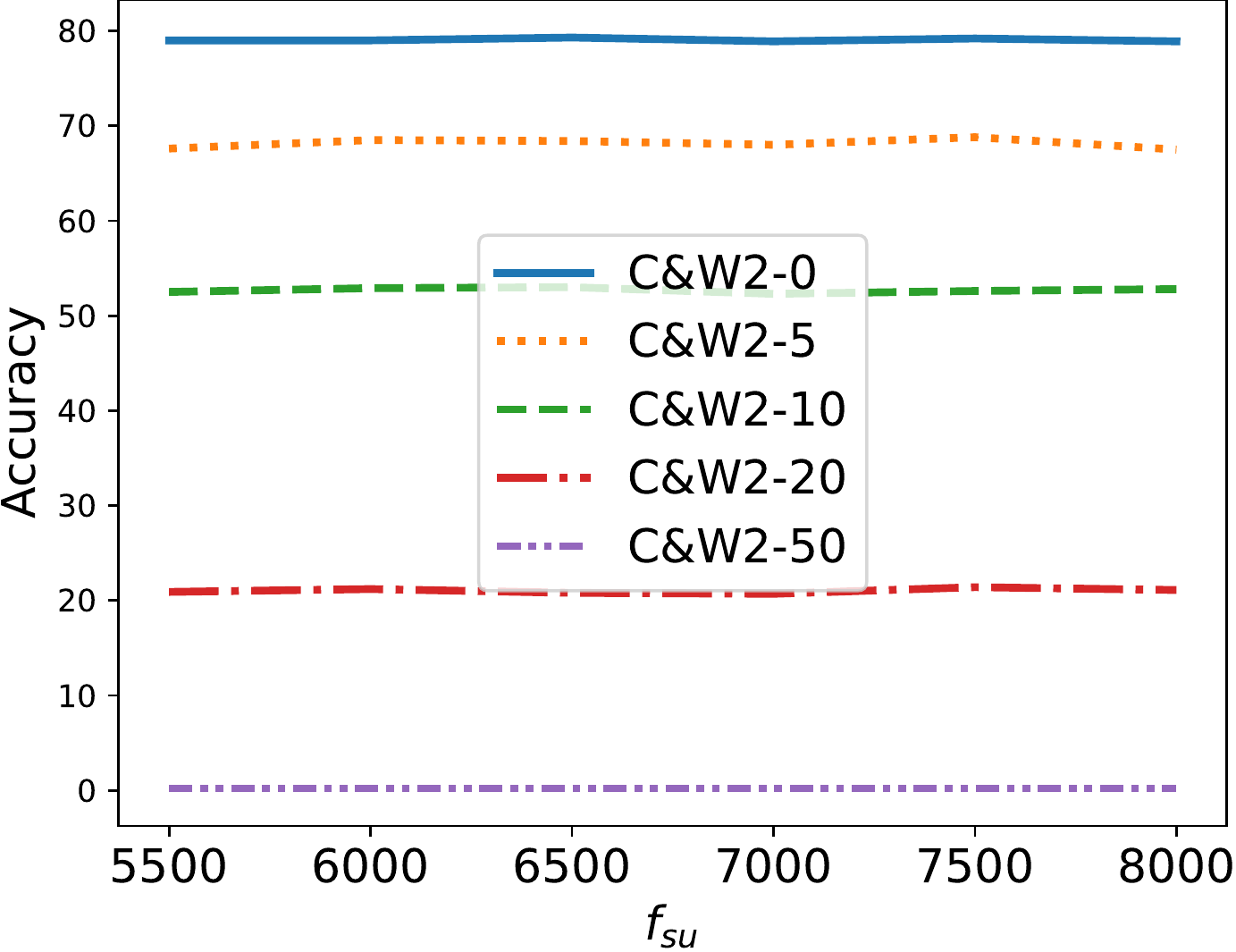}
  \end{minipage}
  }\vspace{-3mm}

  \subfigure[OPUS]{
  \begin{minipage}[t]{0.22\textwidth}
  \includegraphics[width=1.0\textwidth]{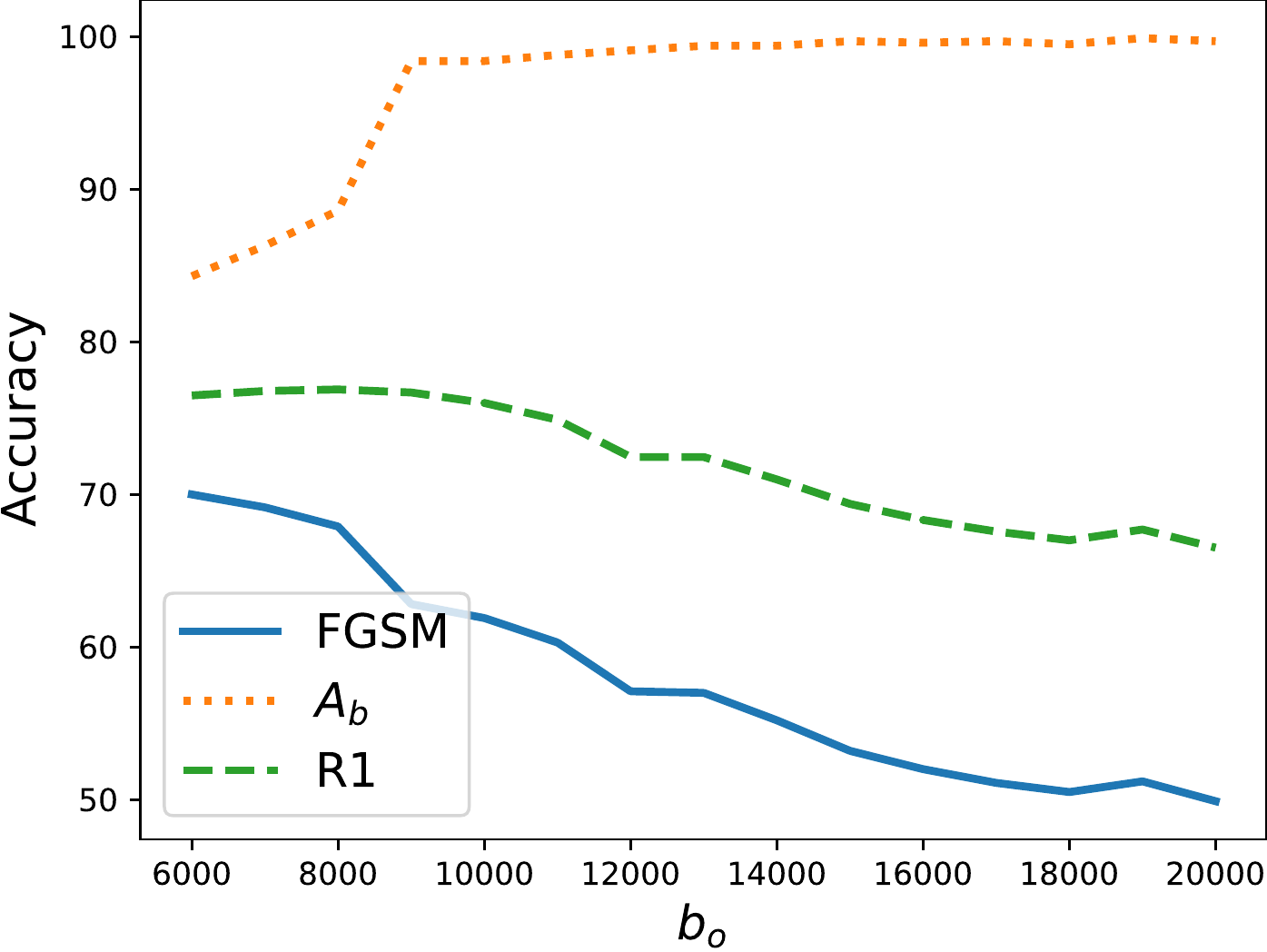}
  \end{minipage}
  \begin{minipage}[t]{0.22\textwidth}
  \includegraphics[width=1.0\textwidth]{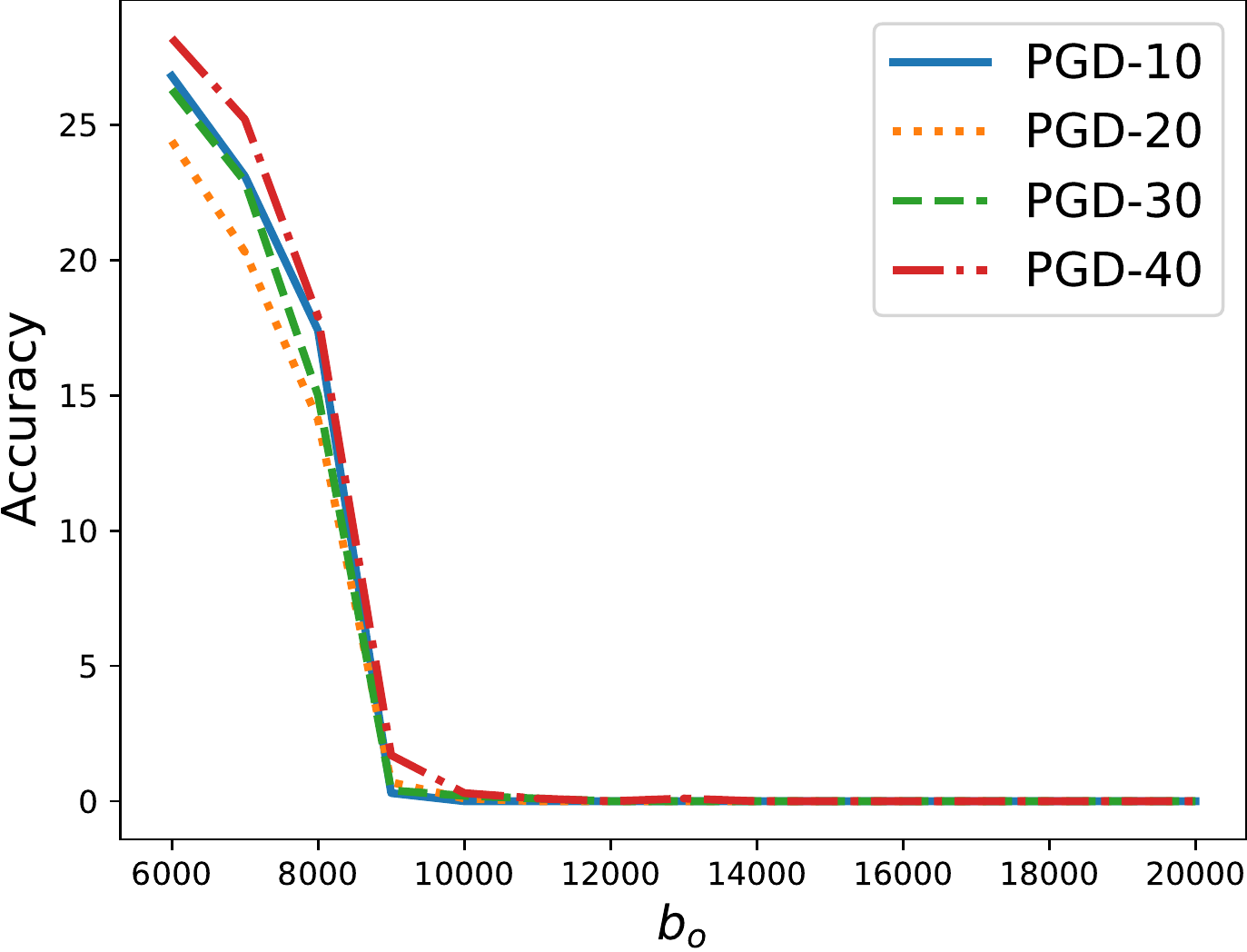}
  \end{minipage}
  \begin{minipage}[t]{0.22\textwidth}
  \includegraphics[width=1.0\textwidth]{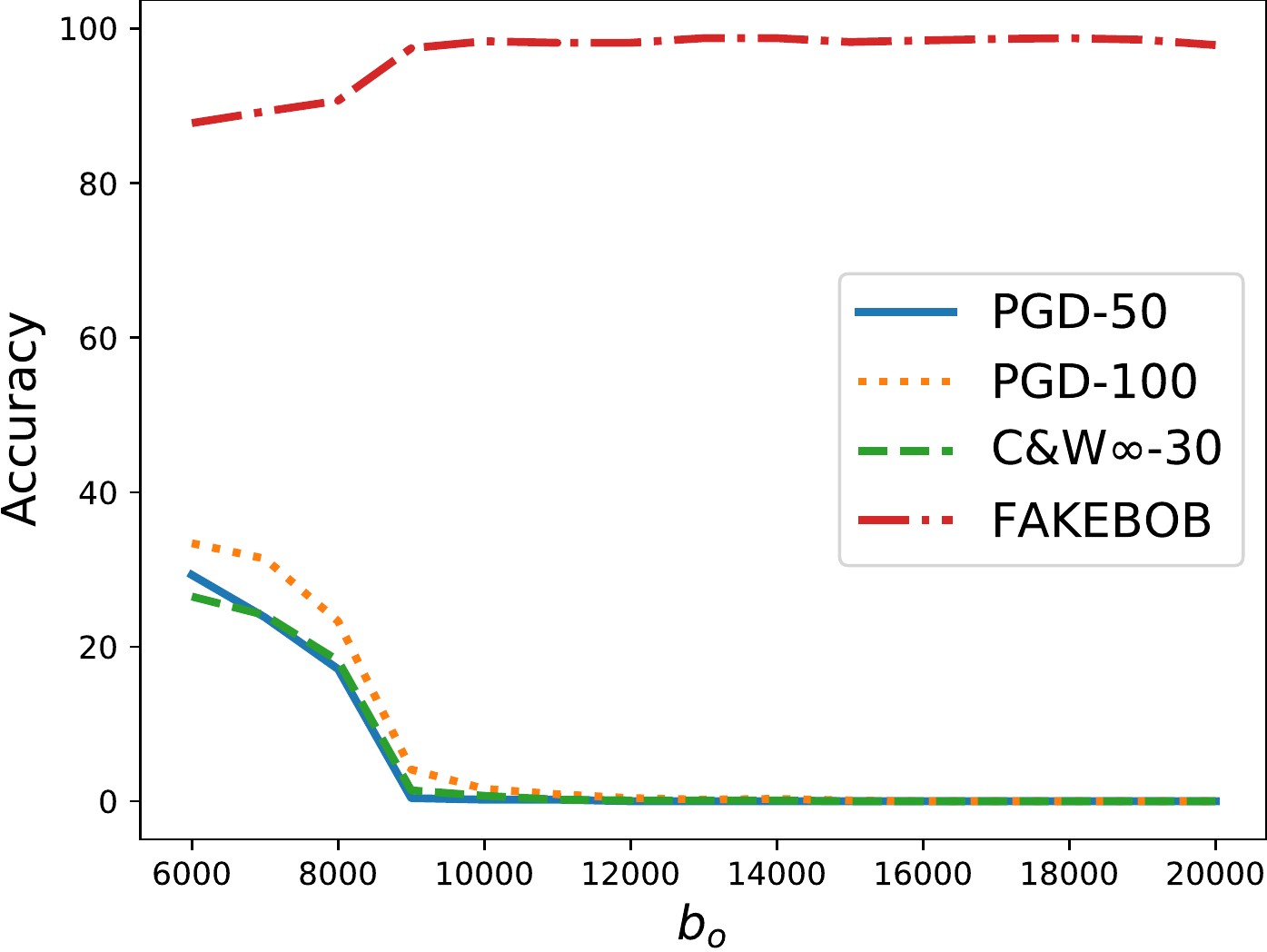}
  \end{minipage}
  \begin{minipage}[t]{0.22\textwidth}
  \includegraphics[width=1.0\textwidth]{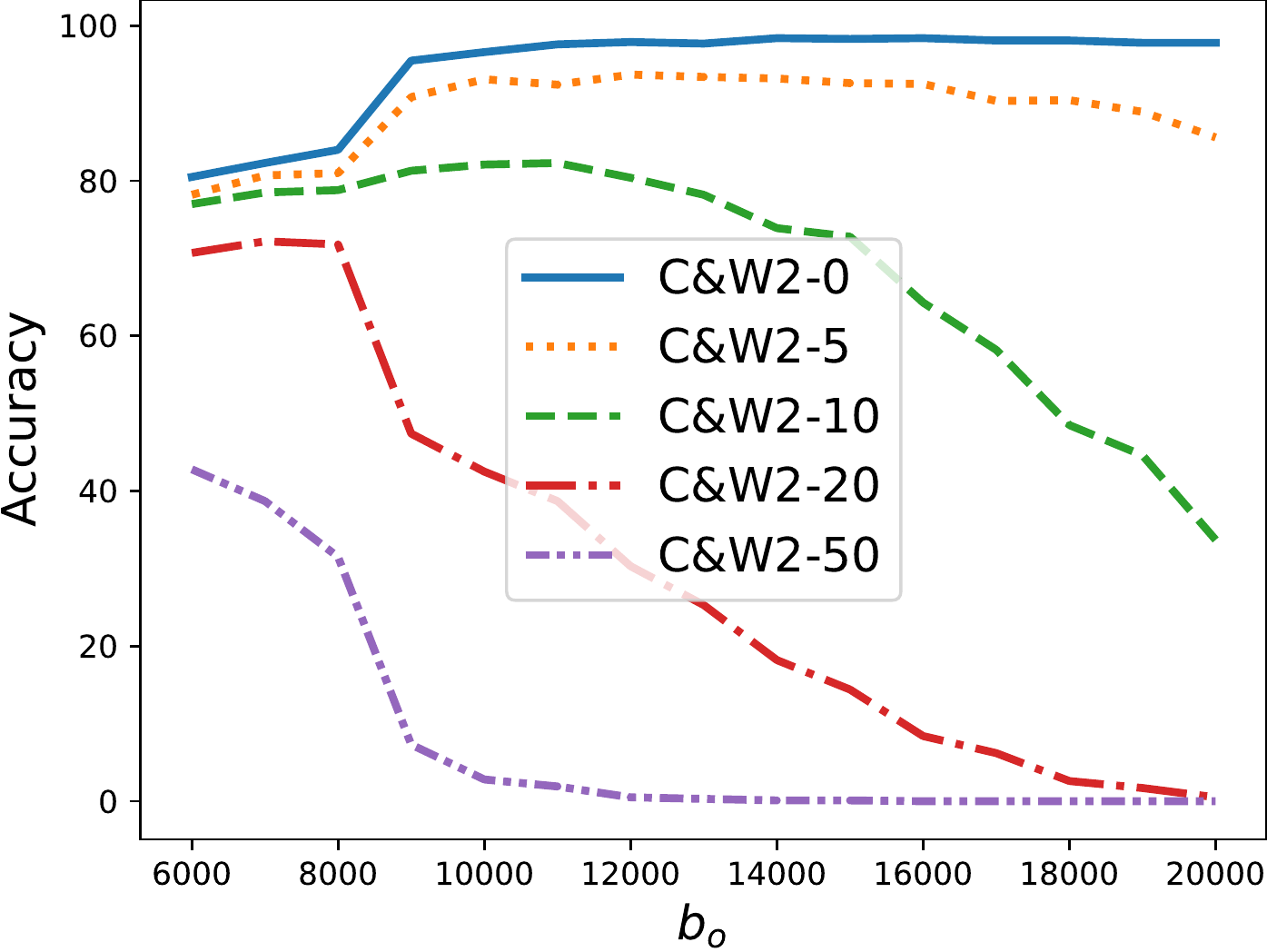}
  \end{minipage}
  }\vspace{-3mm}

  \subfigure[SPEEX]{
  \begin{minipage}[t]{0.22\textwidth}
  \includegraphics[width=1.0\textwidth]{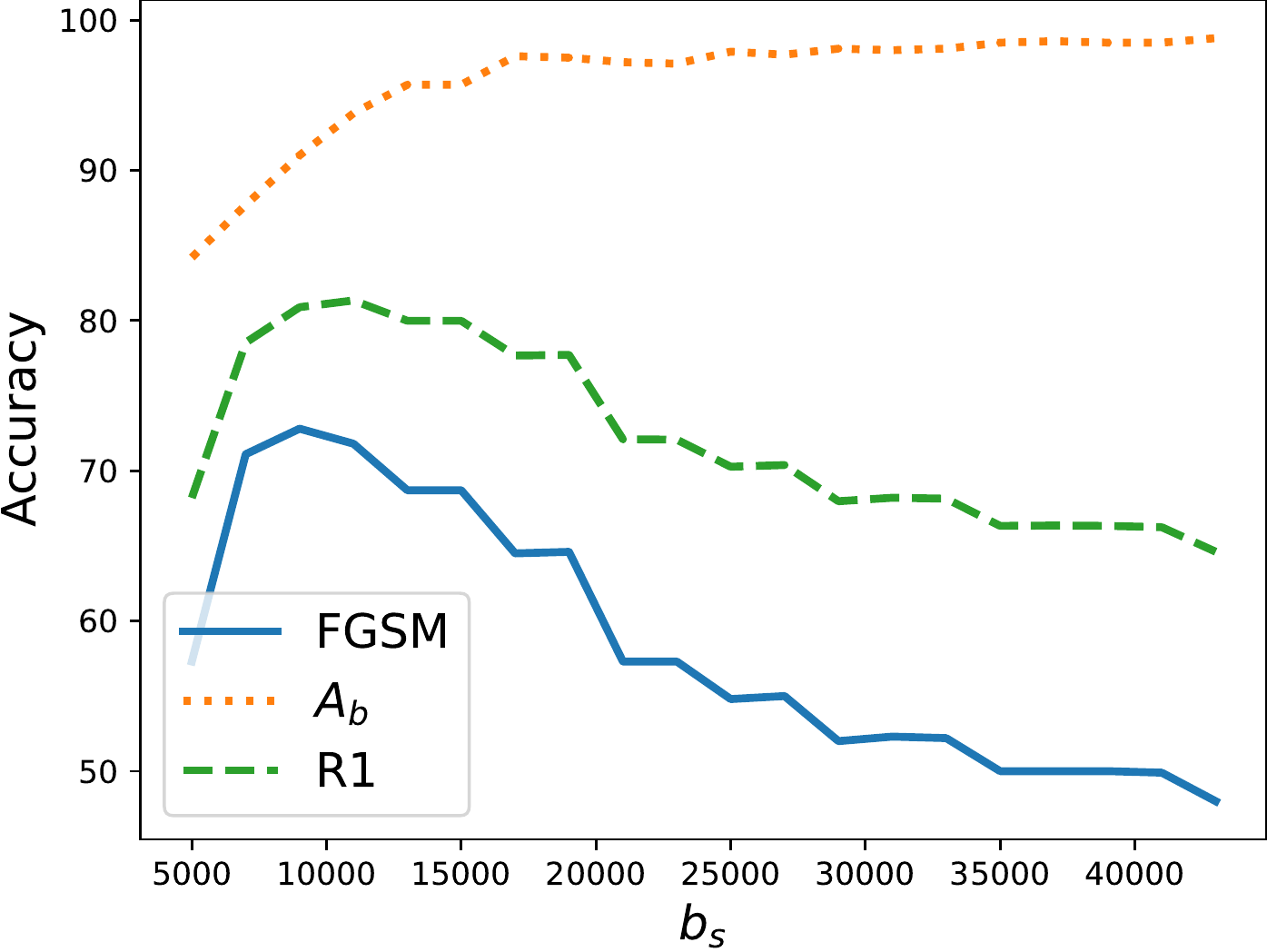}
  \end{minipage}
  \begin{minipage}[t]{0.22\textwidth}
  \includegraphics[width=1.0\textwidth]{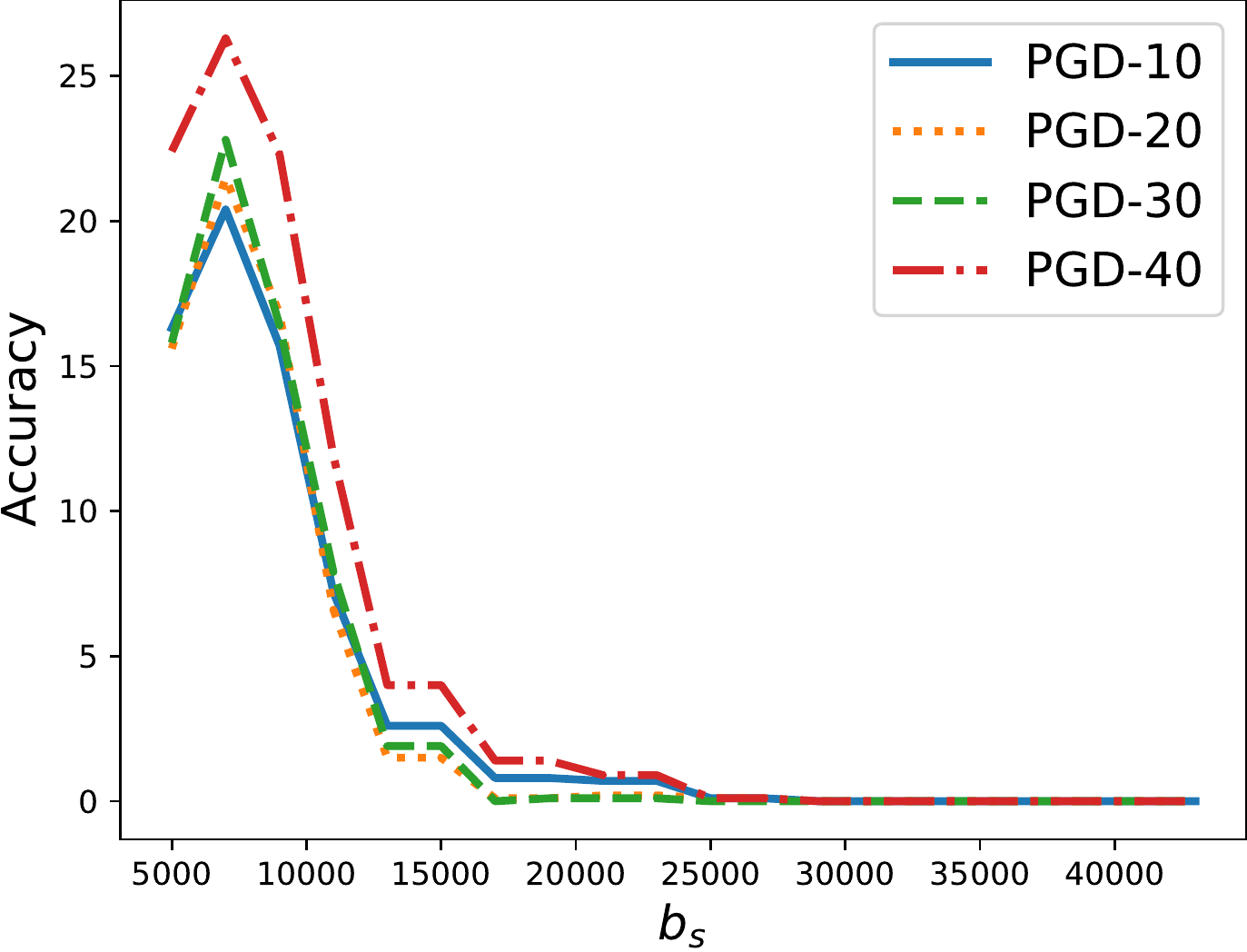}
  \end{minipage}
  \begin{minipage}[t]{0.22\textwidth}
  \includegraphics[width=1.0\textwidth]{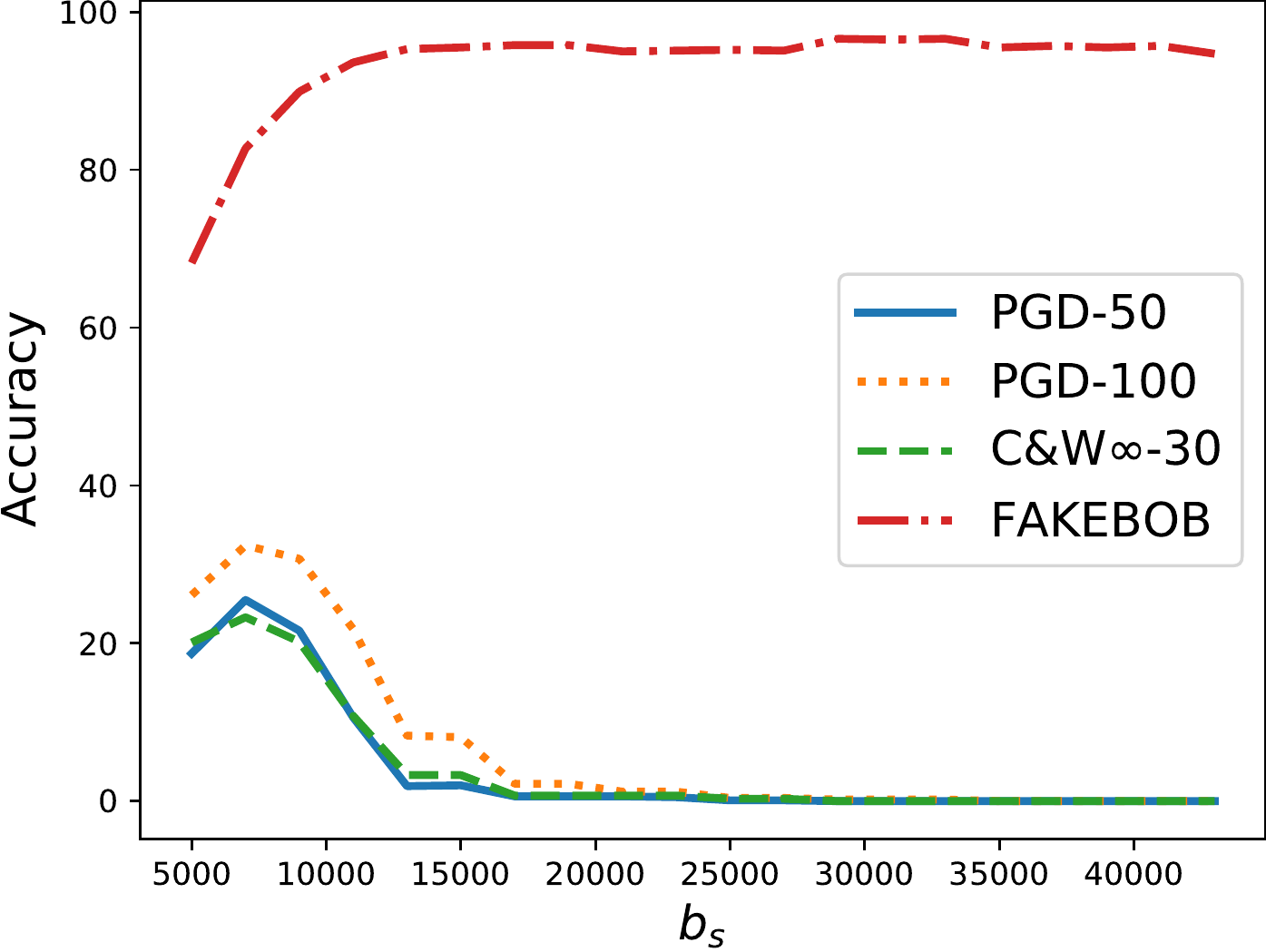}
  \end{minipage}
  \begin{minipage}[t]{0.22\textwidth}
  \includegraphics[width=1.0\textwidth]{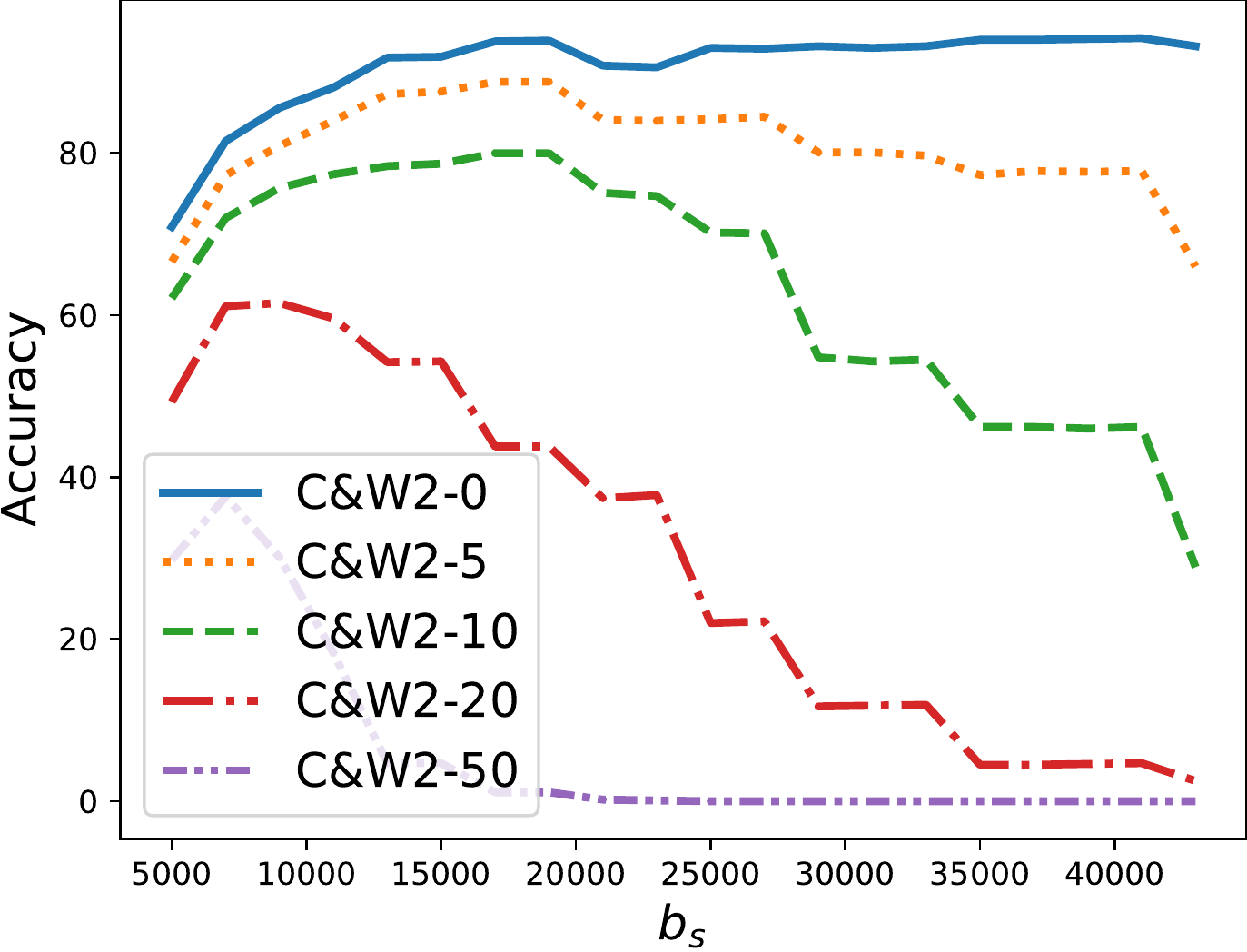}
  \end{minipage}
  }\vspace{-3mm}

  \subfigure[AMR]{
  \begin{minipage}[t]{0.22\textwidth}
  \includegraphics[width=1.0\textwidth]{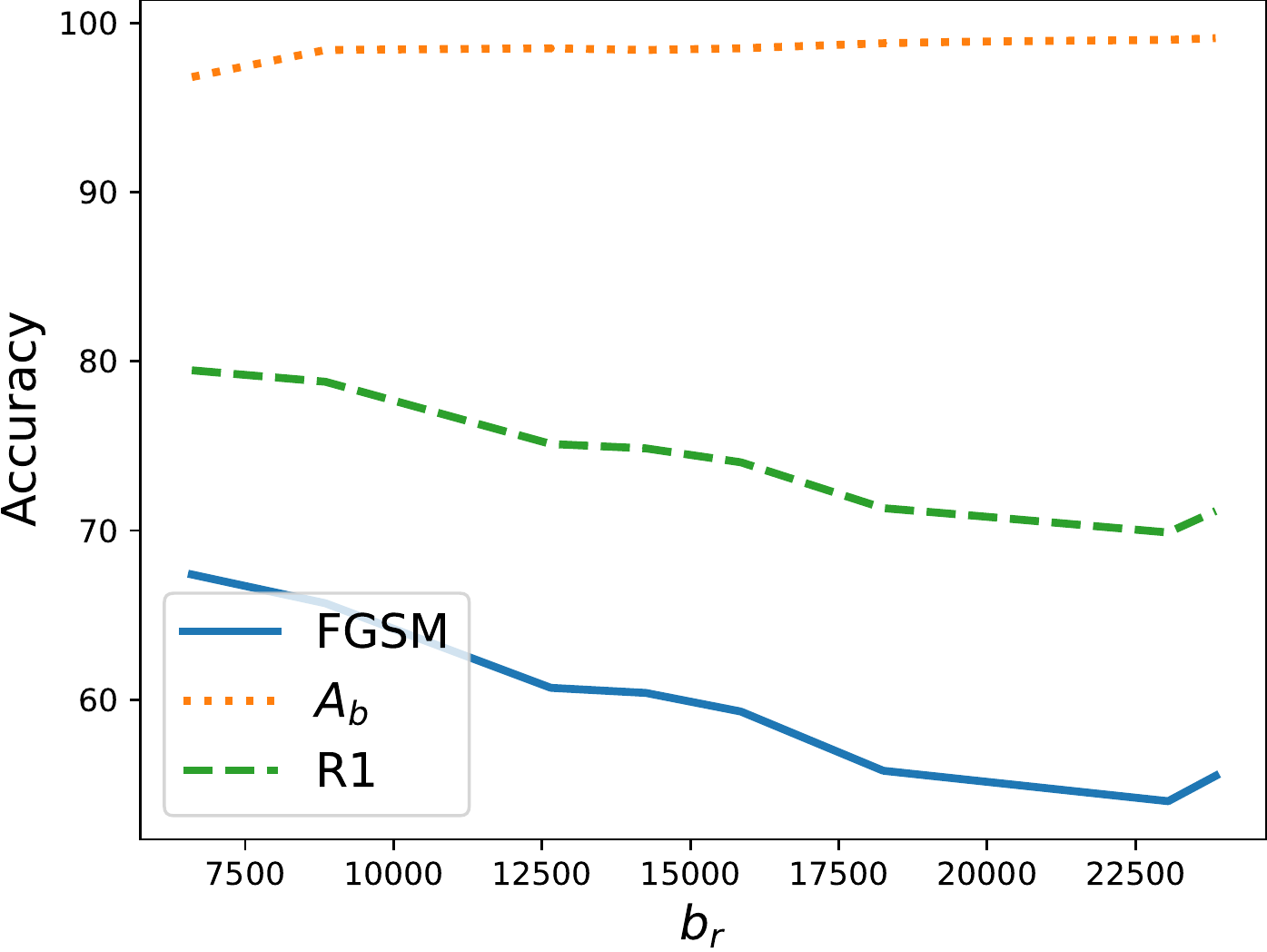}
  \end{minipage}
  \begin{minipage}[t]{0.22\textwidth}
  \includegraphics[width=1.0\textwidth]{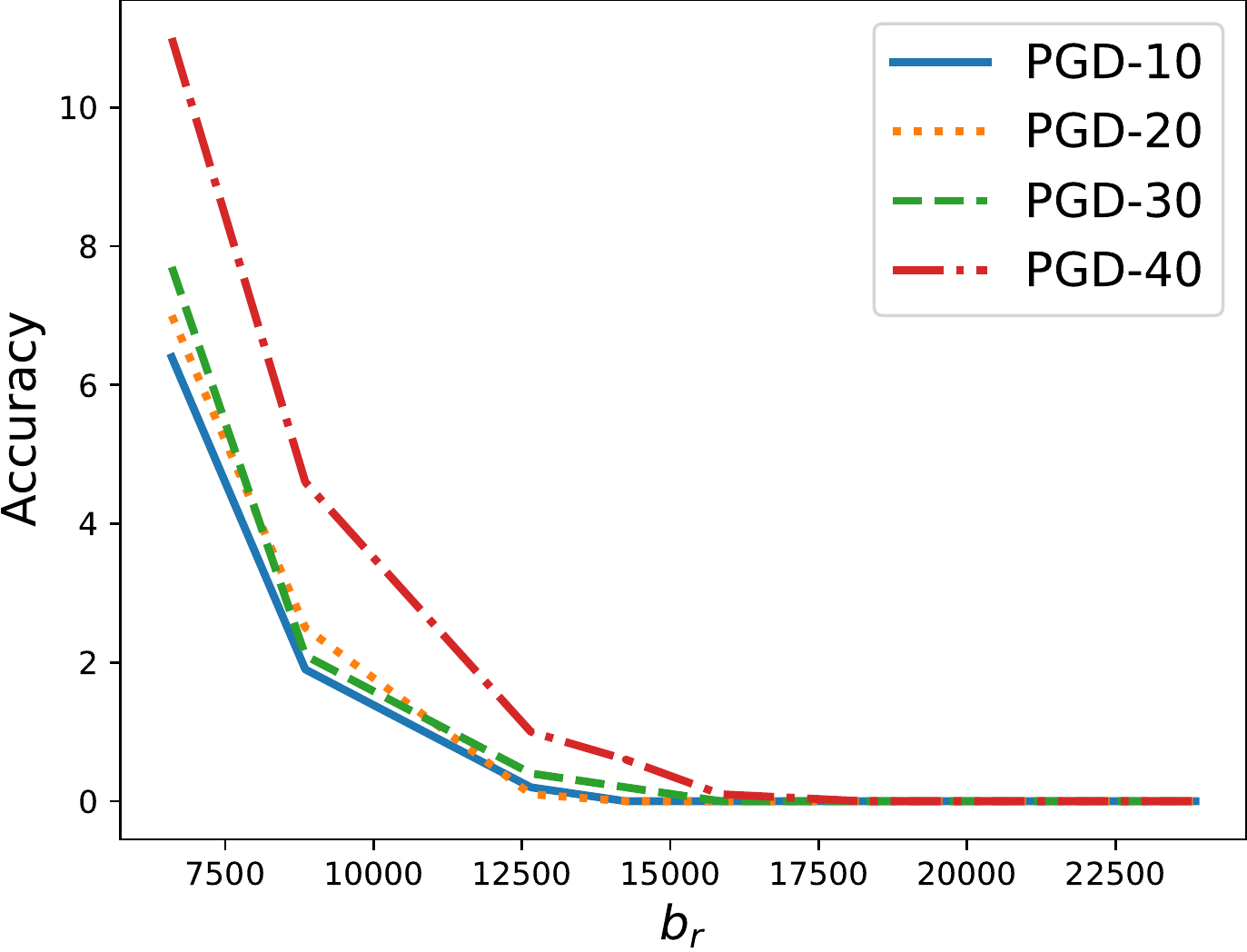}
  \end{minipage}
  \begin{minipage}[t]{0.22\textwidth}
  \includegraphics[width=1.0\textwidth]{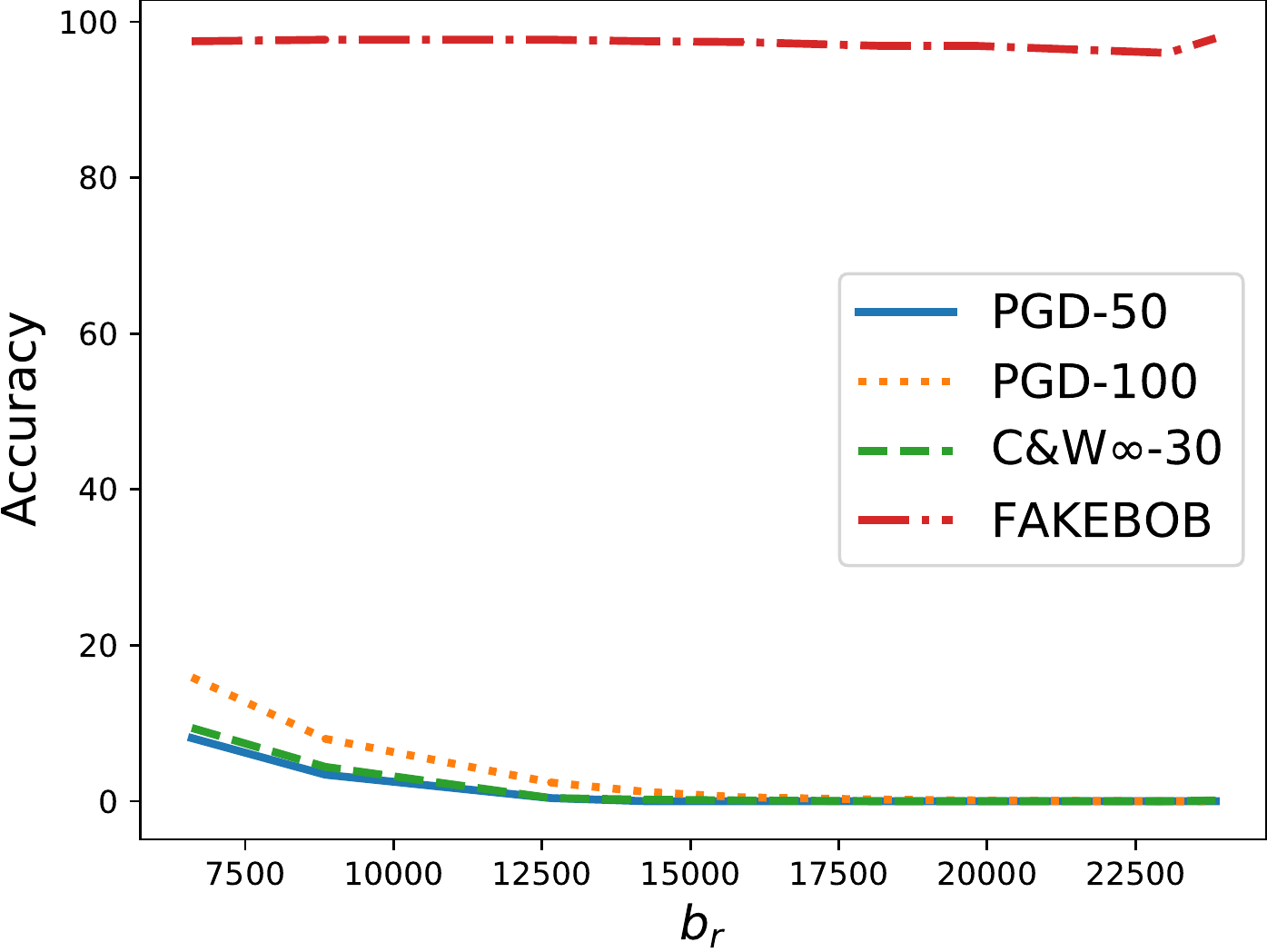}
  \end{minipage}
  \begin{minipage}[t]{0.22\textwidth}
  \includegraphics[width=1.0\textwidth]{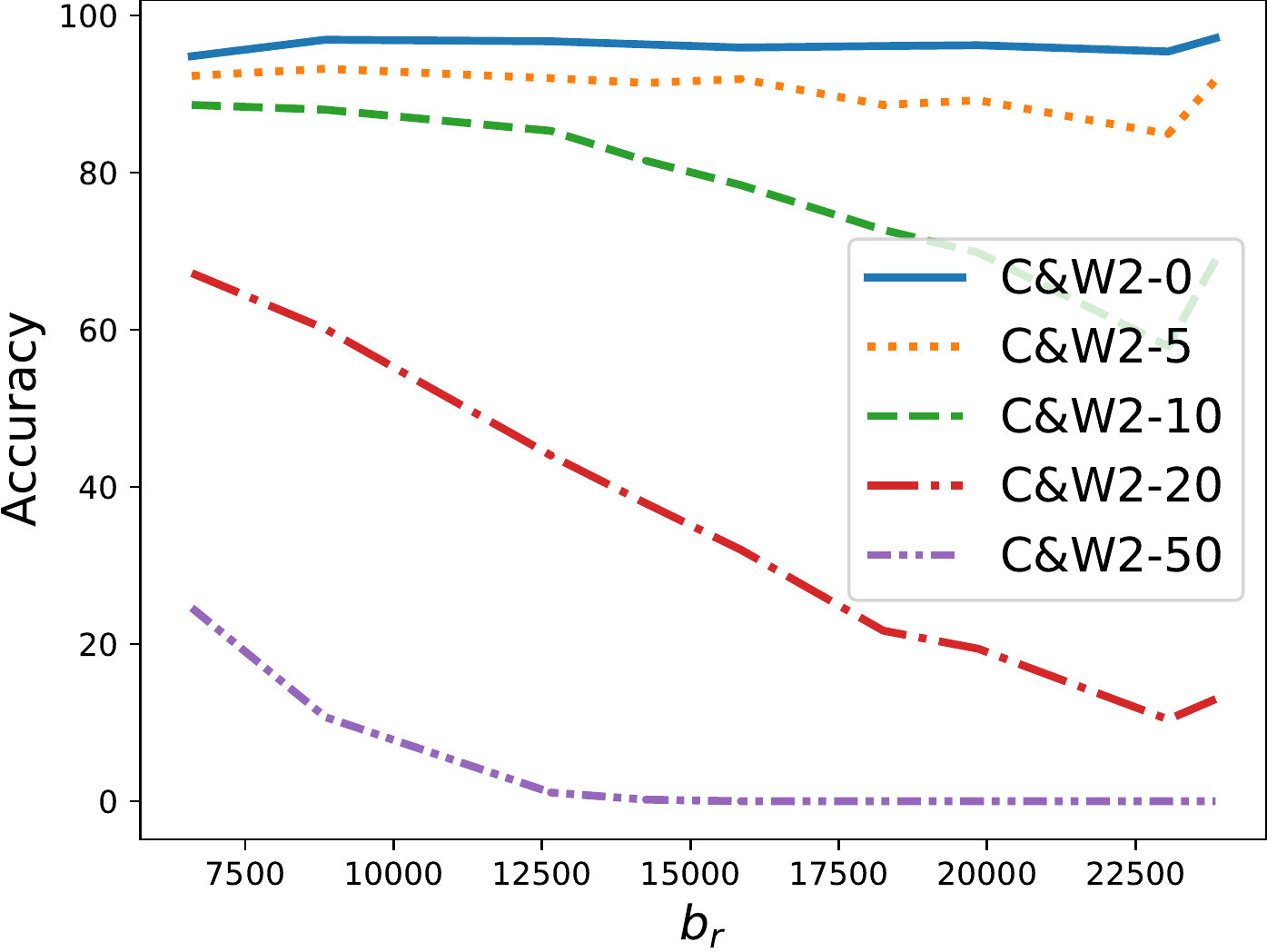}
  \end{minipage}
  }\vspace{-3mm}

   \subfigure[AAC-V]{
  \begin{minipage}[t]{0.22\textwidth}
  \includegraphics[width=1.0\textwidth]{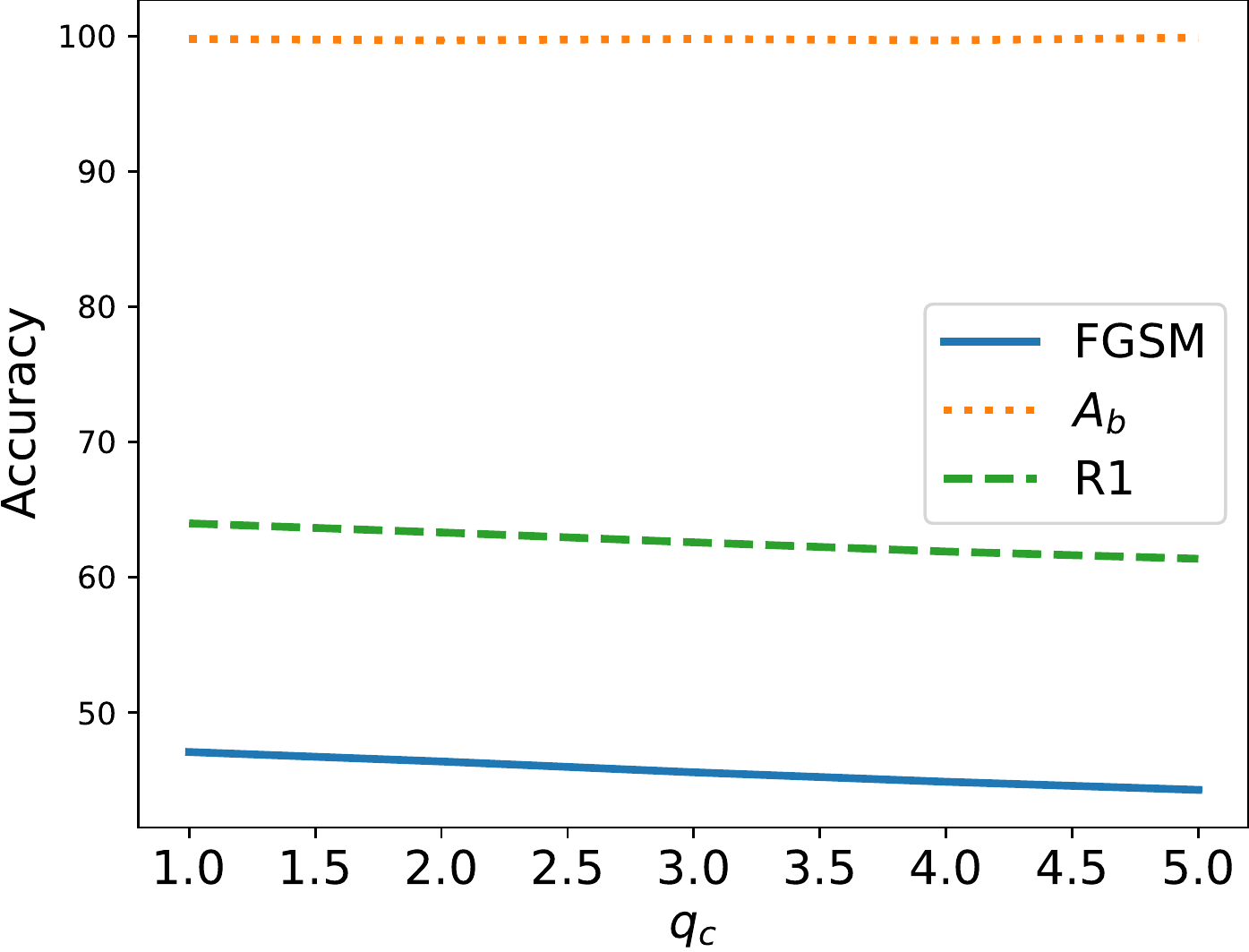}
  \end{minipage}
  \begin{minipage}[t]{0.22\textwidth}
  \includegraphics[width=1.0\textwidth]{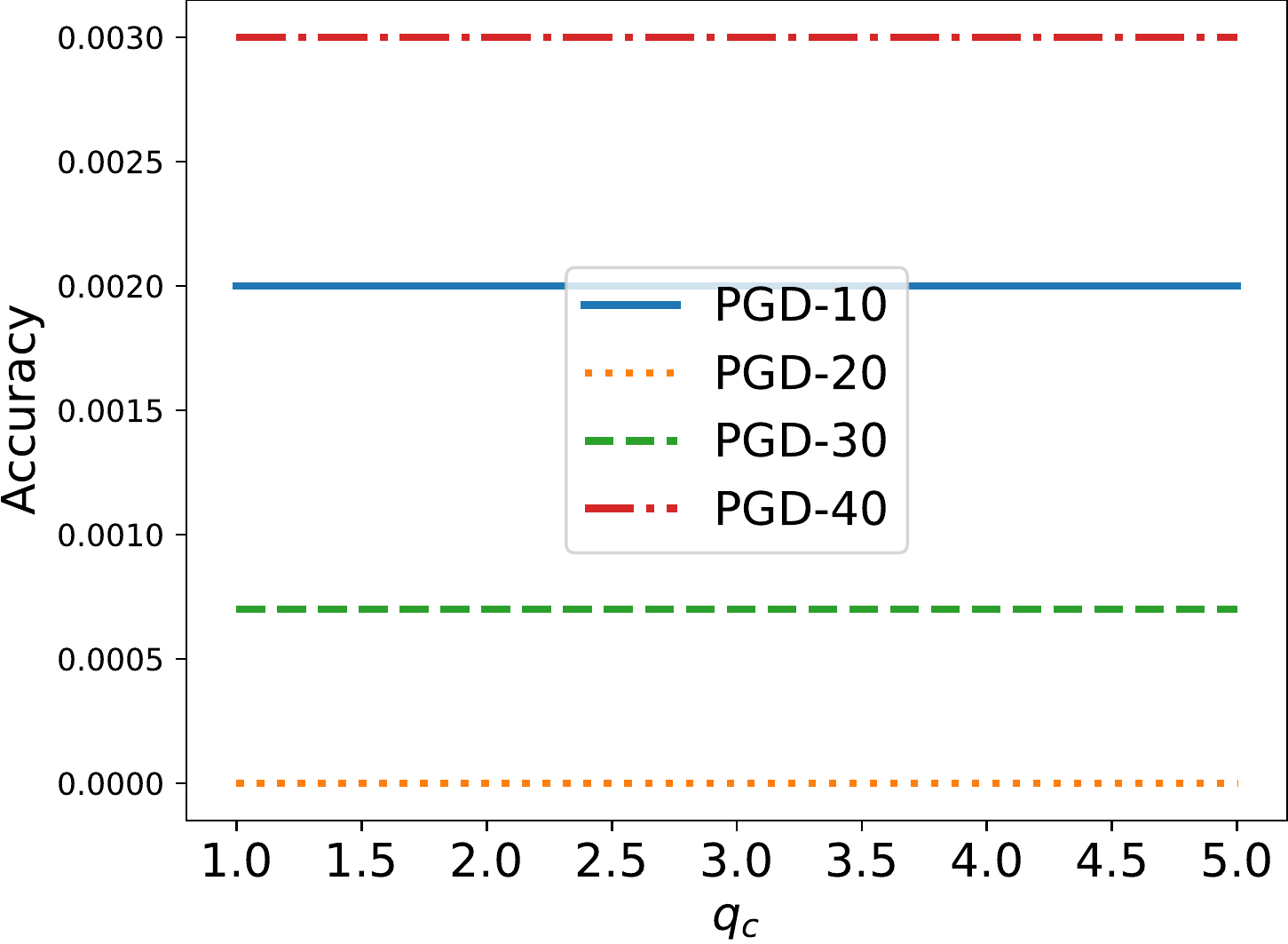}
  \end{minipage}
  \begin{minipage}[t]{0.22\textwidth}
  \includegraphics[width=1.0\textwidth]{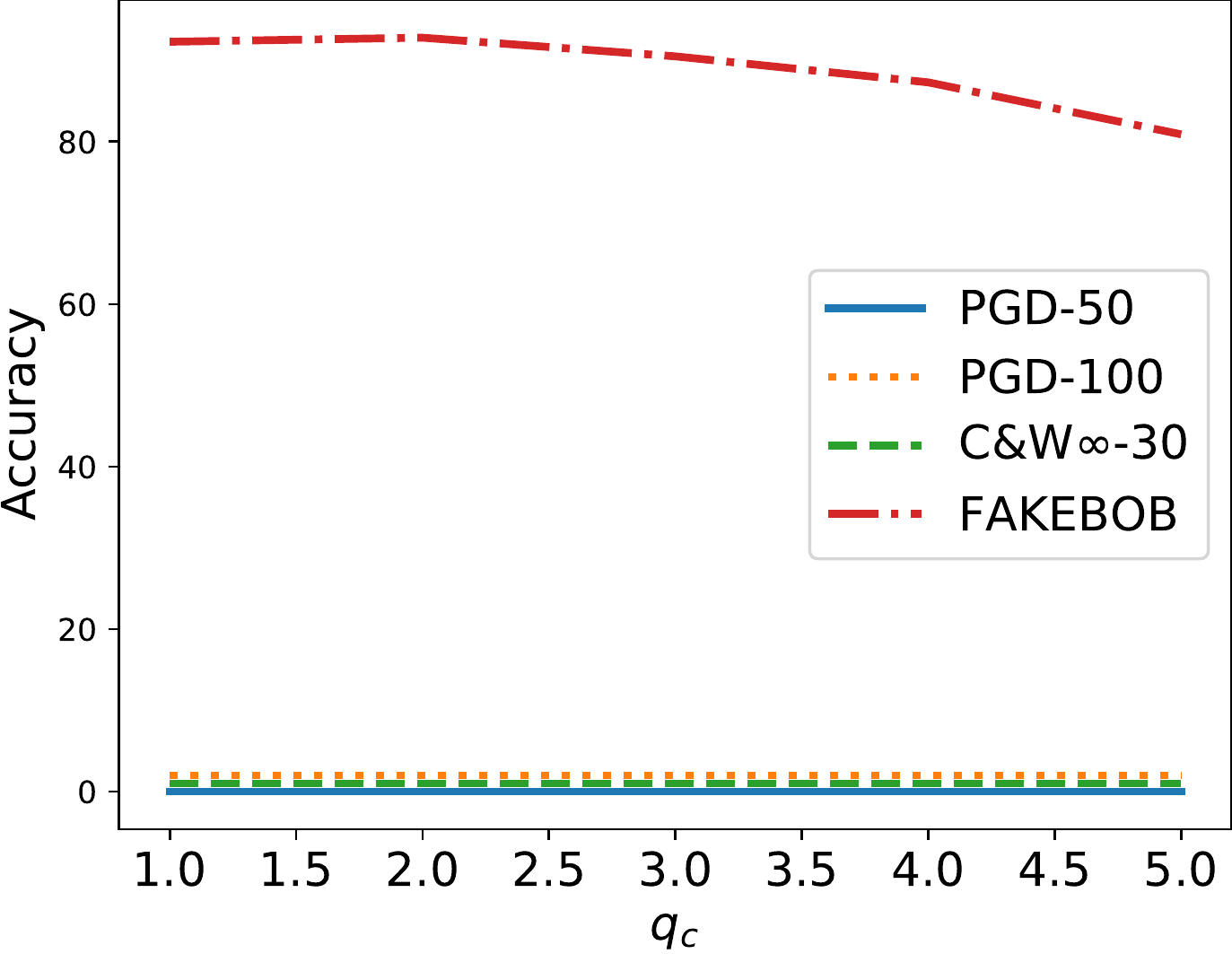}
  \end{minipage}
  \begin{minipage}[t]{0.22\textwidth}
  \includegraphics[width=1.0\textwidth]{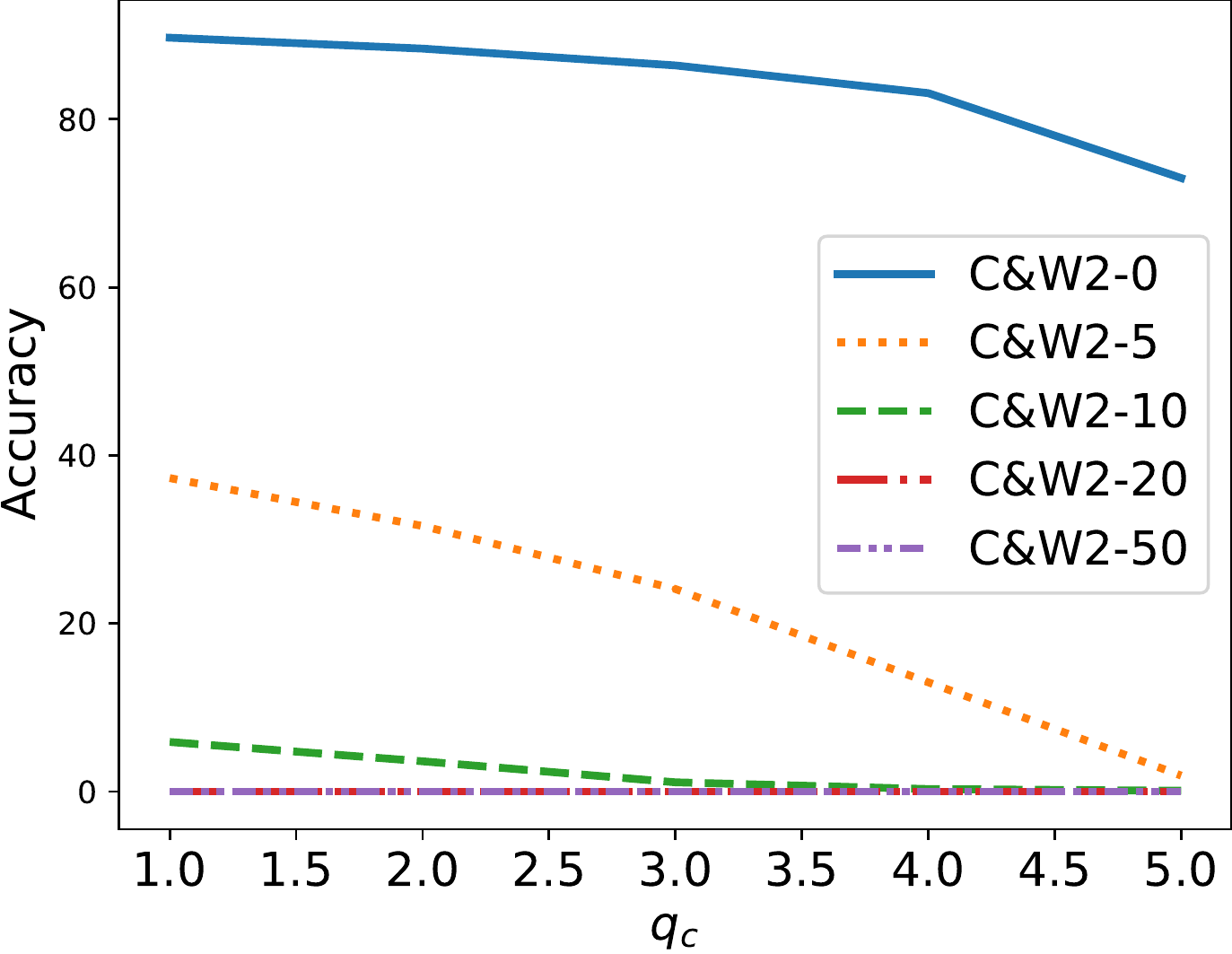}
  \end{minipage}
  }\vspace{-3mm}

  \subfigure[AAC-C]{
  \begin{minipage}[t]{0.22\textwidth}
  \includegraphics[width=1.0\textwidth]{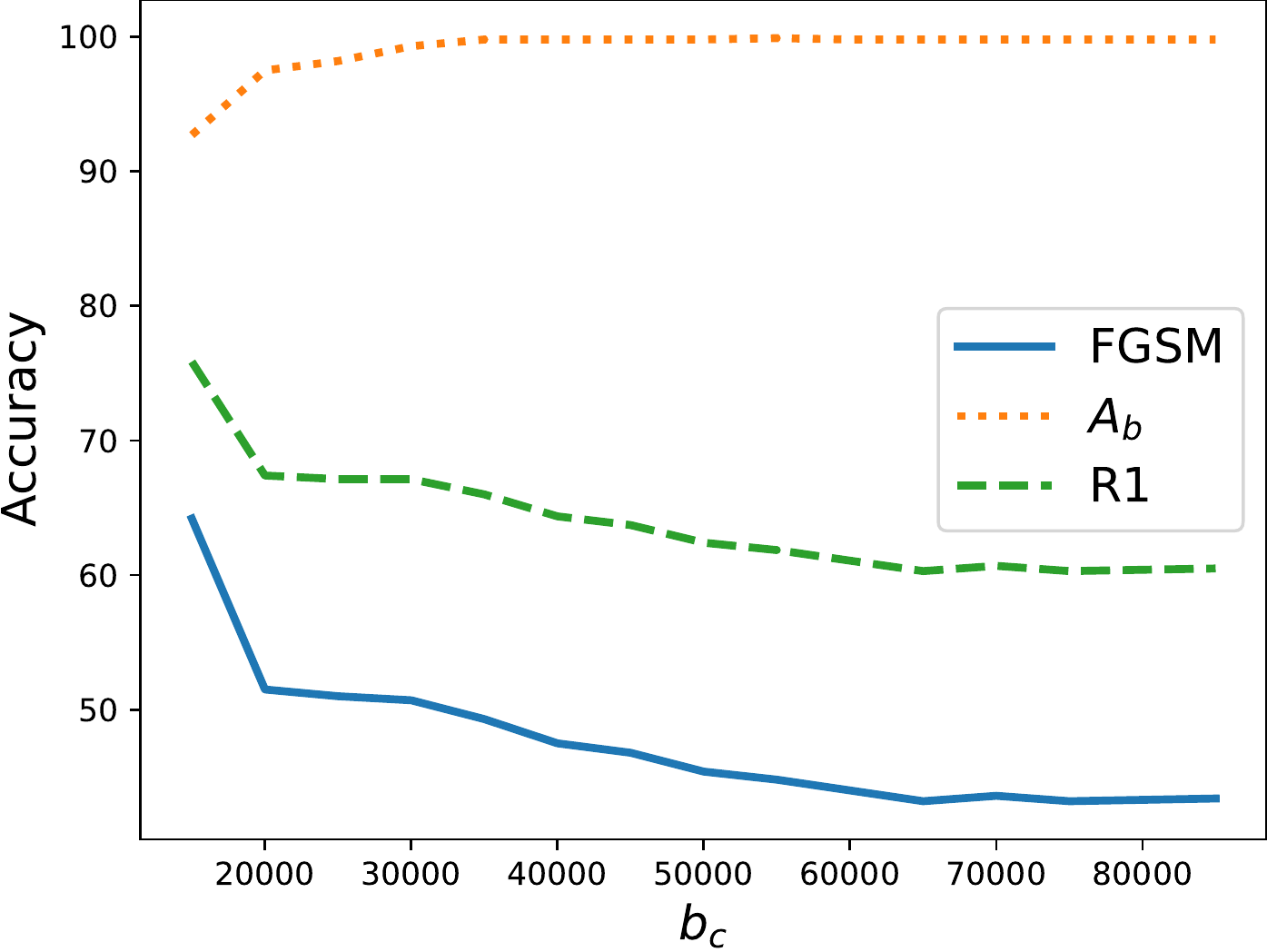}
  \end{minipage}
  \begin{minipage}[t]{0.22\textwidth}
  \includegraphics[width=1.0\textwidth]{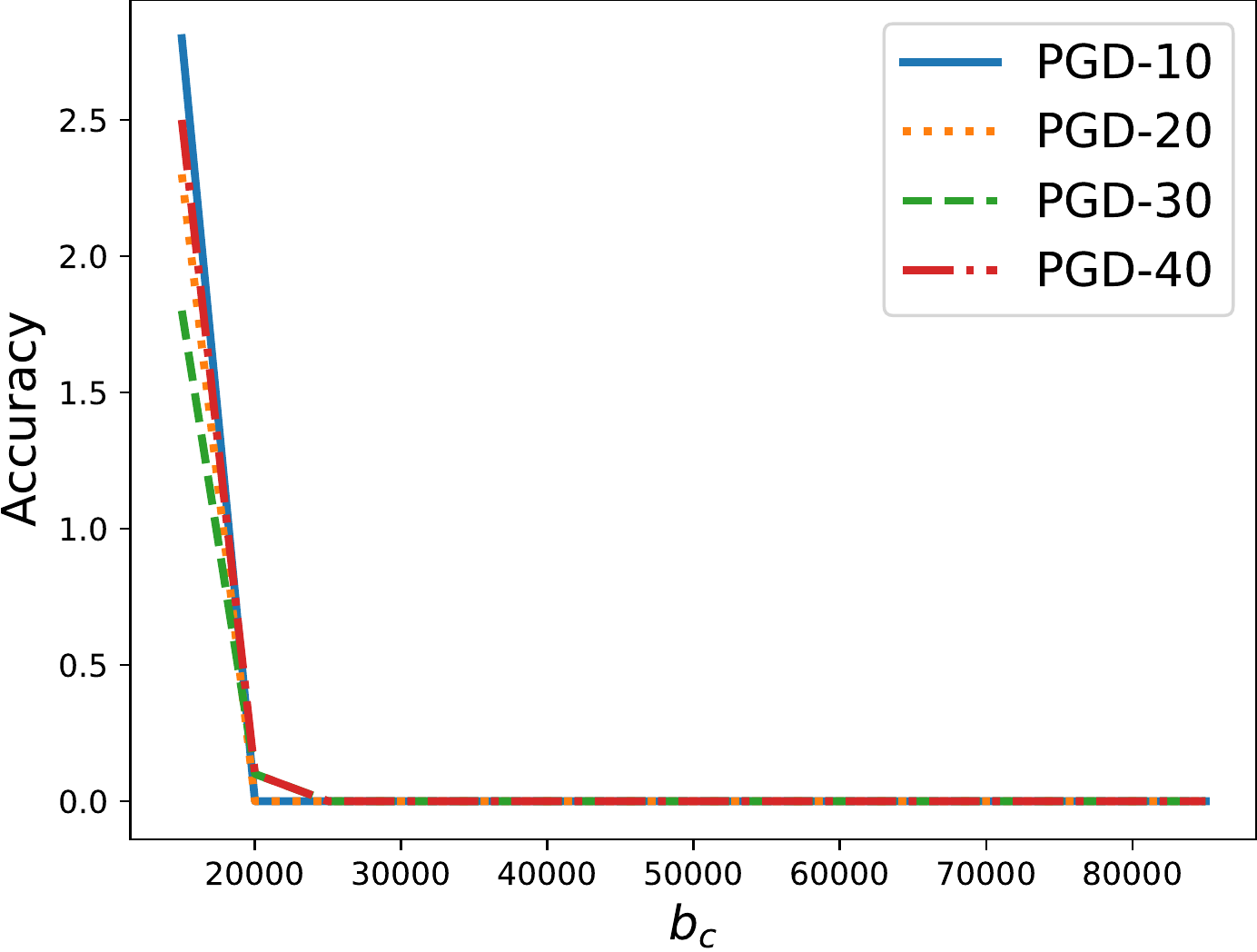}
  \end{minipage}
  \begin{minipage}[t]{0.22\textwidth}
  \includegraphics[width=1.0\textwidth]{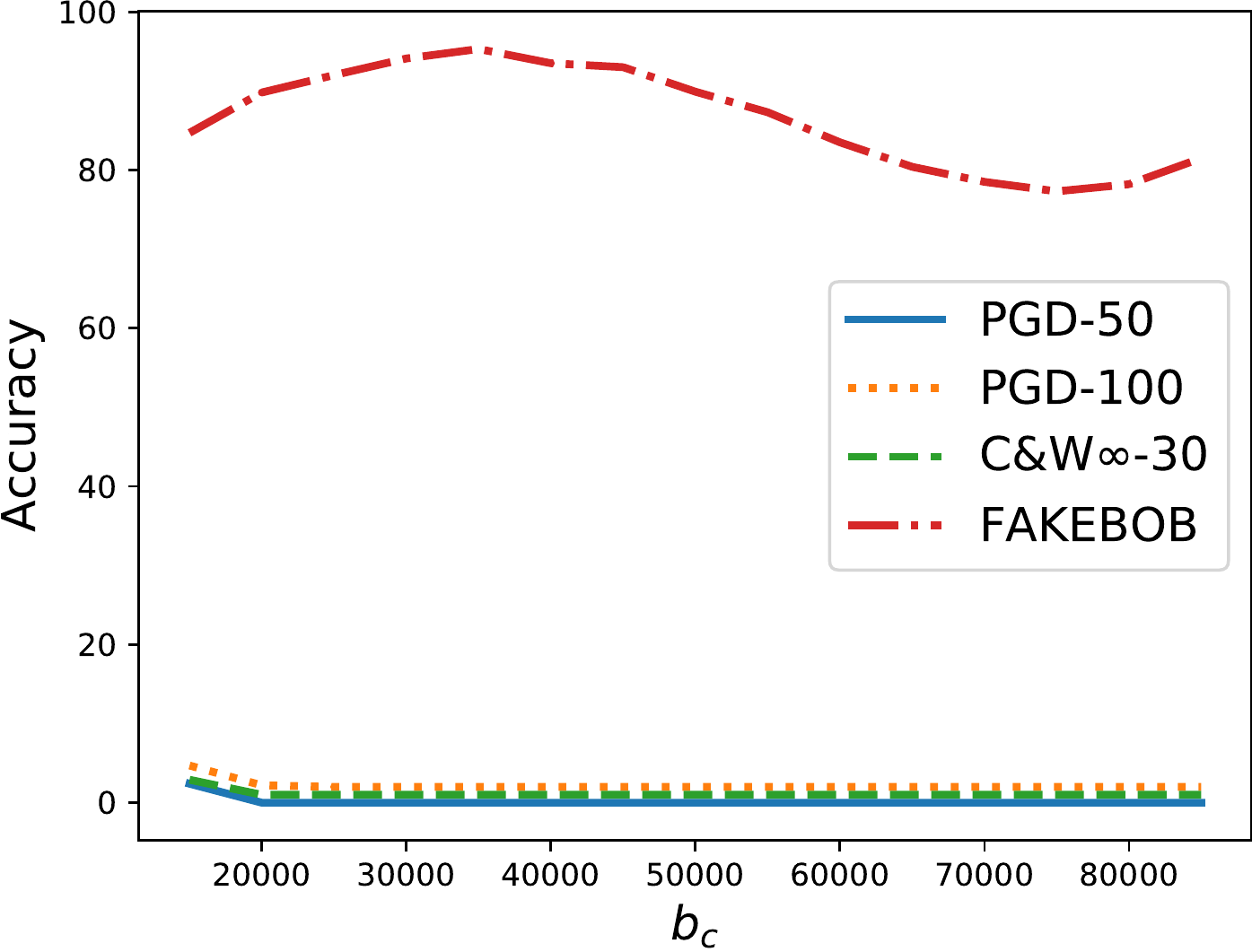}
  \end{minipage}
  \begin{minipage}[t]{0.22\textwidth}
  \includegraphics[width=1.0\textwidth]{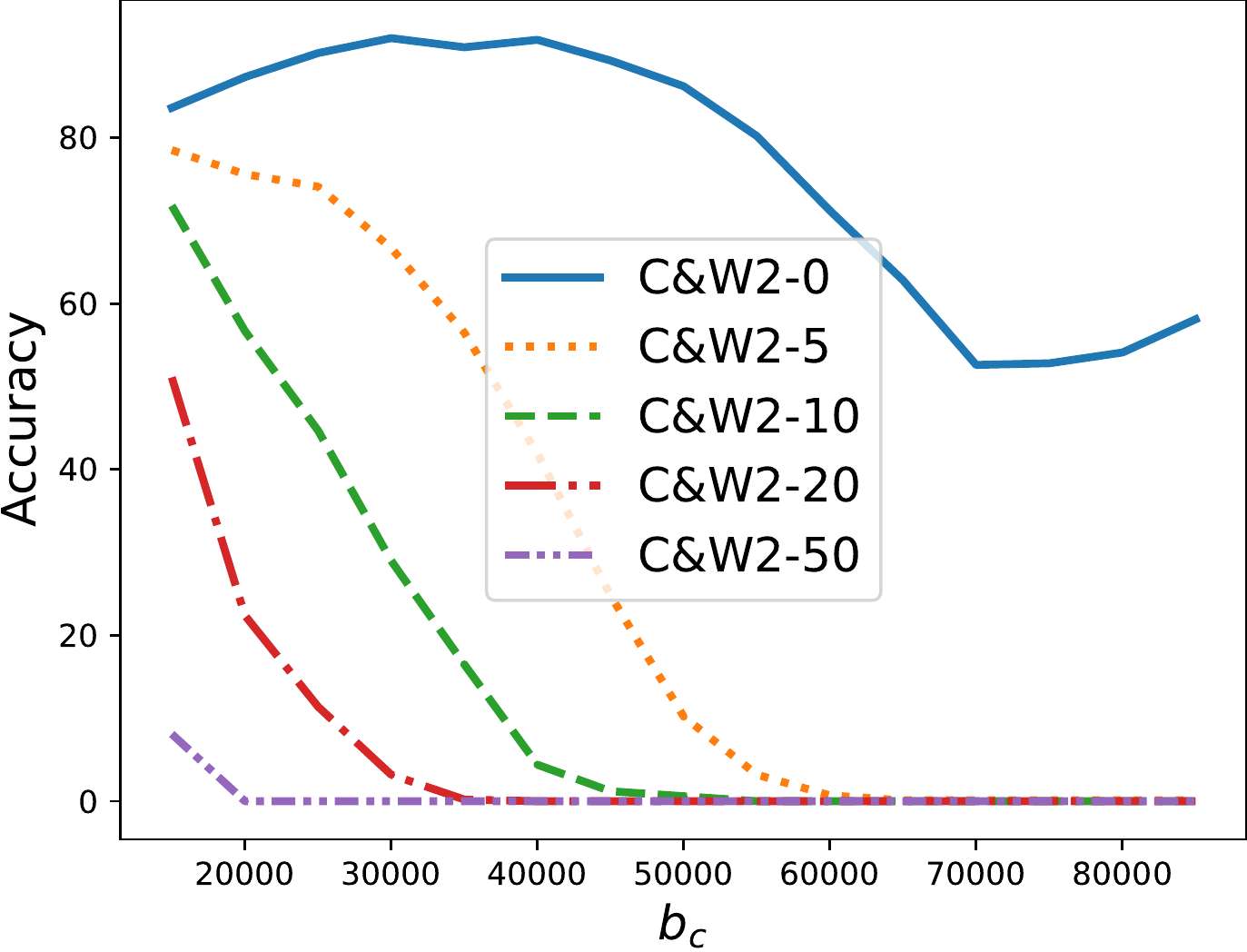}
  \end{minipage}
  }\vspace{-3mm}

  \caption{The performance of input transformations vs. parameter values.
  For better visualization, we fix $f_{sl}=150$ Hz of BPF and shows how its performance varies with $f_{su}$.}
   \label{fig:parameter-2}
\end{figure*}

\begin{figure*}
  \centering

  \subfigure[MP3-V]{
  \begin{minipage}[t]{0.22\textwidth}
  \includegraphics[width=1.0\textwidth]{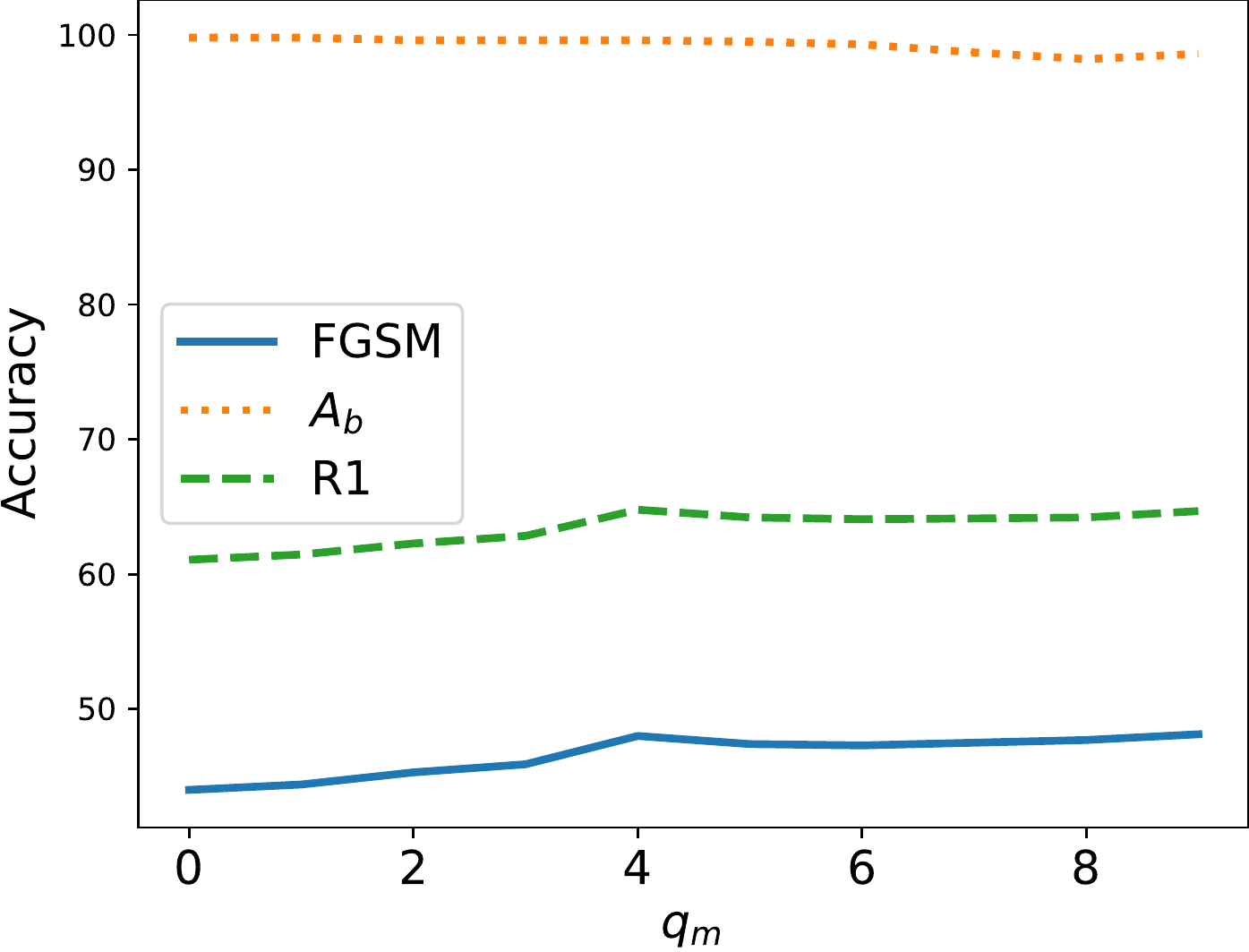}
  \end{minipage}
  \begin{minipage}[t]{0.22\textwidth}
  \includegraphics[width=1.0\textwidth]{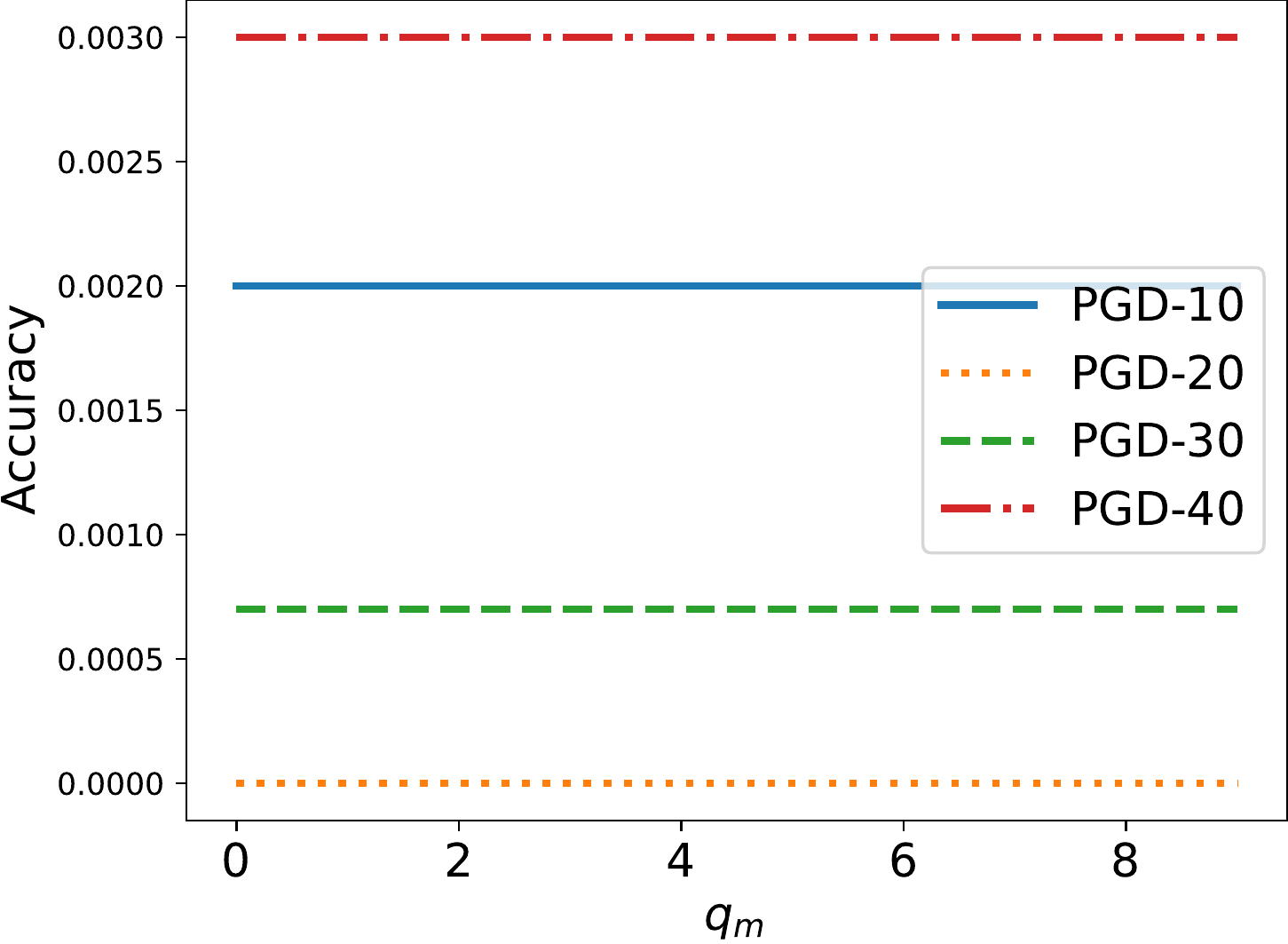}
  \end{minipage}
  \begin{minipage}[t]{0.22\textwidth}
  \includegraphics[width=1.0\textwidth]{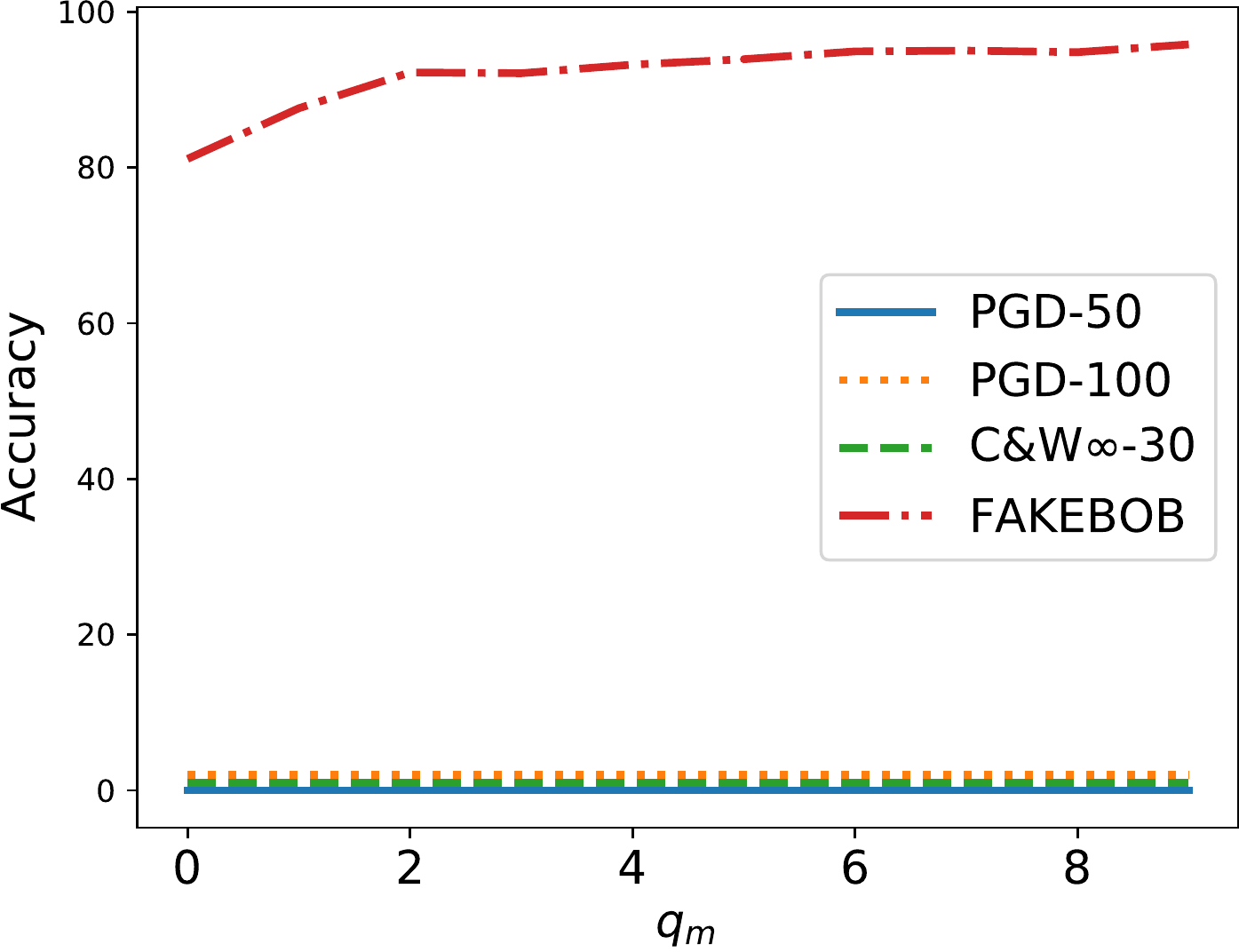}
  \end{minipage}
  \begin{minipage}[t]{0.22\textwidth}
  \includegraphics[width=1.0\textwidth]{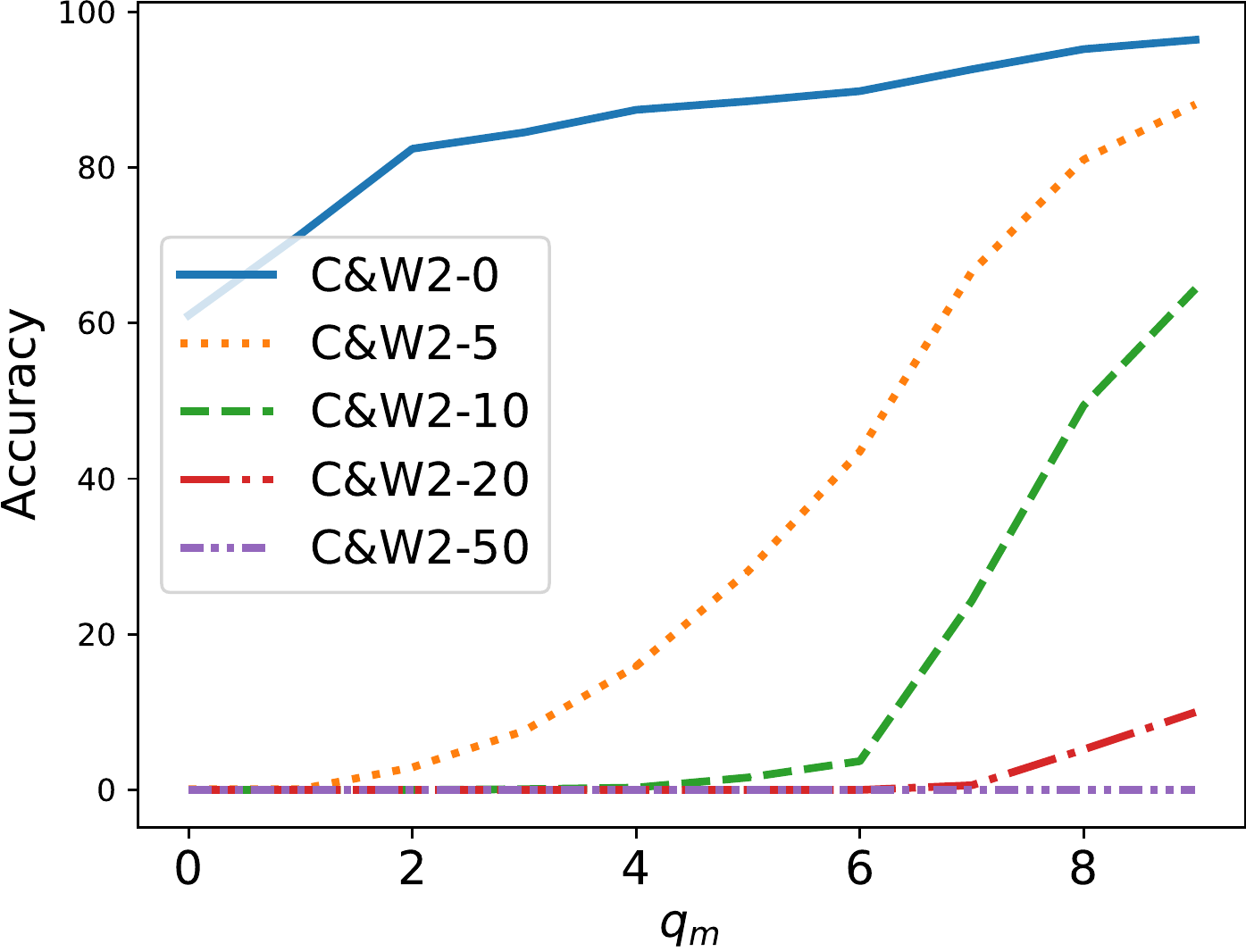}
  \end{minipage}
  }\vspace{-3mm}


  \subfigure[MP3-C]{
  \begin{minipage}[t]{0.22\textwidth}
  \includegraphics[width=1.0\textwidth]{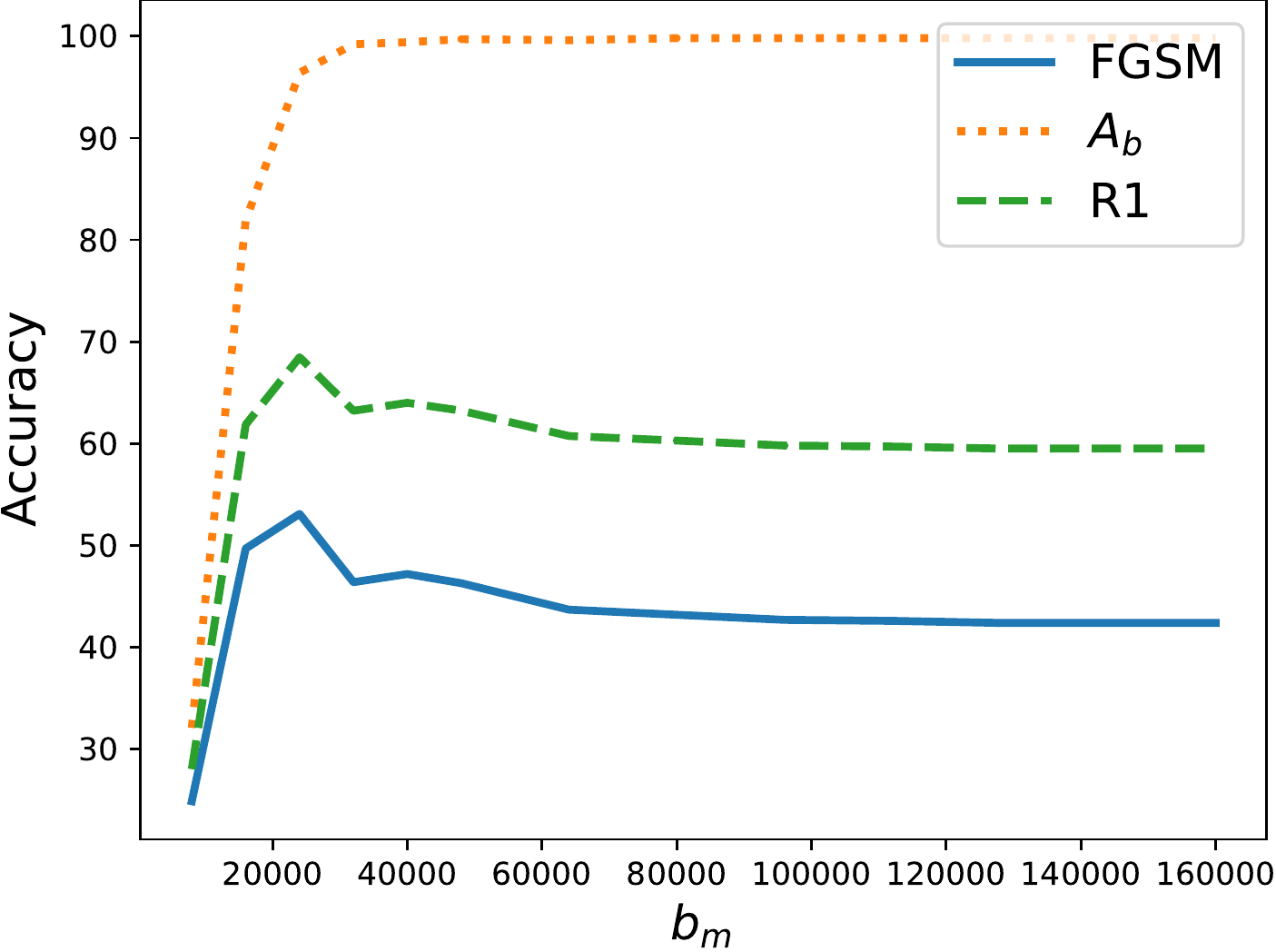}
  \end{minipage}
  \begin{minipage}[t]{0.22\textwidth}
  \includegraphics[width=1.0\textwidth]{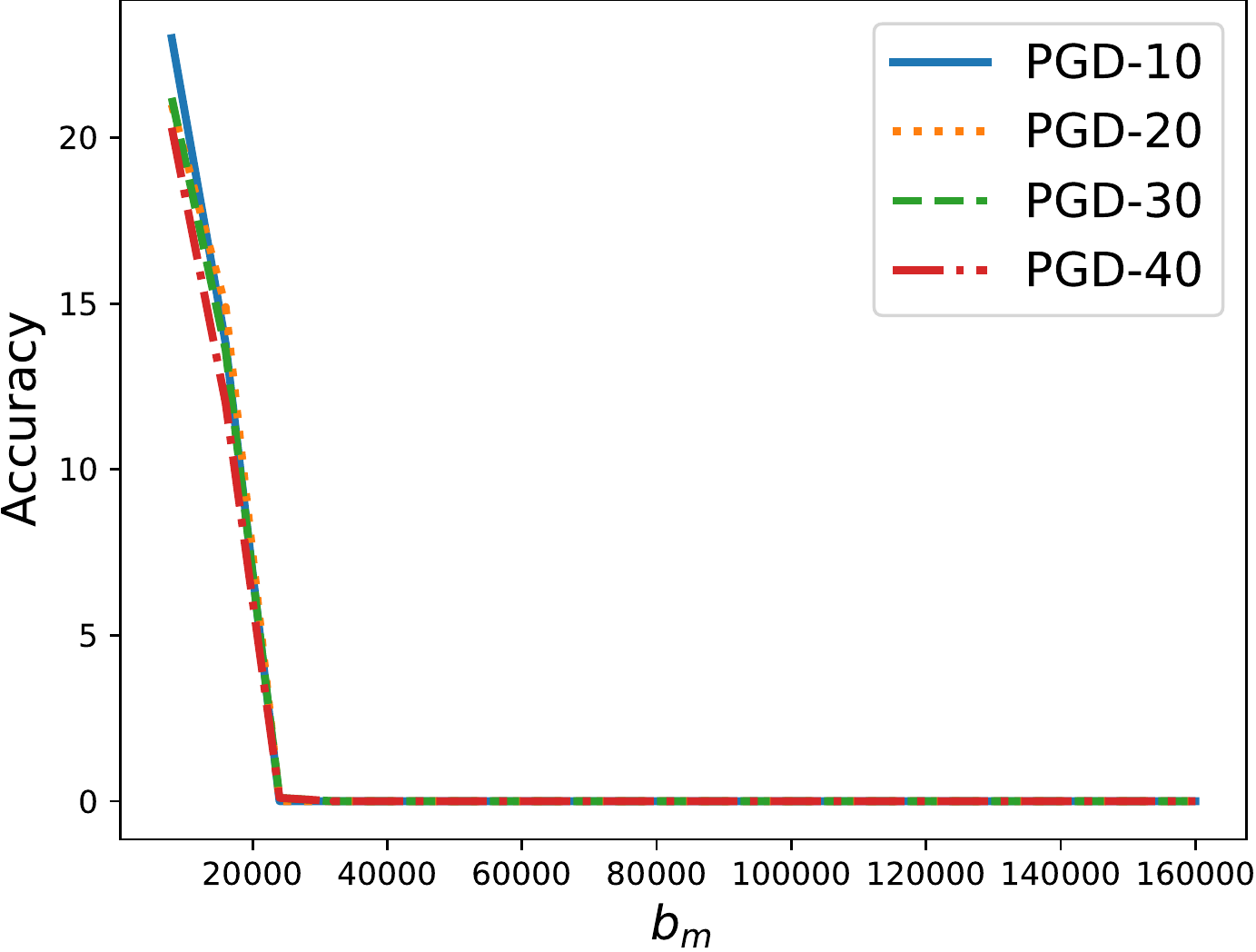}
  \end{minipage}
  \begin{minipage}[t]{0.22\textwidth}
  \includegraphics[width=1.0\textwidth]{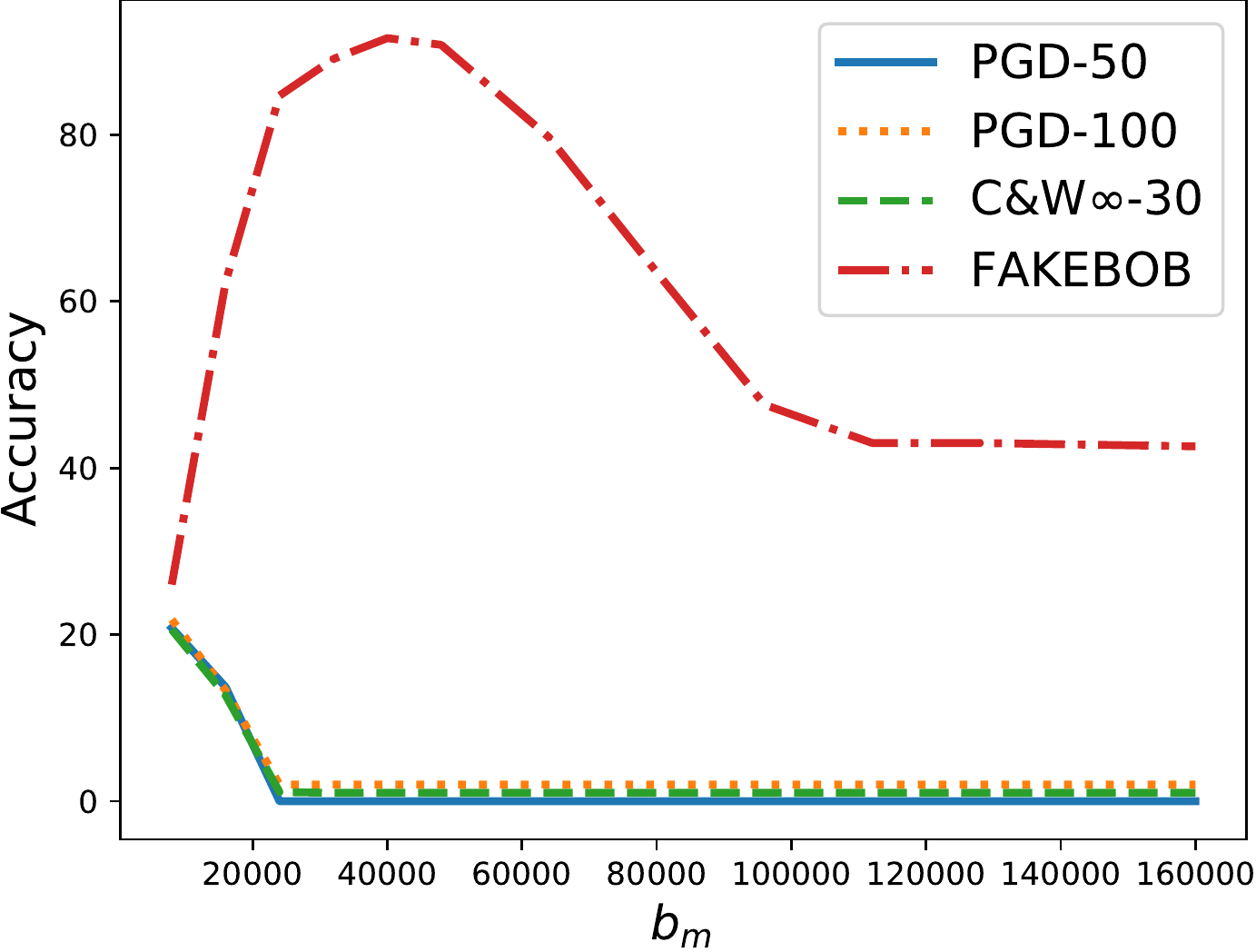}
  \end{minipage}
  \begin{minipage}[t]{0.22\textwidth}
  \includegraphics[width=1.0\textwidth]{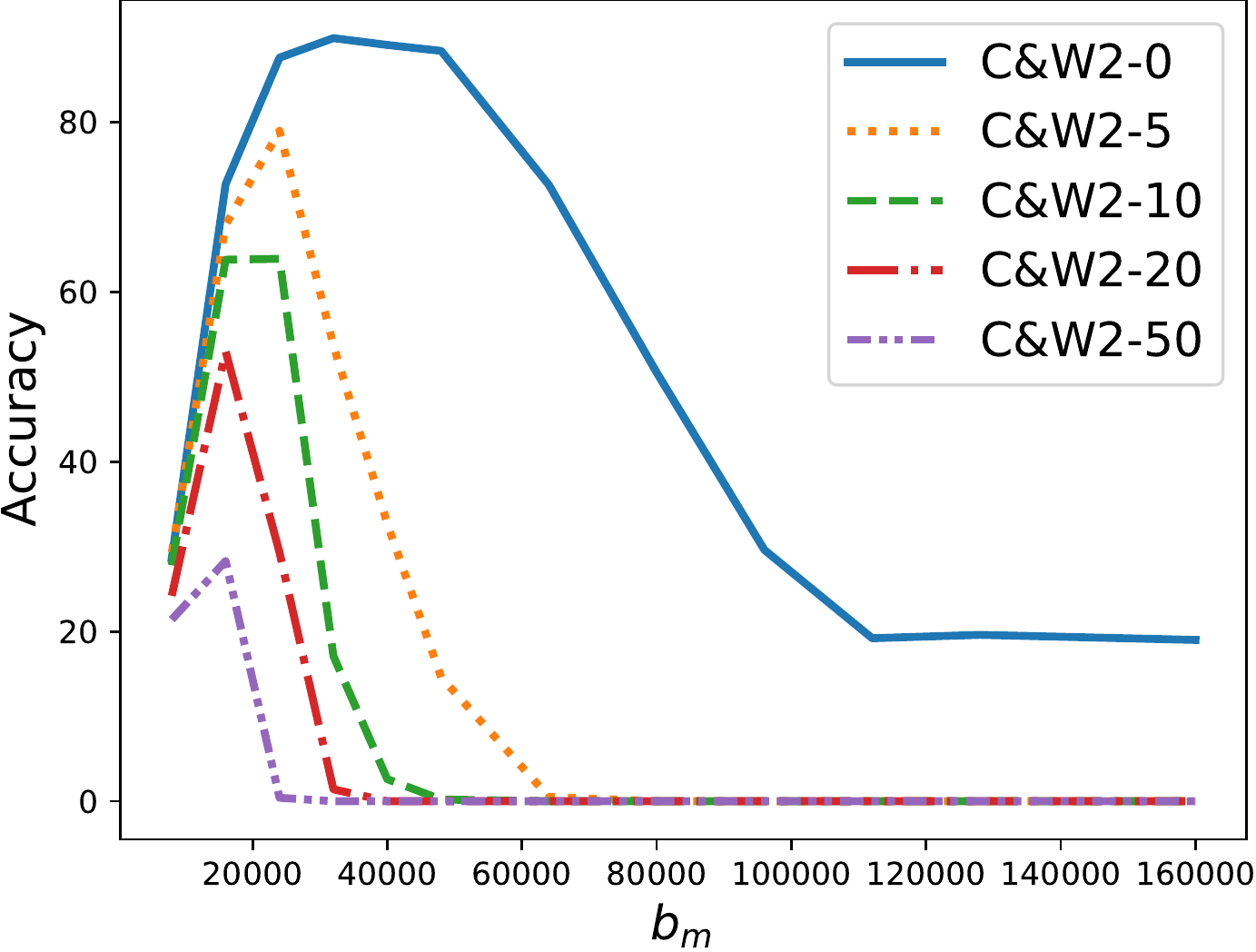}
  \end{minipage}
  }\vspace{-3mm}

  \subfigure[\defensenameabbr-o]{
  \begin{minipage}[t]{0.22\textwidth}
  \includegraphics[width=1.0\textwidth]{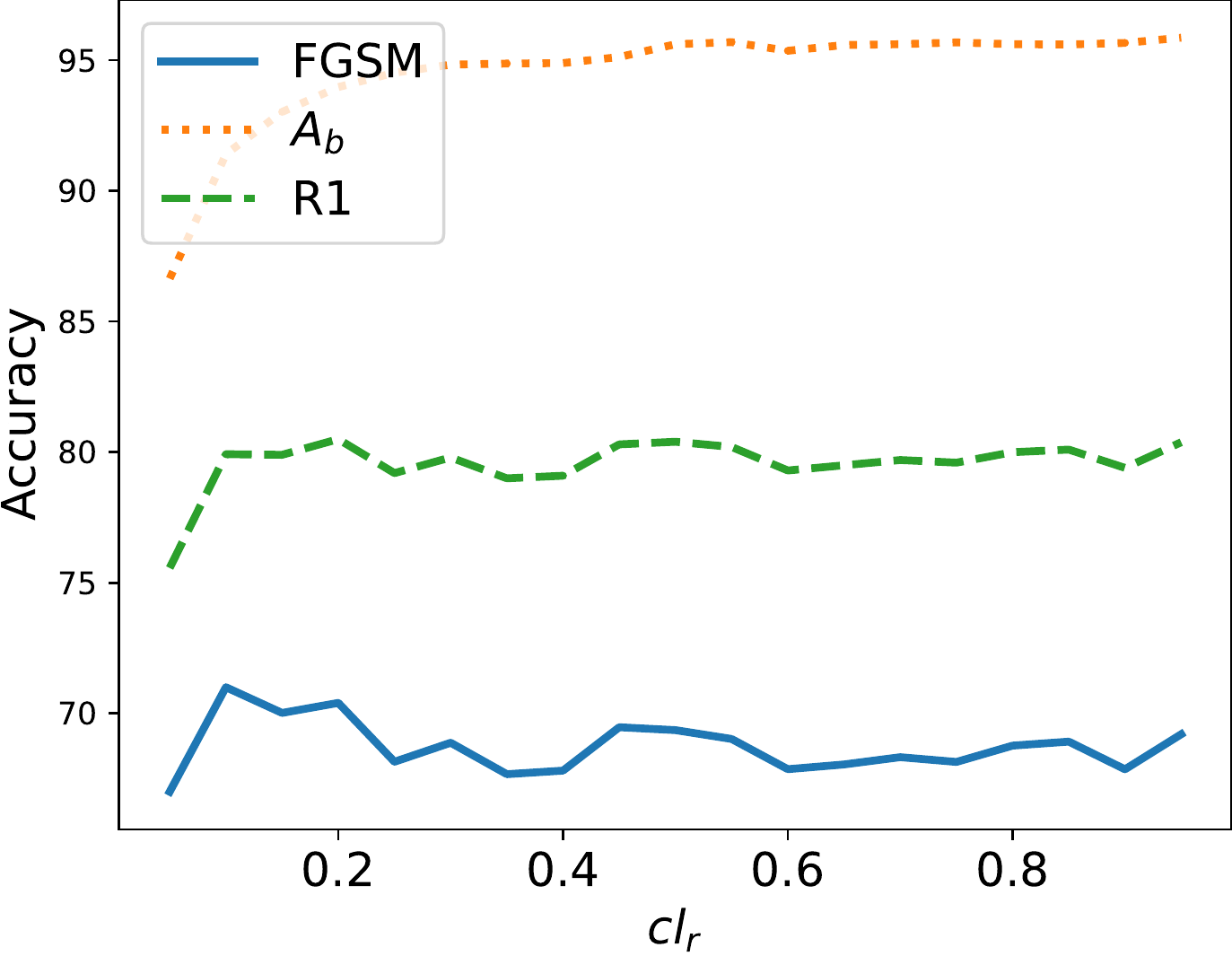}
  \end{minipage}
  \begin{minipage}[t]{0.22\textwidth}
  \includegraphics[width=1.0\textwidth]{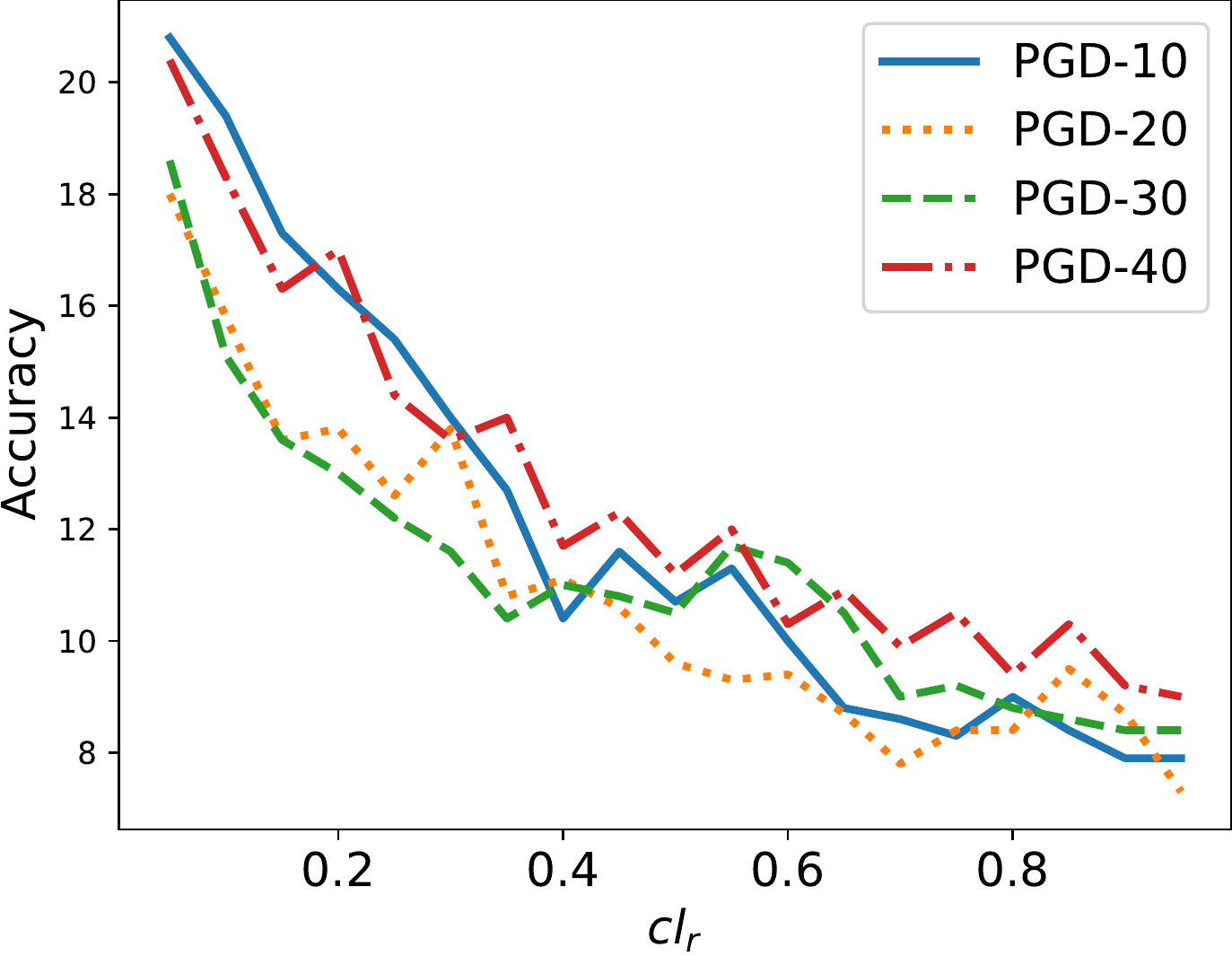}
  \end{minipage}
  \begin{minipage}[t]{0.22\textwidth}
  \includegraphics[width=1.0\textwidth]{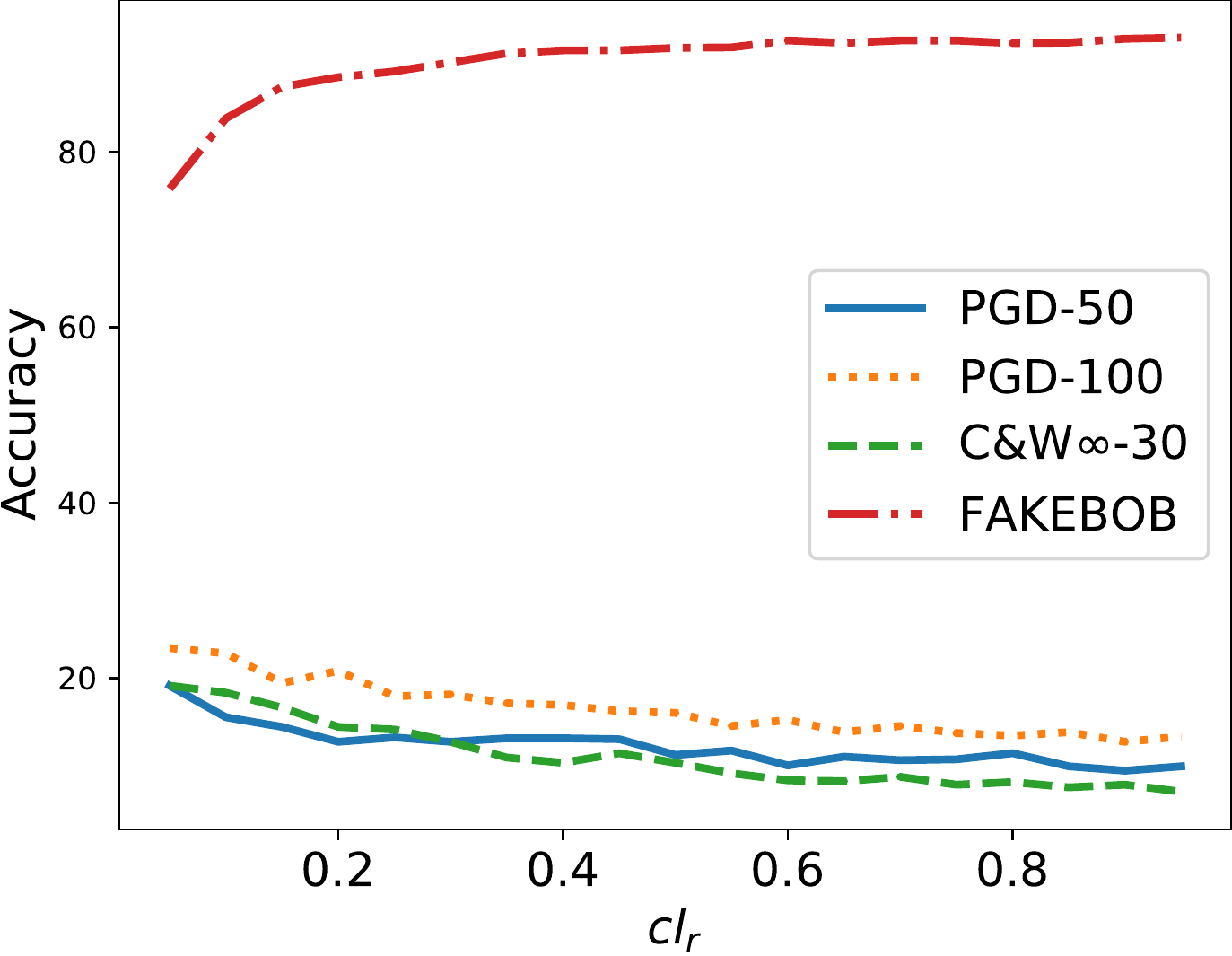}
  \end{minipage}
  \begin{minipage}[t]{0.22\textwidth}
  \includegraphics[width=1.0\textwidth]{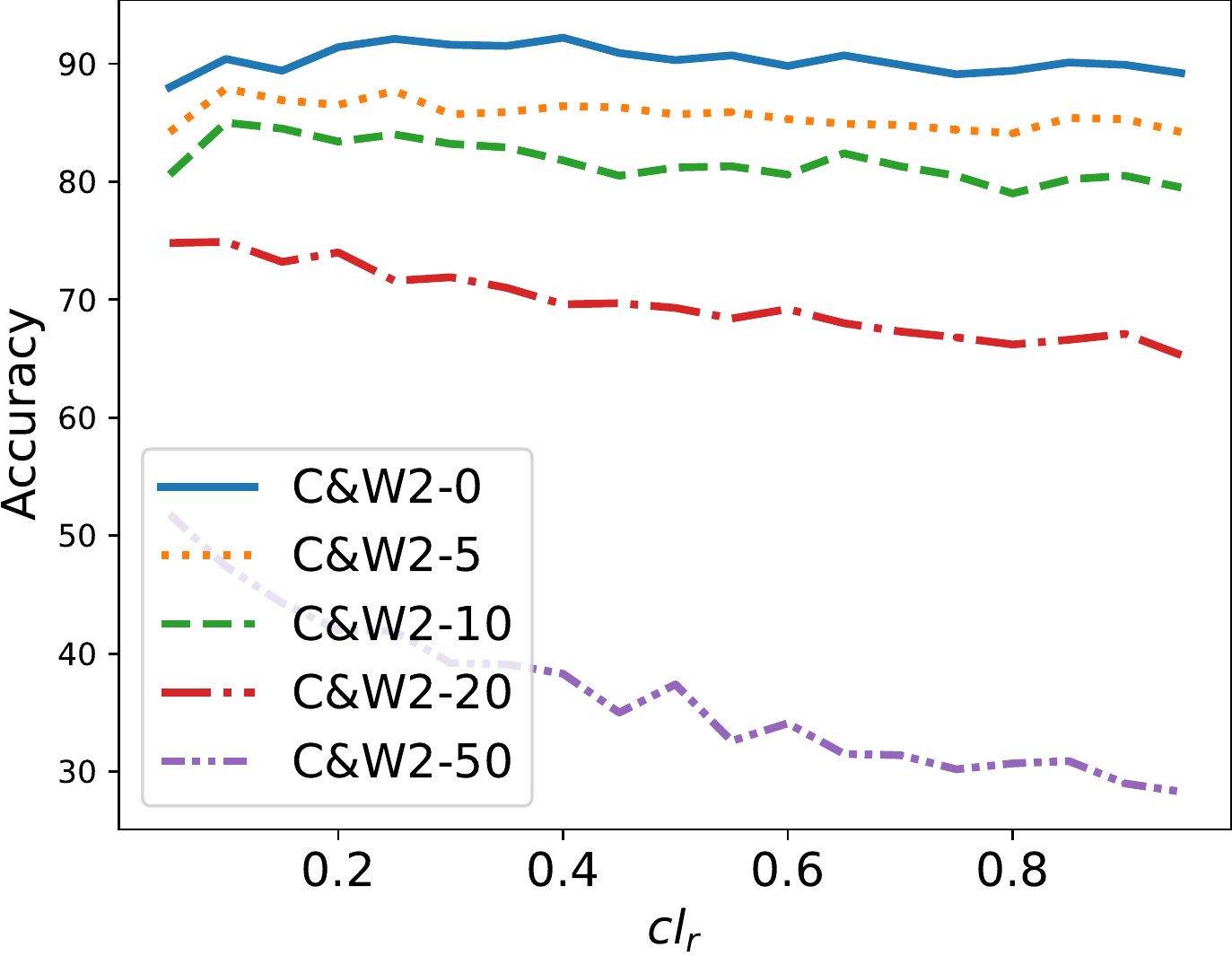}
  \end{minipage}
  \label{fig:p-ff-o}
  }\vspace{-3mm}

   \subfigure[\defensenameabbr-d]{
  \begin{minipage}[t]{0.22\textwidth}
  \includegraphics[width=1.0\textwidth]{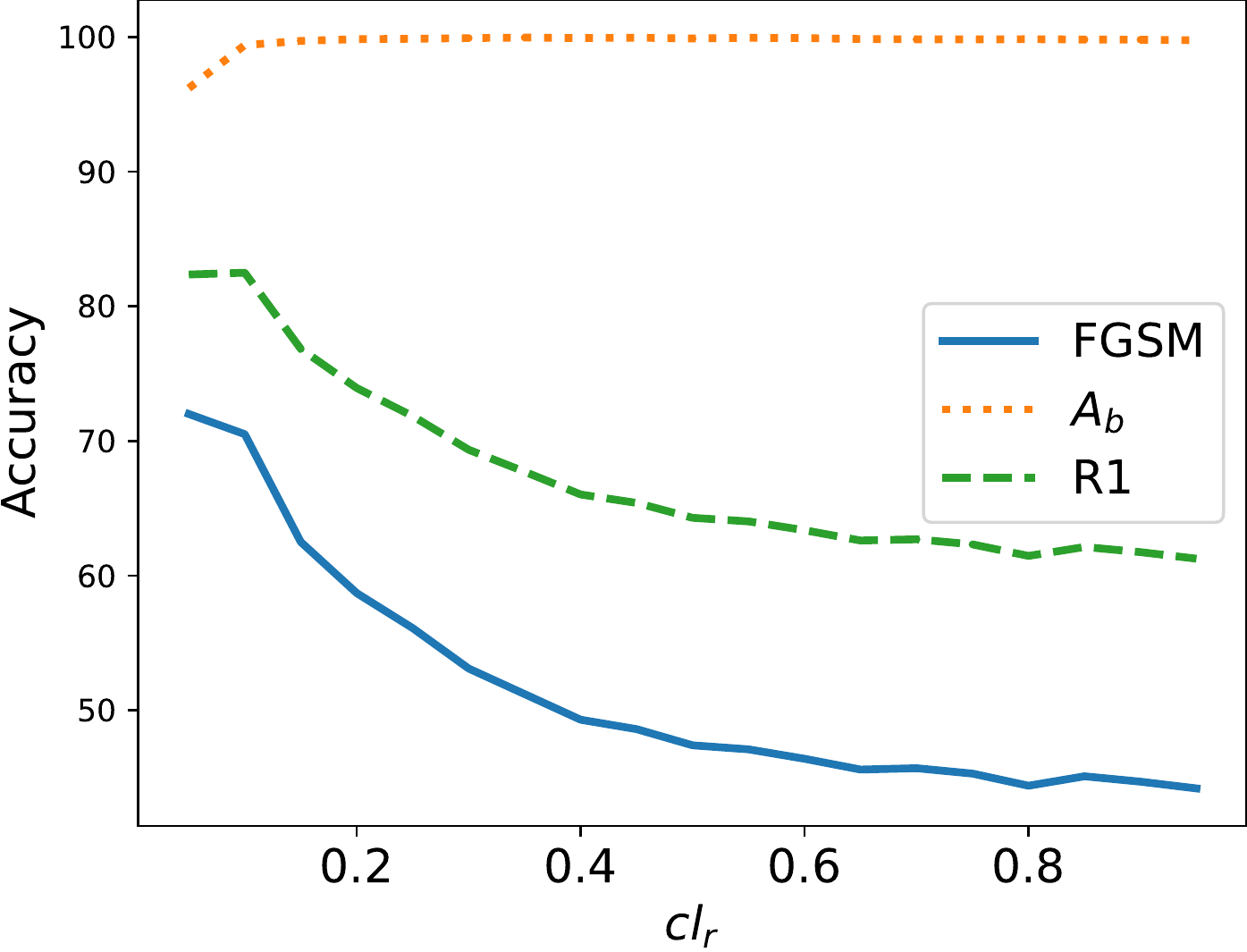}
  \end{minipage}
  \begin{minipage}[t]{0.22\textwidth}
  \includegraphics[width=1.0\textwidth]{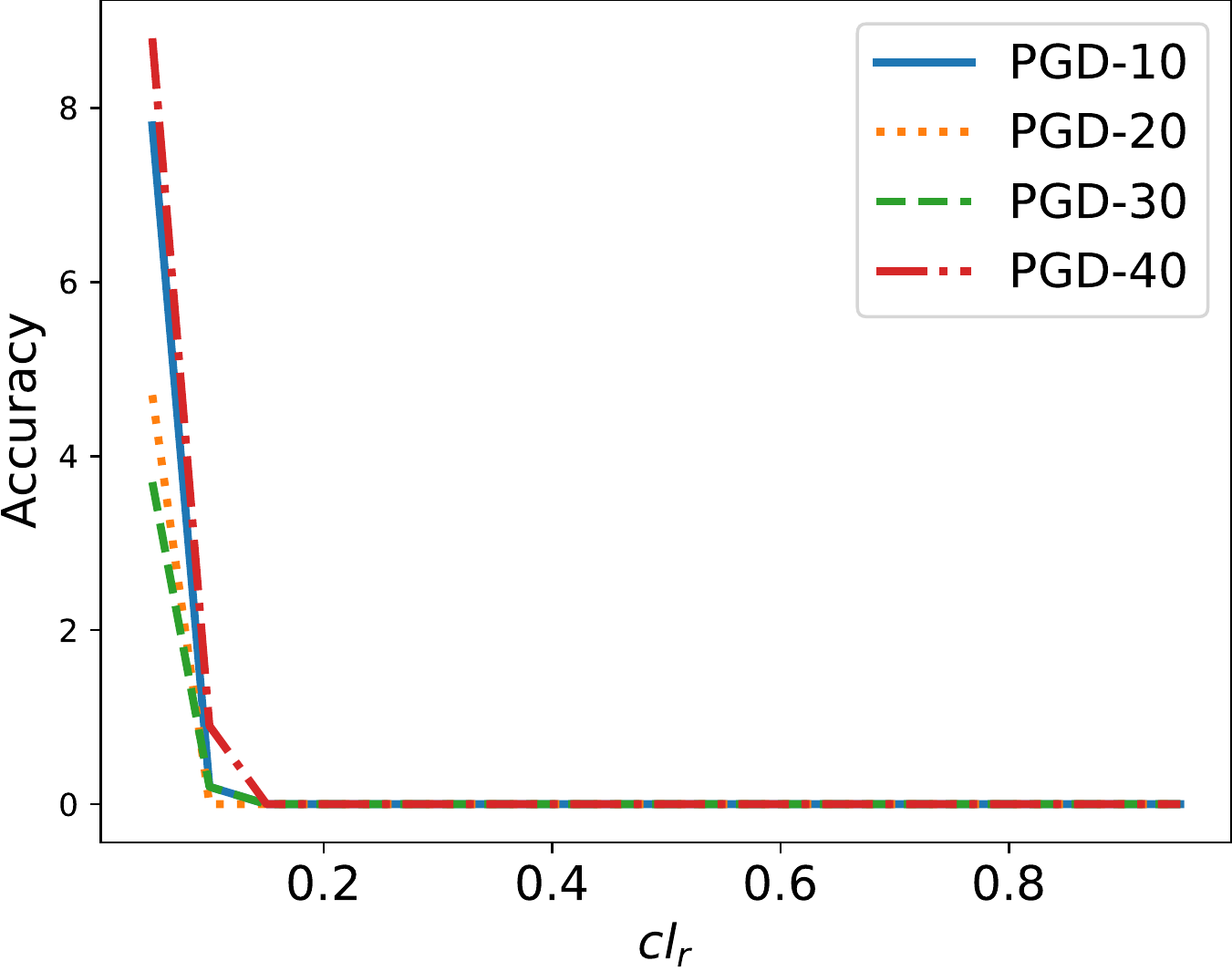}
  \end{minipage}
  \begin{minipage}[t]{0.22\textwidth}
  \includegraphics[width=1.0\textwidth]{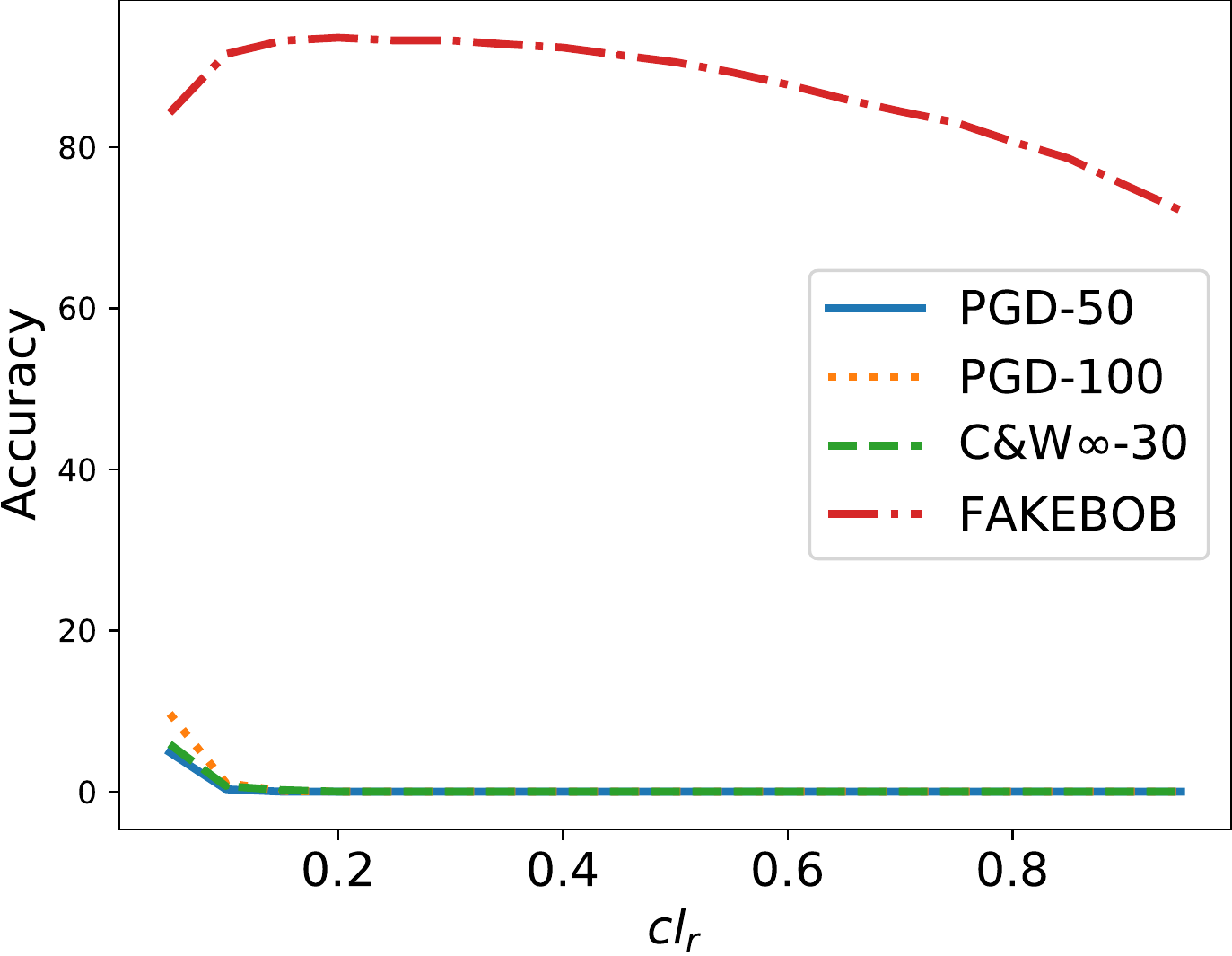}
  \end{minipage}
  \begin{minipage}[t]{0.22\textwidth}
  \includegraphics[width=1.0\textwidth]{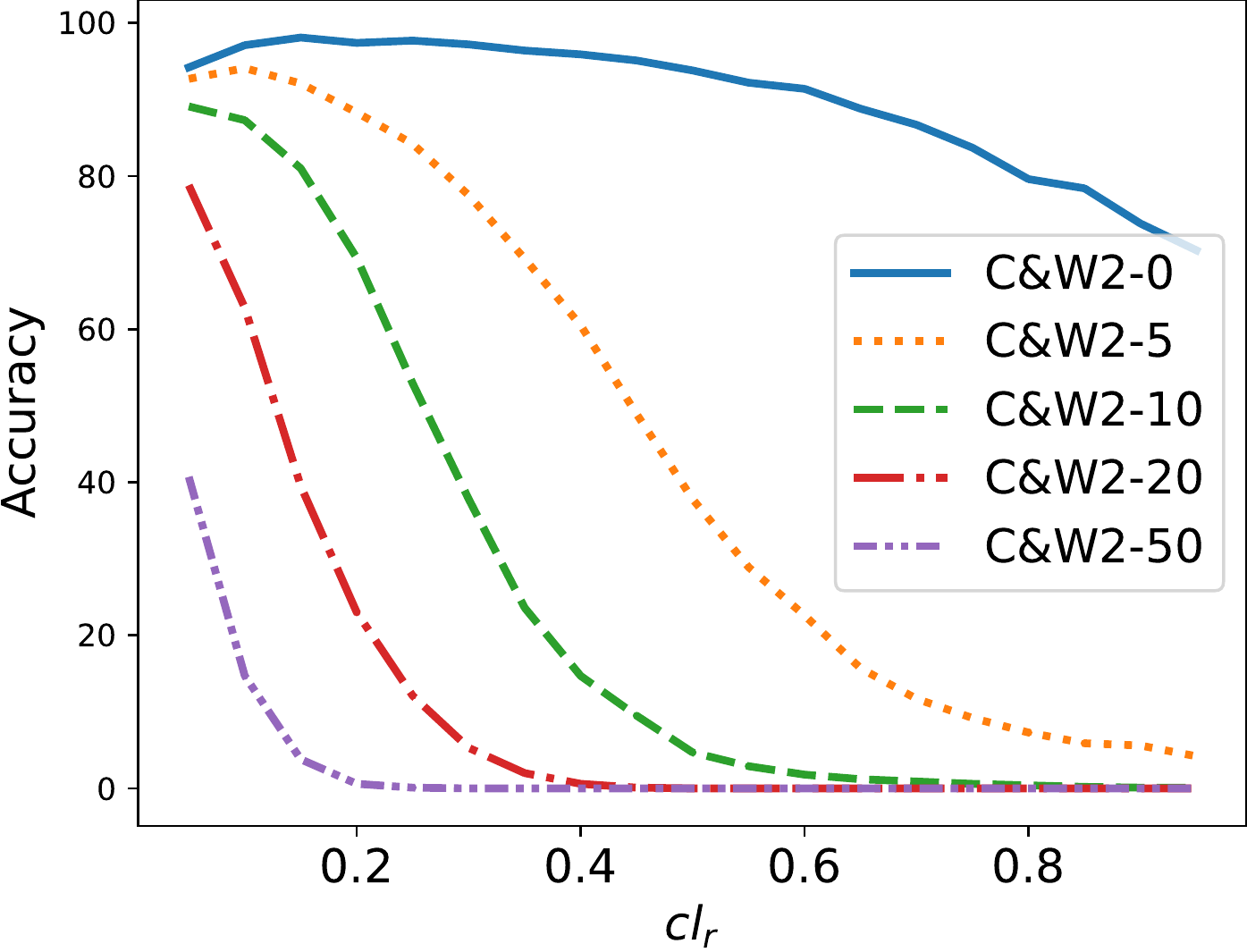}
  \end{minipage}
  \label{fig:p-ff-d}
  }\vspace{-3mm}

   \subfigure[\defensenameabbr-c]{
  \begin{minipage}[t]{0.22\textwidth}
  \includegraphics[width=1.0\textwidth]{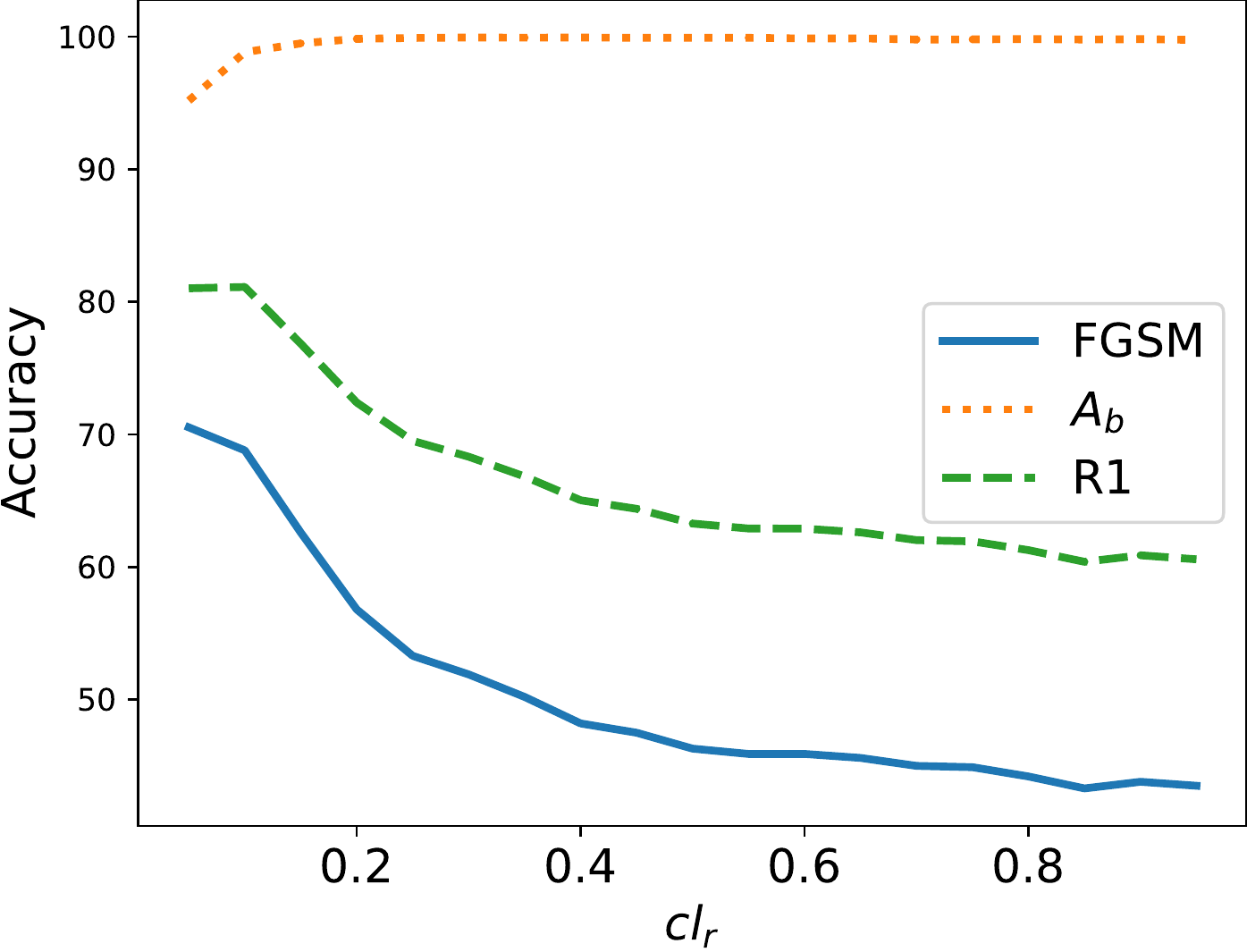}
  \end{minipage}
  \begin{minipage}[t]{0.22\textwidth}
  \includegraphics[width=1.0\textwidth]{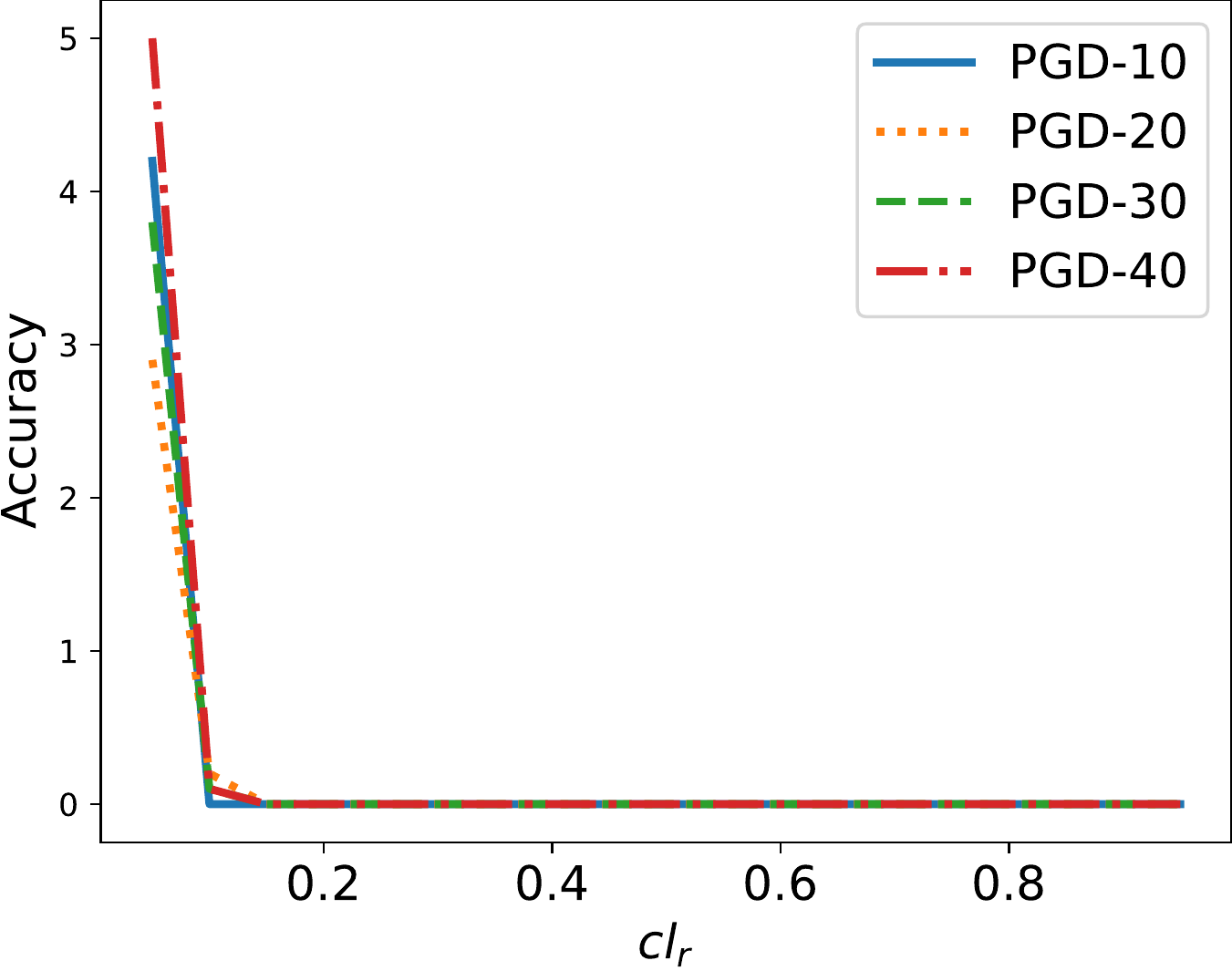}
  \end{minipage}
  \begin{minipage}[t]{0.22\textwidth}
  \includegraphics[width=1.0\textwidth]{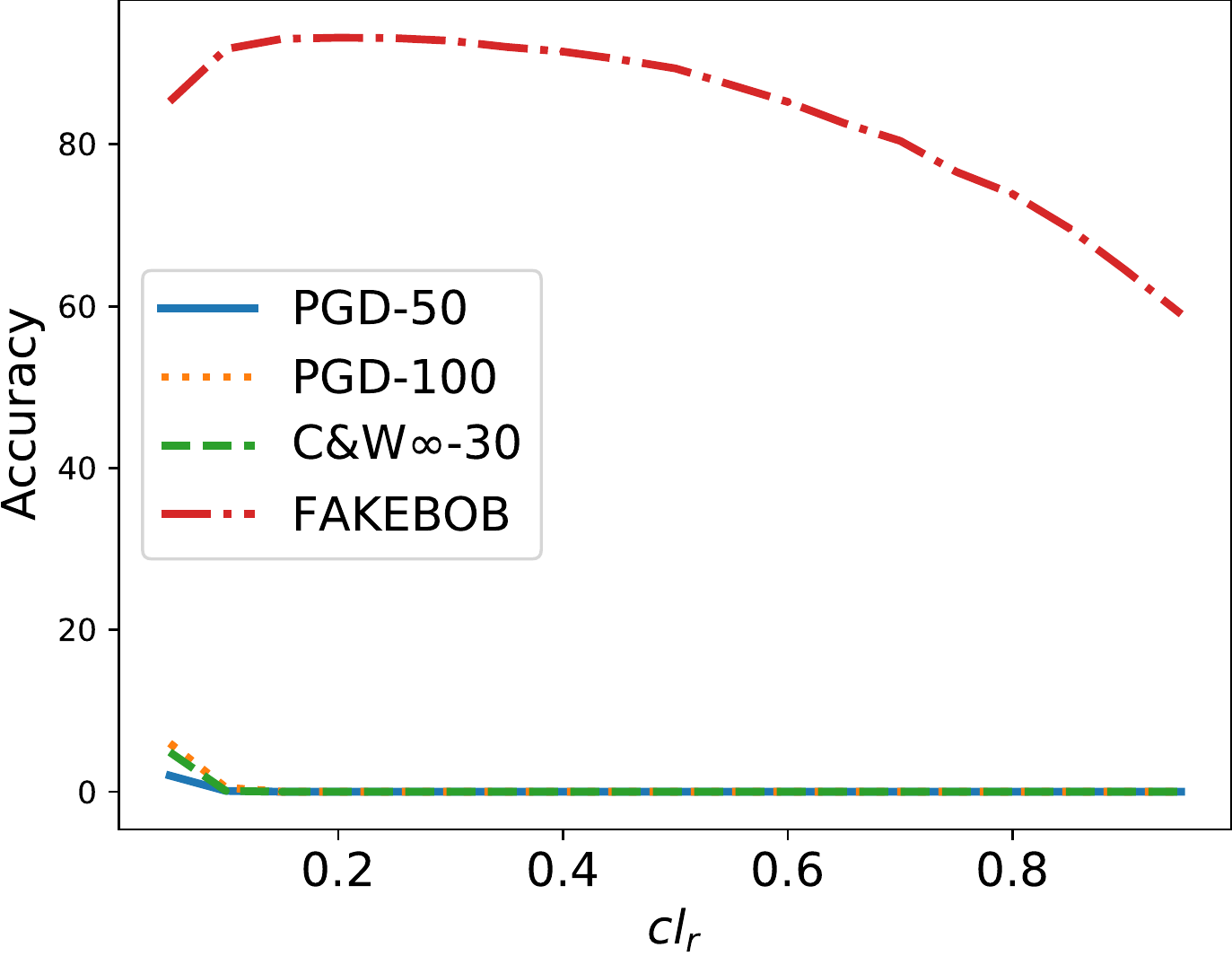}
  \end{minipage}
  \begin{minipage}[t]{0.22\textwidth}
  \includegraphics[width=1.0\textwidth]{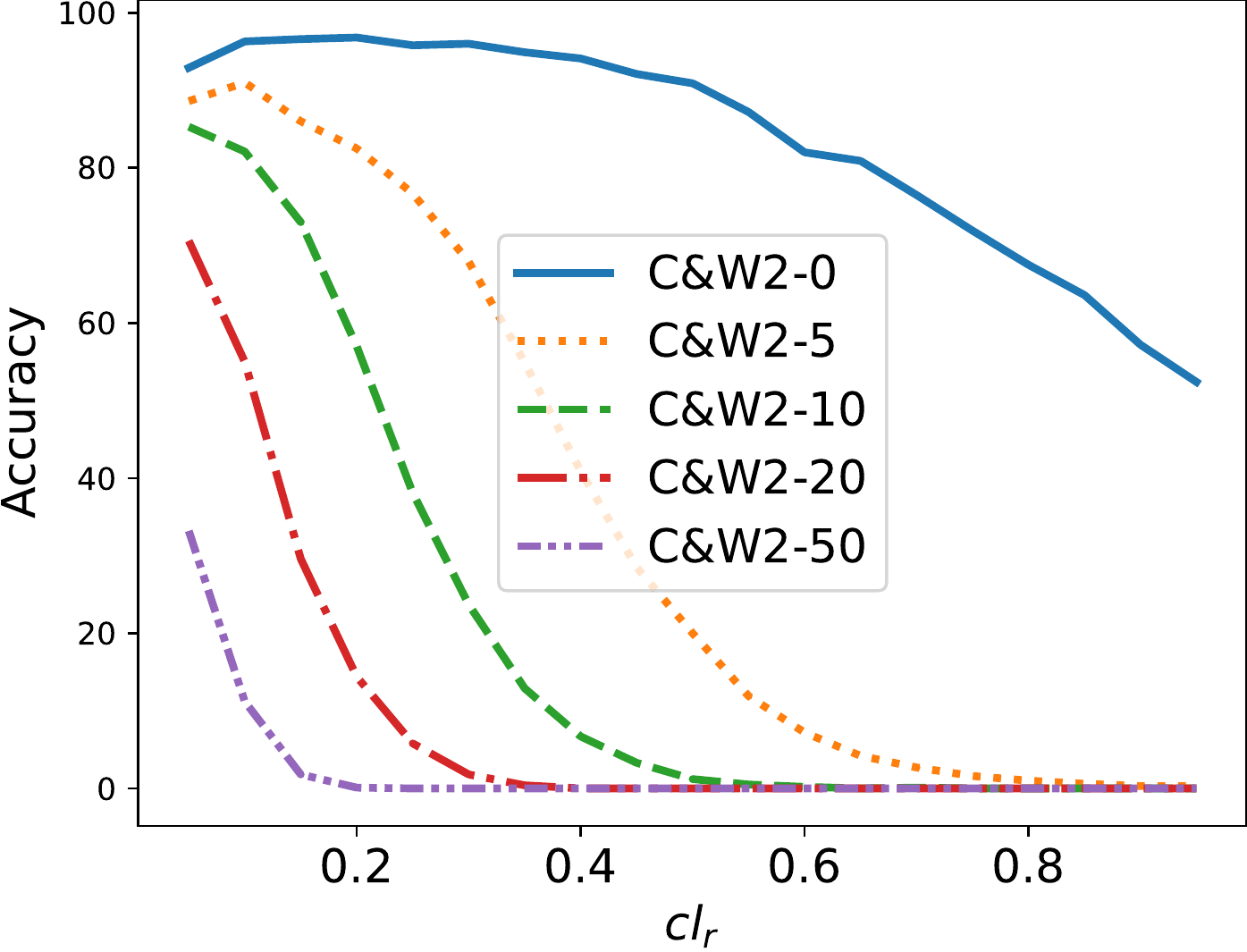}
  \end{minipage}
  \label{fig:p-ff-c}
  }\vspace{-3mm}

  \subfigure[\defensenameabbr-f]{
  \begin{minipage}[t]{0.22\textwidth}
  \includegraphics[width=1.0\textwidth]{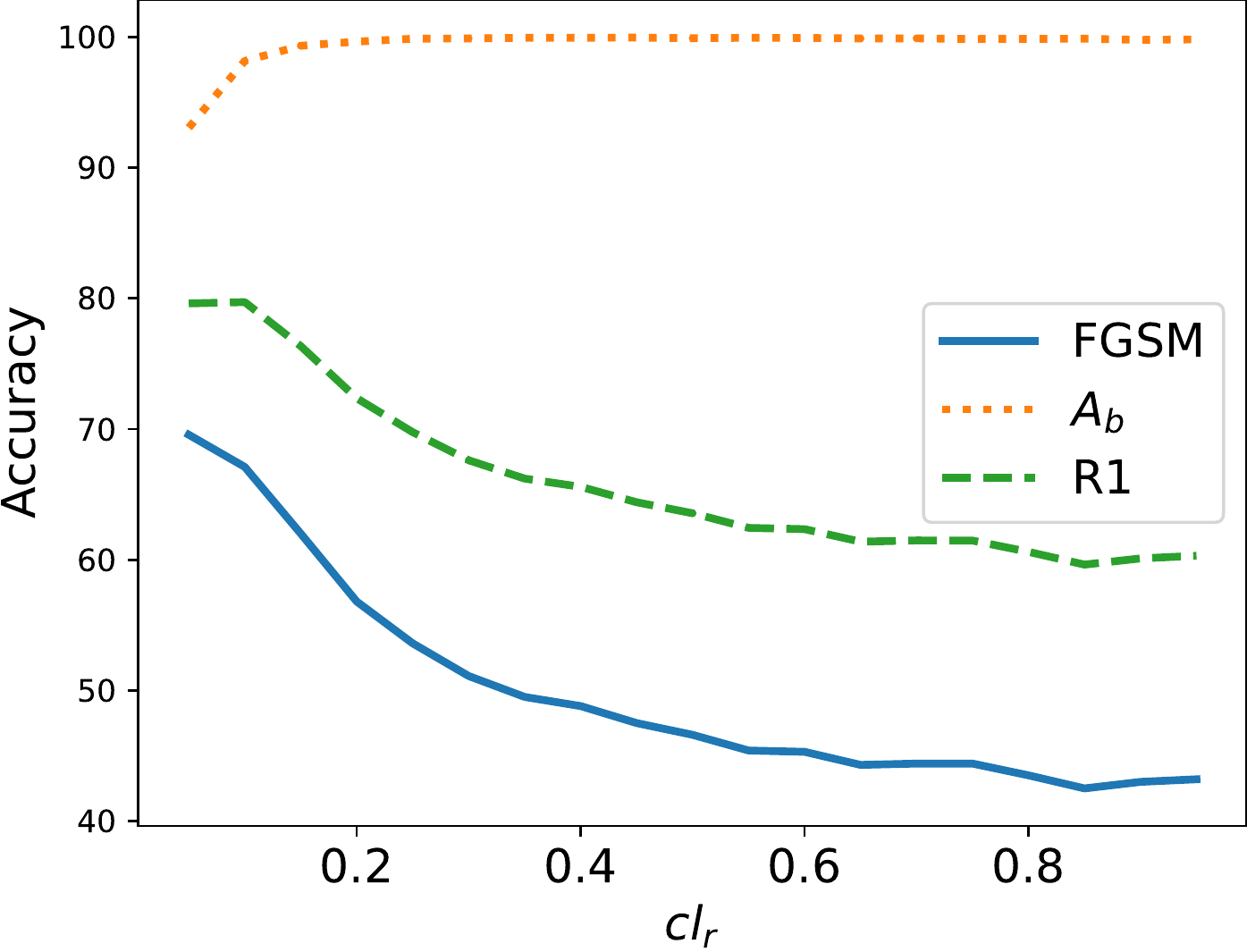}
  \end{minipage}
  \begin{minipage}[t]{0.22\textwidth}
  \includegraphics[width=1.0\textwidth]{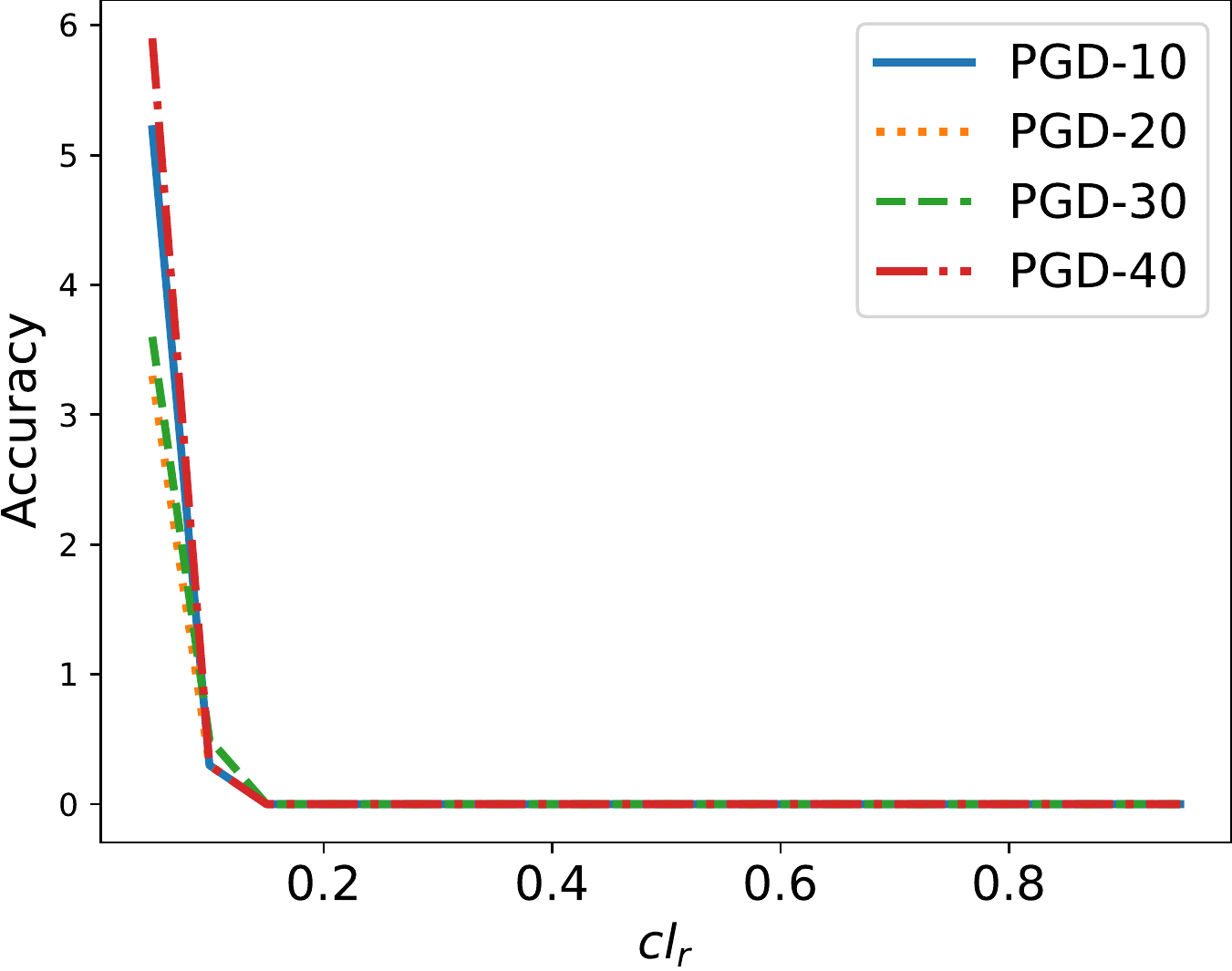}
  \end{minipage}
  \begin{minipage}[t]{0.22\textwidth}
  \includegraphics[width=1.0\textwidth]{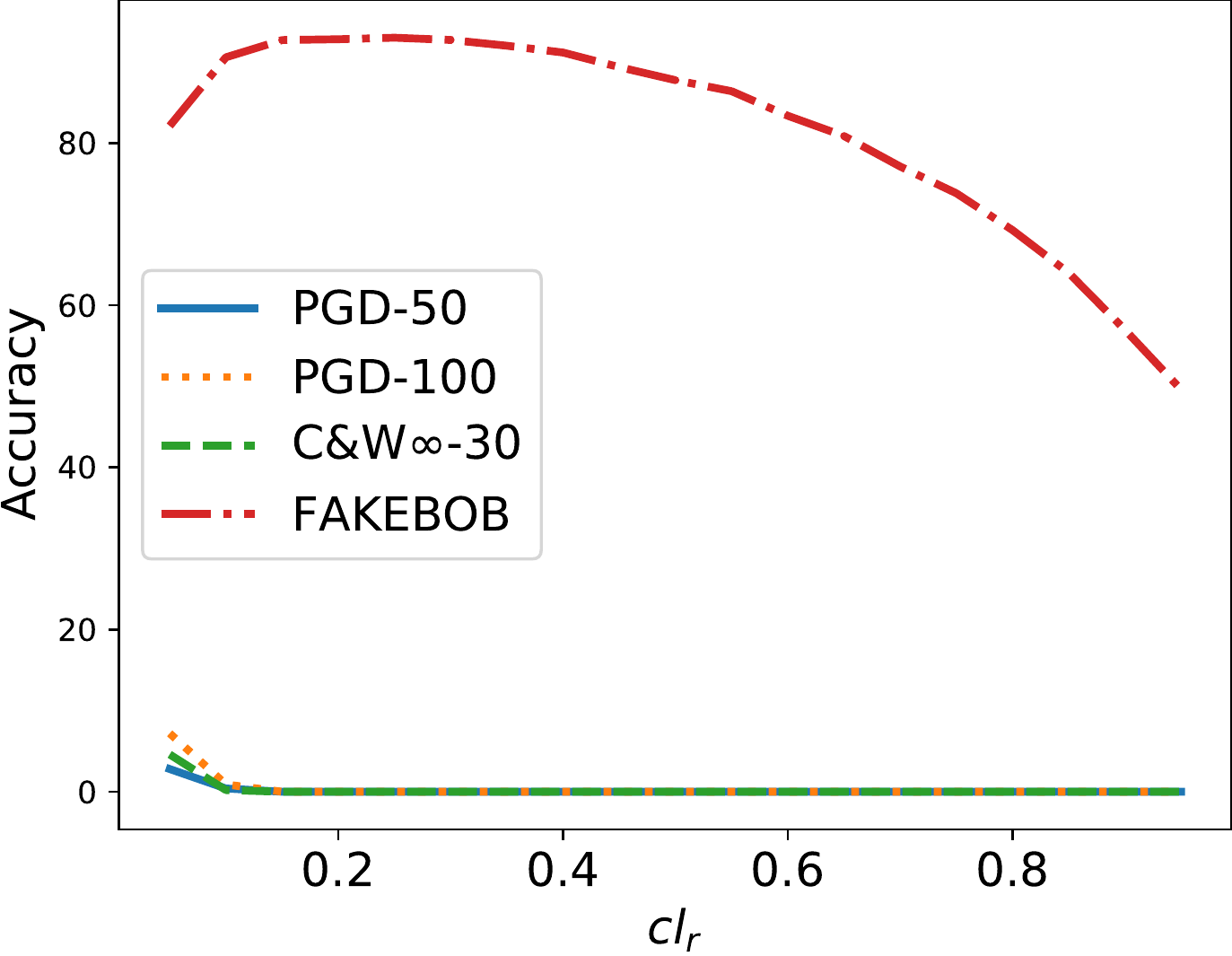}
  \end{minipage}
  \begin{minipage}[t]{0.22\textwidth}
  \includegraphics[width=1.0\textwidth]{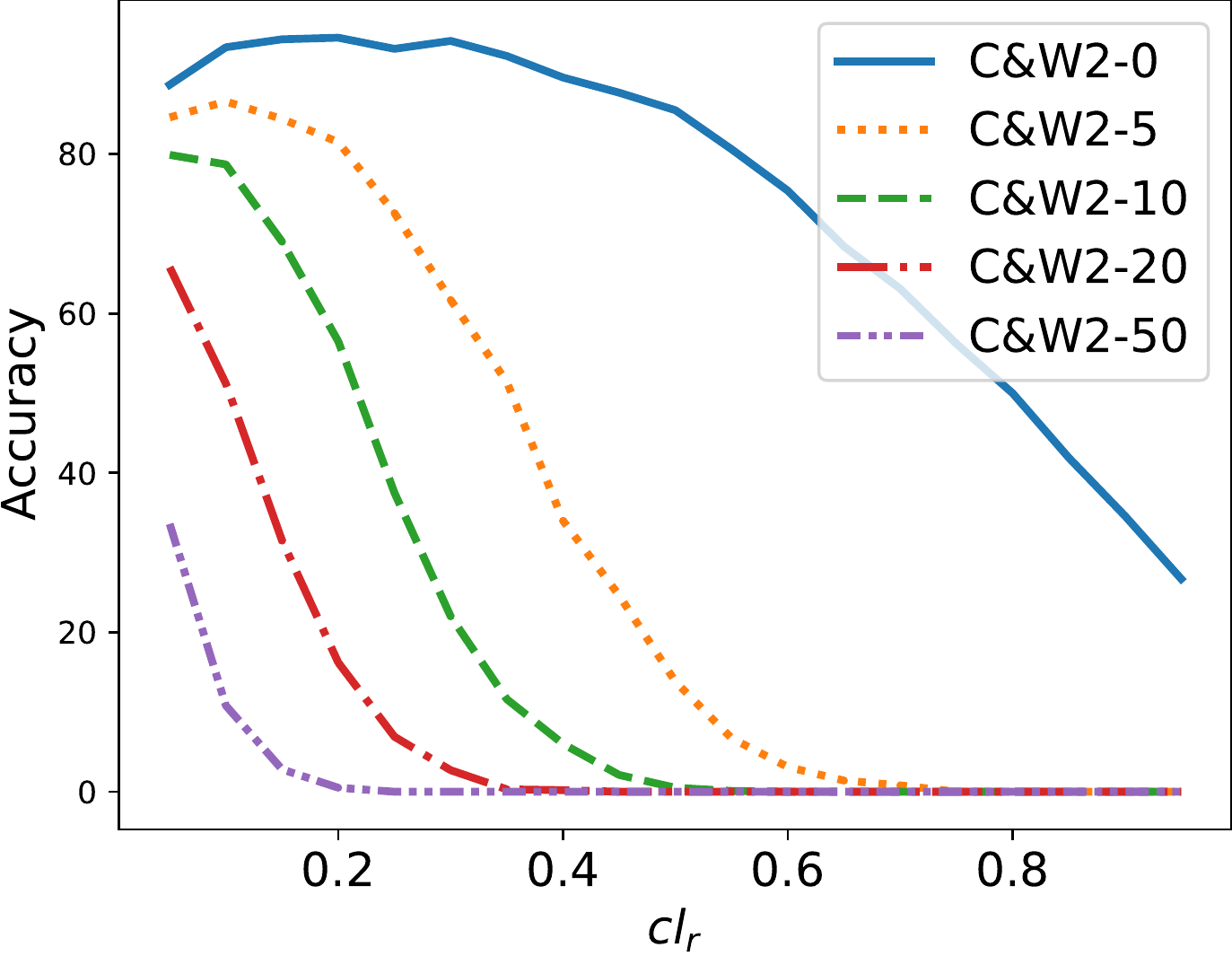}
  \end{minipage}
  \label{fig:p-ff-f}
  }\vspace{-3mm}
  \caption{The performance of input transformations vs. parameter values.}
  \label{fig:parameter-3}
\end{figure*}

%% file: main.bbl
\begin{thebibliography}{100}
\providecommand{\url}[1]{#1}
\csname url@samestyle\endcsname
\providecommand{\newblock}{\relax}
\providecommand{\bibinfo}[2]{#2}
\providecommand{\BIBentrySTDinterwordspacing}{\spaceskip=0pt\relax}
\providecommand{\BIBentryALTinterwordstretchfactor}{4}
\providecommand{\BIBentryALTinterwordspacing}{\spaceskip=\fontdimen2\font plus
\BIBentryALTinterwordstretchfactor\fontdimen3\font minus
  \fontdimen4\font\relax}
\providecommand{\BIBforeignlanguage}[2]{{%
\expandafter\ifx\csname l@#1\endcsname\relax
\typeout{** WARNING: IEEEtran.bst: No hyphenation pattern has been}%
\typeout{** loaded for the language `#1'. Using the pattern for}%
\typeout{** the default language instead.}%
\else
\language=\csname l@#1\endcsname
\fi
#2}}
\providecommand{\BIBdecl}{\relax}
\BIBdecl

\bibitem{Homayoon11}
H.~Beigi, \emph{Fundamentals of Speaker Recognition}.\hskip 1em plus 0.5em
  minus 0.4em\relax Springer, 12 2011.

\bibitem{kaldi}
``Kaldi toolkit,'' \url{https://github.com/kaldi-asr/kaldi}, 2022.

\bibitem{microsoft-azure-vpr}
``Microsoft azure speaker recognition,''
  \url{https://azure.microsoft.com/en-us/services/cognitive-services/speaker-recognition},
  2022.

\bibitem{Alexa}
(2022) {Amazon Alexa}. \url{https://developer.amazon.com/en-US/alexa}.

\bibitem{TD-Bank}
{TD Bank voiceprint}, \url{https://www.tdbank.com/bank/tdvoiceprint.html},
  2022.

\bibitem{RenSYS16}
H.~Ren, Y.~Song, S.~Yang, and F.~Situ, ``Secure smart home: {A} voiceprint and
  internet based authentication system for remote accessing,'' in \emph{ICCSE},
  2016.

\bibitem{abs-1801-03339}
F.~Kreuk, Y.~Adi, M.~Ciss{\'{e}}, and J.~Keshet, ``Fooling end-to-end speaker
  verification with adversarial examples,'' in \emph{{ICASSP}}, 2018.

\bibitem{li2020adversarial}
X.~Li, J.~Zhong, X.~Wu, J.~Yu, X.~Liu, and H.~Meng, ``Adversarial attacks on
  gmm i-vector based speaker verification systems,'' in \emph{ICASSP}, 2020.

\bibitem{jati2021adversarial}
A.~Jati, C.-C. Hsu, M.~Pal, R.~Peri, W.~AbdAlmageed, and S.~Narayanan,
  ``Adversarial attack and defense strategies for deep speaker recognition
  systems,'' \emph{Computer Speech \& Language}, vol.~68, p. 101199, 2021.

\bibitem{zhang2021attack}
W.~Zhang, S.~Zhao, L.~Liu, J.~Li, X.~Cheng, T.~F. Zheng, and X.~Hu, ``Attack on
  practical speaker verification system using universal adversarial
  perturbations,'' in \emph{ICASSP}, 2021.

\bibitem{LiZJXZWM020}
J.~Li, X.~Zhang, C.~Jia, J.~Xu, L.~Zhang, Y.~Wang, S.~Ma, and W.~Gao,
  ``Universal adversarial perturbations generative network for speaker
  recognition,'' in \emph{{ICME}}, 2020.

\bibitem{xie2020enabling}
Y.~Xie, Z.~Li, C.~Shi, J.~Liu, Y.~Chen, and B.~Yuan, ``Enabling fast and
  universal audio adversarial attack using generative model,'' in \emph{AAAI},
  2021.

\bibitem{WangGX20}
Q.~Wang, P.~Guo, and L.~Xie, ``Inaudible adversarial perturbations for targeted
  attack in speaker recognition,'' in \emph{INTERSPEECH}, 2020.

\bibitem{shamsabadi2021foolhd}
A.~S. Shamsabadi, F.~S. Teixeira, A.~Abad, B.~Raj, A.~Cavallaro, and
  I.~Trancoso, ``Foolhd: Fooling speaker identification by highly imperceptible
  adversarial disturbances,'' in \emph{ICASSP}, 2021.

\bibitem{chen2019real}
G.~Chen, S.~Chen, L.~Fan, X.~Du, Z.~Zhao, F.~Song, and Y.~Liu, ``Who is real
  {Bob}? adversarial attacks on speaker recognition systems,'' in \emph{S\&P},
  2021.

\bibitem{du2020sirenattack}
T.~Du, S.~Ji, J.~Li, Q.~Gu, T.~Wang, and R.~Beyah, ``Sirenattack: Generating
  adversarial audio for end-to-end acoustic systems,'' in \emph{{ASIACCS}},
  2020.

\bibitem{DBLP:journals/corr/abs-1711-03280}
Y.~Gong and C.~Poellabauer, ``Crafting adversarial examples for speech
  paralinguistics applications,'' \emph{CoRR}, vol. abs/1711.03280, 2017.

\bibitem{voting-defense}
H.~Wu, Y.~Zhang, Z.~Wu, D.~Wang, and H.~Lee, ``Voting for the right answer:
  Adversarial defense for speaker verification,'' in \emph{INTERSPEECH}, 2021.

\bibitem{HPF-defense}
R.~Olivier, B.~Raj, and M.~Shah, ``High-frequency adversarial defense for
  speech and audio,'' in \emph{{ICASSP}}, 2021.

\bibitem{tramer2020adaptive}
F.~Tram{\`{e}}r, N.~Carlini, W.~Brendel, and A.~Madry, ``On adaptive attacks to
  adversarial example defenses,'' in \emph{NeurIPS}, 2020.

\bibitem{ASGBWYST19}
H.~Abdullah, M.~S. Rahman, W.~Garcia, L.~Blue, K.~Warren, A.~S. Yadav,
  T.~Shrimpton, and P.~Traynor, ``Hear "no evil", see "kenansville": Efficient
  and transferable black-box attacks on speech recognition and voice
  identification systems,'' in \emph{IEEE S\&P}, 2021.

\bibitem{athalye2018obfuscated}
A.~Athalye, N.~Carlini, and D.~A. Wagner, ``Obfuscated gradients give a false
  sense of security: Circumventing defenses to adversarial examples,'' in
  \emph{ICML}, 2018.

\bibitem{athalye2018synthesizing}
A.~Athalye, L.~Engstrom, A.~Ilyas, and K.~Kwok, ``Synthesizing robust
  adversarial examples,'' in \emph{ICML}, 2018.

\bibitem{wierstra2014natural}
D.~Wierstra, T.~Schaul, T.~Glasmachers, Y.~Sun, J.~Peters, and J.~Schmidhuber,
  ``Natural evolution strategies,'' \emph{The Journal of Machine Learning
  Research}, vol.~15, no.~1, pp. 949--980, 2014.

\bibitem{wang2020simulation}
D.~Wang, ``A simulation study on optimal scores for speaker recognition,''
  \emph{EURASIP Journal on Audio, Speech, and Music Processing}, vol. 2020,
  no.~1, pp. 1--23, 2020.

\bibitem{ivector-2011}
N.~Dehak, P.~Kenny, R.~Dehak, P.~Dumouchel, and P.~Ouellet, ``Front-end factor
  analysis for speaker verification,'' \emph{{IEEE} Trans. Speech Audio
  Process.}, 2011.

\bibitem{reynolds2000speaker}
D.~A. Reynolds, T.~F. Quatieri, and R.~B. Dunn, ``Speaker verification using
  adapted gaussian mixture models,'' \emph{Digit. Signal Process.}, 2000.

\bibitem{becker2018interpreting}
S.~Becker, M.~Ackermann, S.~Lapuschkin, K.-R. M\"uller, and W.~Samek,
  ``Interpreting and explaining deep neural networks for classification of
  audio signals,'' \emph{CoRR}, vol. abs/1807.03418, 2018.

\bibitem{snyder2018x}
D.~Snyder, D.~Garcia-Romero, G.~Sell, D.~Povey, and S.~Khudanpur, ``X-vectors:
  Robust dnn embeddings for speaker recognition,'' in \emph{ICASSP}, 2018.

\bibitem{PLDA}
S.~J.~D. Prince and J.~H. Elder, ``Probabilistic linear discriminant analysis
  for inferences about identity,'' in \emph{{ICCV}}, 2007.

\bibitem{dehak2010cosine}
N.~Dehak, R.~Dehak, J.~R. Glass, D.~A. Reynolds, P.~Kenny \emph{et~al.},
  ``Cosine similarity scoring without score normalization techniques,'' in
  \emph{Odyssey}, 2010.

\bibitem{muda2010voice}
L.~Muda, M.~Begam, and I.~Elamvazuthi, ``Voice recognition algorithms using mel
  frequency cepstral coefficient {(MFCC)} and dynamic time warping {(DTW)}
  techniques,'' \emph{CoRR}, vol. abs/1003.4083, 2010.

\bibitem{FilterBanks}
H.~F. {Pardede}, V.~{Zilvan}, D.~{Krisnandi}, A.~{Heryana}, and R.~B.~S.
  {Kusumo}, ``Generalized filter-bank features for robust speech recognition
  against reverberation,'' in \emph{IC3INA}, 2019.

\bibitem{yang2018characterizing}
Z.~Yang, B.~Li, P.~Chen, and D.~Song, ``Characterizing audio adversarial
  examples using temporal dependency,'' in \emph{ICLR}, 2019.

\bibitem{Xu0Q18}
W.~Xu, D.~Evans, and Y.~Qi, ``Feature squeezing: Detecting adversarial examples
  in deep neural networks,'' in \emph{NDSS}, 2018.

\bibitem{MengC17}
D.~Meng and H.~Chen, ``Magnet: {A} two-pronged defense against adversarial
  examples,'' in \emph{CCS}, 2017.

\bibitem{data-aug-advt-effective-image}
S.~Rebuffi, S.~Gowal, D.~A. Calian, F.~Stimberg, O.~Wiles, and T.~Mann, ``Data
  augmentation can improve robustness,'' in \emph{{NeurIPS}}, 2021.

\bibitem{voice-feature-review}
D.~Prabakaran and R.~Shyamala, ``A review on performance of voice feature
  extraction techniques,'' in \emph{Proceedings of the 3rd International
  Conference on Computing and Communications Technologies}, 2019.

\bibitem{xiao2016speech}
X.~Xiao, S.~Zhao, D.~H.~H. Nguyen, X.~Zhong, D.~L. Jones, E.~S. Chng, and
  H.~Li, ``Speech dereverberation for enhancement and recognition using dynamic
  features constrained deep neural networks and feature adaptation,''
  \emph{EURASIP Journal on Advances in Signal Processing}, vol. 2016, no.~1,
  p.~4, 2016.

\bibitem{PurwinsLVSCS19}
H.~Purwins, B.~Li, T.~Virtanen, J.~Schl{\"{u}}ter, S.~Chang, and T.~N. Sainath,
  ``Deep learning for audio signal processing,'' \emph{{IEEE} J. Sel. Top.
  Signal Process.}, vol.~13, no.~2, pp. 206--219, 2019.

\bibitem{paszke2017automatic}
A.~Paszke, S.~Gross, S.~Chintala, G.~Chanan, E.~Yang, Z.~DeVito, Z.~Lin,
  A.~Desmaison, L.~Antiga, and A.~Lerer, ``Automatic differentiation in
  pytorch,'' in \emph{NIPS-W}, 2017.

\bibitem{goodfellow2014explaining}
I.~J. Goodfellow, J.~Shlens, and C.~Szegedy, ``Explaining and harnessing
  adversarial examples,'' in \emph{ICLR}, 2015.

\bibitem{madry2017towards}
A.~Madry, A.~Makelov, L.~Schmidt, D.~Tsipras, and A.~Vladu, ``Towards deep
  learning models resistant to adversarial attacks,'' in \emph{ICLR}, 2018.

\bibitem{kwon2019poster}
H.~Kwon, H.~Yoon, and K.-W. Park, ``Poster: Detecting audio adversarial example
  through audio modification,'' in \emph{CCS}, 2019.

\bibitem{rajaratnam2018speech}
K.~Rajaratnam, B.~Alshemali, and J.~Kalita, ``Speech coding and audio
  preprocessing for mitigating and detecting audio adversarial examples on
  automatic speech recognition,''
  \url{http://cs.uccs.edu/~jkalita/work/reu/REU2018/07Rajaratnam.pdf}, 2018.

\bibitem{vos2013voice}
K.~Vos, K.~V. S{\o}rensen, S.~S. Jensen, and J.-M. Valin, ``Voice coding with
  opus,'' in \emph{Audio Engineering Society Convention}, 2013.

\bibitem{valin2016speex}
J.~Valin, ``Speex: {A} free codec for free speech,'' \emph{CoRR}, vol.
  abs/1602.08668, 2016.

\bibitem{ekudden1999adaptive}
E.~Ekudden, R.~Hagen, I.~Johansson, and J.~Svedberg, ``The adaptive multi-rate
  speech coder,'' in \emph{Workshop on Speech Coding}, 1999.

\bibitem{bosi1997iso}
M.~Bosi, K.~Brandenburg, S.~Quackenbush, L.~Fielder, K.~Akagiri, H.~Fuchs, and
  M.~Dietz, ``Iso/iec mpeg-2 advanced audio coding,'' \emph{Journal of the
  Audio engineering society}, vol.~45, no.~10, pp. 789--814, 1997.

\bibitem{hacker2000mp3}
S.~Hacker, \emph{MP3: The definitive guide}.\hskip 1em plus 0.5em minus
  0.4em\relax O'Reilly Sebastopol, 2000.

\bibitem{speech-processing-book}
J.~Benesty, \emph{Springer handbook of speech processing}, ser. Springer
  Handbooks, 2008.

\bibitem{sohn1999statistical}
J.~Sohn, N.~S. Kim, and W.~Sung, ``A statistical model-based voice activity
  detection,'' \emph{IEEE signal processing letters}, vol.~6, no.~1, pp. 1--3,
  1999.

\bibitem{hartigan1979algorithm}
J.~A. Hartigan and M.~A. Wong, ``Algorithm as 136: A k-means clustering
  algorithm,'' \emph{Journal of the Royal Statistical Society}, vol.~28, no.~1,
  pp. 100--108, 1979.

\bibitem{mackay2003information}
D.~J. MacKay and D.~J. Mac~Kay, \emph{Information theory, inference and
  learning algorithms}.\hskip 1em plus 0.5em minus 0.4em\relax Cambridge
  university press, 2003.

\bibitem{LeivaV13}
L.~A. Leiva and E.~Vidal, ``Warped k-means: An algorithm to cluster
  sequentially-distributed data,'' \emph{Inf. Sci.}, vol. 237, pp. 196--210,
  2013.

\bibitem{kaldi-ivector-plda}
``Ivector-plda model released by kaldi,''
  \url{https://kaldi-asr.org/models/m7}, 2022.

\bibitem{panayotov2015librispeech}
V.~Panayotov, G.~Chen, D.~Povey, and S.~Khudanpur, ``Librispeech: an asr corpus
  based on public domain audio books,'' in \emph{ICASSP}, 2015.

\bibitem{carlini2017towards}
N.~Carlini and D.~A. Wagner, ``Towards evaluating the robustness of neural
  networks,'' in \emph{S\&P}, 2017.

\bibitem{bu2019taking}
L.~Bu, Z.~Zhao, Y.~Duan, and F.~Song, ``Taking care of the discretization
  problem: A comprehensive study of the discretization problem and a black-box
  adversarial attack in discrete integer domain,'' \emph{IEEE TDSC}, 2021.

\bibitem{yuan2018commandersong}
X.~Yuan, Y.~Chen, Y.~Zhao, Y.~Long, X.~Liu, K.~Chen, S.~Zhang, H.~Huang,
  X.~Wang, and C.~A. Gunter, ``Commandersong: {A} systematic approach for
  practical adversarial voice recognition,'' in \emph{{USENIX} Security}, 2018.

\bibitem{rix2001perceptual}
A.~W. Rix, J.~G. Beerends, M.~P. Hollier, and A.~P. Hekstra, ``Perceptual
  evaluation of speech quality (pesq)-a new method for speech quality
  assessment of telephone networks and codecs,'' in \emph{ICASSP}, 2001.

\bibitem{xiang2017digital}
Y.~Xiang, G.~Hua, and B.~Yan, \emph{Digital audio watermarking: fundamentals,
  techniques and challenges}.\hskip 1em plus 0.5em minus 0.4em\relax Springer,
  2017.

\bibitem{AWBPT20}
H.~Abdullah, K.~Warren, V.~Bindschaedler, N.~Papernot, and P.~Traynor, ``Sok:
  The faults in our asrs: An overview of attacks against automatic speech
  recognition and speaker identification systems,'' in \emph{{S\&P}}, 2021.

\bibitem{GL-algo}
D.~W. Griffin and J.~S. Lim, ``Signal estimation from modified short-time
  fourier transform,'' in \emph{ICASSP}.\hskip 1em plus 0.5em minus 0.4em\relax
  {IEEE}, 1983.

\bibitem{Carlini018}
N.~Carlini and D.~A. Wagner, ``Audio adversarial examples: Targeted attacks on
  speech-to-text,'' in \emph{SPW}, 2018.

\bibitem{kingma2014adam}
D.~P. Kingma and J.~Ba, ``Adam: {A} method for stochastic optimization,'' in
  \emph{ICLR}, 2015.

\bibitem{eckle2019comparison}
K.~Eckle and J.~Schmidt-Hieber, ``A comparison of deep networks with relu
  activation function and linear spline-type methods,'' \emph{Neural Networks},
  vol. 110, pp. 232--242, 2019.

\bibitem{DTW}
M.~M{\"u}ller, ``Dynamic time warping,'' \emph{Information retrieval for music
  and motion}, pp. 69--84, 2007.

\bibitem{NATTACK}
Y.~Li, L.~Li, L.~Wang, T.~Zhang, and B.~Gong, ``{NATTACK:} learning the
  distributions of adversarial examples for an improved black-box attack on
  deep neural networks,'' in \emph{ICML}, 2019.

\bibitem{boundary-attack}
W.~Brendel, J.~Rauber, and M.~Bethge, ``Decision-based adversarial attacks:
  Reliable attacks against black-box machine learning models,'' in
  \emph{{ICLR}}, 2018.

\bibitem{CMA-ES}
Y.~Dong, H.~Su, B.~Wu, Z.~Li, W.~Liu, T.~Zhang, and J.~Zhu, ``Efficient
  decision-based black-box adversarial attacks on face recognition,'' in
  \emph{{CVPR}}, 2019.

\bibitem{suitibility-lp}
M.~Sharif, L.~Bauer, and M.~K. Reiter, ``On the suitability of lp-norms for
  creating and preventing adversarial examples,'' in \emph{{CVPR} Workshops},
  2018.

\bibitem{Occam}
B.~Zheng, P.~Jiang, Q.~Wang, Q.~Li, C.~Shen, C.~Wang, Y.~Ge, Q.~Teng, and
  S.~Zhang, ``Black-box adversarial attacks on commercial speech platforms with
  minimal information,'' in \emph{{CCS}}, 2021.

\bibitem{zhao2021attack}
Z.~Zhao, G.~Chen, J.~Wang, Y.~Yang, F.~Song, and J.~Sun, ``Attack as defense:
  Characterizing adversarial examples using robustness,'' in \emph{ISSTA},
  2021.

\bibitem{AS2T}
G.~Chen, Z.~Zhao, F.~Song, S.~Chen, L.~Fan, and Y.~Liu, ``{AS2T}: Arbitrary
  source-to-target adversarial attack on speaker recognition systems,''
  ShanghaiTech University, Tech. Rep., 2022,
  \url{https://faculty.sist.shanghaitech.edu.cn/faculty/songfu/publications/AS2T.pdf}.

\bibitem{liveness-detection-wisec}
L.~Blue, L.~Vargas, and P.~Traynor, ``Hello, is it me you're looking for?:
  Differentiating between human and electronic speakers for voice interface
  security,'' in \emph{WiSec}, 2018.

\bibitem{liveness-detection-usenix}
Y.~Meng, J.~Li, M.~Pillari, A.~Deopujari, L.~Brennan, H.~Shamsie, H.~Zhu, and
  Y.~Tian, ``Your microphone array retains your identity: A robust voice
  liveness detection system for smart speaker,'' in \emph{USENIX Security},
  2022.

\bibitem{AbdullahGPTBW19}
H.~Abdullah, W.~Garcia, C.~Peeters, P.~Traynor, K.~R.~B. Butler, and J.~Wilson,
  ``Practical hidden voice attacks against speech and speaker recognition
  systems,'' in \emph{NDSS}, 2019.

\bibitem{wenger2021hello}
E.~Wenger, M.~Bronckers, C.~Cianfarani, J.~Cryan, A.~Sha, H.~Zheng, and B.~Y.
  Zhao, ````hello, it's me": Deep learning-based speech synthesis attacks in
  the real world,'' in \emph{CCS}, 2021.

\bibitem{chen2021sok}
Y.~Chen, J.~Zhang, X.~Yuan, S.~Zhang, K.~Chen, X.~Wang, and S.~Guo, ``Sok: {A}
  modularized approach to study the security of automatic speech recognition
  systems,'' \emph{CoRR}, vol. abs/2103.10651, 2021.

\bibitem{qin2019imperceptible}
Y.~Qin, N.~Carlini, G.~W. Cottrell, I.~J. Goodfellow, and C.~Raffel,
  ``Imperceptible, robust, and targeted adversarial examples for automatic
  speech recognition,'' in \emph{ICML}, 2019.

\bibitem{psychoacoustic-hiding-attack}
L.~Sch{\"{o}}nherr, K.~Kohls, S.~Zeiler, T.~Holz, and D.~Kolossa, ``Adversarial
  attacks against automatic speech recognition systems via psychoacoustic
  hiding,'' in \emph{{NDSS}}, 2019.

\bibitem{247642}
Y.~Chen, X.~Yuan, J.~Zhang, Y.~Zhao, S.~Zhang, K.~Chen, and X.~Wang,
  ``Devil{\textquoteright}s whisper: A general approach for physical
  adversarial attacks against commercial black-box speech recognition
  devices,'' in \emph{{USENIX} Security}, 2020.

\bibitem{LiW00020}
Z.~Li, Y.~Wu, J.~Liu, Y.~Chen, and B.~Yuan, ``Advpulse: Universal,
  synchronization-free, and targeted audio adversarial attacks via subsecond
  perturbations,'' in \emph{{CCS}}, 2020.

\bibitem{TaoriKCV19}
R.~Taori, A.~Kamsetty, B.~Chu, and N.~Vemuri, ``Targeted adversarial examples
  for black box audio systems,'' in \emph{SPW}, 2019.

\bibitem{dolphin-attack}
G.~Zhang, C.~Yan, X.~Ji, T.~Zhang, T.~Zhang, and W.~Xu, ``Dolphinattack:
  Inaudible voice commands,'' in \emph{CCS}, 2017.

\bibitem{Dompteur}
T.~Eisenhofer, L.~Sch{\"o}nherr, J.~Frank, L.~Speckemeier, D.~Kolossa, and
  T.~Holz, ``Dompteur: Taming audio adversarial examples,'' in \emph{USENIX
  Security}, 2021.

\bibitem{hautamaki2013vectors}
R.~G. Hautam{\"a}ki, T.~Kinnunen, V.~Hautam{\"a}ki, T.~Leino, and A.-M.
  Laukkanen, ``I-vectors meet imitators: on vulnerability of speaker
  verification systems against voice mimicry,'' in \emph{INTERSPEECH}, 2013.

\bibitem{shirvanian2019quantifying}
M.~Shirvanian, S.~Vo, and N.~Saxena, ``Quantifying the breakability of voice
  assistants,'' in \emph{{PerCom}}, 2019.

\bibitem{mukhopadhyay2015all}
D.~Mukhopadhyay, M.~Shirvanian, and N.~Saxena, ``All your voices are belong to
  us: Stealing voices to fool humans and machines,'' in \emph{ESORICS}, 2015.

\bibitem{shirvanian2014wiretapping}
M.~Shirvanian and N.~Saxena, ``Wiretapping via mimicry: Short voice imitation
  mitm attacks on crypto phones,'' in \emph{{CCS}}, 2014.

\bibitem{shirvanian2018short}
M.~Shirvanian, N.~Saxena, and D.~Mukhopadhyay, ``Short voice imitation
  man-in-the-middle attacks on crypto phones: Defeating humans and machines,''
  \emph{Journal of Computer Security}, 2018.

\bibitem{Defend-GAN}
P.~Samangouei, M.~Kabkab, and R.~Chellappa, ``Defense-gan: Protecting
  classifiers against adversarial attacks using generative models,'' in
  \emph{{ICLR}}, 2018.

\bibitem{TIFS-defense}
S.~Joshi, J.~Villalba, P.~Zelasko, L.~Moro{-}Vel{\'{a}}zquez, and N.~Dehak,
  ``Study of pre-processing defenses against adversarial attacks on
  state-of-the-art speaker recognition systems,'' \emph{TIFS}, 2021.

\bibitem{XuLidetection}
X.~Li, N.~Li, J.~Zhong, X.~Wu, X.~Liu, D.~Su, D.~Yu, and H.~Meng,
  ``Investigating robustness of adversarial samples detection for automatic
  speaker verification,'' in \emph{INTERSPEECH}, 2020.

\bibitem{pairing-weak-strong}
Z.~Peng, X.~Li, and T.~Lee, ``Pairing weak with strong: Twin models for
  defending against adversarial attack on speaker verification,'' in
  \emph{INTERSPEECH}, 2021.

\bibitem{CW17a}
N.~Carlini and D.~A. Wagner, ``Adversarial examples are not easily detected:
  Bypassing ten detection methods,'' in \emph{{AISec@CCS}}, 2017.

\bibitem{ParkYKKA17}
S.~J. Park, G.~Yeung, J.~Kreiman, P.~A. Keating, and A.~Alwan, ``Using voice
  quality features to improve short-utterance, text-independent speaker
  verification systems,'' in \emph{INTERSPEECH}, 2017.

\bibitem{app9183697}
I.~Vi{\~{n}}als, A.~Ortega, A.~Miguel, and E.~Lleida, ``An analysis of the
  short utterance problem for speaker characterization,'' \emph{Applied
  Sciences}, 2019.

\bibitem{akhtar2018threat}
N.~Akhtar and A.~Mian, ``Threat of adversarial attacks on deep learning in
  computer vision: A survey,'' \emph{{IEEE} Access}, vol.~6, pp.
  14\,410--14\,430, 2018.

\bibitem{PSO}
R.~Eberhart and J.~Kennedy, ``A new optimizer using particle swarm theory,'' in
  \emph{MHS}, 1995.

\bibitem{kmeans++}
D.~Arthur and S.~Vassilvitskii, ``k-means++: the advantages of careful
  seeding,'' in \emph{SODA}, 2007.

\end{thebibliography}
